%% file: main.tex
\documentclass[twocolumn]{aastex63}

\usepackage{float,graphicx,amsmath,multirow}
\usepackage{color}
\usepackage{chemformula}
\newcommand{\ce}[1]{\ch{#1}}
\usepackage{booktabs}

\usepackage{xcolor}
\usepackage{xifthen}
\usepackage[normalem]{ulem}
\newcommand{\stkout}[1]{\ifmmode\text{\sout{\ensuremath{#1}}}\else\sout{#1}\fi}
\newcommand{\edited}[2]{\ifthenelse{\isempty{#1}}{\textbf{#2}}{\ifthenelse{\isempty{#2}}{\textcolor{gray}{\stkout{#1}}}{\textcolor{gray}{\stkout{#1}} \textcolor{red}{#2}}}}

\usepackage{xstring,xifthen}
\newcommand{\replacedecimal}[2]{\ifthenelse{\isin{.}{#1}}{\text{\StrBefore{#1}{.}}\ensuremath{\overset{#2}{.}}\text{\StrBehind{#1}{.}}}{#1\ensuremath{^{#2}}}}
\newcommand{\hms}[3]{\ensuremath{#1\overset{\text{h}}{\phantom{.}}#2\overset{\text{m}}{\phantom{.}}\replacedecimal{#3}{\text{s}}}}
\newcommand{\dms}[3]{\ensuremath{#1\overset{\circ}{\phantom{.}}#2\overset{\prime}{\phantom{.}}\replacedecimal{#3}{\prime\prime}}}

\makeatletter
\newcommand*{\rom}[1]{\expandafter\@slowromancap\romannumeral #1@}
\makeatother
\newlength{\VSpaceBeforeTabBib}
\newlength{\VSpaceBeforeTabFoot}
\newcommand{\posteriorfignum}{8}
\newcounter{posteriorsubfigure}
\newcommand{\posteriorsubfignum}{\stepcounter{posteriorsubfigure}Fig.~\posteriorfignum.\theposteriorsubfigure}
\newcounter{cachedposteriorsubfigure}
\newcommand{\cacheposteriorsubfigure}{\setcounter{cachedposteriorsubfigure}{\theposteriorsubfigure}}
\newcommand{\loadcachedposteriorsubfigure}{\setcounter{posteriorsubfigure}{\thecachedposteriorsubfigure}}
\begin{document}


\title{The Molecular Inventory of TMC-1 with GOTHAM Observations}

\author[0000-0003-2760-2119]{Ci Xue}
\affiliation{Department of Chemistry, Massachusetts Institute of Technology, Cambridge, MA 02139, USA}

\author[0000-0002-4593-518X]{Alex N. Byrne}
\affiliation{Department of Chemistry, Massachusetts Institute of Technology, Cambridge, MA 02139, USA}

\author[0000-0001-8708-5593]{Larry Morgan}
\affiliation{Green Bank Observatory, 155 Observatory Rd, Green Bank, West Virginia, WV 24944, USA}

\author[0000-0002-0332-2641]{Gabi Wenzel}
\affiliation{Department of Chemistry, Massachusetts Institute of Technology, Cambridge, MA 02139, USA}
\affiliation{Center for Astrophysics \textbar{} Harvard \& Smithsonian, Cambridge, MA 02138, USA}

\author[0000-0003-0304-9814]{P. Bryan Changala}
\affiliation{Center for Astrophysics \textbar{} Harvard \& Smithsonian, Cambridge, MA 02138, USA}
\affiliation{JILA, University of Colorado Boulder and National Institute of Standards and Technology, Boulder, CO 80309, USA}
\affiliation{Department of Physics, University of Colorado Boulder, Boulder CO 80309, USA}

\author[0000-0001-5020-5774]{Zachary T.P. Fried}
\affiliation{Department of Chemistry, Massachusetts Institute of Technology, Cambridge, MA 02139, USA}

\author[0000-0002-8932-1219]{Ryan A. Loomis}
\affiliation{National Radio Astronomy Observatory, Charlottesville, VA 22903, USA}

\author[0000-0001-9479-9287]{Anthony Remijan}
\affiliation{National Radio Astronomy Observatory, Charlottesville, VA 22903, USA}

\author[0000-0003-4179-6394]{Edwin A. Bergin}
\affiliation{Department of Astronomy, University of Michigan, Ann Arbor, MI 48109, USA}

\author[0000-0002-0850-7426]{Ilsa R. Cooke}
\affiliation{Department of Chemistry, University of British Columbia, 2036 Main Mall, Vancouver, BC V6T 1Z1, Canada}

\author[0000-0003-1924-1122]{David Frayer}
\affiliation{Green Bank Observatory, 155 Observatory Rd, Green Bank, West Virginia, WV 24944, USA}

\author[0000-0003-0799-0927]{Andrew M. Burkhardt}
\affiliation{Department of Earth, Environment, and Physics, Worcester State University, Worcester, MA 01602, USA}

\author[0000-0001-6752-5109]{Steven B. Charnley}
\affiliation{Astrochemistry Laboratory and the Goddard Center of Astrobiology, Solar System Exploration Division, NASA Goddard Space Flight Center,  8800 Greenbelt Road, Greenbelt, MD 20771, USA.}

\author[0000-0001-8233-2436]{Martin A. Cordiner}
\affiliation{Astrochemistry Laboratory and the Goddard Center of Astrobiology, Solar System Exploration Division, NASA Goddard Space Flight Center,  8800 Greenbelt Road, Greenbelt, MD 20771, USA.}
\affiliation{Department of Physics, Catholic University of America, Washington, DC 20064, USA.}

\author[0000-0002-6667-7773]{Andrew Lipnicky}
\affiliation{National Radio Astronomy Observatory, Charlottesville, VA 22903, USA}

\author[0000-0001-9142-0008]{Michael C. McCarthy}
\affiliation{Center for Astrophysics \textbar{} Harvard \& Smithsonian, Cambridge, MA 02138, USA}

\author[0000-0003-1254-4817]{Brett A. McGuire}
\affiliation{Department of Chemistry, Massachusetts Institute of Technology, Cambridge, MA 02139, USA}
\affiliation{National Radio Astronomy Observatory, Charlottesville, VA 22903, USA}

\correspondingauthor{Ci Xue, Brett A. McGuire}
\email{cixue@mit.edu, brettmc@mit.edu}

\begin{abstract}
Spectral line surveys of the Taurus Molecular Cloud-1 (TMC-1) have led to the detection of more than 100 new molecular species, making it the most prolific source of interstellar molecular discoveries. These wide-band, high-sensitivity line surveys have been enabled by advances in telescope and receiver technology, particularly at centimeter and millimeter wavelengths. In this work, we present a statistical analysis of the molecular inventory of TMC-1 as probed by the GOTHAM large program survey from 3.9 to 36.4 GHz. To fully unlock the potential of the $\sim$29 GHz spectral bandwidth, we developed an automated pipeline for data reduction and calibration. We applied a Bayesian approach with Markov-Chain Monte Carlo fitting to the calibrated spectra and constrained column densities for 102 molecular species detected in TMC-1, including 75 main isotopic species, 20 carbon-13 substituted species, and seven deuterium-substituted species. This list of the detected gas-phase molecules is populated by unsaturated hydrocarbons, in stark contrast to the oxygen-rich organics found in sublimated ices around protostars. Of note, ten individual aromatic molecules were identified in the GOTHAM observations, contributing 0.011\% of the gas-phase carbon budget probed by detected molecules when including \ce{CO} and 6\% when excluding \ce{CO}. This work provides a reference set of observed gas-phase molecular abundances for interstellar clouds, offering a new benchmark for astrochemical theoretical models.

\end{abstract}


\section{Introduction \label{sec:introd}}
TMC-1 is a filamentary condensation within the extended Taurus Molecular Cloud. It has a mass of ${\sim}8\,\mathrm{M_\odot}$, a central density of ${\sim}2\times10^4\,\mathrm{cm^{-3}}$, and is at a distance of ${\sim}130\,\mathrm{pc}$ \citep{2008MNRAS.384..755N}. The molecular richness of TMC-1 makes it a prototypical target in the quest to understand dark cloud chemistry \citep{2019AnA...624A.105F}. Numerous astrochemical models have been benchmarked against the observed molecular abundances in TMC-1 \citep[e.g.,][]{2016MNRAS.459.3756R,2021AnA...652A..63W,2023ApJ...957...88B,2024AnA...682A.109M}. A large fraction of TMC-1's molecular inventory can be explained as a result of binary gas-phase interstellar chemical processes that occur after UV shielding but before grain-surface chemistry becomes dominant \citep{2011AnA...530A..61H, 2017MNRAS.470.4075L, 2014MNRAS.437..930L, 2019AnA...625A..91V}. It sets the reference initial conditions for the later prestellar and protostellar phases. During later phases, when gravitational collapse has proceeded further, freeze-out of the gas onto dust grains occurs more efficiently, which requires more elaborate yet less well-constrained gas-grain models to describe the chemistry \citep{hin13,cou20}. 

TMC-1 is the most prolific source for interstellar molecular discoveries\footnote{\footnotesize 77 new detections have been made in Sgr B2, making it the second most prolific source of molecular discoveries.}, contributing 109 new molecular detections as of February 2025 \citep[${\sim}31\%$ of the interstellar molecular inventory,][]{2022ApJS..259...30M}, from the detection of \ce{C3N} in 1980 to \ce{H2CCCN} and \ce{CH3CHS} in 2025 \citep{1980ApJ...241L..99F, 2025AnA...693L..20A}. TMC-1's sparse spectra and narrow linewidths allow for spectral stacking techniques to be utilized to increase the sensitivity of new molecular searches \citep{2021NatAs...5..188L}, in contrast to the other more turbulent and line-dense sources (e.g., Sgr B2, Orion KL, IRAS+16293).

The recent advances in high-sensitivity wide-band observations have greatly accelerated the rate of molecular discoveries. Molecular species peak in abundance at different regions along the TMC-1 filament. The most commonly observed region is the “cyanopolyyne peak” (CP) located in the southeastern end of the filament \citep{1983ApJ...267..163S}. Several spectral surveys have been carried out toward the CP position with radio observations to establish the molecular inventory in TMC-1 \citep[e.g.,][]{2004ApJ...610..329K, 2004PASJ...56...69K, 2016ApJS..225...25G, 2018ApJ...854..116S}. In particular, the most recent spectral line surveys, GBT Observations of TMC-1: Hunting Aromatic Molecules \citep[GOTHAM,][]{2020ApJ...900L..10M} and Q-band Ultrasensitive Inspection Journey to the Obscure TMC-1 Environment \citep[QUIJOTE,][]{2021A&A...652L...9C}, have reported the detection of a combined 78 new molecular species since 2020; $22\%$ of all interstellar molecular discoveries. Unsaturated species are strongly represented in these detections, which include carbon-chain cyanopolyynes and aromatic molecules \citep{2001ApJ...552..168F, 2018Sci...359..202M, 2021A&A...652L...9C, 2023ApJ...944L..45R}.

While unbiased molecular line surveys constrain the chemical conditions in TMC-1, a challenge faced by these surveys is the efficient processing of the vast amount of data and its subsequent presentation to the scientific community. For example, the GOTHAM project provides highly sensitive datasets essential for detecting faint molecular signals, yet the amount of observational raw data necessitates automated processing. Previous data reductions for this large project were based on the first-generation pipeline \citep{2020ApJ...900L..10M}. This pipeline has several limitations, including difficulties in extending the \texttt{IDL}-based packages and reliance solely on visual inspection for data cleaning, resulting in non-reproducible results and potential human errors. In order to automate data processing, we developed a \texttt{Python}-based pipeline to calibrate, reduce, and analyze spectral data with robust algorithms for signal identification and interference mitigation, which enables accurate downstream statistical analyses.

In this work, we present a statistical study of the molecular inventory of TMC-1 with a Bayesian approach to the spectral analysis. We describe the spectral survey calibrated using the newly developed pipeline in Section~\ref{sec:obs-data} and the spectroscopic data in Section~\ref{sec:spec-data}. A description of the methodology employed to constrain the excitation conditions is presented in Section~\ref{sec:methods}. Section~\ref{sec:rslt} presents a summary of the molecular column densities based on Bayesian statistical analysis and the derived abundance ratios for each isotopologue family. Section~\ref{sec:disc} contains discussions on elemental abundances probed by the gas-phase molecules and the treatment of velocity components in spectral modeling.

\section{Observational Data \label{sec:obs-data}}

\begin{deluxetable*}{cclc}
    \tablecaption{Summary of GOTHAM Spectral Setups \label{tab:projects}}
    \tablehead{
        \colhead{Project Code} & \colhead{Band(s)} & \colhead{Frequency Range} & \colhead{Observing Time}\\
        ~    &         & \colhead{(MHz)}     & \colhead{(hr)}
    }
    \startdata
        AGBT17A\nobreakdash-164 & K &   18306.2 -- 18493.8, 19076.2 -- 19263.8, 19996.2 -- 20183.8, 20726.2 -- 20913.8, & 14.83\\
        && 21766.2 -- 21953.8, 22836.2 -- 23193.8&\\
        AGBT17A\nobreakdash-434 & K & 20181.2 -- 20546.8, 20666.2 -- 20853.8, 21171.2 -- 21358.8, 21706.2 -- 21893.8, & 16.78\\
        && 22126.2 -- 22313.8, 22831.2 -- 23018.8, 23081.2 -- 23268.8 &\\
        AGBT18A\nobreakdash-333 & X, Ka & 7906.2 -- 11608.8, 26879.2 -- 29101.8  & 138.54 \\
        AGBT18B\nobreakdash-007 & K, Ka & 22031.2 -- 27503.8, 27606.2 -- 28348.8, 29086.2 -- 29828.8 & 129.14  \\
        AGBT19B\nobreakdash-047 & K, Ka & 22031.2 -- 33528.8 & 438.13  \\
        AGBT20A\nobreakdash-516 & C & 3943.2 -- 8132.8 & 74.51 \\
        AGBT21A\nobreakdash-414 & Ka & 33296.2 -- 33483.8, 33602.2 -- 33789.8, 34253.2 -- 34440.8, 35246.2 -- 35433.8 & 0.79  \\
        AGBT21B\nobreakdash-210 & Ku, K, Ka & 12700.2 -- 15587.8, 18000.2 -- 22327.8, 33525.2 -- 36412.8 & 432.57\\
        AGBT24A\nobreakdash-124 & X & 9386.2 -- 10962.5 & 192.79
 \\
    \enddata
    \tablecomments{Data were recorded with a uniform bandwidth of 187.5\,MHz per window at a frequency resolution of 1.43\,kHz. The total observing time per project includes on-source observations, pointing scans, flux calibration scans, and overhead slewing time.}
\end{deluxetable*}

The GOTHAM program is a dedicated spectral line observing program of TMC-1 CP covering almost 30~GHz of bandwidth at high sensitivity and spectral resolution. Four data reductions (DR) have been performed and analyzed thus far, comprising observations obtained between May 2017 -- May 2019 (DR1), June 2020 (DR2), April 2021 (DR3), and May 2022 (DR4); each DR includes the prior data plus additional observations. Detailed information concerning the previous data calibration methods can be found in \citet{2020ApJ...900L..10M, 2021Sci...371.1265M, 2022ApJ...938L..12S}. In this work, we present the fifth DR of the GOTHAM program, hereafter referred to as DR\rom{5}. Compared to the previous DR4, DRV additionally includes new C-band observations (project code AGBT20A-516) and substantially deeper X-band observations (project code AGBT24A-124). All included data were calibrated and reduced using our newly developed in-house data processing pipeline to maintain consistency. A detailed description of the pipeline is presented in Appendix~\ref{apx:reduction}. The change from using Arabic numbers to Roman numerals for version numbering highlights the implementation of different data calibration and reduction strategies.

All observations were performed with the VErsatile GBT Astronomical Spectrometer \citep[VEGAS,][]{VEGAS} spectral line backend on the Robert C. Byrd 100-m Green Bank Telescope (GBT). The targeted position was the TMC-1 CP, centered at $\alpha_\text{J2000}=\hms{04}{41}{42.5}$, $\delta_\text{J2000}=\dms{25}{41}{26.8}$, with the blank-sky `off' position at an azimuth offset of $-1^{\circ}/\cos(\mathrm{el.})$. Pointing and focus corrections were measured and applied every 3–4 hours during the 24A semester and every 1–2 hours during other semesters, primarily on the calibrator J0530+1331. Pointing solutions across all observed frequencies averaged to $\sim7\%$ of the relevant beam size. The lowest pointing corrections were for the Ku-Band observations (with a median solution value of $2.2\arcsec$, corresponding to $4.1\%$ of the Ku-Band beam) and the highest were for the X-Band observations (with a median value of $9.5\arcsec$, corresponding to $11.4\%$ of the X-Band beam). Flux calibration was achieved with switched noise-diode measurements, resulting in an estimated antenna temperature accuracy of $10-20\%$ (see e.g., \citealp{Braatz_2009}).

Details of individual spectral setups are summarized in Table~\ref{tab:projects}. The data set covers frequency ranges set by the individual receiver bands, C, X, Ku, K and Ka, with nearly continuous coverage from 3.9 to 11.6 GHz, 12.7 to 15.6 GHz, and 18.0 to 36.4 GHz (29 GHz of total bandwidth). The total observing time amounts to 1438 hours. Observations were made with a uniform spectral resolution of 1.43\,kHz, corresponding to $0.11–0.01\,\mathrm{km\,s^{-1}}$ in velocity. The Full Width at Half Maximum (FWHM) of Gaussian beamwidths ($\theta_{B}$) varied from $\sim$194\arcsec\ at 3.9 GHz to $\sim$21\arcsec\ at 36.4 GHz. 

\begin{figure*}[ht]
    \centering
    \includegraphics[width=\textwidth]{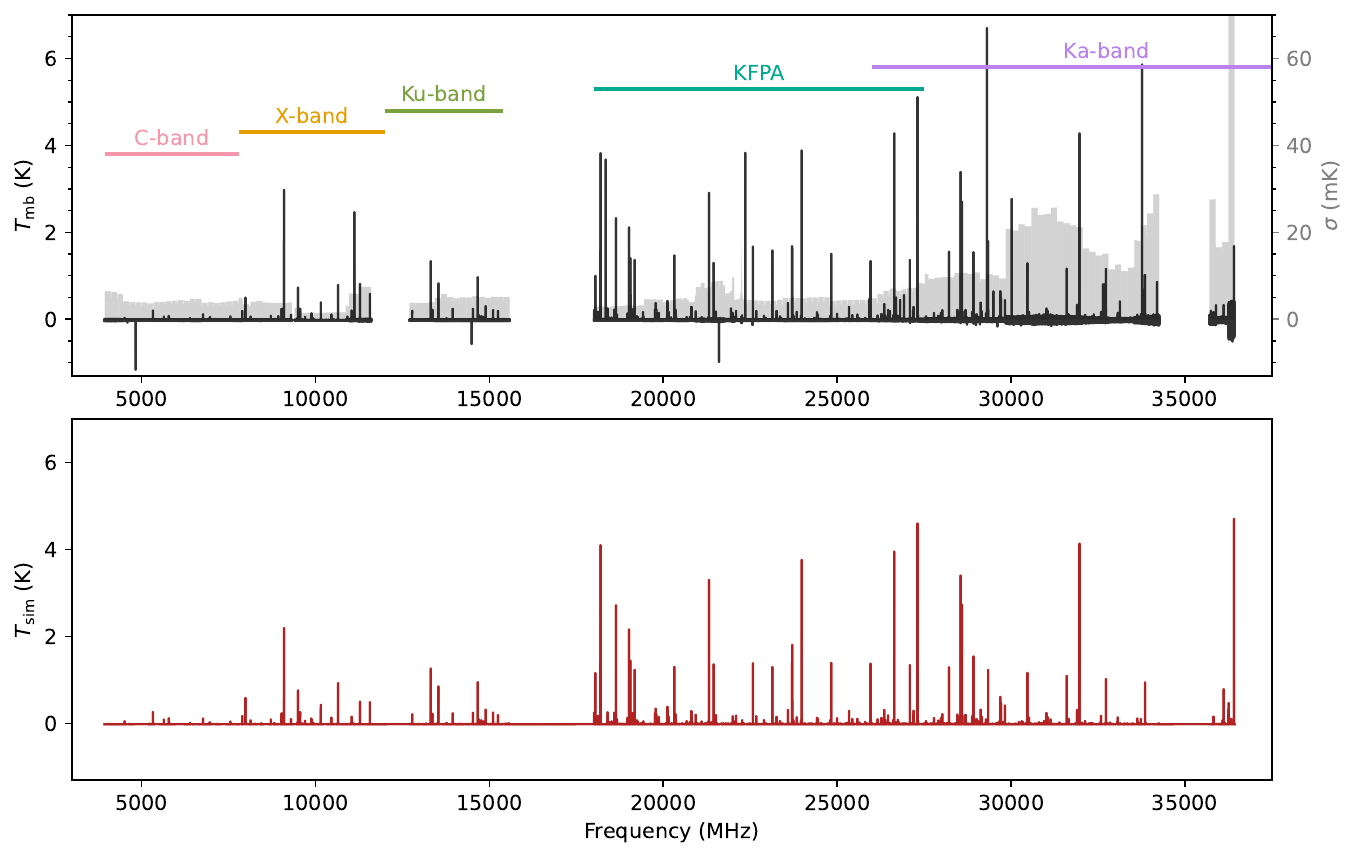}
    \caption{\label{fig:spectra} GOTHAM DRV observations and simulations. The top panel shows the calibrated observed spectra in black in K (left axis) and the noise level per channel in grey in mK (right axis). The receivers used in each frequency range are shown as colored bars. Data between 34250 and 35686 MHz in the Ka band are flagged due to abnormally high $T_{\mathrm{sys}}$ (see Appendix~\ref{apx:reduction} for details). The bottom panel shows the simulated spectrum in red in K, generated from the MCMC fitting of 102 analyzed molecules. The absorption lines and some bright emission lines contributed by molecules that need special treatment are not included in the simulation. See the main text for details.}
\end{figure*}

Channel-to-channel root-mean-square (RMS) noise was determined at $\approx$4--15\,mK across most of the observed frequency range, as shown with the grey-shaded area in Figure~\ref{fig:spectra}. For each integration, we use the radiometer equation \citep{2004tra..book.....R} to compute average noise level, 
\begin{equation} \label{eqa:sigma}
    \sigma = \frac{T_\mathrm{sys}}{\sqrt{t \times \Delta f}},
\end{equation}
where $T_{\mathrm{sys}}$ is the system temperature averaged over a single spectral window, $t$ is the effective exposure time\footnote{for position-switched observations, it is the harmonic mean of the exposure time of the four sampling phases in each On-Off scan pair.} ($\sim$15 seconds), and $\Delta f$ is the channel width (i.e., 1.43\,kHz). Calculated noise levels offer a more practical approach than measured noise levels for processing the GOTHAM data. Accurate direct measurements require finding spectral regions free of both astronomical signal and radio-frequency interference (RFI). However, signals may be buried in the noise in short integrations and RFI may be sporadic and unpredictable. Instead of relying on exhaustive and often unfeasible visual inspection, the radiometer equation shows that $T_\mathrm{sys}$ is a reliable proxy for the noise level, allowing straightforward and consistent estimation. Moreover, the algorithms implemented for automated signal identification rely on local noise levels as input. We find that, in noise-only spectral regions, the calculated noise levels are consistent with the measured standard deviation, differing by less than $2\%$. Noise levels across all integrations are derived using error propagation through inverse-variance-weighted averaging of the calibrated spectrum of each integration. The RMS noise varies depending on frequencies, mainly due to different total integration times and zenith opacity correction at certain channels. 

\section{Spectroscopic Data}\label{sec:spec-data}
For this work, spectroscopic information was taken from the CDMS database whenever available \citep{2005JMoSt.742..215M}. Considering the cold nature of TMC-1 \citep[$T_{\textrm{ex}}=5-10\,\mathrm{K}$,][]{2016ApJS..225...25G}, the energy in these regions is generally not sufficient to populate the vibrationally excited states and hence we only considered rotational transitions in the ground vibrational state. Note that one vibrationally excited state of \ce{C6H} has been detected towards TMC-1 \citep{2023AnA...680L...4C}. For nitrogen(N)-bearing species, we used catalogs that include the appropriate hyperfine splitting due to the $^{14}$N nuclear electric quadrupolar coupling.

The CDMS database provides tabulated values of partition functions ($Q$) computed by state counting, with most entries containing values from a rotational temperature of 300 K down to 2.725 K. To approximate the underlying $Q(T)$ at the lower rotational temperature in TMC-1, we fitted a low-order (3--5) polynomial function to the tabulated values on a logarithmic scale. This approach enabled us to achieve the desired accuracy of $Q(T)$, with the deviation between the fitted and tabulated values constrained to be within 5\%. The $Q(T)$ value at the corresponding rotational temperature used in the analysis of each species is presented in Tables~\ref{tab:main-mols}, \ref{tab:13C-mols}, and \ref{tab:D-mols}. The $Q(T)$ values, fitted polynomial expressions, and the maximum percentage difference for each species are publicly available in the Harvard Dataverse Repository \citep{GOTHAMDRV}.

We found that the line frequencies for some species reported in the CDMS database were insufficiently accurate to match the astronomically observed features due to the extremely narrow linewidths in TMC-1 ($<$0.3\,km\,s$^{-1}$) and the ultra-high spectral resolution of VEGAS. Examples included \ce{H2CCN}, \ce{C4H}, \ce{C6H}, \ce{CH3C_{2n+1}N}, \ce{CH3C_{2n}H}, \ce{C5N-}, and the \ce{^{13}C} isotopologues of \ce{HC5N} and \ce{HC7N}. The frequency mismatch was significant and prevented reliable spectral fitting. As such, we adapted the remeasured frequencies or/and recalculated spectroscopic parameters from recent individual publications for these species \citep{2020AnA...641L...9C, 2022ApJ...924...21S, 2023MNRAS.525.2154T, 2023ApJ...944L..45R}. In addition, we remeasured the ground state transition frequencies for \ce{C3N}, $c$-\ce{C3H}, and \ce{CH3C6H} ourselves specifically for this work using a Balle-Flygare-type cavity Fourier transform microwave (FTMW) spectrometer~\citep{2016JChPh.144l4201C, 2005RScI...76i3106G}. Appendix~\ref{apx:lab} summarizes the spectroscopic properties of the rotational transitions generated from these measurements.

\section{Methods \label{sec:methods}}
\subsection{Single Excitation Temperature Model}
Most of the molecular species in TMC-1 exhibit a Boltzmann distribution of the populations across the rotational energy levels at centimeter wavelengths \citep{2004PASJ...56...69K}. The population of these levels can thus be well-described at a single, uniform excitation temperature. As such, we assume a single excitation temperature for deriving the molecular excitation conditions and use it to describe the distribution of rotational levels (hereafter $T_{\textrm{rot}}$). However, the assumption of complete thermalization fails for $A/E$ spin isomers of $C_{3v}$ symmetric tops (e.g., methylpolyynes \ce{CH3(CC)_nH} and methylcyanopolynnes \ce{CH3(CC)_nCN}). Because the two spin isomers are radiatively decoupled from each other, we can treat them as separate chemical species by simply partitioning their radio catalog into separate $A$- and $E$-transitions.\footnote{The reference catalog intensities are calculated assuming complete $A$/$E$ thermalization, so the retrieved column densities must be corrected by the $A$/$E$ population fractions at the excitation temperature derived for each spin isomer, respectively, which we calculate by re-summing the $A$ and $E$ Boltzmann factors separately.} 

We note that $T_{\textrm{rot}}$ is not necessarily the same as the kinetic temperature ($T_k$) of the gas, since the gas density of TMC-1 \citep[$1-2\times10^4\,\mathrm{cm}^{-3}$,][]{1982ApJ...255..149S, 2016A&A...590A..75F} might not be high enough to thermalize the observed energy levels. Considering the limited availability of collisional rate coefficients and the high computational cost, calculating the possible radiative transfer processes for all detected species is beyond the scope of this paper.

We used the \texttt{molsim} \citep{molsim} program to model molecular line profiles. To determine $T_{\textrm{rot}}$ and column densities ($N_T$), we followed the convention of \citet{1991ApJS...76..617T} which includes corrections for optical depth. The local optical depth is formulated as
\begin{multline}
    \tau(\nu) = \frac{16\pi^3}{3h}\sqrt{\frac{\ln{2}}{\pi}}\frac{S_{ij}\mu^2N_T}{\Delta V}\frac{e^{-E_u/kT_{\rm rot}}}{Q(T_{\rm rot})}\\
    \times(e^{h\nu/kT_{\rm rot}}-1),
    \label{eqn:tau}
\end{multline}
where $S_{ij}\mu^2$ is the transition intensity expressed as the product of the relevant component of the dipole moment squared times the intrinsic line strength, $\Delta V$ is the FWHM linewidth, $Q(T_{\rm rot})$ is the partition function at $T_{\rm rot}$, and the other symbols carry their usual meaning. This $\tau$ profile is converted to the brightness temperature ($T_b$) using the source function \citep{1991ApJS...76..617T},
\begin{multline}
    \Delta T_b(\nu) = \biggl[\frac{h\nu}{k(e^{h\nu/kT_{\rm rot}}-1)} - \frac{h\nu}{k(e^{h\nu/kT_{\rm bg}}-1)}\biggr]\\
    \times (1-e^{-\tau}).
    \label{eqn:tb}
\end{multline}
Line profiles are multiplied by the beam filling factor ($B$) to correct for beam dilution and then compared with the observed main-beam brightness temperatures ($T_{\rm mb}$). The beam filling factor $B$ is defined as
\begin{equation}
    B = \frac{\theta_s^2}{\theta_{B}^2 + \theta_s^2}\ \ ,
\end{equation} where $\theta_s$ is the source emission extent and $\theta_{B}$ is the beam
width.

\subsection{Bayesian Inference via MCMC}
We used a Bayesian approach for line fitting to compare models with varying parameters and calculate their probability distributions. This approach provides an inference by conditioning the data on priors and sampling posterior distributions. Following \citet{2021NatAs...5..188L} and \citet{2021ApJ...910L...2L}, we use an affine-invariant Markov-Chain Monte Carlo (MCMC) ensemble sampler, a wrapper function in \texttt{molsim} to \texttt{emcee} \citep{2013PASP..125..306F}, to explore parameter space.

We assume that there are multiple velocity components that may be blended, but otherwise do not attempt to treat line blending between molecules due to the low line density. Recent observations performed with the 45\,m telescope at Nobeyama Radio Observatory \citep{2018ApJ...864...82D}, as well as the previous GOTHAM DR data \citep{2020ApJ...900L...9X, 2021Sci...371.1265M, 2024ApJ...976..105R}, revealed four distinct velocity components contributing to the overall signal for molecular emission toward TMC-1 CP (see Section~\ref{sec:vel-components} for further discussion of the velocity components). Following the strategy established in \citet{2021NatAs...5..188L}, we model the emission using four velocity components under two simplified spatial distribution scenarios: either separated or co-spatial. In the separated scenario, emission originates from spatially distinct velocity components that are radiatively decoupled, allowing line intensities to be added linearly. In contrast, the co-spatial scenario involves emission from velocity components that spatially overlap and align along the line of sight, requiring line profiles to be added in opacity space before converting to intensity. We found the co-spatial scenario better describes molecules with extended distribution and higher optical depth, while the separated scenario better fits molecules with compact distribution or optically thin molecules, consistent with \citet{2021NatAs...5..188L}. We acknowledge that some species are probably somewhere in-between, but we assume these two simplifying scenarios to make the modeling tractable. As examples, the spectra of \ce{C3N} for the co-spatial model and $c$-\ce{C6H5CN} for the separate model are shown in Appendix~\ref{apx:lab} and Appendix~\ref{apx:mcmc}.

The four velocity components are represented by independent source velocity ($V_\mathrm{LSR}$), $\theta_s$, and $N_T$ parameters but assumed to share a common $T_{\textrm{rot}}$ and $\Delta V$. For the separated model, a forward model of 14 free parameters (see Tables~\ref{tab:priors_separate_nonaromatic} and \ref{tab:priors_separate_aromatic}) is applied to iteratively generate model spectra that are then compared with the observations. In addition, for the co-spatial model, we assume a common source size across four components so that the total number of free parameters is reduced by three, i.e., 11 free parameters (see Tables~\ref{tab:priors_cospatial}).

We use a logarithmic likelihood distribution with:
\begin{equation}
    l = \sum_{i} \left[-\ln \left(\sigma_i \sqrt{2\pi}\right) - \frac{1}{2}\frac{(x_i-m_i)^2}{\sigma_i^2}\right],
    \label{eqn:log_likelihood}
\end{equation}
where $x$ is the observed spectral line intensity, $m$ is the model line intensity, and $\sigma$ is the uncertainty of observations including the local RMS noise and an additional 20\% uncertainty accounting for the systematic errors.

The posterior probability distributions for each parameter and their covariances were generated using 100 Markov chains, often referred to as walkers. Each Markov chain consists of multiple samples, with each sample being randomly drawn from within the parameter space. The number of samples was chosen to ensure convergence for each species, respectively, ranging from 10,000 to 50,000. To aid in convergence, we construct prior distributions for different models based on the posteriors from the previous MCMC analyses as detailed in Appendix~\ref{apx:mcmc}. Convergence of the MCMC was assessed using the $\hat{R}$ diagnostics \citep{1992StaSc...7..457G, Stephen1998GeneralMF}, a wrapper function in \texttt{molsim} to \texttt{arviz} \citep{arviz_2019}, and by visually inspecting the posterior traces.

The statistics on physical parameters derived from the converged posteriors as well as the marginalized posterior probability distribution are presented in Appendix~\ref{apx:mcmc}. The fitted results are represented by 50$^{th}$, 16$^{th}$, and 84$^{th}$ percentiles as the representation value and the range of uncertainty, also known as the 68\% confidence interval. This choice is equivalent to choosing the mean and $+/-1\sigma$ for a normal distribution, which is the case for most of our posterior distributions. 

\section{Results} \label{sec:rslt}
More than 170 molecular species have been identified in TMC-1 from previous observations. However, limited by the frequency coverage of the GOTHAM survey ($3.9-32.0$ GHz)\footnote{See Appendix~\ref{apx:reduction} for details on the exclusion of Ka-band data above 32 GHz from spectral analyses.}, some of these molecular species cannot be observed or analyzed. For example, some small molecules with a large $B$ constant have rotational transitions that all lie above 32.0 GHz, including \ce{CX} (where \ce{X}=\ce{O}, \ce{N}, \ce{S}), \ce{HCO+}, \ce{HCN}, and \ce{SiO} \citep{1989ApJ...343..201Z, 2019AnA...624A.105F}. While \ce{CH} was detected with its ground-state $\Lambda$-doubling transition at 3.335 GHz toward TMC-1 \citep{2012AnA...546A.103S}, the transitions covered by GOTHAM are between energy levels too high to be populated under TMC-1 conditions. Similarly, some species, such as \ce{CH2CCH} and \ce{HCOOCH3}, have emission features that are too weak within our frequency coverage, despite being observed with bright features at millimeter wavelengths \citep{2021AnA...649L...4A, 2022AnA...657A..96A}.

We performed MCMC fitting for a total of 102 molecular species, all of which have been previously detected in TMC-1. This work does not include upper-limit analyses for undetected species. In the following sections, we present the MCMC fitted results for 75 main isotopic species, 20 \ce{^{13}C}-substituted species, and seven \ce{D}-substituted species. Figure~\ref{fig:spectra} shows the simulated spectra for all molecules analyzed using the MCMC fitting. There are no corresponding simulations for some strong lines in the observed spectra because those lines are contributed by molecules that either have only one transition covered (e.g., \ce{H2CS}; \ce{SO}; \ce{OCS}) or need special treatment of excitation conditions (e.g., \ce{CCS}; \ce{H2CO}; $c$-\ce{C3H2}).

In particular, the population of \ce{CCS} transitions in the GOTHAM observations significantly deviate from a Boltzmann distribution. Potential explanations for this deviation include geometric effects related to the compact spatial distribution of \ce{CCS} and its offset from the TMC-1 CP, as revealed by \citet{2019ApJ...879...88D}. The standard beam-dilution correction, which assumes Gaussian sources centered within the beam, may not be applicable in this case. Moreover, the observed line profiles of individual \ce{CCS} transitions vary notably with frequencies, suggesting that the four velocity components of \ce{CCS} have different excitation conditions. This invalidates the assumption of a common $T_{\rm rot}$. Therefore, we exclude \ce{CCS} from the MCMC analysis to avoid fitting a non-descriptive model to the data.

\subsection{Main Isotopic Species}
Table~\ref{tab:main-mols} summarizes the physical parameters derived from the marginalized posteriors for the main isotopic molecular species. It includes 7 cations, 6 anions, and 62 neutral molecules, with their chemical formulae detailed in the first column. The third, fourth, and fifth columns list the 50$^{th}$ percentile values for $N_T$, $T_{\rm rot}$, and $\Delta V$, respectively, along with uncertainties. The sixth column lists the fitted partition function value at the corresponding $T_{\rm rot}$, the seventh column lists the model of velocity components considered in MCMC fits, and the last column specifies the figure number where individual species' marginalized posterior distributions are presented.

$N_T$ ranges from $\sim1 \times 10^{10}$~cm$^{-2}$ to $\sim3 \times 10^{14}$~cm$^{-2}$, with \ce{NH3} being the most abundant molecule and \ce{C7N^-} and \ce{NC4NH+} the least abundant. There is no clear trend between molecular abundance and number of atoms in general, contrary to the trend of a decrease in abundance with increasing molecular size within a specific molecular family \citep[e.g.,][]{1997ApJ...480L..63L, 2022ApJ...924...21S}.

$T_{\rm rot}$ has a median value of 7.5 K within a range of 3$-$13 K. We found that a few species have very low $T_{\rm rot}$, namely \ce{HCCS+}, \ce{HCCCHO}, and $c$-\ce{H2C3O}, indicating sub-thermal excitation. In particular, the QUIJOTE survey found that the transitions of \ce{HCCS+} at 32-99 GHz need to be fitted by two distinct $T_{\rm rot}$, $4.5\pm1.0$ K and $7.0\pm1.5$ K, and the transitions of \ce{HCCCHO} at 37-47 GHz were fitted by 4 K and 5 K, depending on rotational ladders \citep{2022AnA...657L...4C, 2021AnA...650L..14C}. Potential absorption activity has also been proposed as an explanation for the low $T_{\rm rot}$ of \ce{HCCCHO} \citep{2024ApJ...976..105R}. Although we also found $c$-\ce{H2C3O} with low $T_{\rm rot}$, the observed transitions in GOTHAM span only a modest range of rotational energy levels (upper-state energy, $E_u=0.7-3$ K). Additional observations of higher-energy transitions are needed to confirm its sub-thermal excitation.

\cacheposteriorsubfigure
\startlongtable
\begin{deluxetable*}{cccccclc}
    \setlength{\tabcolsep}{2.5pt}
    \tablecaption{Summary statistics of the marginalized posteriors for main isotopic species, sorted by molecular mass \label{tab:main-mols}}
    \tabletypesize{\footnotesize}
    \tablewidth{0pt}
    \tablehead{
        \multicolumn{2}{c}{Molecules} &\colhead{$N_T$(total)\tablenotemark{$\ast$}} &\colhead{$T_\mathrm{rot}$} &\colhead{$\Delta V$\tablenotemark{$\dagger$}} &\colhead{$Q(T_{\rm rot})$\tablenotemark{$\ddagger$}} &\colhead{Model} &\colhead{Posterior}\\
        Formula &IUPAC Names             &\colhead{($\mathrm{cm}^{-2}$)} &\colhead{(K)} &\colhead{(km\,s$^{-1}$)} & & &
    }
    \startdata
    \loadcachedposteriorsubfigure
    \input{tables/mols_main_mcmc_table}

    \enddata
    \tablenotetext{}{Uncertainties are represented by the 16$^\mathrm{th}$ and 84$^\mathrm{th}$ percentiles, also known as the $68\,\%$ confidence interval, which corresponds to 1$\,\sigma$ for a Gaussian distribution.  $^\ast$~Total column density and its uncertainties, derived by marginalizing over the posterior distributions of four velocity components. $^\dagger$~Velocity linewidth of a single component, assumed to be common across the four components. Considering the radial velocity separation of the four components, the typical width of an observed emission feature not involving hyperfine splitting is $\sim0.4~{\rm km\,s}^{-1}+\Delta V$. $^\ddagger$~Partition function at the median $T_\mathrm{rot}$. $^\mathsection$~Constrained priors on $\Delta V$ were applied as described in Appendix~\ref{apx:mcmc}; otherwise, uninformative priors were applied. $^\Vert$~Constrained priors on $T_\mathrm{rot}$ were applied as described in Appendix~\ref{apx:mcmc}; otherwise, uninformative priors were applied.}
\end{deluxetable*}

\subsection{Carbon-13 Isotopic Species}
Table~\ref{tab:13C-mols} summarizes the physical parameters derived from the marginalized posteriors for the singly-\ce{^{13}C}-substituted isotopologues. We detected emission from all four singly-\ce{^{13}C}-substituted isotopologues of \ce{C4H}, each having similar column densities with the exception of $\ce{^{13}CCCCH}$, which is the least abundant. To evaluate the $\ce{^{12}C}$/$\ce{^{13}C}$ ratio, we sampled the posterior distribution of $N_T$ for the main isotopic species and each \ce{^{13}C}-substituted species with 100,000 iterations to generate a distribution of the $N_T$ ratio. From this $N_T$ ratio distribution, we derived the $\ce{^{12}C}$/$\ce{^{13}C}$ value and its uncertainty by reporting the 50$^\mathrm{th}$, 16$^\mathrm{th}$ and 84$^\mathrm{th}$ percentiles. The $N_T$ ratios of \ce{C4H}/$\ce{^{13}CCCCH}$, \ce{C4H}/$\ce{C^{13}CCCH}$, \ce{C4H}/$\ce{CC^{13}CCH}$, and \ce{C4H}/$\ce{CCC^{13}CH}$ are determined to be $191^{+32}_{-43}$, $106^{+14}_{-19}$, $79^{+11}_{-19}$, and $100^{+28}_{-28}$, respectively, with a mean value of $117^{+13}_{-15}$. These relative ratios are consistent with the $\ce{^{12}C}$/$\ce{^{13}C}$ ratios derived from the integrated intensity ratios of the $N=5-4$ lines reported by \citet{2013JPCA..117.9831S}.

We detected one isotopologue of \ce{CCS}, \ce{C^{13}CS}. However, the $N_T$ of the main isotopic \ce{CCS} cannot be constrained with a single-excitation model in this work. Previous studies report $N_T$ values ranging from $9\times10^{12}$ to $1\times10^{14}\,\mathrm{cm}^{-2}$ \citep{2016ApJS..225...25G, 2019ApJ...879...88D, 2021AnA...646L...3C}, resulting in a $\ce{^{12}C}$/$\ce{^{13}C}$ isotopic ratio of 3$-$53. In addition, there is no detectable emission from \ce{^{13}CCS} in the GOTHAM observations, which was marginally observed in \citet{2007ApJ...663.1174S}.

We detected emission from all three singly-\ce{^{13}C}-substituted isotopologues of \ce{HC3N}, five of \ce{HC5N}, and seven of \ce{HC7N}. Using catalogs with consistent treatment of $\ce{^{14}N}$ hyperfine splitting, we derive mean $\ce{^{12}C}$/$\ce{^{13}C}$ ratios of $56^{+4}_{-6}$ for \ce{HC3N}, $68.7^{+1.4}_{-1.8}$ for \ce{HC5N}, $68^{+20}_{-8}$ for \ce{HC7N}, averaged across the $N_T$ ratios of all isotopomers for each molecule. The large uncertainties in the $\ce{^{12}C}$/$\ce{^{13}C}$ ratio for \ce{HC7N} prevents determining if $\ce{^{13}C}$ becomes more diluted with increasing chain length in \ce{HC_{2n+1}N}. More sensitive observations are needed to assess this trend. We also compared the relative column densities for the different $\ce{^{13}C}$ isotopomers. However, no significant enrichment of $^{13}$C is found at the C atom adjacent to the N atom in \ce{HC_{2n+1}N}, previously found in \ce{HC3N} \citep{1998AnA...329.1156T}.

\cacheposteriorsubfigure
\begin{deluxetable*}{ccccclc}
    \tablecaption{Summary statistics of the marginalized posteriors for $\ce{^{13}C}$-substituted isotopologues, sorted by molecular mass \label{tab:13C-mols}}
    \tablehead{
        \colhead{Molecules} &\colhead{$N_T$(total)\tablenotemark{$\ast$}} &\colhead{$T_\mathrm{rot}$} &\colhead{$\Delta V$\tablenotemark{$\dagger$}} &\colhead{$Q(T_{\rm rot})$\tablenotemark{$\ddagger$}} &\colhead{Model} &\colhead{Posterior}\\
                          &\colhead{($\mathrm{cm}^{-2}$)} &\colhead{(K)} &\colhead{(km\,s$^{-1}$)} & & &
    }
    \startdata
    \loadcachedposteriorsubfigure
    \input{tables/mols_13C_mcmc_table}
    \enddata
    \tablenotetext{}{See Table~\ref{tab:main-mols} for table notes $\ast, \dagger, \ddagger, \mathsection, \Vert$.}
\end{deluxetable*}

\subsection{Deuterium Isotopic Species}
Table~\ref{tab:D-mols} summarizes the physical parameters derived from the marginalized posteriors for the \ce{D}-substituted molecular species. We applied the same method used to derive $\ce{^{12}C}$/$\ce{^{13}C}$ ratios to derive H/D ratios by generating a distribution of the $N_T$ ratios.

We derived an H/D isotopic ratio of $89^{+8}_{-11}$ from \ce{C4H} and \ce{C4D}. For the cyanopolyynes family, we derived the H/D isotopic ratio to be $42^{+2}_{-2}$ for \ce{HC3N}, $50^{+4}_{-8}$ for \ce{HC5N}, and $82^{+11}_{-21}$ for \ce{HC7N}, all with resolved hyperfine components. In addition, we detected the singly-\ce{D}-substituted isotopologues, \ce{CH2DC4H} and \ce{CH2DC3N}, resulting the H/D isotopic ratio to be $42^{+16}_{-13}$ for \ce{CH3C4H} and $11^{+3}_{-4}$ for \ce{CH3C3N}.

Combining the polyyne radical, cyanopolyynes, methylpolyynes, methylcyanopolynnes families, the average H/D isotopic ratio from the GOTHAM observations is $52^{+5}_{-5}$. These ratios are comparable with the H/D isotopic ratios derived from the other carbon chain molecules (20--90) but lower than that from \ce{NH2D} ($\sim$1000) and \ce{N2D+} (156) in TMC-1 \citep{2001ApJS..136..579T}. In contrast, \citet{2023A&A...679A.120R} performed a survey of cold starless cores and reported \ce{H2S}/\ce{HDS} abundance ratios ranging from 3 to 10.

\cacheposteriorsubfigure
\begin{deluxetable*}{ccccclc}
    \tablecaption{Summary statistics of the marginalized posteriors for $\ce{D}$-substituted isotopologues, sorted by molecular mass \label{tab:D-mols}}
    \tablehead{
        \colhead{Molecules} &\colhead{$N_T$(total)\tablenotemark{$\ast$}} &\colhead{$T_\mathrm{rot}$} &\colhead{$\Delta V$\tablenotemark{$\dagger$}} &\colhead{$Q(T_{\rm rot})$\tablenotemark{$\ddagger$}} &\colhead{Model} &\colhead{Posterior}\\
                          &\colhead{($\mathrm{cm}^{-2}$)} &\colhead{(K)} &\colhead{(km\,s$^{-1}$)} & & &
    }
    \startdata
    \loadcachedposteriorsubfigure
    \input{tables/mols_D_mcmc_table}
    \enddata
    \tablenotetext{}{See Table~\ref{tab:main-mols} for table notes $\ast, \dagger, \ddagger, \mathsection, \Vert$.}
\end{deluxetable*}

\section{Discussion \label{sec:disc}}

\subsection{Elemental Compositions}
The known molecular inventory of TMC-1 is mainly constructed from five elements: hydrogen, carbon, nitrogen, oxygen, and sulfur. The inventory presented in this work enables us to revisit the elemental budget in dark clouds, particularly the budget probed by gas-phase molecules. We computed first molecular abundances (relative to H nuclei) by dividing the molecular column densities by $N_H$, where $N_H$ = $N$(H) + 2$N$(\ce{H2}) and atomic H is assumed to be negligible compared to \ce{H2} in TMC-1. We assume a \ce{H2} column density of $N_{T}$(\ce{H2})$=1.82\times10^{22}\,\mathrm{cm}^{-2}$ \citep{2019AnA...624A.105F, 2024MNRAS.532.4661K}. To derive the elemental abundances, we multiply the molecular abundance by the number of nuclei of that element in each molecule. To illustrate this, take for example \ce{C4H^-}, which has a column density of $\sim2.0\times10^{10}\,\mathrm{cm}^{-2}$. The molecular abundance of \ce{C4H^-} relative to H nuclei is therefore $5.5\times10^{-13}$ and the elemental abundance of carbon in \ce{C4H^-} is thus four times this value, i.e., $2.2\times10^{-12}$.

As some abundant molecules, such as \ce{CO}, are not covered by GOTHAM, we included additional species from complementary observations as referenced in Table~\ref{tab:literature-mols}. We considered the species listed in both Tables~\ref{tab:main-mols} and~\ref{tab:literature-mols} when computing the elemental abundance for each metal element, as summarized in Table~\ref{tab:elements}. Since GOTHAM mainly covers larger, less abundant species, particularly in comparison to dominant carbon carriers such as CO and CH, the contribution from this GOTHAM analysis to the total gas-phase carbon budget is limited ($\sim$0.12\%). Similarly, the GOTHAM contributions to the oxygen and sulfur budgets are also modest, at approximately $\sim$0.0015\% and $\sim$5\%, respectively. In contrast, the GOTHAM observations amount to $\sim$80\% of the total gas-phase nitrogen budget. This substantial contribution is due to the prevalence of \ce{NH3} and large CN-functionalized molecules, while simple species such as CN and HCN are relatively less abundant.

\begin{deluxetable}{ccc}
    \tablecaption{Molecular Column Densities from Literature, Sorted by Molecular Mass \label{tab:literature-mols}}
    \tablehead{
        \colhead{Species} & \colhead{$N_T$} & \colhead{Reference}\\
                          & \colhead{($\mathrm{cm}^{-2}$)}
    }
    \startdata
    \input{tables/mols_literature}
    \enddata
    \tablenotetext{\ast}{Derived from molecular abundance and $N_{\ce{H2}}$ of $1.82\times10^{22}$\,cm$^{-2}$.}
    \tablenotetext{\dagger}{Total $N_T$ derived with a $ortho$-to-$para$ ratio of 3.}
\end{deluxetable}

\begin{deluxetable}{cccl}
    \tablecaption{Elemental Abundances ($n/n_{\ce{H}}$) of TMC-1 CP \label{tab:elements}}
    \tablehead{
        \colhead{} & \colhead{Gas-phase Abundance} & \colhead{Initial Abundance in} \\
                          & \colhead{Probed by Detected Mols\tablenotemark{$\ast$}} & \colhead{Astrochemical Models\tablenotemark{$\dagger$}}
    }
    \startdata
    C &$5.0 \times 10^{-5}$ &$1.7 \times 10^{-4}$\,\citep{2011AnA...530A..61H}\\
    N &$2.1 \times 10^{-8}$ &$6.2 \times 10^{-5}$\,\citep{2009ApJ...700.1299J}\\
    O &$4.9 \times 10^{-5}$ &$2.4 \times 10^{-4}$\,\citep{2011AnA...530A..61H}\\
    S &$1.1 \times 10^{-8}$ &$8.0 \times 10^{-8}$\,\citep{1982ApJS...48..321G}\\
    \enddata
    \tablenotetext{\ast}{this work, including the molecular species observed in GOTHAM DRV and additional species from other observations not covered by GOTHAM, as referenced in Table~\ref{tab:literature-mols}.}
    \tablenotetext{\dagger}{ both in the gas and on the dust grains, as used in \citet{2011AnA...530A..61H,2016MNRAS.459.3756R,2017MNRAS.469..435V,2024AnA...689A..63W}.}
\end{deluxetable}

\begin{figure}[ht]
    \centering
    \includegraphics[width=0.5\textwidth]{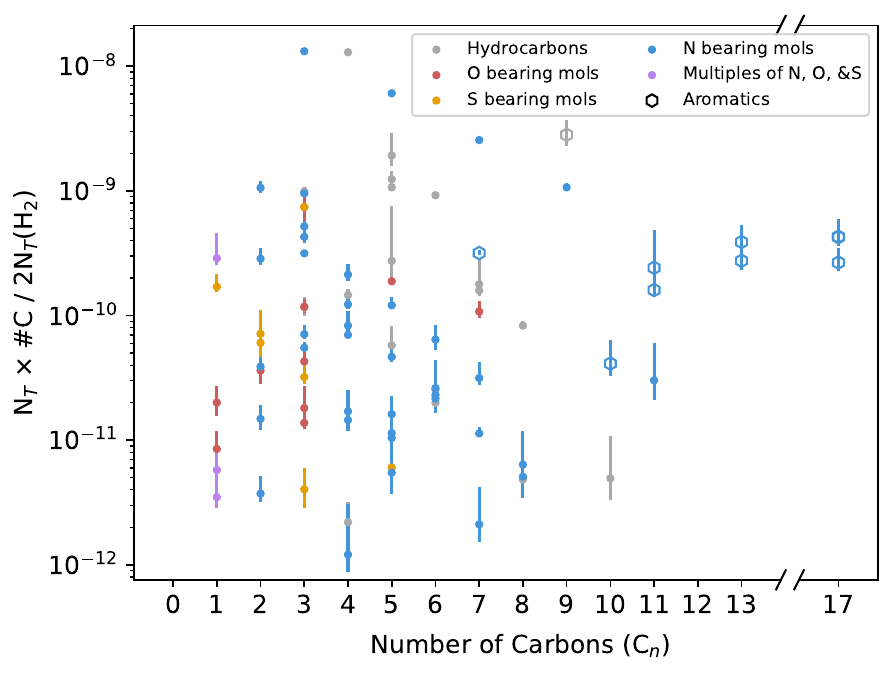}
    \caption{\label{fig:mols_as_scatters} 
    Carbon abundances probed by each molecule detected in GOTHAM (Table~\ref{tab:main-mols}) as a function of the number of carbon atoms ($\#$C), colored by elemental composition. Hydrocarbons are shown in grey, nitrogen-bearing species in blue, oxygen-bearing species in red, and sulfur-bearing species in yellow. Molecules containing more than one heteroatom are marked in purple.
 }
\end{figure}

\paragraph{Carbon} We obtained a total gas-phase carbon abundance ($n_{\ce{C}}/n_{\ce{H}}$) of $5.0\times 10^{-5}$, being about one third of the value of $1.7\times 10^{-4}$ from astrochemical models that accounts for nuclei in both the gas and on dust grains \citep{2011AnA...530A..61H}. Figure~\ref{fig:mols_as_scatters} displays the carbon abundances locked into the molecules listed in Table~\ref{tab:main-mols}, sorted by the number of carbon nuclei per molecule, and colored according to the metal elements contained. All the carbon-chain species\footnote{Linear molecules with two or more adjacent carbon atoms} contribute to $\sim$0.11\% of the gas-phase carbon budget of TMC-1 probed by detected molecules when including \ce{CO} and $\sim$65\% when excluding \ce{CO}.

Of note, among molecules with seven or more carbon atoms, the more abundant species are those containing aromatic rings. A major milestone in the detection of interstellar molecules is the recent discovery of individual aromatic molecules toward TMC-1 CP. Ten aromatic molecules were identified in the GOTHAM DRV, ranging from a single benzene ring \citep[benzonitrile,][]{2018Sci...359..202M} to four-ring molecules, \citep[cyanopyrene,][]{2024Sci...386..810W, 2024NatAs.tmp..264W}. Pure aromatic hydrocarbons typically possess weak or non-existent electric dipole moments and therefore cannot be directly observed via rotational spectra. Instead, CN-functionalized aromatics can serve as observational proxies. Nine of the 10 identified species are CN-functionalized. The total abundance of these 10 molecules amounts to $\sim$0.011\% of the gas-phase carbon budget when \ce{CO} is included and $\sim$6\% when \ce{CO} is excluded. While the X-H to X-CN ratio can vary between molecules, astrochemical modeling has recently constrained this ratio to $\sim$10--40 for aromatic hydrocarbons \citep{2024NatAs.tmp..264W}. This theoretical value depends on the overall rate coefficient for CN addition and the number of CN addition sites. In addition, a ratio of 76$^{+36}_{-29}$ is derived from observations of \ce{C9H8} and 2-\ce{C9H7CN}. It is evident that ``radio invisible'' pure aromatic molecules, along with their functionalized derivatives species, represent a non-trivial reservoir of carbon in TMC-1.

The composition of the gas phase in TMC-1 stands in stark contrast to that of sublimated ices observed towards hot cores and hot corinos. A similar summary of the elemental carbon content of sublimated ices is given by \citet{Bergin2014} using the Orion KL line survey (from 180 $\mu$m to 3\,mm). They find that the organic ice carbon abundance is $\sim3\times10^{-6}$, which is $\sim$2\% of the elemental carbon content using the normalization in Table~\ref{tab:elements}. A key distinction here is that the sublimated ices are characterized by oxygen-rich organics that are theorized to form from \ce{CO} (e.g., \ce{CH3OH}). This is different from the gas-phase carbon content seen in TMC-1 characterized by large unsaturated hydrocarbons.

The Orion KL analysis did not include some of the low-frequency measurements, which are better coupled to emission of large molecules. However, a Q-band survey of Orion KL by \citet{Liu2022} does not detect hydrocarbons in abundance, and this spectrum is also dominated by oxygen-rich organic emission. We acknowledge that, Orion KL, a region associated with high-mass star formation, may not be a direct comparison to the prestellar TMC-1, which is more readily associated with low-mass star formation. However, another inventory study of the low-mass source IRAS16293-2422 also finds that the sublimated organics are oxygen-rich \citep{Manigand2020}. While \citet{Manigand2021} did detect the emission of propyne (\ce{C3H6}) towards IRAS16293-2422, this hydrocarbon is found in an abundance that is less than $1\%$ of that of \ce{CH3OH} and this hydrocarbon is not a major elemental carbon reservoir.

The hydrocarbons detected in TMC-1 should also accrete onto grains \citep[][]{2024Sci...386..810W}, it is curious that the products of this chemistry are not observed in hot corinos or hot cores. Chemically, this suggests that the hydrocarbon-rich chemistry observed in TMC-1 produces hydrocarbon ices that are less volatile than the simple oxygen-rich organics found in ices \citep{doi:10.1021/je7005133, 2019ApJ...875...73B, 2022ESC.....6..597M}. One possibility is that the molecules in TMC-1 {\em may} eventually grow to larger sizes (e.g., aromatics) before they deplete. Aromatics are less volatile than water ice \citep[see sublimation enthalpy's and discussion in][]{2012AIChE..58.2875T, 2020ApJ...897L..38V} and can remain hidden as semi-refractory material when water and organic ice sublimate.

Another possibility is that light hydrocarbons could react with water when accreting onto grains to form oxygen-rich organic ices \citep[e.g., \ce{CCH},][]{2022ESC.....6..496P}. As heavier molecules have longer accretion timescales than lighter ones, the accretion of large hydrocarbons and aromatics could be limited by gas-phase destruction, thereby suppressing the production of hydrocarbon ices. However, extensive deuterium enrichment is found in the meteoritic organic material \citep{2010GeCoA..74.4417A}. Such material is expected to form at cold temperatures and be implanted into hydrogen-rich grains \citep{2024come.book....3B}. A similar action is required to obtain the carbon-13 isotopic signature isolated by \citet{2023Sci...382.1411Z}. Thus, some hydrocarbon ices (including aromatics) should have formed or been processed in the gas at cold temperatures where water is depleted. Such conditions may be plausible in TMC-1.

\paragraph{Nitrogen}
Another notable feature shown in Figure~\ref{fig:mols_as_scatters} is the prevalence of N-bearing species in TMC-1, which are dominated by nitriles. The \ce{CN} radical plays an important role in the formation of these molecules. At low temperatures, addition–elimination reactions between \ce{CN} radicals and unsaturated hydrocarbons are efficient and barrierless in the entrance channels \citep{2001CPL...344..310C}. In addition, experimental measurements have shown \ce{CN} attacks aromatic $\pi$-bonds with a large rate coefficient \citep[$\sim10^{-10}\,\mathrm{cm}^3 \mathrm{s}^{-1}$, ][]{2020ApJ...891L..41C, 2020JPCA..124.7950M, 2000JChPh.113.8643B}. These \ce{CN} reactions contribute to the formation of CN-functionalized unsaturated carbon chains (e.g., \ce{HC_nN} and \ce{CH3C_nN}) and aromatics (e.g., $c$-\ce{C6H5CN}), both of which are abundant in TMC-1.

From the observed gas-phase N-bearing molecules, we obtained a nitrogen abundance ($n_{\ce{N}}/n_{\ce{H}}$) of 2.1$\times 10^{-8}$. The discrepancy with the value of $6.2\times 10^{-5}$ \citep{2009ApJ...700.1299J} used by astrochemical models shows that gas-phase organics are not major nitrogen carriers. The gas-phase nitrogen content carried by sublimated ices in Orion KL is $\sim8.6\pm3.4\times10^{-7}$ \citep{2018ApJ...866..156R}.  Thus, the gas-phase nitriles seen in TMC-1 are under abundant compared to sublimated ices. This implies that much of the nitrogen in gas is either locked in some form that is involved in the buildup of nitriles (e.g., \ce{N2}) or is not present in the gas-phase \citep{2006Natur.442..425M,2010AnA...513A..41H}. In particular, icy ammonia or ammonia salts have been proposed as a major reservoir \citep{2012PNAS..10910233D, 2020Sci...367.7462P}. Follow-up infrared observations will be critical to further explore the nitrogen content in TMC-1.

\subsection{Treatment of Velocity Components \label{sec:vel-components}}
As a representative object of the early quiescence stage of star formation, TMC-1 is often modeled under the assumption of physical conditions both temporally constant and spatially isotropic \citep{2011AnA...530A..61H, 2014MNRAS.437..930L, 2016MNRAS.459.3756R}. However, high spectral resolution observations have revealed multiple velocity components with distinct $V_\mathrm{LSR}$. The term ``component" refers to a spatially localized region with distinct chemical and physical conditions. Observations with NASA's Deep Space Network 70 m antenna revealed the multiple-peaked line profiles of \ce{CCS} and \ce{HC7N} with a spectral resolution of $0.008\,\mathrm{km\,s^{-1}}$, fitted with a combination of three Gaussians \citep{1995ApJ...453..293L, 2001ApJ...558..693D}. Observations with the Nobeyama Radio Observatory 45 m telescope resolved four velocity components for \ce{CCS} and \ce{HC3N} with a spectral resolution of $0.0004\,\mathrm{km\,s^{-1}}$ \citep{2018ApJ...864...82D}. The GOTHAM observations further confirmed four components in the line profiles of five cyanopolyynes (\ce{HC_nN}, n=3, 5, 7, 9, and 11) with a spectral resolution of $0.014–0.054\,\mathrm{km\,s^{-1}}$ \citep{2021NatAs...5..188L}. Additionally, the Green Bank Ammonia Survey conducted mapping covering the TMC-1 filament in \ce{HC5N} and decomposed the emission into three components with a spectral resolution of $0.072\,\mathrm{km\,s^{-1}}$ \citep{2023MNRAS.519..285S}.

From these analyses, it is clear that molecular emission in TMC-1 certainly shows multiple velocity components, yet no consensus has been reached on whether TMC-1 contains three or four velocity components. Differences in spectral resolution, beam size, observed transition frequencies, and fitting methodologies contribute to the varying interpretations, probing different spatial scales of structures. Depending on the molecular tracer used, these velocity components have been attributed to different sub-filamentary structures (or so-called `fibres') \citep{1998ApJ...497..842P, 2016A&A...590A..75F}. The formation and hierarchical structure of these sub-filaments, however, remain uncertain. The recent study of \ce{HC5N} by \citet{2023MNRAS.519..285S} proposed that these components trace different layers of the TMC-1 filament, where low-density outer layers are inflowing toward the high-density inner layers. It is also quite likely that the relative abundances between velocity components are not consistent across all molecular species \citep[e.g.,][]{2001ApJ...558..693D}. Thus, some may present with fewer than four components as a result of the observations not being sensitive enough to detect the weak signal in one or more components. In this work, we will not attempt to interpret the differences in physical and chemical properties among these velocity components; instead, we present the statistical results of the molecular column densities for each component.

Nevertheless, for the analysis here using our MCMC model, decisions had to be made regarding the assumed number of velocity components. Previous GOTHAM studies have found that one or more of the four components exhibit(s) no significant detection for propargyl cyanide \citep[\ce{HCCCH2CN},][]{2020ApJ...900L..10M} and E-1-cyano-1,3-butadiene \citep[(E-1-\ce{C4H5CN}, ][]{2023ApJ...948..133C}. In this work, similar results are found for other molecules in our expanded list, such as \ce{HCNO}. \ce{t-HCOOH}, \ce{HNC3}, \ce{NCCNH+}, \ce{HC3O+}, \ce{l-HC4N}, \ce{HC5O}, \ce{C10H-}, \ce{CH3CN}, \ce{HOCO+}, \ce{C2H3C3N}, and \ce{DC7N}. These findings reinforce the idea that TMC-1 exhibits chemical heterogeneity among velocity components. 

The key objectives for a statistical census include maintaining consistent data series and ensuring a sufficient number of variables. Uniform treatment of the number of velocity components ensures comparability across all species and maintains data consistency. However, this approach involves trade-offs between the need for consistency and improving data accuracy. One possible adjustment is to introduce the number of components as an additional variable. To assess this adjustment, we compared the performance of MCMC fitting and the derived column densities when using a three-component versus four-component model for species showing significant detections in only three or fewer components. We found the difference is negligible, as the fourth component contributes minimally in the four-component model. Therefore, for the sake of consistency, we chose to conduct all MCMC analyses using a four-velocity-component model in this inventory study, as the MCMC procedure appropriately nulls some components as needed. 

\subsection{Anomalous Absorption}
The majority of the molecular species detected in GOTHAM exhibit emission features following a Boltzmann distribution of level populations. However, a few outlier molecules exhibit non-Boltzmann population distributions. In particular, we observed anomalous absorption by \ce{H2CO}, $c$-\ce{C3H2}, and $c$-\ce{C3HD} against the cosmic microwave background of 2.73 K toward TMC-1. We observed three \ce{H2CO} absorption lines, $1_{1,0}-1_{1,1}$ at 4829.6 MHz, $2_{1,1}-2_{1,2}$ at 14488.4 MHz, and $3_{1,2}-3_{1,3}$ at 28975.8 MHz.

We observed both absorption and emission features from $c$-\ce{C3H2} and its deuterium isotopologue, $c$-\ce{C3HD}. For $c$-\ce{C3H2}, the $para$ transition $2_{2,0}-2_{1,1}$ at 21587.4 MHz exhibits anomalous absorption, while two $ortho$ transitions, $1_{1,0}-1_{0,1}$ at 18343.1 MHz and $3_{3,0}-3_{2,1}$ at 27084.3 MHz, show in emission. $c$-\ce{C3HD} exhibit similar behavior: the $2_{2,0}-2_{1,1}$ transition at 29593.2 MHz is in absorption, while two transitions, $1_{1,0}-1_{0,1}$ at 19418.7 MHz and $2_{1,1}-2_{1,2}$ at 35600.5 MHz, are in emission. The collision rate coefficients for $ortho$ $c$-\ce{C3H2} transitions are more sensitive to the environmental temperature and are generally higher compared to those for $para$ transitions \citep{2019PCCP...21.1443B}. As such, in cold regions, $ortho$ transitions can be efficiently excited by collisions, turning into emission, whereas $para$ transitions remain in absorption. In addition, theoretical calculations suggest that the $c$-\ce{C3H2} $1_{1,0}-1_{0,1}$ transition might be a maser transition, exhibiting yet another type of anomalous phenomenon \citep{2022MNRAS.514.2116S}.

A detailed non-local thermodynamic equilibrium (non-LTE) radiative transfer model is required to understand the excitation of \ce{H2CO}, $c$-\ce{C3H2} and $c$-\ce{C3HD}. Collisional rate coefficients are available for \ce{H2CO} with \ce{H2} \citep{2013MNRAS.432.2573W}, and for $c$-\ce{C3H2} and $c$-\ce{C3HD} with \ce{He} \citep{2019PCCP...21.1443B}. Such non-LTE modeling will be presented in future work (Ben Khalifa et al. in prep.).

\section{Summary \label{sec:sum}}
The GOTHAM program is a wide-band spectral line survey of TMC-1 CP, covering almost 30~GHz with high sensitivity and spectral resolution. Alongside the QUIJOTE survey, these observations have made TMC-1 the most prolific source of interstellar molecular discoveries. In this work, we present a statistical analysis of the gas-phase molecular inventory of TMC-1, using high-spectral-resolution radio observations and Bayesian MCMC spectral fitting that explores multi-dimensional likelihood distributions. The inventory of molecular species of TMC-1 places crucial constraints on the chemical content of molecular clouds in the early stages of star formation.

We developed a \texttt{Python}-based pipeline for data calibration and reduction, generalized for single-beam position-switched observations. This pipeline introduces an innovative interference identification and mitigation technique using the $\chi^2$ test, enhancing its capabilities and improving the legacy of previous pipelines. Key features include receiver performance assessment for data quality control, automated interference detection and removal, per-channel Doppler tracking correction, and dynamic zenith opacity correction. These advancements ensure automation, accuracy, and reproducibility in the data reduction process.

All GOTHAM observations were calibrated and reduced using this new pipeline to maintain consistency with the DRV data reduction. Using data from GOTHAM DRV, we performed MCMC fitting to over 102 molecular species identified in TMC-1, including 75 main isotopic species, 20 $^{13}$C-substituted, and 7 D-substituted isotopologues. Molecular column densities range from $\sim1 \times 10^{10}$~cm$^{-2}$ to $\sim4 \times 10^{14}$~cm$^{-2}$ with no clear trend between abundance and molecular size. We derived $^{12}$C/$^{13}$C isotopic ratios in the range of $50-223$ and H/D ratios in the range of $7-97$, varying among species.

We revisited the elemental abundances in TMC-1 as probed by gas-phase molecules, including species detected by both GOTHAM and complementary observations. The (non-aromatic) carbon-chain species contribute to 0.11\% of the gas-phase carbon content when including \ce{CO} and 65\% when excluding \ce{CO}. In addition, ten detected aromatic molecules contribute to 0.011\% when including \ce{CO} and 6\% when excluding \ce{CO}. The CN-functionalized aromatic species serve as observational proxies for otherwise undetectable pure aromatics. The elemental carbon content of gas-phase molecules in the prestellar TMC-1 stands in stark contrast to that in hot cores and hot corinos. Hydrocarbons detected in TMC-1 are likely to accrete onto grains, but their products are not observed in hot corinos or hot cores, suggesting that TMC-1 produces hydrocarbon ices that may grow into larger, less volatile aromatic species.

At centimeter and millimeter wavelength, while most of TMC-1's molecules exhibit emission features following a Boltzmann distribution of level populations, the spectra of several outliers cannot be described by a single-excitation-temperature model. Example molecules include \ce{CCS}, \ce{H2CO}, and $c$-\ce{C3H2}, which require a different radiative transfer model to understand their excitation conditions. The line profiles of these molecules differ substantially from most other species, likely due to underlying kinematic variations between velocity components. Modeling these spectral profiles with special treatment of the velocity structure and non-LTE effects is subject to future work.

\section{Data access \& code}
The raw data and the calibrated data products of the GOTHAM observations are publicly available in the GBT Legacy Data Archive\footnote{\footnotesize \url{https://greenbankobservatory.org/portal/gbt/gbt-legacy-archive/gotham-data/}}. The code used in the data calibration and reduction, \texttt{GOTHAM Spectral Pipeline}, is open source and available at \citet{gotham-spectral-pipeline}. The code used in the MCMC fitting analysis is part of the \texttt{molsim} open-source package; an archival version of the code can be accessed at \citet{molsim}. The complete MCMC fitting datasets for each individual molecule, including observational data around relevant transitions, spectroscopic properties, partition functions, and full posterior samples, are publicly available in the Harvard Dataverse Repository \citep{GOTHAMDRV}.

\facilities{GBT}

\software{
    \texttt{GOTHAM Spectral Pipeline} \citep{gotham-spectral-pipeline},
    \texttt{Molsim} \citep{molsim},
    \texttt{Emcee} \citep{foreman-mackey_emcee_2013},
    \texttt{Numba} \citep{2015llvm.confE...1L},
          }

\acknowledgments
B. A. M. and C. X. gratefully acknowledge support of National Science Foundation grant AST-2205126. We thank H. Gupta for providing additional flux calibration data obtained as part of the MEDIUM large program. We thank M. Rodr{\'\i}guez-Baras for providing the observed column densities for the relevant molecules. P. B. C. was supported by the National Science Foundation (AST-2307137) and NIST. S. B. C. was supported by the Goddard Center for Astrobiology and the NASA Planetary Science Division Internal Scientist Funding Program through the Fundamental Laboratory Research work package (FLaRe). The National Radio Astronomy Observatory is a facility of the National Science Foundation operated under cooperative agreement by Associated Universities, Inc. ALMA is a partnership of ESO (representing its member states), NSF (USA) and NINS (Japan), together with NRC (Canada), MOST and ASIAA (Taiwan), and KASI (Republic of Korea), in cooperation with the Republic of Chile. The Joint ALMA Observatory is operated by ESO, AUI/NRAO and NAOJ.   This paper makes use of the following GBT data: AGBT17A-164, AGBT17A-434, AGBT18A-333, AGBT18B-007, AGBT19B-047, AGBT20A-516, AGBT21A-414, 21B-210, and AGBT24A-124.

\bibliographystyle{aasjournal}

\newpage

\appendix
\section{Data Calibration and Reduction \label{apx:reduction}}
\begin{figure*}[ht]
    \centering
    \includegraphics[width=0.7\textwidth]{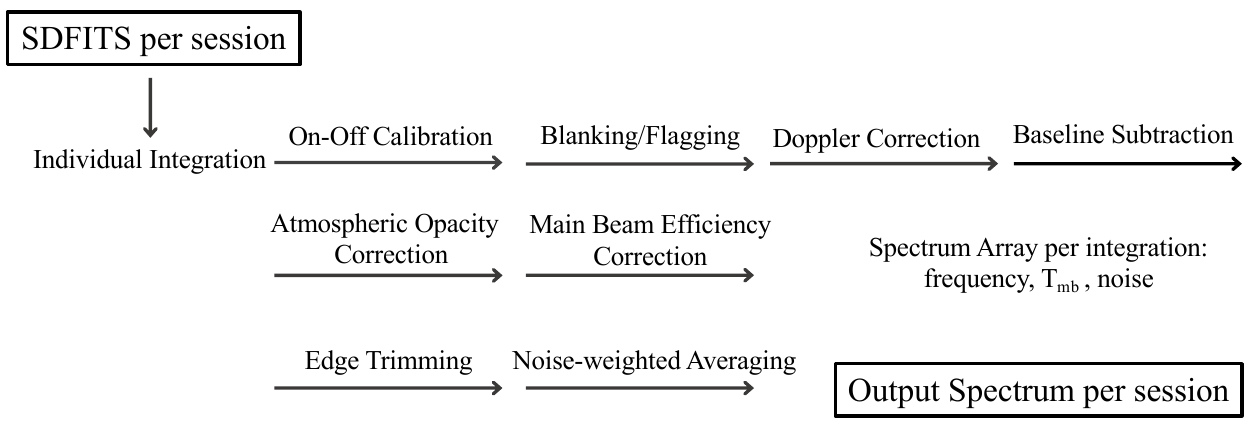}
    \caption{\label{fig:flowchart} Schematic diagram for data flow of the  DRV data calibration and reduction pipeline. }
\end{figure*}

To unlock the full potential of the GOTHAM data, we developed a \texttt{Python}-based pipeline for data calibration and reduction \citep[\texttt{GOTHAM Spectral Pipeline, }][]{gotham-spectral-pipeline}, expanding in capabilities and improving the legacy of our first-generation pipeline based on \texttt{GBTIDL} \citep{2013ascl.soft03019M}. The current version is generalized for single-beam position-switched observations but does not yet support frequency-switched or beam-switched observations. Calibration and reduction were conducted on a per-integration basis, in contrast to the per-scan basis implemented in the previous pipeline. Since each On-Off scan pair consists of four integrations, this approach increases data granularity by a factor of four. This data segmentation enables us to maximize data retention in the presence of abnormal observation conditions, such as when transient RFI occurs or the observing system is unbalanced.

A flowchart providing an overview of the pipeline is shown in Fig~\ref{fig:flowchart}. Each On-Off position switched integration (consisting of four sampling phases, 15 seconds each) was calibrated to the internal noise diodes, flagged for interference, corrected for Doppler Tracking caused by the Earth's rotation, baseline subtracted, and then placed on an intensity scale corrected for atmospheric attenuation and telescope efficiencies. The antenna temperature ($T_A$) was corrected to the main-beam brightness temperature ($T_\mathrm{mb}$) using 
\begin{equation}
    T_{mb} = T_A \times \frac{e^{\tau_0/\sin(\mathrm{el})}}{\eta_B}, \label{eq:T_A*}
\end{equation} where $\tau_0$ is zenith sky opacity (see Appx~\ref{sec:zenith_opacity}), $\mathrm{el}$ is elevation of the observation, and $\eta_B$ is frequency-dependent main-beam efficiency. GBT's $\eta_B$ is approximately $1.37\,\times$ aperture efficiency ($\eta_{ap}$), which is computed following the Ruze equation, 
\begin{equation}
    \eta_{ap} = 0.71\,\exp\left[-\left(\frac{4\pi\,\epsilon\,\nu}{c}\right)^2\right],
\end{equation}
with a surface error $\epsilon$ of 290 microns. During GOTHAM observations, $\eta_B$ varies from 0.94 at 3.9 GHz to 0.77 at 36.4 GHz. All integrations for each observing session were averaged using inverse-variance weighting, and the output data were exported on a per-observing-session basis. A final spectrum was then generated by performing inverse-variance weighted averages across all sessions of the GOTHAM observations. During spectral resampling, nearest neighbor interpolation was used to retain the intrinsic noise distribution, resulting a $\sim96\%$ reduction in noise covariance compared to the previous implementation. The key features of the pipeline are discussed below.

\subsection{Receiver performance assessment \label{sec:T_sys}}
\begin{figure*}[ht]
    \centering
    \includegraphics[width=\textwidth]{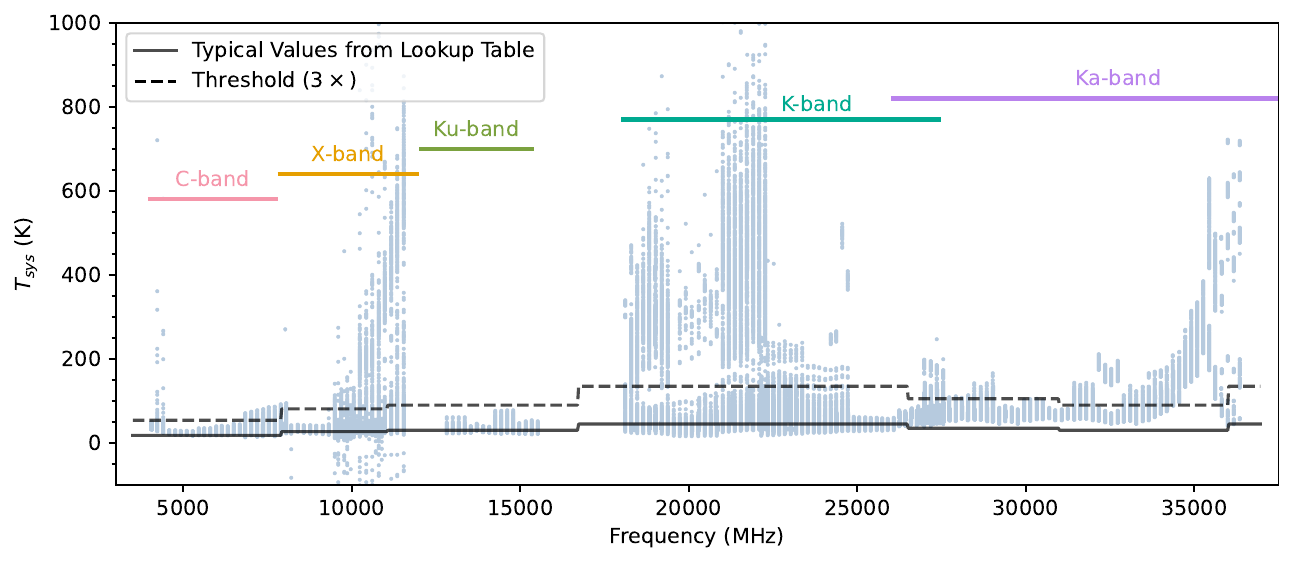}
    \caption{\label{fig:tsys} Statistic of system temperatures of all GOTHAM observations, grouped by spectral windows and polarization. The black solid line represents the typical $T_{\mathrm{sys}}$ reported in the GBT Proposer's Guide. The dashed lines show a threshold of three times the typical values.}
\end{figure*}

The GOTHAM observations cover several frequency bands, including C-, X-, Ku-, K-, and Ka-bands. The typical $T_{\mathrm{sys}}$ for these GBT receivers range from 18 to 45 K, as reported in the GBT Proposer's Guide \footnote{\footnotesize Table 2.2 in \url{https://www.gb.nrao.edu/scienceDocs/GBTog.pdf}}. However, a significant portion of the GOTHAM observations had abnormal $T_{\mathrm{sys}}$ values, such as negative values or values around 1000~K, as depicted in Figure~\ref{fig:tsys}. These abnormal $T_{\mathrm{sys}}$ values indicate that a proper balance and stabilization in the system were not achieved during these observations so the corresponding integrations become unusable. As such, a criterion is needed to systematically eliminate data with abnormal $T_{\mathrm{sys}}$ values. In this reduction, we adapted a threshold of three times the typical values of $T_{\mathrm{sys}}$. Any integration with a negative $T_{\mathrm{sys}}$ or a $T_{\mathrm{sys}}$ higher than this threshold was eliminated. For C-, X-, Ku-, K- and the lower end of Ka-bands, we used the values from the lookup table in the GBT Proposer's Guide (black traces in Figure~\ref{fig:tsys}). However, a consistent increase in $T_{\mathrm{sys}}$ has been observed in most of the recent GBT Ka-band observations (particularly in the 32-40 GHz range, private communication with GBO staff). This increase was also observed in the GOTHAM observations (Figure~\ref{fig:tsys}). Therefore, we exclude the Ka-band data above 32 GHz from the quantitative spectral analyses, although we still include them as part of the data release.

\subsection{Interference identification and flagging \label{sec:interference}}
\begin{figure*}[ht]
    \centering
    \includegraphics[width=\textwidth]{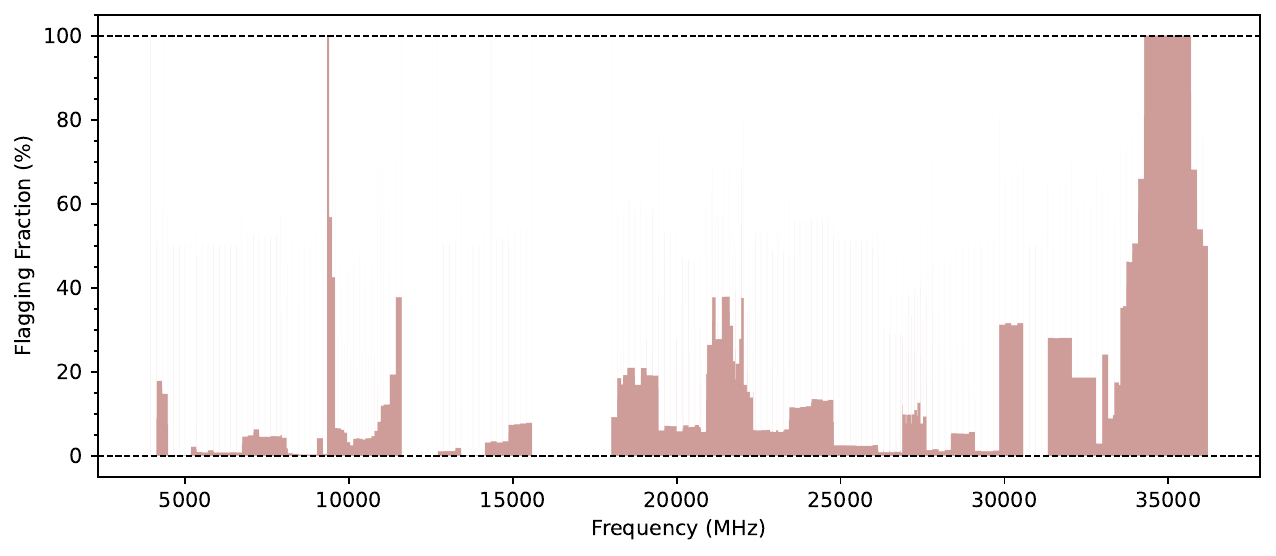}
    \caption{\label{fig:flagging} Fractions of the amount of flagged data per channel of GOTHAM DRV. The reasons for flagging include channels with interference, integrations with abnormal $T_{\mathrm{sys}}$, and spectral window edges.}
\end{figure*}

We developed several techniques to automatically identify and mitigate interference, which are detailed below. These in-house interference mitigation techniques identified more artifacts compared to the previous GOTHAM reductions while achieving a low false-positive rate. Figure~\ref{fig:flagging} shows the flagging rates per channel of the calibrated data of the GOTHAM observations.

\paragraph{Transient interference \label{sec:time-interference}}
The VEGAS spectrometer records raw data as the power spectral densities ($P$) in arbitrary units (i.e., counts). Differences in P between on-source and off-source observations may contain contributions from both astronomical and terrestrial signals. Terrestrial signals include typically stable contributions, such as receiver noise (incorporating reflections, feed resonances, etc.) and atmospheric contributions, which are usually incorporated in the measured system temperatures and are assumed to contribute equally to the on-source and off-source sampling phases. Conversely, transient interference signals may vary on short timescales (e.g., minutes) and thus provide significant variations between the on-source and off-source sampling phases. Any integrations showing significant differences in the average power level between the on-source and off-source sampling phases have therefore been flagged.

\paragraph{Frequency-domain interference \label{sec:freq-interference}}

To identify non-noise channels in a spectrum, we developed a technique based on the $\chi^2$ test. The $\chi^2$ distribution, with $k$ degrees of freedom, describes the probability density distribution of $\chi^2$ values for $k$ independent, normally distributed random samples. This provides a confidence interval within which a $\chi^2$ value for $k$ noise samples is expected to fall, enabling the identification of non-noise outliers. Channels with extreme $\chi^2$ values correspond to low probability. For the benefit of doubt, we compute the minimum $\chi^2$ and compare it with a predefined threshold for significance. This technique is implemented in a way that optimizes performance and can also be generalized easily to detect any signals within a spectral window.

In particular, to identify noise spikes, which are featured with a width of a few channels and heavily deviating in intensity from the adjacent channels, we used $k=7$ degrees of freedom. In our implementation, this technique conceptually fits straight lines to points in each 7-channel spectral chunk. Because of the narrow nature of the noise spikes (with a width of 1--2 channels), the best-fit straight lines typically represent the baseline of these spectral chunks. A $k$-channel chunk has a minimum $\chi^2$ value of
\begin{equation}
    \chi^2_{min} = \sum_{i = 1}^k \frac{ \left[ m\left( \nu_i \right) + c - T_{\mathrm{A},i}^* \right]^2 }{\sigma_i^2}, \label{chi_min}
\end{equation}
where $m$ and $c$ are the coefficients of the linear function minimizing $\chi^2$, $T_{\mathrm{A}}^*$ is the observed antenna temperature, and $\sigma$ is the local noise level. We conducted null hypothesis tests assuming all seven points deviate from the best-fit line with a standard normal distribution and therefore flagged any 7-channel spectral chunks with extreme values of $\chi^2$. To reduce the false-positive rate, we adopted a confidence threshold of 99.9999\% for $k = 7$, corresponding to a $\chi^2$ threshold of 40.5. A single integration of the GOTHAM observations consists of $2^{17}$ channels, meaning that the number of false positive channels is less than 0.15 per integration on average.

\paragraph{Radio Frequency Interference \label{sec:rfi}}
RFI in the X-band has been an increasing problem, such as the external interference from the Garmin aircraft system and the recent TerraSAR-X and TanDEM-X infographic missions. These X-band RFI have been identified in both recent VLA observations\footnote{\footnotesize \url{https://science.nrao.edu/facilities/vla/observing/RFI/feb-2024-c-configuration/X-Band_spectra_202402C}} and the GBT observations of project code 24A-124. RFI can be both time-dependent (i.e., transient) and frequency-dependent. As such, the third technique leverages both the On-Off power level comparison technique and the $\chi^2$-test technique. 

The GBT noise diode calibrator sets the gain factor to calibrate instrumental and atmospheric effects by introducing artificial signals. These signals are stable over short-to-medium time scales so that the difference in the power level per channel between the noise diode turned on and off should be consistent within one integration. As such, any channel showing abnormal values in this difference was identified as RFI and flagged out. 

The classical calibration equation \citep{Braatz_2009,2012A&A...540A.140W} defines the antenna temperature as
\begin{equation}
    T_\mathrm{A} = T_\mathrm{sys} \times (\bar{P}_+^\mathrm{sig} - \bar{P}_+^\mathrm{ref}) / \bar{P}_+^\mathrm{ref}.
\end{equation}
Following this equation, we introduced a quantity that measures the difference in the power level per channel (in Kelvin) between the calibration diode turned on and off as
\begin{equation}
    T_\mathrm{caldiff} = T_\mathrm{sys} \times \frac{\bar{P}_-^\mathrm{sig} - \bar{P}_-^\mathrm{ref}}{\bar{P}_+^\mathrm{ref}}
\end{equation}

where $\bar{P}_\pm^{[~]} = (P_\mathrm{calon}^{[~]} \pm P_\mathrm{caloff}^{[~]}) / 2$, with the superscript $[~]$ denoting either the on-source (sig) or off-source (ref) phase. $P$ is the power level or raw count of each sampling phase. The subscripts calon and caloff denote the calibration diode on and off phase, respectively. For example, $P^\mathrm{sig}_\mathrm{calon}$ is the power level of the on-source phase with calibration diode turned off. $T_\mathrm{caldiff}$ is a frequency-dependent function and its uncertainty is computed using the same radiometer equation used for $T_\mathrm{A}$ (Eq~\ref{eqa:sigma}). To identify the channels with abnormal values in $T_\mathrm{caldiff}$ spectra, we used the $\chi^2$ test described above with a degree of freedom of 127 as the RFI features are relatively broader. We adapted the same confidence threshold of 99.9999\% with Section~\ref{sec:interference}--\textit{Frequency-domain interference}, which corresponds to a $\chi^2$ threshold of 217.6 for $k = 127$.

\subsection{Baseline fitting and subtraction \label{sec:baseline}}
The baseline fitting technique is a hybrid of the polynomial fitting and Lomb-Scargle periodogram techniques. As detailed in \citet{2022GeoRL..4901055C}, the Lomb-Scargle technique identifies periodic spectral components with the highest power determined through the Fourier transformation. We adopt it for the removal of the baseline ripples resulting primarily from temperature-dependent variations in instrument contributions.

Molecular lines were identified and masked using a modified $\chi^2$-test technique. Given the typical line width of tens of channels in GOTHAM observations, $\chi^2$ was computed and assessed with $k = 63$. In practice, strong and weak molecular lines require different treatments when determining the baselines (i.e., $m$ and $c$ in Eq~\ref{chi_min}) for comparison with the $k = 63$ sample points. For strong signals, we determined $m$ and $c$ over a narrow 63-channel spectral chunk, while for weak signals, we determined them over a wider 511-channel chunk. This adjustment is necessary, as more accurate baseline fitting improves the sensitivity required to detect weak signals. A best-fit line determined over a wider chunk results in a higher $\chi^2_{min}$ for the same $k$ points. When applying the same threshold, this increased $\chi^2_{min}$ makes it more sensitive to identifying non-noise channels.

The fitting process to line-free channels was first performed iteratively with up to 20 degrees of polynomial, optimized with the unbounded binary search algorithm. In this pipeline, baselines were determined until the absolute maximum values of the moving average baseline of residual spectra, with a 1025-channel moving window, became less than $10\%$ of the spectral noise level. In the case where polynomial fitting failed with the maximum degree of 20, we subtracted the fitted polynomial function and then proceeded with the Lomb-Scargle fitting, ramping the number of terms up to 40 until the requirement on residual spectra was satisfied. These iterative processes minimized the degree of polynomials and the number of terms of Lomb-Scargle, helping to avoid introducing high-frequency artifacts that mimic real molecular lines.

\subsection{Dynamic zenith opacity correction \label{sec:zenith_opacity}}
\begin{figure*}[ht]
    \centering
    \includegraphics[width=0.48\textwidth]{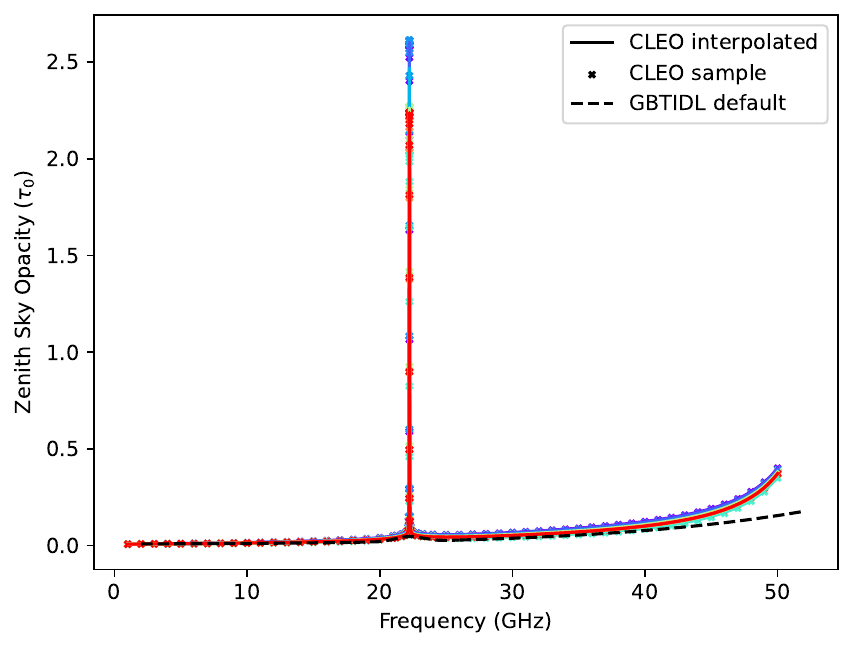}
    \includegraphics[width=0.48\textwidth]{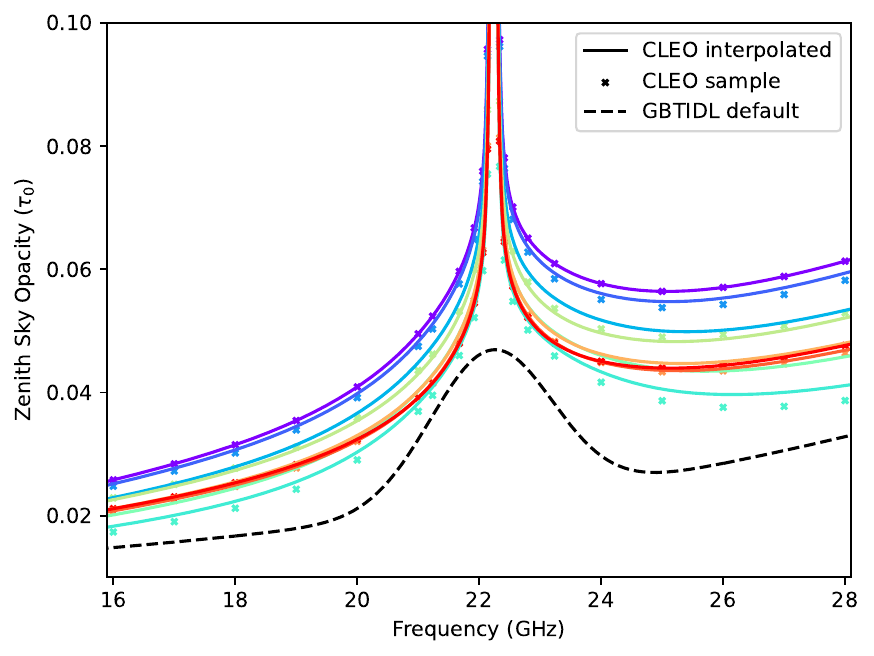}
    \caption{\label{fig:tau} Zenith sky opacity of the default values used in \texttt{GBTIDL} (black dashed line) and the dynamic values retrieved from \texttt{CLEO} associated with the session 08 of AGBT18B-007 as examples (color solid lines). Different colors represent linear interpolation between observation times, while the cubic spline represents interpolation between frequencies. $Left$: frequency range of 0-50 GHz; $Right$: zoom-in frequency range near the 22-GHz water vapor line.}
\end{figure*}

The reduction scheme used in \texttt{GBTIDL} selects representative values of $\tau_0$ from a set of historical data and applies them in the correction of antenna temperature using Eq.~\ref{eq:T_A*}, as default. However, the weather during observations changes in real-time and the actual $\tau_0$ can be significantly different from the default values in \texttt{GBTIDL}. This introduced a non-negligible inaccuracy in the previous GOTHAM data reductions. The effect of the zenith opacity is particularly evident around the 22-GHz water vapor line ($6_{1,6}-5_{2,3}$). Therefore, for this analysis, we retrieved and applied $\tau_0$ values according to the time and frequency of observations, which are more accurate and relevant to our observations. The weather information was obtained from the observing weather forecast tool included in the GBT's \texttt{CLEO} software \citep{2010AAS...21544204M}. We used the CLEO weather archival data forecasted with the ``North American Mesoscale" (NAM) model and averaged among the three sites surrounding the location of the GBT, namely the Elkins, HotSprings and Lewisburg sites. The $\tau_0$ values were retrieved with one-hour time resolution covering the period of GOTHAM observations and interpolated between frequencies and times associated with each GOTHAM observation integration. As shown as an example in Figure~\ref{fig:tau}, the black dashed line represents the default $\tau_0$ value from \texttt{GBTIDL} and the color lines show the retrieved and interpolated $\tau_0$ values associated with some GOTHAM observations. We applied the observation-time-dependent and frequency-dependent $\tau_0$ values to this reduction process.

\section{Laboratory Measurements \label{apx:lab}}
New laboratory measurements were carried out at the Center for Astrophysics, Harvard \& Smithsonian to refine the hyperfine-resolved cm-wave catalogs of \ce{C3N} and c-\ce{C3H} and the $K$-splitting-resolved catalog of \ce{CH3C6H} needed for their robust MCMC analysis in the GOTHAM DRV dataset. The spectroscopic data files 
containing measured frequencies, spectroscopic assignments, and the measurement–prediction difference, are provided in the Harvard Dataverse Repository \citep{GOTHAMDRV}.

\begin{figure*}[b!]
    \centering
    \includegraphics[width=0.75\textwidth]{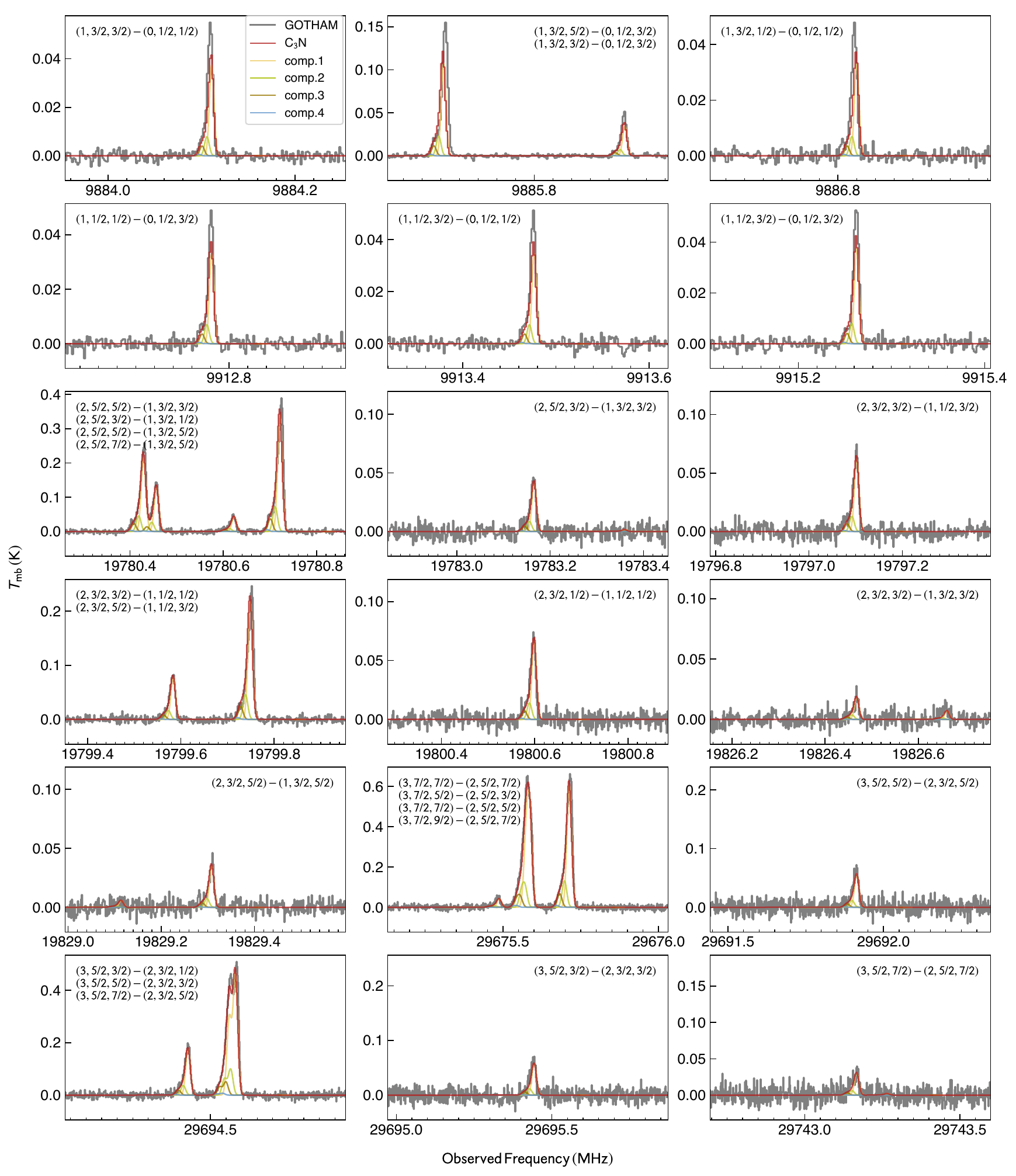}
    \caption{\label{fig:spectra_c3n} Individual line detection of \ce{C3N} transitions in the GOTHAM DRV data. Each panel provides a window context of 7\,km\,s$^{-1}$ and the quantum numbers in a format of $(N, J, F)$ of the covered transitions. The observed spectra are displayed in black while the simulated spectra using the MCMC-derived parameters is overlaid in red. Simulated spectra of the individual velocity components are shown in: yellow (5.64\,km\,s$^{-1}$), green (5.80\,km\,s$^{-1}$), brown (5.96\,km\,s$^{-1}$), and blue (6.07\,km\,s$^{-1}$).}
\end{figure*}

Cyanoacetylene, \ce{HC3N}, was seeded in Ne (0.1\,\%) and injected into the cavity-enhanced FTMW spectrometer in a pulsed supersonic expansion at 5\,Hz with a backing pressure of 2.5\,kTorr. A discharge voltage of 800\,V was applied to the gas pulse to generate \ce{C3N} radicals. We measured 39 individual hyperfine components from 4 rotational transitions between 6 and 40\,GHz, which were fit to a standard Hamiltonian for a linear molecule in a $^2\Sigma^+$ electronic state. Previous mm-wave data ranging from 168 to 198\,GHz were included in the fit~\citep{1983ApJ...275..916G}. The derived spectroscopic constants are listed in Table~\ref{tab:c3n}. These constants contribute to the improved frequency match between the simulated spectra and the GOTHAM observations, as shown in Figure~\ref{fig:spectra_c3n}.

Similarly, acetylene, \ce{C2H2}, was used in a discharge experiment to generate $c$-\ce{C3H} and to measure the hyperfine structure of its $1_{10} - 1_{11}$ transition at 14~GHz. These data were combined with \citet{1992ApJ...399..325L} and \citet{1994JChPh.101.5484Y} to re-derive the spectroscopic parameters of the S-reduced (I$^r$) asymmetric top Hamiltonian listed in Table~\ref{tab:c-c3h}.

The same discharge mixture was used to produce \ce{CH3C6H}. Starting with predictions derived from the original microwave measurements by \citet{Alexander1978:ch3c6h} from 28 to 40\,GHz, we detected 24 new transitions from 6 to 23\,GHz ($J' = 4-15$, $K'_a = 0,1$, where $J'$ and $K'_a$ are the total angular momentum quantum number and $a$-axis projection of the upper state). The high resolution of the cavity FT spectrometer made it possible to resolve the closely spaced doublets ($\sim$30--100\,kHz) of the $K'_a = 0$ and $1$ transitions. The optimized spectroscopic constants from a combined fit of the prior high-frequency and new low-frequency data are summarized in Table~\ref{tab:ch3c6h}. The catalog rest frequencies derived from these data have an uncertainty below 1\,kHz, which is about an order of magnitude more precise than the predictions derived from the high-frequency data alone.

\begin{table}[t!]
    \centering
    \caption{Spectroscopic Constants of \ce{C3N} (in MHz) \label{tab:c3n}}
    \begin{tabular}{ccc}
    \toprule
        Constant & This work$^\ast$ &\citet{1982AnA...109...23G}\\
        \midrule
        $B$ &  4947.620042( 66) & 4947.6190(18) \\
        $D \times 10^3$ &  0.752463(199) & 0.68(12) \\
        $\gamma$ &  -18.74658( 32) & -18.743( 8) \\
        $b$ & -1.19085( 79) & -2.16(10)  \\
        $c$ & 2.86476(230) & 2.85(20)  \\
        $eQq$ & -4.32567(193) & -4.34(25) \\
        $\sigma^\mathrm{norm}_\mathrm{fit}$ & 1.07434 & \\
        \bottomrule
    \end{tabular}\\
    $^\ast$ 1$\sigma$ uncertainties are given in parentheses in units of the last digit.\\
\end{table}

\begin{table}[h!]
    \centering
    \caption{Spectroscopic Constants of c-\ce{C3H} in the S (I$^r$) reduction (in MHz)\label{tab:c-c3h}}
    \begin{tabular}{ccc}
    \toprule
        Constant & This work$^\ast$                & JPL \citep{1998JQSRT..60..883P}/\citet{1994JChPh.101.5484Y}\\
        \midrule
        $A$ &  44528.46(75)                     & 44528.4(99)\\
        $B$ &  34015.4921( 79)                  & 34015.494(17)\\
        $C$ &  19189.4309( 59)                  & 19189.4304(95)\\
        $D_J$ &  -0.01208(186)                  & -0.0119(69)\\
        $D_{JK}$ & 0.9975( 87)                  & 0.997(40)\\
        $D_K$ &  -2.041(109)                    & -2.05(141)\\
        $d_1$ & -0.02232( 89)                   & -0.02240(134)\\
        $d_2$ & -0.03403( 41)                   & -0.0340(37)\\
        $H_J \times 10^3$ &  0.0854(157)        & 0.087(108)\\
        $H_{KJ} \times 10^3$ &  -1.196(171)     & -1.21(172)\\
        $H_K \times 10^3$ &   -35.15(310)       & -35(40)\\
        $h_1 \times 10^3$ &  0.0489(182)        & 0.051(55)\\
        $h_2 \times 10^3$ &  -0.009(23)         & -0.008(40)\\
        $h_3 \times 10^3$ &  -0.0182(135)       & -0.018(39)\\
        $\epsilon_{aa}$ &   113.2839(240)       & 113.2637(253)\\
        $\epsilon_{bb}$ &   59.3849(231)        & 59.3811(239)\\
        $\epsilon_{cc}$ &   -205.7549(246)      & -205.7540(248)\\
        $D^s_N$ & 0.00160(69)                   & 0.00137(79)\\ 
        $D^s_{NK}$ & -0.0269(41)                & -0.0252(45)\\ 
        $D^s_K$ & 0.0229(39)                    & 0.0230(45)\\ 
        $d^s_1 \times 10^3$ & -0.05(44)         & -0.16(55)\\ 
        $d^s_2 \times 10^3$ & -0.84(39)         & -0.84(43)\\
        $a_F$ & -27.23854(209)                  & -27.2459(57)\\ 
        $T_{aa}$ & 16.96933(139)                & 16.980(24)\\ 
        $T_{bb}-T_{cc}$ & 14.90328(504)         & 14.9296(220)\\
        
        $\sigma_\mathrm{fit}$ (norm.) & 0.7 \\
        \bottomrule
    \end{tabular}\\
    $^\ast$ 1$\sigma$ uncertainties are given in parentheses in units of the last digit.
\end{table}

\begin{table}[t!]
    \centering
    \caption{Spectroscopic Constants of \ce{CH3C6H} (in MHz) \label{tab:ch3c6h}}
    \begin{tabular}{ccc}
    \toprule
        Constant & This work$^\ast$ & CDMS/\citet{Alexander1978:ch3c6h}\\
        \midrule
        $A-B$ & [158362.09] & [158362.09]\\
        $B$ &  778.243717(41) & 778.24452(46) \\
        $D_J \times 10^6$ &  8.54(12) & 9.22(47)  \\
        $D_{JK} \times 10^3$ & 4.4391(36) & 4.4417(42) \\ 
        $D_{K}$ & [2.9072] &  [2.9072] \\ 
        $\sigma^\mathrm{norm}_\mathrm{fit}$ & 0.749 & 0.767\\
        \bottomrule
    \end{tabular}\\
    $^\ast$ 1$\sigma$ uncertainties are given in parentheses in units of the last digit.\\
\end{table}

\newpage

\section{MCMC Fitting Detail \label{apx:mcmc}}
\vspace{-1em}
\begin{deluxetable}{cccccccccc}[h!]
    \tablecaption{Priors for the Co-spatial Model in MCMC Analyses \label{tab:priors_cospatial}}
    \tablehead{
    Component &$V_\mathrm{LSR}$\tablenotemark{$\ast$}   &$\theta_s$ &$\mathrm{log_{10}}(N_T)$ &$T_\mathrm{rot}$ &$\Delta V$	\\
    No. &($\mathrm{km\,s^{-1}}$) &($^{\prime\prime}$) &($\mathrm{cm}^{-2}$) &($\mathrm{K}$) &($\mathrm{km\,s^{-1}}$)
    }
    \startdata
    1 &{$N(5.663,0.01)$} &\multirow{4}{*}{$U\{\mathrm{0,500}\}$} &{$U\{\mathrm{11.0,14.0}\}$} &\multirow{4}{*}{$U\{\mathrm{3.0,15.0}\}$\tablenotemark{$\dagger$}} &\multirow{4}{*}{$U\{\mathrm{0.1,0.3}\}$}\\
    2 &{$N(5.817,0.01)$} & &{$U\{\mathrm{11.0,14.0}\}$} & &\\
    3 &{$N(5.935,0.01)$} & &{$U\{\mathrm{11.0,14.0}\}$} & &\\
    4 &{$N(6.065,0.01)$} & &{$U\{\mathrm{11.0,14.0}\}$} & &\\
    \enddata
    \tablenotetext{\ast}{The Gaussian priors for $V_\mathrm{LSR}$ are informed by the posterior distribution of \ce{HC3N} presented in \citet{2021NatAs...5..188L}, with $N(\mu,\sigma)$ denoting a Gaussian distribution with mean, $\mu$, and standard deviation, $\sigma$.}
    \tablenotetext{\dagger}{$U\{min,max\}$ denotes a uniform distribution with minimum, $min$, and maximum, $max$. When only limited transitions were detected for certain species in the GOTHAM observations, Gaussian prior of $N(8.1,2.0)$ for $T_\mathrm{rot}$ was applied, informed by the posterior distribution of \ce{HC3N} \citep{2021NatAs...5..188L}.}
\end{deluxetable}

\begin{deluxetable}{cccccccccc}[h!]
    \tablecaption{Priors for the Separated Model in MCMC Analyses of Linear Molecules \label{tab:priors_separate_nonaromatic}}
    \tablehead{
    Component &$V_\mathrm{LSR}$\tablenotemark{$\ast$}   &$\theta_s$ &$\mathrm{log_{10}}(N_T)$ &$T_\mathrm{rot}$ &$\Delta V$	\\
    No. &($\mathrm{km\,s^{-1}}$) &($^{\prime\prime}$) &($\mathrm{cm}^{-2}$) &($\mathrm{K}$) &($\mathrm{km\,s^{-1}}$)
    }
    \startdata
    1 &{$N(5.624,0.01)$} &{$U\{\mathrm{0,250}\}$} &{$U\{\mathrm{9.5,13.0}\}$} &\multirow{4}{*}{$U\{\mathrm{3.0,15.0}\}$\tablenotemark{$\dagger$}} &\multirow{4}{*}{$U\{\mathrm{0.1,0.3}\}$\tablenotemark{$\dagger$}}\\
    2 &{$N(5.790,0.01)$} &{$U\{\mathrm{0,250}\}$} &{$U\{\mathrm{9.5,13.0}\}$} & &\\
    3 &{$N(5.910,0.01)$} &{$U\{\mathrm{0,250}\}$} &{$U\{\mathrm{9.5,13.0}\}$} & &\\
    4 &{$N(6.033,0.01)$} &{$U\{\mathrm{0,250}\}$} &{$U\{\mathrm{9.5,13.0}\}$} & &\\
    \enddata
    \tablenotetext{\ast}{The Gaussian priors for $V_\mathrm{LSR}$ are informed by the posterior distribution of \ce{HC9N} presented in \citet{2021NatAs...5..188L}, with $N(\mu,\sigma)$ denoting a Gaussian distribution with mean, $\mu$, and standard deviation, $\sigma$.}
    \tablenotetext{\dagger}{$U\{min,max\}$ denotes a uniform distribution with minimum, $min$, and maximum, $max$. When only limited transitions were detected for certain species in the GOTHAM observations, Gaussian priors of $N(6.7,2.0)$ for $T_\mathrm{rot}$ and/or $N(0.117,0.005)$ for $\Delta V$ were applied, informed by the posterior distribution of \ce{HC9N} \citep{2021NatAs...5..188L}.}
\end{deluxetable}

\begin{deluxetable}{cccccccccc}[b!]
    \tablecaption{Priors for the Separated Model in MCMC Analyses of Cyclic Molecules \label{tab:priors_separate_aromatic}}
    \tablehead{
    Component &$V_\mathrm{LSR}$\tablenotemark{$\ast$}   &$\theta_s$ &$\mathrm{log_{10}}(N_T)$ &$T_\mathrm{rot}$ &$\Delta V$	\\
    No. &($\mathrm{km\,s^{-1}}$) &($^{\prime\prime}$) &($\mathrm{cm}^{-2}$) &($\mathrm{K}$) &($\mathrm{km\,s^{-1}}$)
    }
    \startdata
    1 &{$N(5.575,0.01)$} &{$U\{\mathrm{0,250}\}$} &{$U\{\mathrm{9.5,13.0}\}$} &\multirow{4}{*}{$U\{\mathrm{3.0,15.0}\}$\tablenotemark{$\dagger$}} &\multirow{4}{*}{$U\{\mathrm{0.1,0.3}\}$\tablenotemark{$\dagger$}}\\
    2 &{$N(5.767,0.01)$} &{$U\{\mathrm{0,250}\}$} &{$U\{\mathrm{9.5,13.0}\}$} & &\\
    3 &{$N(5.892,0.01)$} &{$U\{\mathrm{0,250}\}$} &{$U\{\mathrm{9.5,13.0}\}$} & &\\
    4 &{$N(6.018,0.01)$} &{$U\{\mathrm{0,250}\}$} &{$U\{\mathrm{9.5,13.0}\}$} & &\\
    \enddata
    \tablenotetext{\ast}{The Gaussian priors for $V_\mathrm{LSR}$ are informed by the posterior distribution of benzonitrile presented in \citet{2021Sci...371.1265M}, with $N(\mu,\sigma)$ denoting a Gaussian distribution with mean, $\mu$, and standard deviation, $\sigma$.}
    \tablenotetext{\dagger}{$U\{min,max\}$ denotes a uniform distribution with minimum, $min$, and maximum, $max$. When only limited transitions were detected for certain species in the GOTHAM observations, Gaussian priors of $N(8.9,2.0)$ for $T_\mathrm{rot}$ and/or $N(0.125,0.005)$ for $\Delta V$ were applied, informed by the posterior distribution of benzonitrile \citep{2021Sci...371.1265M}.}
\end{deluxetable}

\include{figure_set}

We applied two models in the MCMC analyses, the co-spatial model and the separated model. The adopted prior probability distributions are summarized in Table~\ref{tab:priors_cospatial}, Table~\ref{tab:priors_separate_nonaromatic}, and Table~\ref{tab:priors_separate_aromatic}. Except for source velocities for the four components, we initiated fitting with uniformly distributed priors on parameter values with only physical constraints (i.e., positive values). The priors for $V_\mathrm{LSR}$ were determined using three well-characterized molecules, \ce{HC3N}, \ce{HC9N}, and $c$-\ce{C6H5CN}. In the co-spatial case, we found that molecules appeared to share $V_\mathrm{LSR}$ similar to \ce{HC3N}, so we adopted the priors informed by \citet{2021NatAs...5..188L}. In the separated case, linear species share $V_\mathrm{LSR}$ similar to \ce{HC9N} and cyclic species to $c$-\ce{C6H5CN}, so we adopted the priors informed by \citet{2021NatAs...5..188L} and \citet{2021Sci...371.1265M}, respectively.

For species with no individual line detections, the observation data were insufficient to constrain $\Delta V$. In these cases, the prior for $\Delta V$ was adjusted to a Gaussian distribution centered at the median value of the corresponding reference molecule with a standard deviation of $0.05\,\mathrm{km\,s^{-1}}$. In addition, for species with a limited energy range of detected transitions, the prior for $T_\mathrm{rot}$ was adjusted to a Gaussian distribution centered at the median value of the reference molecule with a standard deviation of 2 K. These treatments are specified with markers in Tables~\ref{tab:main-mols}, \ref{tab:13C-mols}, and \ref{tab:D-mols}.

\begin{figure*}[b!]
    \centering
    \includegraphics[width=\textwidth]{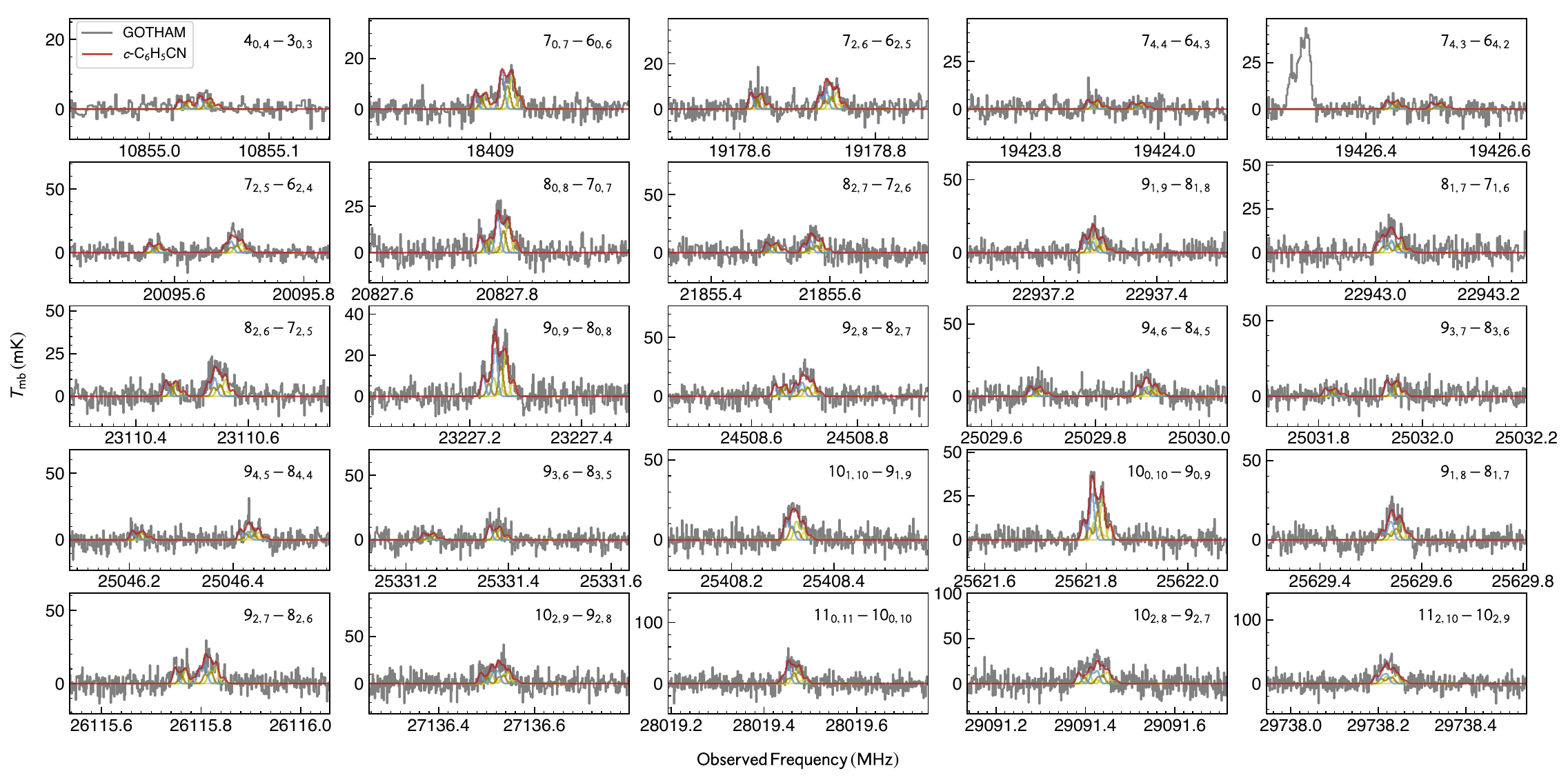}
    \caption{\label{fig:spectra_c-c6h5cn} Individual line detection of $c$-\ce{C6H5CN} transitions in the GOTHAM DRV data. Each panel provides a window context of 4\,km\,s$^{-1}$ and the quantum numbers in a format of $J(K_a, K_c)$ of the covered transitions. The observed spectra are displayed in black while the simulated spectra using the MCMC-derived parameters is overlaid in red. Simulated spectra of the individual velocity components are shown in: yellow (5.61\,km\,s$^{-1}$), green (5.78\,km\,s$^{-1}$), brown (5.90\,km\,s$^{-1}$), and blue (6.03\,km\,s$^{-1}$).}
\end{figure*}

Figure~\ref{fig:corner_c-c6h5cn} presents an example corner plot of the marginalized posterior distribution and parameter covariances for $c$-\ce{C6H5CN}. The complete figure set for 102 molecular species is available online. The 2D posterior distribution is shown in yellow for the co-spatial model and in brown for the separated model. We adopt the 50$^{th}$ percentile value as the representative value of each parameter to simulate the $c$-\ce{C6H5CN} spectra, as shown in Figure~\ref{fig:spectra_c-c6h5cn}.

\end{document}

%% file: tables/mols_main_mcmc_table.tex
NH$_{3}$ & Ammonia & $2.99^{+0.08}_{-0.07} \times 10^{14}$ & $9.1^{+0.1}_{-0.1}$ & $0.283^{+0.006}_{-0.007}$ & 17.3 & Co-spatial & \posteriorsubfignum \\ 
$c$-C$_{3}$H & 2-cyclopropyn-1-yl & $1.20^{+0.11}_{-0.05} \times 10^{13}$ & $4.6^{+0.3}_{-0.4}$ & $0.191^{+0.009}_{-0.008}$ & 40.7 & Co-spatial\tablenotemark{$^\Vert$} & \posteriorsubfignum \\ 
$l$-C$_{3}$H$_{2}$ & Propadienylidene & $1.41^{+0.30}_{-0.21} \times 10^{12}$ & $6.5^{+0.9}_{-1.0}$ & $0.121^{+0.002}_{-0.002}$ & 23.1 & Separated\tablenotemark{$^\mathsection$}\tablenotemark{$^\Vert$} & \posteriorsubfignum \\ 
HCCN & Cyanomethylene & $7.1^{+1.3}_{-0.9} \times 10^{11}$ & $6.4^{+1.0}_{-1.0}$ & $0.118^{+0.003}_{-0.003}$ & 220.9 & Separated\tablenotemark{$^\mathsection$}\tablenotemark{$^\Vert$} & \posteriorsubfignum \\ 
H$_{2}$CCN & Cyanomethyl & $1.93^{+0.24}_{-0.19} \times 10^{13}$ & $8.2^{+1.4}_{-1.4}$ & $0.158^{+0.004}_{-0.003}$ & 349.2 & Co-spatial & \posteriorsubfignum \\ 
CH$_{3}$CN & Acetonitrile & $5.2^{+1.1}_{-0.6} \times 10^{12}$ & $6.9^{+0.9}_{-0.9}$ & $0.286^{+0.010}_{-0.010}$ & 127.2 & Separated\tablenotemark{$^\Vert$} & \posteriorsubfignum \\ 
HCCO & Ketenyl & $6.6^{+2.5}_{-1.5} \times 10^{11}$ & $6.7^{+1.0}_{-1.0}$ & $0.117^{+0.002}_{-0.003}$ & 53.2 & Separated\tablenotemark{$^\mathsection$}\tablenotemark{$^\Vert$} & \posteriorsubfignum \\ 
CH$_{3}$NC & Isocyanomethane & $2.7^{+0.8}_{-0.5} \times 10^{11}$ & $6.7^{+1.0}_{-1.0}$ & $0.159^{+0.018}_{-0.016}$ & 71.8 & Separated\tablenotemark{$^\mathsection$}\tablenotemark{$^\Vert$} & \posteriorsubfignum \\ 
HCNO & Fulminic acid & $1.27^{+0.56}_{-0.22} \times 10^{11}$ & $6.7^{+1.0}_{-1.0}$ & $0.118^{+0.002}_{-0.002}$ & 37.5 & Separated\tablenotemark{$^\mathsection$}\tablenotemark{$^\Vert$} & \posteriorsubfignum \\ 
HOCN & Cyanic acid & $2.1^{+0.8}_{-0.4} \times 10^{11}$ & $6.7^{+1.0}_{-1.0}$ & $0.118^{+0.002}_{-0.002}$ & 41.3 & Separated\tablenotemark{$^\mathsection$}\tablenotemark{$^\Vert$} & \posteriorsubfignum \\ 
HNCO & Isocyanic acid & $1.05^{+0.61}_{-0.13} \times 10^{13}$ & $6.5^{+1.1}_{-1.2}$ & $0.120^{+0.002}_{-0.002}$ & 37.0 & Separated\tablenotemark{$^\mathsection$}\tablenotemark{$^\Vert$} & \posteriorsubfignum \\ 
HOCO$^{+}$ & Oxohydroxymethylium & $3.1^{+1.2}_{-0.5} \times 10^{11}$ & $6.7^{+1.0}_{-1.0}$ & $0.118^{+0.003}_{-0.002}$ & 13.4 & Separated\tablenotemark{$^\mathsection$}\tablenotemark{$^\Vert$} & \posteriorsubfignum \\ 
$t$-HCOOH & Trans-formic acid & $7.3^{+2.7}_{-1.6} \times 10^{11}$ & $6.7^{+1.0}_{-1.0}$ & $0.118^{+0.002}_{-0.003}$ & 30.2 & Separated\tablenotemark{$^\mathsection$}\tablenotemark{$^\Vert$} & \posteriorsubfignum \\ 
C$_{4}$H$^{-}$ & Butadiynyl anion & $2.0^{+0.9}_{-0.3} \times 10^{10}$ & $6.7^{+1.0}_{-1.0}$ & $0.118^{+0.003}_{-0.002}$ & 30.3 & Separated\tablenotemark{$^\mathsection$}\tablenotemark{$^\Vert$} & \posteriorsubfignum \\ 
C$_{4}$H & Butadiynyl & $1.18^{+0.03}_{-0.03} \times 10^{14}$ & $9.1^{+0.7}_{-0.6}$ & $0.162^{+0.002}_{-0.002}$ & 160.5 & Co-spatial & \posteriorsubfignum \\ 
C$_{3}$N & Cyanoethynyl & $1.16^{+0.04}_{-0.05} \times 10^{13}$ & $10.7^{+0.8}_{-0.8}$ & $0.178^{+0.005}_{-0.004}$ & 271.2 & Co-spatial\tablenotemark{$^\Vert$} & \posteriorsubfignum \\ 
HC$_{3}$N & Prop-2-ynenitrile & $1.60^{+0.02}_{-0.02} \times 10^{14}$ & $7.5^{+0.2}_{-0.2}$ & $0.138^{+0.002}_{-0.002}$ & 103.8 & Co-spatial & \posteriorsubfignum \\ 
HNC$_{3}$ & Hydrogen isocyanide & $6.7^{+0.7}_{-0.3} \times 10^{11}$ & $5.0^{+1.0}_{-0.7}$ & $0.231^{+0.009}_{-0.008}$ & 68.3 & Co-spatial & \posteriorsubfignum \\ 
HCCNC & Isocyanoethyne & $5.2^{+1.4}_{-0.6} \times 10^{12}$ & $10.8^{+2.6}_{-2.4}$ & $0.145^{+0.007}_{-0.006}$ & 137.4 & Separated & \posteriorsubfignum \\ 
C$_{3}$O & 3-oxopropadienylidene & $1.43^{+0.21}_{-0.15} \times 10^{12}$ & $10.0^{+3.1}_{-2.8}$ & $0.161^{+0.008}_{-0.008}$ & 43.7 & Separated & \posteriorsubfignum \\ 
HC$_{3}$NH$^{+}$ & Prop-2-ynenitrilium & $8.6^{+1.6}_{-0.8} \times 10^{11}$ & $6.7^{+0.9}_{-1.0}$ & $0.120^{+0.002}_{-0.002}$ & 98.4 & Separated\tablenotemark{$^\mathsection$}\tablenotemark{$^\Vert$} & \posteriorsubfignum \\ 
C$_{2}$H$_{3}$CN & Vinyl cyanide & $6.3^{+0.6}_{-0.4} \times 10^{12}$ & $4.4^{+0.1}_{-0.1}$ & $0.147^{+0.004}_{-0.004}$ & 139.6 & Separated & \posteriorsubfignum \\ 
NCCNH$^{+}$ & Oxalonitrilium & $6.8^{+2.6}_{-1.0} \times 10^{10}$ & $6.7^{+1.0}_{-1.0}$ & $0.18^{+0.06}_{-0.04}$ & 285.4 & Separated\tablenotemark{$^\Vert$} & \posteriorsubfignum \\ 
HC$_{3}$O$^{+}$ & 3-oxopropadienium & $1.67^{+0.40}_{-0.18} \times 10^{11}$ & $6.7^{+1.0}_{-1.0}$ & $0.117^{+0.002}_{-0.002}$ & 31.6 & Separated\tablenotemark{$^\mathsection$}\tablenotemark{$^\Vert$} & \posteriorsubfignum \\ 
$c$-H$_{2}$C$_{3}$O & Cyclopropenone & $5.2^{+2.3}_{-0.8} \times 10^{11}$ & $4.3^{+1.4}_{-0.8}$ & $0.163^{+0.019}_{-0.016}$ & 76.8 & Separated & \posteriorsubfignum \\ 
HCCCHO & Propynal & $9^{+4}_{-3} \times 10^{12}$ & $3.1^{+0.1}_{-0.1}$ & $0.175^{+0.013}_{-0.012}$ & 24.2 & Separated & \posteriorsubfignum \\ 
trans-C$_{2}$H$_{3}$CHO & Trans-prop-2-enal & $2.2^{+1.1}_{-0.3} \times 10^{11}$ & $5.7^{+1.1}_{-1.1}$ & $0.118^{+0.003}_{-0.002}$ & 75.7 & Separated\tablenotemark{$^\mathsection$}\tablenotemark{$^\Vert$} & \posteriorsubfignum \\ 
HCCS & Thioketenyl & $1.1^{+0.9}_{-0.3} \times 10^{12}$ & $6.7^{+1.0}_{-1.0}$ & $0.117^{+0.002}_{-0.002}$ & 104.7 & Separated\tablenotemark{$^\mathsection$}\tablenotemark{$^\Vert$} & \posteriorsubfignum \\ 
HCCS$^{+}$ & Thioxoethenylium & $1.30^{+0.61}_{-0.24} \times 10^{12}$ & $3.7^{+0.5}_{-0.5}$ & $0.118^{+0.003}_{-0.003}$ & 34.1 & Separated\tablenotemark{$^\mathsection$}\tablenotemark{$^\Vert$} & \posteriorsubfignum \\ 
C$_{5}$H & Pentadiynylidyne & $2.0^{+3.5}_{-0.6} \times 10^{12}$ & $6.9^{+0.9}_{-0.8}$ & $0.141^{+0.006}_{-0.005}$ & 242.5 & Separated & \posteriorsubfignum \\ 
$l$-HC$_{4}$N & Isocyanopropadiene-1,3-diyl & $1.55^{+0.77}_{-0.25} \times 10^{11}$ & $6.5^{+1.0}_{-1.0}$ & $0.118^{+0.003}_{-0.002}$ & 980.5 & Separated\tablenotemark{$^\mathsection$}\tablenotemark{$^\Vert$} & \posteriorsubfignum \\ 
H$_{2}$C$_{3}$HCCH & 1,2-pentadiyn-4-ene & $1.40^{+0.74}_{-0.25} \times 10^{13}$ & $9.2^{+2.0}_{-1.6}$ & $0.157^{+0.023}_{-0.019}$ & 370.9 & Separated & \posteriorsubfignum \\ 
\multirow{2}{*}{CH$_{3}$C$_{4}$H} & \multirow{2}{*}{Penta-1,3-diyne} & \multirow{2}{*}{$1.71^{+0.20}_{-0.12} \times 10^{13}$} & $9.3^{+2.2}_{-2.0}$ ($A$) & $0.125^{+0.004}_{-0.004}$ ($A$) & 559.5 & \multirow{2}{*}{Separated} & \posteriorsubfignum \\ 
                                  &                                  &                     & $8.6^{+0.4}_{-0.4}$ ($E$) & $0.119^{+0.004}_{-0.004}$ ($E$) & 504.5 &     & \posteriorsubfignum \\ 
H$_{2}$CCCHCN & Cyanoallene & $1.94^{+0.43}_{-0.23} \times 10^{12}$ & $4.8^{+0.2}_{-0.2}$ & $0.137^{+0.002}_{-0.002}$ & 403.2 & Separated\tablenotemark{$^\mathsection$} & \posteriorsubfignum \\ 
HCCCH$_{2}$CN & Propargyl cyanide & $1.12^{+0.16}_{-0.09} \times 10^{12}$ & $6.1^{+1.0}_{-0.8}$ & $0.143^{+0.024}_{-0.017}$ & 632.0 & Separated & \posteriorsubfignum \\ 
\multirow{2}{*}{CH$_{3}$C$_{3}$N} & \multirow{2}{*}{But-2-ynenitrile} & \multirow{2}{*}{$1.41^{+0.25}_{-0.11} \times 10^{12}$} & $6.0^{+1.2}_{-1.0}$ ($A$) & $0.158^{+0.010}_{-0.007}$ ($A$) & 937.7 & \multirow{2}{*}{Separated} & \posteriorsubfignum \\ 
                                  & & & $7.6^{+0.5}_{-0.5}$ ($E$) & $0.153^{+0.007}_{-0.006}$ ($E$) & 1273.8 & & \posteriorsubfignum \\ 
C$_{3}$S & Thioxopropadieneylidene & $9.0^{+0.4}_{-0.3} \times 10^{12}$ & $6.0^{+0.4}_{-0.3}$ & $0.152^{+0.004}_{-0.004}$ & 43.7 & Co-spatial & \posteriorsubfignum \\ 
H$_{2}$C$_{3}$S & Propadienethione & $3.9^{+1.0}_{-0.5} \times 10^{11}$ & $10.4^{+0.7}_{-0.7}$ & $0.118^{+0.002}_{-0.002}$ & 217.4 & Separated\tablenotemark{$^\mathsection$}\tablenotemark{$^\Vert$} & \posteriorsubfignum \\ 
$c$-H$_{2}$C$_{3}$S & Cyclopropenthione & $4.9^{+2.4}_{-1.4} \times 10^{10}$ & $8.9^{+1.0}_{-1.0}$ & $0.125^{+0.002}_{-0.003}$ & 412.6 & Separated\tablenotemark{$^\mathsection$}\tablenotemark{$^\Vert$} & \posteriorsubfignum \\ 
C$_{6}$H$^{-}$ & Hexatriynyl anion & $1.21^{+0.26}_{-0.09} \times 10^{11}$ & $7.0^{+0.9}_{-0.8}$ & $0.144^{+0.007}_{-0.007}$ & 106.7 & Separated & \posteriorsubfignum \\ 
C$_{6}$H & Hexatriynyl & $5.6^{+0.3}_{-0.3} \times 10^{12}$ & $5.4^{+0.1}_{-0.1}$ & $0.131^{+0.002}_{-0.002}$ & 338.8 & Separated & \posteriorsubfignum \\ 
C$_{5}$N & Cyanobutadiynyl & $3.4^{+0.9}_{-0.3} \times 10^{11}$ & $12.0^{+2.1}_{-2.7}$ & $0.119^{+0.002}_{-0.002}$ & 1071.2 & Separated\tablenotemark{$^\mathsection$} & \posteriorsubfignum \\ 
$l$-C$_{6}$H$_{2}$ & Hexapentaenylidene & $1.54^{+0.07}_{-0.06} \times 10^{11}$ & $13.3^{+0.5}_{-0.5}$ & $0.126^{+0.008}_{-0.006}$ & 622.7 & Separated\tablenotemark{$^\Vert$} & \posteriorsubfignum \\ 
C$_{5}$N$^{-}$ & Cyanobutadiynyl anion & $7.6^{+2.2}_{-0.8} \times 10^{10}$ & $8.7^{+3.0}_{-2.1}$ & $0.123^{+0.029}_{-0.014}$ & 392.9 & Separated & \posteriorsubfignum \\ 
HC$_{4}$NC & Isocyanobutadiyne & $1.18^{+0.31}_{-0.14} \times 10^{11}$ & $8.5^{+3.7}_{-2.7}$ & $0.117^{+0.002}_{-0.002}$ & 378.6 & Separated\tablenotemark{$^\mathsection$} & \posteriorsubfignum \\ 
HC$_{5}$N & Penta-2,4-diynenitrile & $4.42^{+0.04}_{-0.05} \times 10^{13}$ & $8.6^{+0.2}_{-0.2}$ & $0.127^{+0.002}_{-0.001}$ & 406.8 & Co-spatial & \posteriorsubfignum \\ 
C$_{2}$H$_{3}$C$_{3}$N & Pent-4-en-2-ynenitrile & $8.3^{+8.3}_{-2.5} \times 10^{10}$ & $5.6^{+3.7}_{-1.7}$ & $0.117^{+0.002}_{-0.002}$ & 785.6 & Separated\tablenotemark{$^\mathsection$} & \posteriorsubfignum \\ 
HC$_{5}$O & Butadiynylformyl & $1.37^{+0.11}_{-0.07} \times 10^{12}$ & $8.0^{+0.9}_{-0.8}$ & $0.156^{+0.009}_{-0.008}$ & 542.0 & Separated & \posteriorsubfignum \\ 
NC$_{4}$NH$^{+}$ & 4-iminobut-2-ynenitrile & $1.1^{+1.7}_{-0.3} \times 10^{10}$ & $6.1^{+4.6}_{-2.1}$ & $0.117^{+0.003}_{-0.002}$ & 881.0 & Separated\tablenotemark{$^\mathsection$} & \posteriorsubfignum \\ 
E-1-C$_{4}$H$_{5}$CN & E-1-cyano-1,3-butadiene & $4.0^{+2.6}_{-1.3} \times 10^{10}$ & $6.9^{+1.0}_{-1.0}$ & $0.117^{+0.003}_{-0.002}$ & 1289.7 & Separated\tablenotemark{$^\mathsection$}\tablenotemark{$^\Vert$} & \posteriorsubfignum \\ 
\multirow{2}{*}{CH$_{3}$C$_{6}$H} & \multirow{2}{*}{Hepta-1,3,5-triyne} & \multirow{2}{*}{$1.83^{+0.64}_{-0.26} \times 10^{12}$} & $11.2^{+2.3}_{-2.3}$ ($A$) & $0.118^{+0.002}_{-0.002}$ ($A$) & 1904.0 & \multirow{2}{*}{Separated\tablenotemark{$^\mathsection$}} & \posteriorsubfignum \\ 
                                  & & & $10.4^{+1.6}_{-1.4}$ ($E$) & $0.118^{+0.002}_{-0.002}$ ($E$) & 1716.0 & & \posteriorsubfignum \\ 
\multirow{2}{*}{CH$_{3}$C$_{5}$N} & \multirow{2}{*}{Hexa-2,4-diynenitrile} & \multirow{2}{*}{$3.0^{+1.4}_{-0.3} \times 10^{11}$} & $9.4^{+1.4}_{-1.2}$ ($A$) & $0.126^{+0.019}_{-0.014}$ ($A$) & 4482.5 & \multirow{2}{*}{Separated} & \posteriorsubfignum \\ 
                                  & & & $7.6^{+0.7}_{-0.7}$ ($E$) & $0.138^{+0.021}_{-0.012}$ ($E$) & 3377.7 & & \posteriorsubfignum \\ 
$c$-1-C$_{5}$H$_{5}$CN & 1-cyanocyclopentadiene & $3.9^{+1.2}_{-0.7} \times 10^{11}$ & $9.8^{+3.1}_{-2.5}$ & $0.125^{+0.002}_{-0.002}$ & 3098.3 & Separated\tablenotemark{$^\mathsection$} & \posteriorsubfignum \\ 
$c$-2-C$_{5}$H$_{5}$CN & 2-cyanocyclopentadiene & $1.4^{+1.2}_{-0.4} \times 10^{11}$ & $7.3^{+4.4}_{-2.7}$ & $0.125^{+0.002}_{-0.002}$ & 2030.5 & Separated\tablenotemark{$^\mathsection$} & \posteriorsubfignum \\ 
C$_{5}$S & Thioxopentatetraenylidene & $4.4^{+4.0}_{-1.6} \times 10^{10}$ & $4.6^{+1.8}_{-0.9}$ & $0.17^{+0.04}_{-0.03}$ & 104.8 & Separated & \posteriorsubfignum \\ 
C$_{8}$H & Octatetraynyl & $3.78^{+0.36}_{-0.26} \times 10^{11}$ & $7.3^{+0.4}_{-0.3}$ & $0.142^{+0.006}_{-0.005}$ & 1073.5 & Separated & \posteriorsubfignum \\ 
C$_{8}$H$^{-}$ & Octatetraynyl anion & $2.2^{+1.7}_{-0.5} \times 10^{10}$ & $8.0^{+1.4}_{-1.2}$ & $0.112^{+0.010}_{-0.007}$ & 285.2 & Separated & \posteriorsubfignum \\ 
C$_{7}$N$^{-}$ & Cyanohexatriynyl anion & $1.1^{+1.1}_{-0.3} \times 10^{10}$ & $10.9^{+2.7}_{-2.9}$ & $0.117^{+0.002}_{-0.003}$ & 389.2 & Separated\tablenotemark{$^\mathsection$} & \posteriorsubfignum \\ 
HC$_{7}$N & Hepta-2,4,6-triynenitrile & $1.33^{+0.02}_{-0.01} \times 10^{13}$ & $10.5^{+0.1}_{-0.2}$ & $0.119^{+0.007}_{-0.001}$ & 1164.8 & Co-spatial & \posteriorsubfignum \\ 
HC$_{7}$NH$^{+}$ & Hepta-2,4,6-triynenitrilium & $5.9^{+0.7}_{-0.4} \times 10^{10}$ & $11.2^{+1.5}_{-1.3}$ & $0.120^{+0.002}_{-0.002}$ & 423.0 & Separated\tablenotemark{$^\mathsection$} & \posteriorsubfignum \\ 
HC$_{7}$O & Hexatriynylformyl & $5.6^{+1.2}_{-0.6} \times 10^{11}$ & $9.3^{+1.8}_{-1.4}$ & $0.143^{+0.028}_{-0.019}$ & 1505.0 & Separated & \posteriorsubfignum \\ 
$c$-C$_{6}$H$_{5}$CN &  & $1.65^{+0.09}_{-0.06} \times 10^{12}$ & $10.0^{+0.7}_{-0.6}$ & $0.121^{+0.005}_{-0.004}$ & 19530.6 & Separated & \posteriorsubfignum \\ 
\multirow{2}{*}{CH$_{3}$C$_{7}$N} & \multirow{2}{*}{Octa-2,4,6-triynenitrile} & \multirow{2}{*}{$5.8^{+5.6}_{-1.5} \times 10^{10}$} & $8.8^{+2.4}_{-1.7}$ ($A$) & $0.117^{+0.002}_{-0.002}$ ($A$) & 8455.4 & \multirow{2}{*}{Separated\tablenotemark{$^\mathsection$}} & \posteriorsubfignum \\ 
                                  & & & $9.0^{+3.1}_{-2.3}$ ($E$) & $0.117^{+0.002}_{-0.003}$ ($E$) & 8772.2 & & \posteriorsubfignum \\ 
C$_{9}$H$_{8}$ & Indene & $1.14^{+0.37}_{-0.22} \times 10^{13}$ & $6.2^{+3.1}_{-1.7}$ & $0.125^{+0.002}_{-0.003}$ & 999.5 & Separated\tablenotemark{$^\mathsection$} & \posteriorsubfignum \\ 
C$_{10}$H$^{-}$ & Deca-1,3,5,7,9-pentayne & $1.8^{+2.1}_{-0.6} \times 10^{10}$ & $6.6^{+0.9}_{-0.8}$ & $0.118^{+0.002}_{-0.002}$ & 458.6 & Separated\tablenotemark{$^\mathsection$} & \posteriorsubfignum \\ 
HC$_{9}$N & Nona-2,4,6,8-tetraynenitrile & $4.32^{+0.39}_{-0.28} \times 10^{12}$ & $8.7^{+0.1}_{-0.1}$ & $0.118^{+0.002}_{-0.002}$ & 1870.3 & Separated & \posteriorsubfignum \\ 
2-C$_{9}$H$_{7}$CN & 2-cyanoindene & $1.5^{+0.8}_{-0.3} \times 10^{11}$ & $8.4^{+3.3}_{-2.3}$ & $0.125^{+0.002}_{-0.002}$ & 10078.5 & Separated\tablenotemark{$^\mathsection$} & \posteriorsubfignum \\ 
HC$_{11}$N & Undeca-2,4,6,8,10-pentaynenitrile & $1.0^{+1.0}_{-0.3} \times 10^{11}$ & $7.2^{+1.5}_{-1.4}$ & $0.117^{+0.002}_{-0.002}$ & 886.7 & Separated\tablenotemark{$^\mathsection$} & \posteriorsubfignum \\ 
1-C$_{10}$H$_{7}$CN & 1-cyanonaphthalene & $8.0^{+8.1}_{-1.6} \times 10^{11}$ & $9.2^{+1.0}_{-0.9}$ & $0.133^{+0.018}_{-0.014}$ & 15370.6 & Separated & \posteriorsubfignum \\ 
2-C$_{10}$H$_{7}$CN & 2-cyanonaphthalene & $5.3^{+1.4}_{-0.5} \times 10^{11}$ & $9.4^{+0.9}_{-0.8}$ & $0.121^{+0.014}_{-0.010}$ & 15963.0 & Separated & \posteriorsubfignum \\ 
1-C$_{12}$H$_{7}$CN & 1-cyanoacenaphthylene & $1.09^{+0.40}_{-0.18} \times 10^{12}$ & $7.9^{+0.6}_{-0.5}$ & $0.132^{+0.002}_{-0.002}$ & 18666.2 & Separated\tablenotemark{$^\mathsection$} & \posteriorsubfignum \\ 
5-C$_{12}$H$_{7}$CN & 5-cyanoacenaphthylene & $7.7^{+3.8}_{-1.2} \times 10^{11}$ & $9.0^{+0.9}_{-0.8}$ & $0.131^{+0.002}_{-0.002}$ & 21689.3 & Separated\tablenotemark{$^\mathsection$} & \posteriorsubfignum \\ 
1-C$_{16}$H$_{9}$CN & 1-cyanopyrene & $9.1^{+1.9}_{-0.8} \times 10^{11}$ & $8.5^{+0.5}_{-0.4}$ & $0.126^{+0.002}_{-0.002}$ & 43049.5 & Separated\tablenotemark{$^\mathsection$} & \posteriorsubfignum \\ 
2-C$_{16}$H$_{9}$CN & 2-cyanopyrene & $5.7^{+1.8}_{-0.8} \times 10^{11}$ & $8.3^{+0.7}_{-0.7}$ & $0.125^{+0.002}_{-0.002}$ & 690081.3 & Separated\tablenotemark{$^\mathsection$} & \posteriorsubfignum \\ 
4-C$_{16}$H$_{9}$CN & 4-cyanopyrene & $9.2^{+3.5}_{-1.4} \times 10^{11}$ & $8.4^{+0.7}_{-0.6}$ & $0.126^{+0.002}_{-0.002}$ & 42836.3 & Separated\tablenotemark{$^\mathsection$} & \posteriorsubfignum \\ 

%% file: tables/mols_13C_mcmc_table.tex
$^{13}$CCCCH & $6.2^{+1.7}_{-0.9} \times 10^{11}$ & $6.6^{+1.0}_{-1.0}$ & $0.190^{+0.031}_{-0.028}$ & 241.3 & Separated\tablenotemark{$^\Vert$} & \posteriorsubfignum \\ 
C$^{13}$CCCH & $1.11^{+0.23}_{-0.12} \times 10^{12}$ & $6.8^{+1.0}_{-1.0}$ & $0.195^{+0.025}_{-0.021}$ & 241.6 & Separated\tablenotemark{$^\Vert$} & \posteriorsubfignum \\ 
CC$^{13}$CCH & $1.50^{+0.46}_{-0.18} \times 10^{12}$ & $6.8^{+1.0}_{-1.0}$ & $0.181^{+0.018}_{-0.015}$ & 241.7 & Separated\tablenotemark{$^\Vert$} & \posteriorsubfignum \\ 
CCC$^{13}$CH & $1.18^{+0.45}_{-0.26} \times 10^{12}$ & $6.8^{+1.0}_{-1.0}$ & $0.180^{+0.019}_{-0.018}$ & 248.2 & Separated\tablenotemark{$^\Vert$} & \posteriorsubfignum \\ 
H$^{13}$CCCN & $2.75^{+0.52}_{-0.26} \times 10^{12}$ & $4.9^{+0.6}_{-0.6}$ & $0.163^{+0.007}_{-0.007}$ & 69.7 & Separated\tablenotemark{$^\Vert$} & \posteriorsubfignum \\ 
HC$^{13}$CCN & $2.39^{+0.69}_{-0.20} \times 10^{12}$ & $5.6^{+1.4}_{-0.9}$ & $0.147^{+0.005}_{-0.005}$ & 77.6 & Separated & \posteriorsubfignum \\ 
HCC$^{13}$CN & $3.41^{+0.39}_{-0.22} \times 10^{12}$ & $4.8^{+0.7}_{-0.5}$ & $0.142^{+0.004}_{-0.003}$ & 67.3 & Separated & \posteriorsubfignum \\ 
C$^{13}$CS & $2.3^{+0.7}_{-0.4} \times 10^{12}$ & $3.43^{+0.27}_{-0.21}$ & $0.158^{+0.012}_{-0.011}$ & 29.8 & Separated\tablenotemark{$^\Vert$} & \posteriorsubfignum \\ 
H$^{13}$CC$_{4}$N & $6.35^{+0.30}_{-0.24} \times 10^{11}$ & $11.3^{+1.7}_{-1.4}$ & $0.135^{+0.010}_{-0.007}$ & 547.4 & Separated & \posteriorsubfignum \\ 
HC$^{13}$CC$_{3}$N & $6.07^{+0.20}_{-0.20} \times 10^{11}$ & $12.9^{+1.3}_{-1.5}$ & $0.132^{+0.009}_{-0.007}$ & 613.6 & Separated & \posteriorsubfignum \\ 
HC$_{2}$$^{13}$CC$_{2}$N & $6.18^{+0.20}_{-0.18} \times 10^{11}$ & $11.3^{+1.2}_{-1.1}$ & $0.128^{+0.008}_{-0.006}$ & 530.3 & Separated & \posteriorsubfignum \\ 
HC$_{3}$$^{13}$CCN & $6.54^{+0.32}_{-0.24} \times 10^{11}$ & $10.7^{+1.3}_{-1.1}$ & $0.135^{+0.008}_{-0.007}$ & 504.9 & Separated & \posteriorsubfignum \\ 
HC$_{4}$$^{13}$CN & $7.00^{+0.28}_{-0.24} \times 10^{11}$ & $10.8^{+1.4}_{-1.1}$ & $0.142^{+0.009}_{-0.007}$ & 512.9 & Separated & \posteriorsubfignum \\ 
H$^{13}$CC$_{6}$N & $2.3^{+2.7}_{-0.7} \times 10^{11}$ & $8.8^{+0.9}_{-0.8}$ & $0.120^{+0.002}_{-0.002}$ & 993.7 & Separated\tablenotemark{$\dagger$} & \posteriorsubfignum \\ 
HC$^{13}$CC$_{5}$N & $2.8^{+3.1}_{-1.0} \times 10^{11}$ & $7.2^{+0.8}_{-0.8}$ & $0.119^{+0.002}_{-0.002}$ & 806.1 & Separated\tablenotemark{$\dagger$} & \posteriorsubfignum \\ 
HC$_{2}$$^{13}$CC$_{4}$N & $3.1^{+3.5}_{-1.3} \times 10^{11}$ & $8.4^{+0.8}_{-0.7}$ & $0.120^{+0.002}_{-0.002}$ & 939.0 & Separated\tablenotemark{$\dagger$} & \posteriorsubfignum \\ 
HC$_{3}$$^{13}$CC$_{3}$N & $1.49^{+0.11}_{-0.07} \times 10^{11}$ & $10.6^{+1.0}_{-0.9}$ & $0.123^{+0.014}_{-0.011}$ & 1180.3 & Separated & \posteriorsubfignum \\ 
HC$_{4}$$^{13}$CC$_{2}$N & $1.45^{+0.15}_{-0.08} \times 10^{11}$ & $11.6^{+1.2}_{-1.1}$ & $0.129^{+0.012}_{-0.010}$ & 1285.9 & Separated & \posteriorsubfignum \\ 
HC$_{5}$$^{13}$CCN & $1.72^{+0.29}_{-0.15} \times 10^{11}$ & $8.1^{+0.7}_{-0.7}$ & $0.119^{+0.012}_{-0.009}$ & 905.3 & Separated & \posteriorsubfignum \\ 
HC$_{6}$$^{13}$CN & $2.2^{+0.7}_{-0.4} \times 10^{11}$ & $7.9^{+0.9}_{-0.8}$ & $0.134^{+0.014}_{-0.011}$ & 889.0 & Separated & \posteriorsubfignum \\ 

%% file: tables/mols_D_mcmc_table.tex
HDCS & $6.2^{+1.7}_{-0.6} \times 10^{12}$ & $5.7^{+0.8}_{-0.8}$ & $0.141^{+0.013}_{-0.011}$ & 11.2 & Separated\tablenotemark{$^\Vert$} & \posteriorsubfignum \\ 
C$_{4}$D & $1.33^{+0.17}_{-0.10} \times 10^{12}$ & $6.7^{+1.0}_{-1.0}$ & $0.159^{+0.028}_{-0.020}$ & 190.4 & Separated\tablenotemark{$^\Vert$} & \posteriorsubfignum \\ 
DC$_{3}$N & $3.83^{+0.16}_{-0.12} \times 10^{12}$ & $6.9^{+0.9}_{-0.9}$ & $0.145^{+0.008}_{-0.006}$ & 310.8 & Separated\tablenotemark{$^\Vert$} & \posteriorsubfignum \\ 
CH$_{2}$DC$_{4}$H & $4.2^{+1.8}_{-1.0} \times 10^{11}$ & $6.9^{+0.9}_{-1.0}$ & $0.117^{+0.003}_{-0.003}$ & 144.0 & Separated\tablenotemark{$^\mathsection$}\tablenotemark{$^\Vert$} & \posteriorsubfignum \\ 
CH$_{2}$DC$_{3}$N & $1.32^{+0.86}_{-0.24} \times 10^{11}$ & $7.0^{+1.5}_{-1.2}$ & $0.118^{+0.003}_{-0.003}$ & 432.3 & Separated\tablenotemark{$^\mathsection$} & \posteriorsubfignum \\ 
DC$_{5}$N & $8.8^{+1.5}_{-0.6} \times 10^{11}$ & $10.8^{+2.0}_{-1.7}$ & $0.158^{+0.010}_{-0.009}$ & 534.2 & Separated & \posteriorsubfignum \\ 
DC$_{7}$N & $1.64^{+0.58}_{-0.19} \times 10^{11}$ & $10.7^{+1.4}_{-1.3}$ & $0.157^{+0.015}_{-0.011}$ & 3673.6 & Separated & \posteriorsubfignum \\ 

%% file: tables/mols_literature.tex
CH &$1.98\pm0.17 \times 10^{14}$ &\citet{2012AnA...546A.103S} \\
CCH & $7.24\pm0.67 \times 10^{13}$ &\citet{1997ApJ...486..862P}\\
CN & $7.76\pm1.43 \times 10^{12}$ &\citet{1997ApJ...486..862P}\\
HCN &$9.64\pm0.11 \times 10^{13}$ &\citet{2021AnA...648A.120R}\\
HNC &$2.97\pm0.26 \times 10^{13}$ &\citet{2021AnA...648A.120R}\\
DCN &$2.3\pm0.5 \times 10^{12}$ &\citet{2021AnA...653A..15N}\\
DNC &$2.0\pm0.5 \times 10^{12}$ &\citet{2021AnA...653A..15N}\\
CO &$1.77 \times 10^{18}$\tablenotemark{$\ast$} &\citet{2019AnA...624A.105F}\\
$^{13}$CO &$2.94 \times 10^{16}$\tablenotemark{$\ast$} &\citet{2019AnA...624A.105F}\\
HCO$^{+}$ &$1.82 \times 10^{14}$\tablenotemark{$\ast$} &\citet{2019AnA...624A.105F}\\
N$_{2}$H$^{+}$ &$1.40 \times 10^{13}$\tablenotemark{$\ast$} &\citet{2019AnA...624A.105F}\\
C$^{18}$O &$2.94 \times 10^{15}$\tablenotemark{$\ast$} &\citet{2019AnA...624A.105F}\\
CH$_{3}$OH &$2.0^{+0.6}_{-0.3} \times 10^{13}$ &\citet{2022AnA...657A..10S}\\
H$_{2}$S &$4.1\pm1.2 \times 10^{13}$\tablenotemark{$\dagger$} &\citet{2020AnA...637A..39N}\\
CS &$2.9\pm1.8 \times 10^{14}$ &\citet{2021AnA...653A..15N}\\
$^{13}$CS &$3.9\pm1.0 \times 10^{12}$ &\citet{2020AnA...637A..39N}\\
HCS$^{+}$ &$2.55 \times 10^{12}$\tablenotemark{$\ast$} &\citet{2019AnA...624A.105F}\\
SO &$3.28 \times 10^{13}$\tablenotemark{$\ast$} &\citet{2019AnA...624A.105F}\\
OCS &$4.29\pm0.78 \times 10^{12}$ &\citet{2021AnA...648A.120R}\\


%% file: figure_set.tex
\figsetstart
\figsetnum{8}
\figsettitle{Parameter covariances and marginalized posterior distributions for the MCMC fit in GOTHAM DR}

\figsetgrpstart
\figsetgrpnum{8.1}
\figsetgrptitle{Corner plot for NH$_{3}$.}
\figsetplot{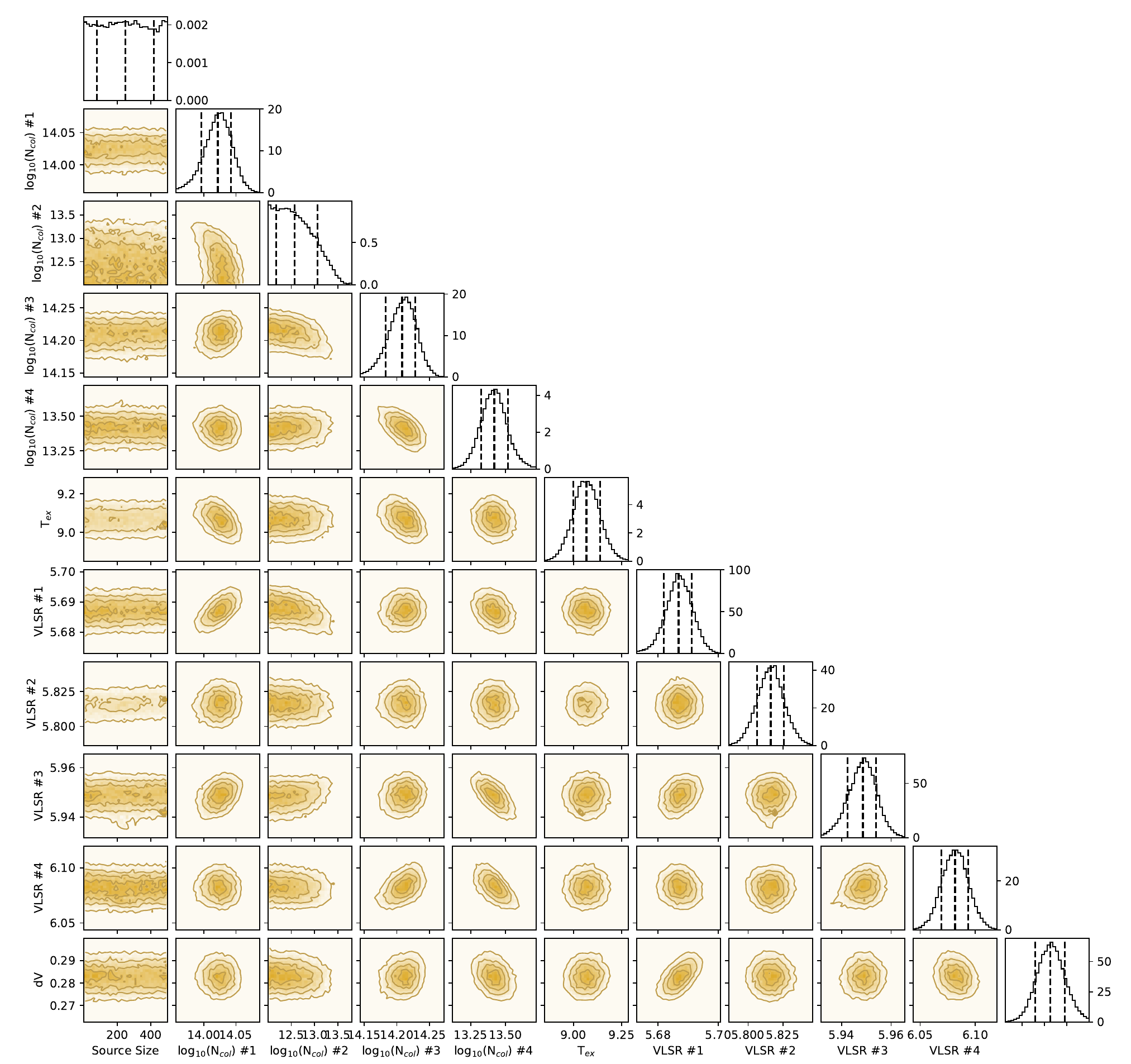}
\figsetgrpnote{The 16$^{th}$, 50$^{th}$, and 84$^{th}$ confidence intervals (corresponding to $\pm$1 sigma for a Gaussian posterior distribution) are shown as vertical lines. The contour lines are posterior probability levels, starting at $20\%$ of the maximum a posteriori estimate, with evenly spaced intervals of $20\%$ up to the peak density.}
\figsetgrpend

\figsetgrpstart
\figsetgrpnum{8.2}
\figsetgrptitle{Corner plot for $c$-C$_{3}$H.}
\figsetplot{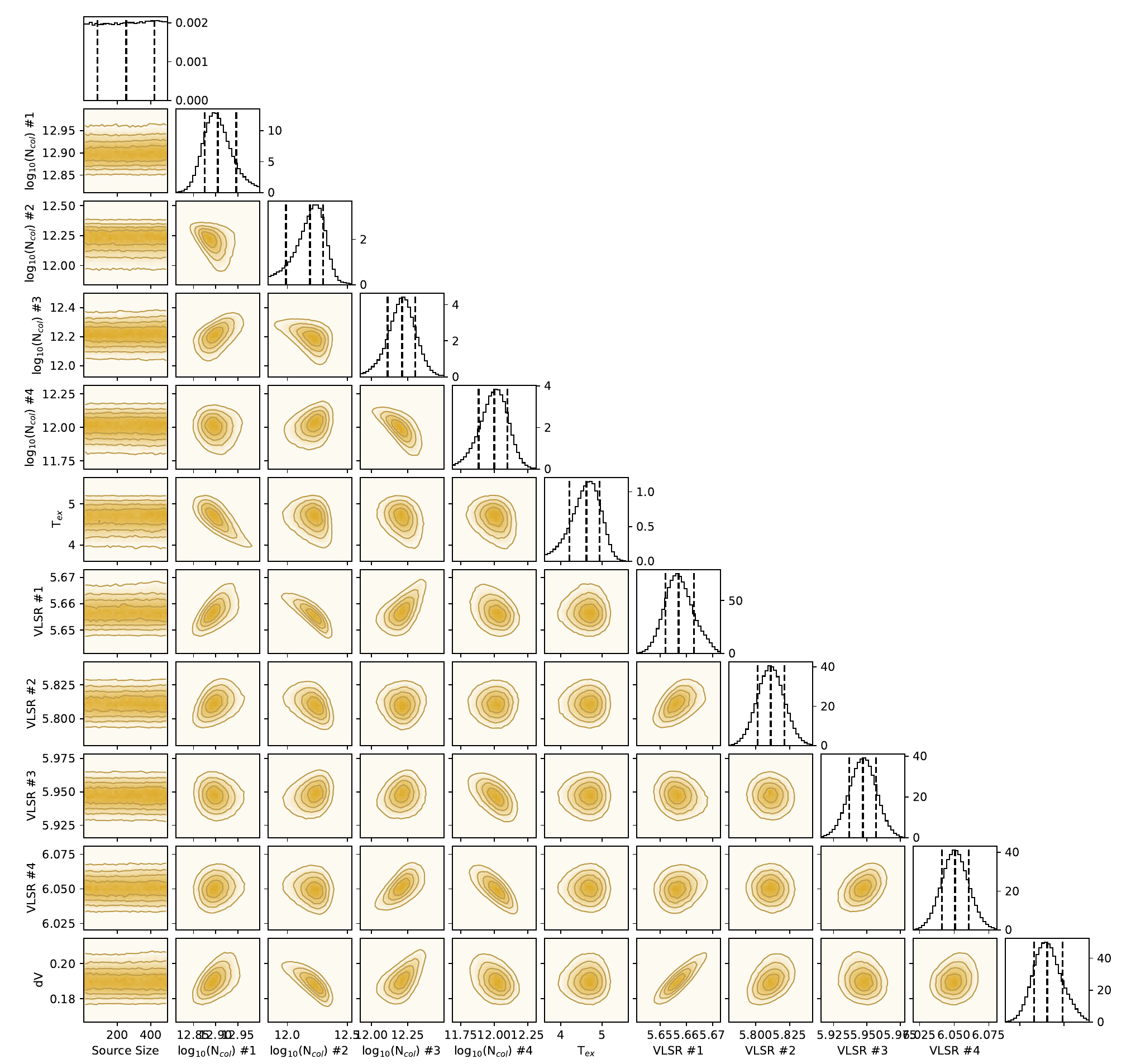}
\figsetgrpnote{The 16$^{th}$, 50$^{th}$, and 84$^{th}$ confidence intervals (corresponding to $\pm$1 sigma for a Gaussian posterior distribution) are shown as vertical lines. The contour lines are posterior probability levels, starting at $20\%$ of the maximum a posteriori estimate, with evenly spaced intervals of $20\%$ up to the peak density.}
\figsetgrpend

\figsetgrpstart
\figsetgrpnum{8.3}
\figsetgrptitle{Corner plot for $l$-C$_{3}$H$_{2}$.}
\figsetplot{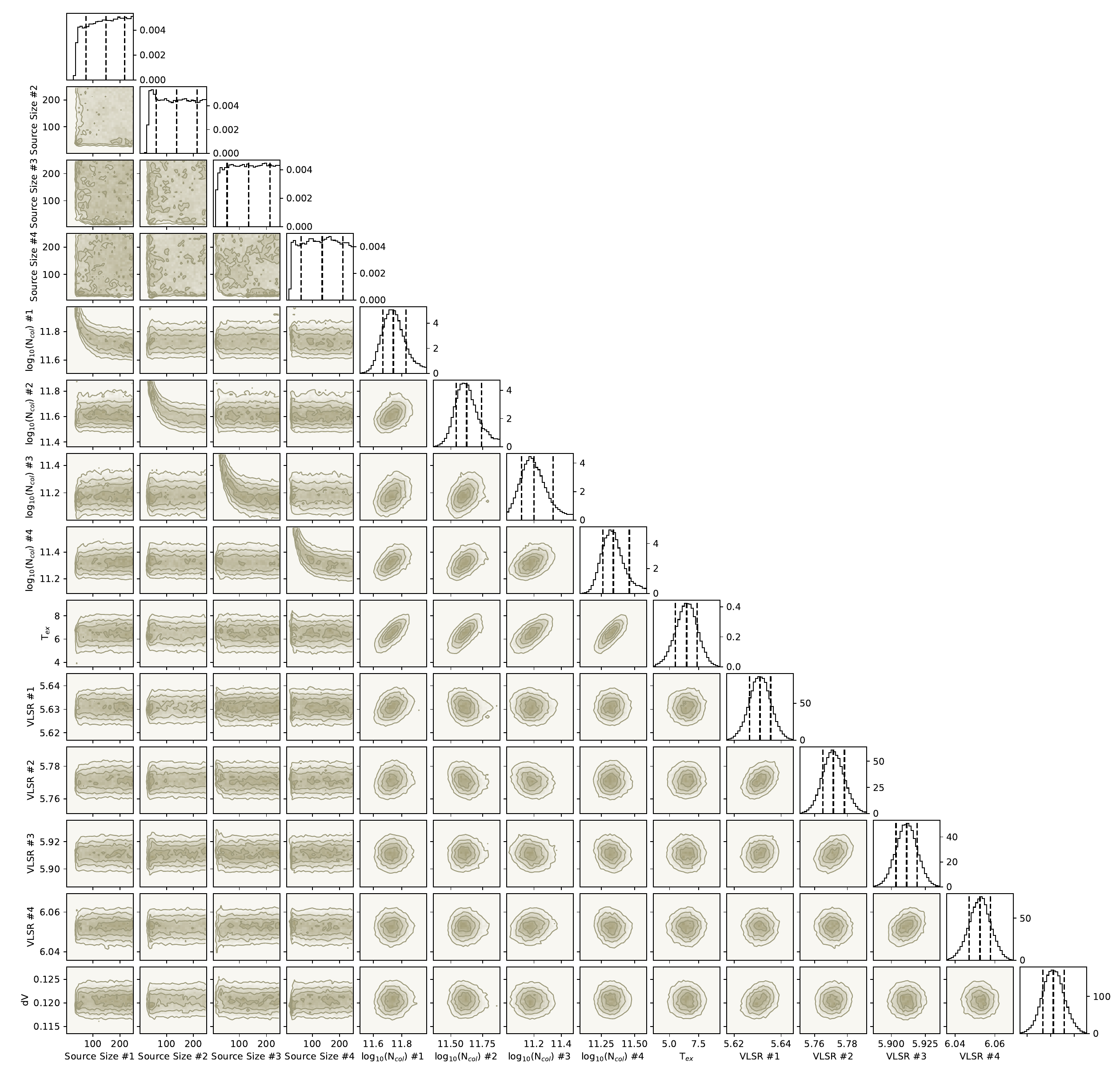}
\figsetgrpnote{The 16$^{th}$, 50$^{th}$, and 84$^{th}$ confidence intervals (corresponding to $\pm$1 sigma for a Gaussian posterior distribution) are shown as vertical lines. The contour lines are posterior probability levels, starting at $20\%$ of the maximum a posteriori estimate, with evenly spaced intervals of $20\%$ up to the peak density.}
\figsetgrpend

\figsetgrpstart
\figsetgrpnum{8.4}
\figsetgrptitle{Corner plot for HCCN.}
\figsetplot{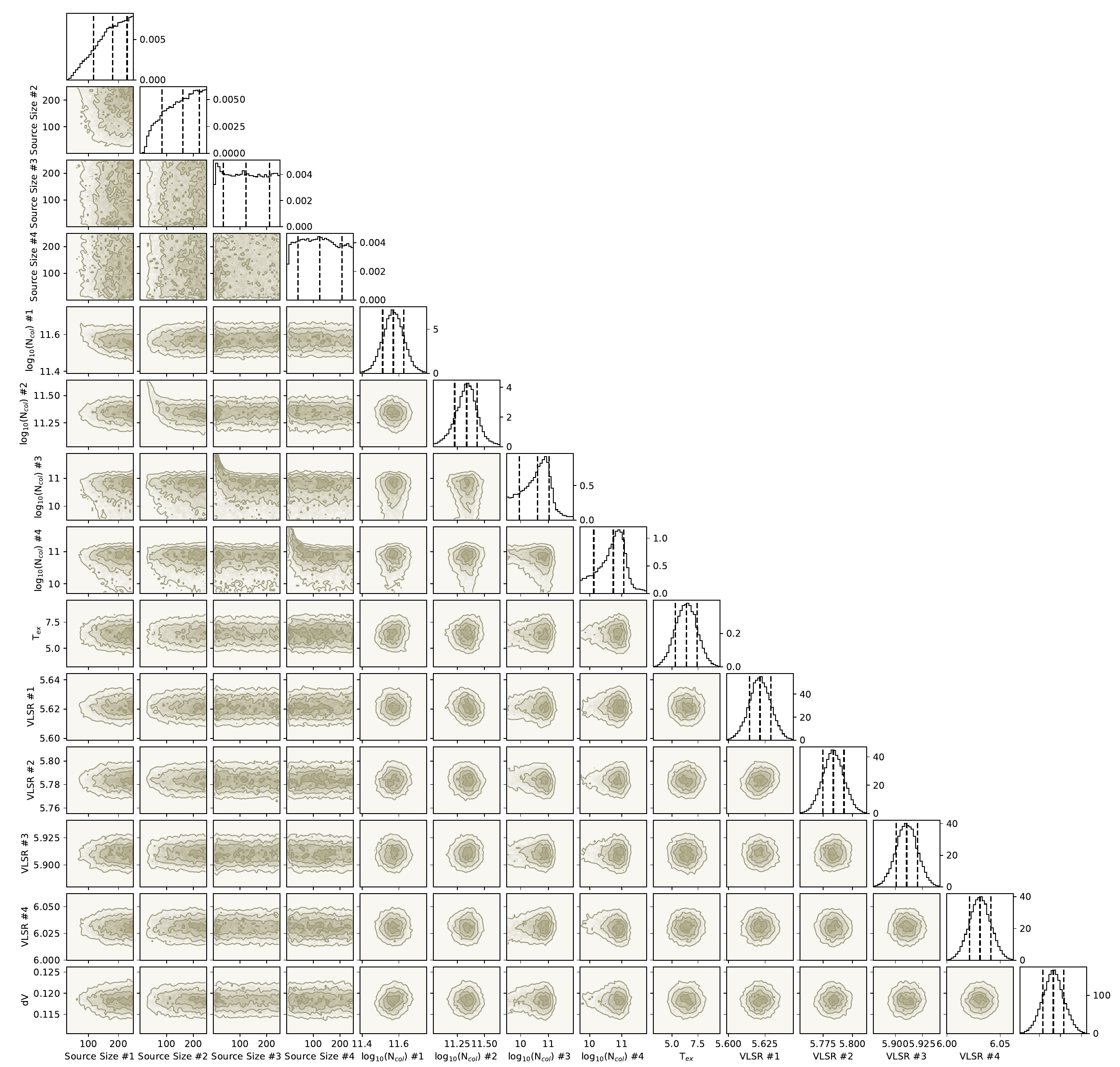}
\figsetgrpnote{The 16$^{th}$, 50$^{th}$, and 84$^{th}$ confidence intervals (corresponding to $\pm$1 sigma for a Gaussian posterior distribution) are shown as vertical lines. The contour lines are posterior probability levels, starting at $20\%$ of the maximum a posteriori estimate, with evenly spaced intervals of $20\%$ up to the peak density.}
\figsetgrpend

\figsetgrpstart
\figsetgrpnum{8.5}
\figsetgrptitle{Corner plot for H$_{2}$CCN.}
\figsetplot{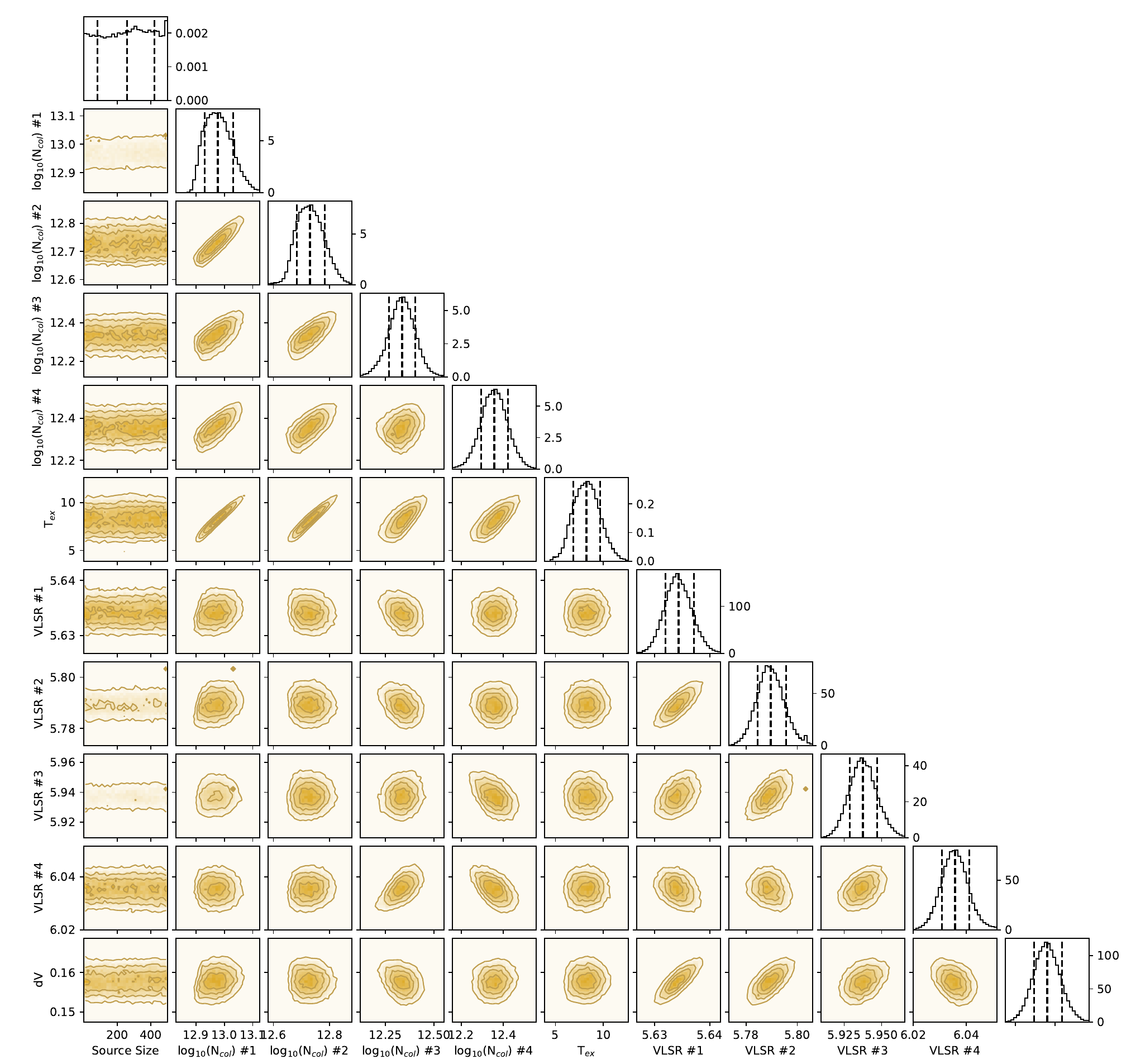}
\figsetgrpnote{The 16$^{th}$, 50$^{th}$, and 84$^{th}$ confidence intervals (corresponding to $\pm$1 sigma for a Gaussian posterior distribution) are shown as vertical lines. The contour lines are posterior probability levels, starting at $20\%$ of the maximum a posteriori estimate, with evenly spaced intervals of $20\%$ up to the peak density.}
\figsetgrpend

\figsetgrpstart
\figsetgrpnum{8.6}
\figsetgrptitle{Corner plot for CH$_{3}$CN.}
\figsetplot{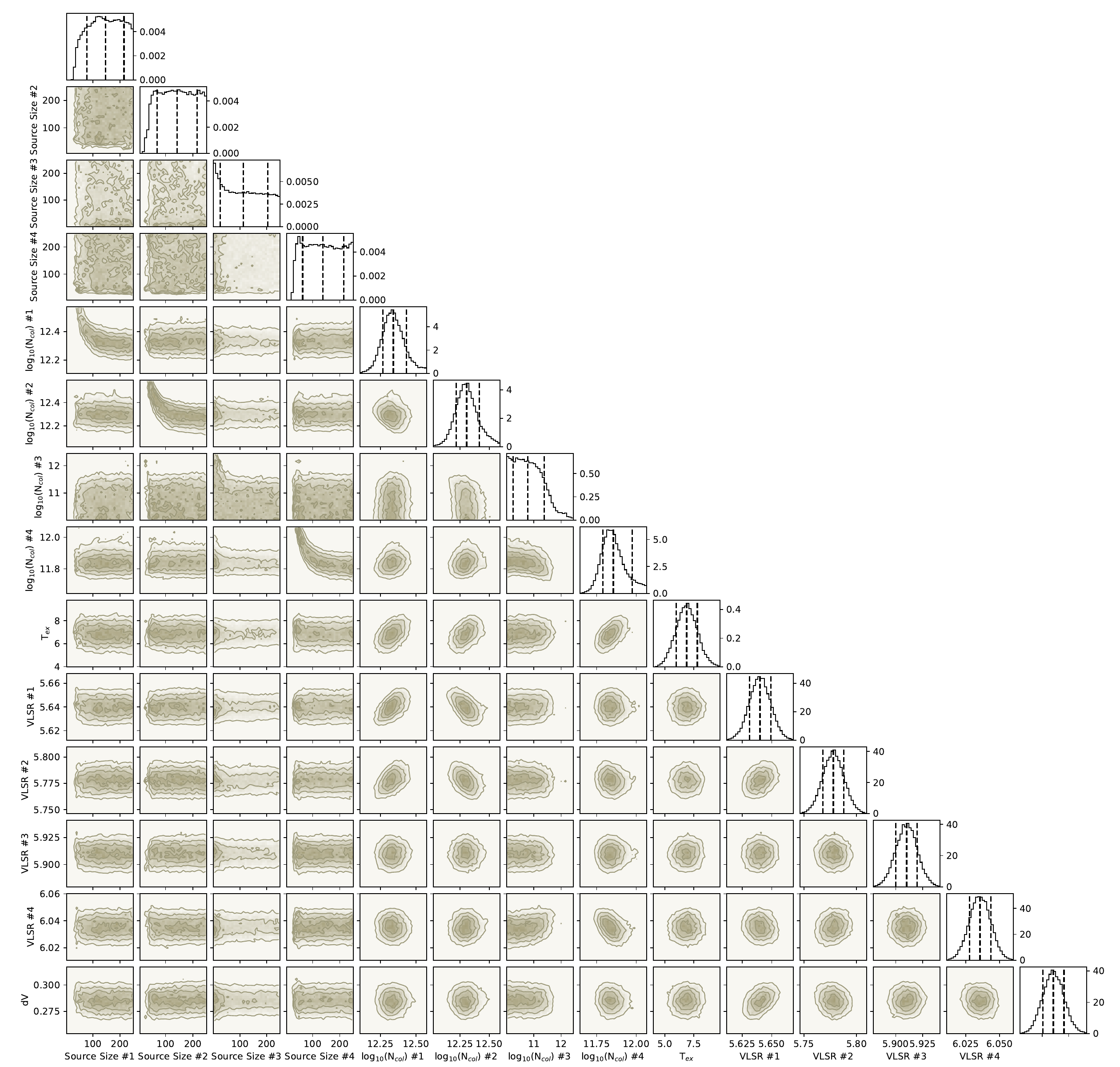}
\figsetgrpnote{The 16$^{th}$, 50$^{th}$, and 84$^{th}$ confidence intervals (corresponding to $\pm$1 sigma for a Gaussian posterior distribution) are shown as vertical lines. The contour lines are posterior probability levels, starting at $20\%$ of the maximum a posteriori estimate, with evenly spaced intervals of $20\%$ up to the peak density.}
\figsetgrpend

\figsetgrpstart
\figsetgrpnum{8.7}
\figsetgrptitle{Corner plot for HCCO.}
\figsetplot{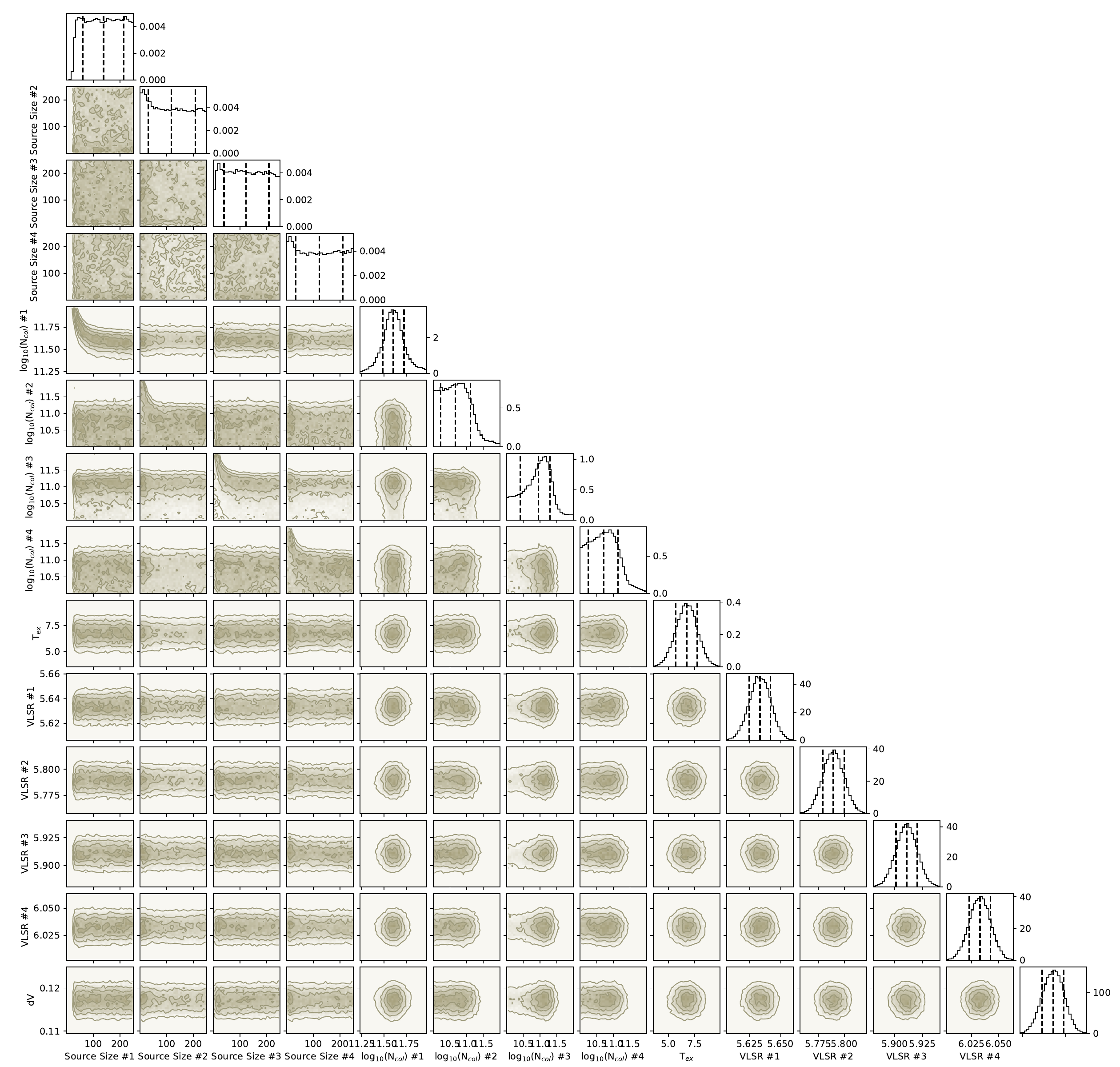}
\figsetgrpnote{The 16$^{th}$, 50$^{th}$, and 84$^{th}$ confidence intervals (corresponding to $\pm$1 sigma for a Gaussian posterior distribution) are shown as vertical lines. The contour lines are posterior probability levels, starting at $20\%$ of the maximum a posteriori estimate, with evenly spaced intervals of $20\%$ up to the peak density.}
\figsetgrpend

\figsetgrpstart
\figsetgrpnum{8.8}
\figsetgrptitle{Corner plot for CH$_{3}$NC.}
\figsetplot{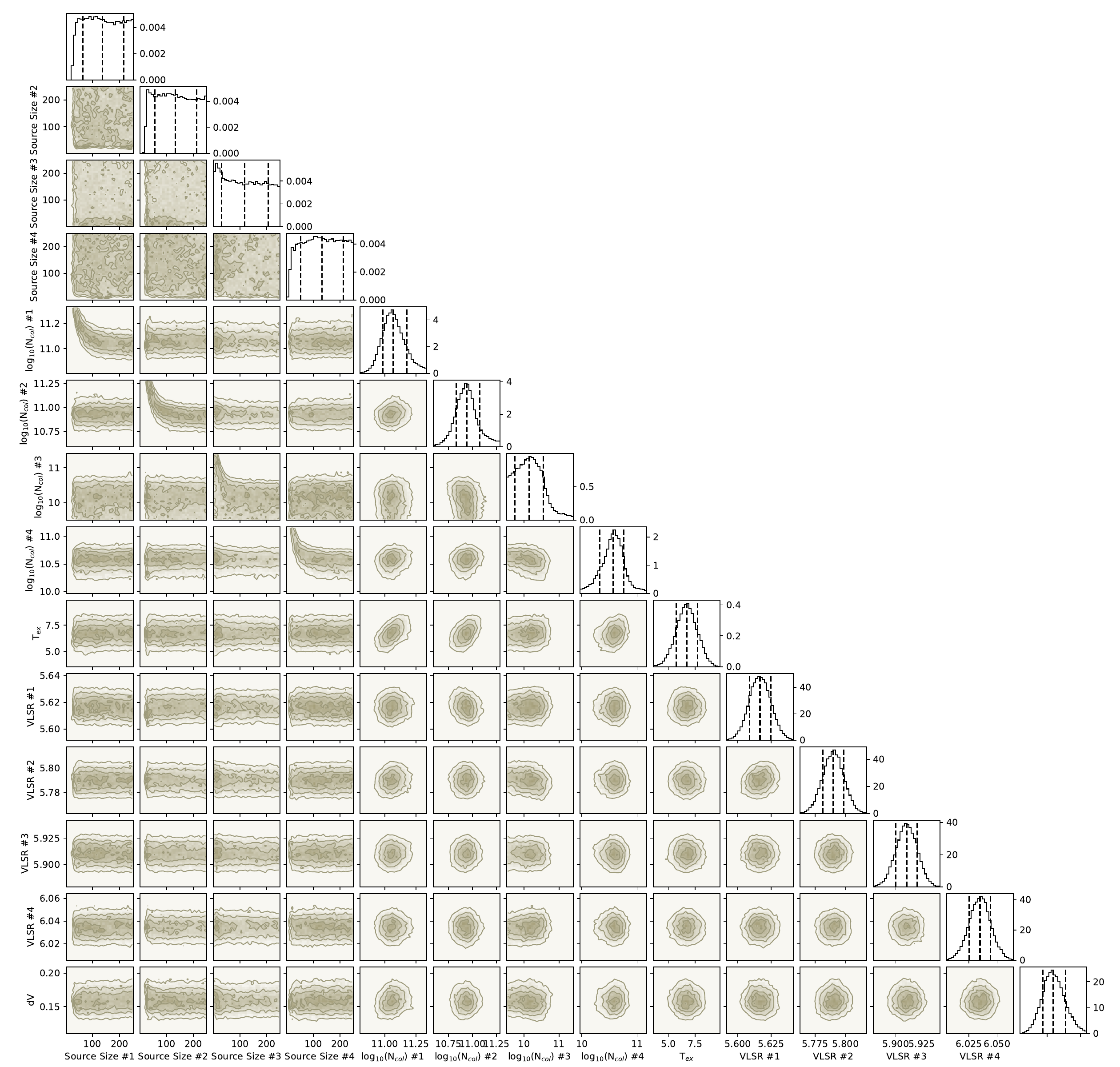}
\figsetgrpnote{The 16$^{th}$, 50$^{th}$, and 84$^{th}$ confidence intervals (corresponding to $\pm$1 sigma for a Gaussian posterior distribution) are shown as vertical lines. The contour lines are posterior probability levels, starting at $20\%$ of the maximum a posteriori estimate, with evenly spaced intervals of $20\%$ up to the peak density.}
\figsetgrpend

\figsetgrpstart
\figsetgrpnum{8.9}
\figsetgrptitle{Corner plot for HCNO.}
\figsetplot{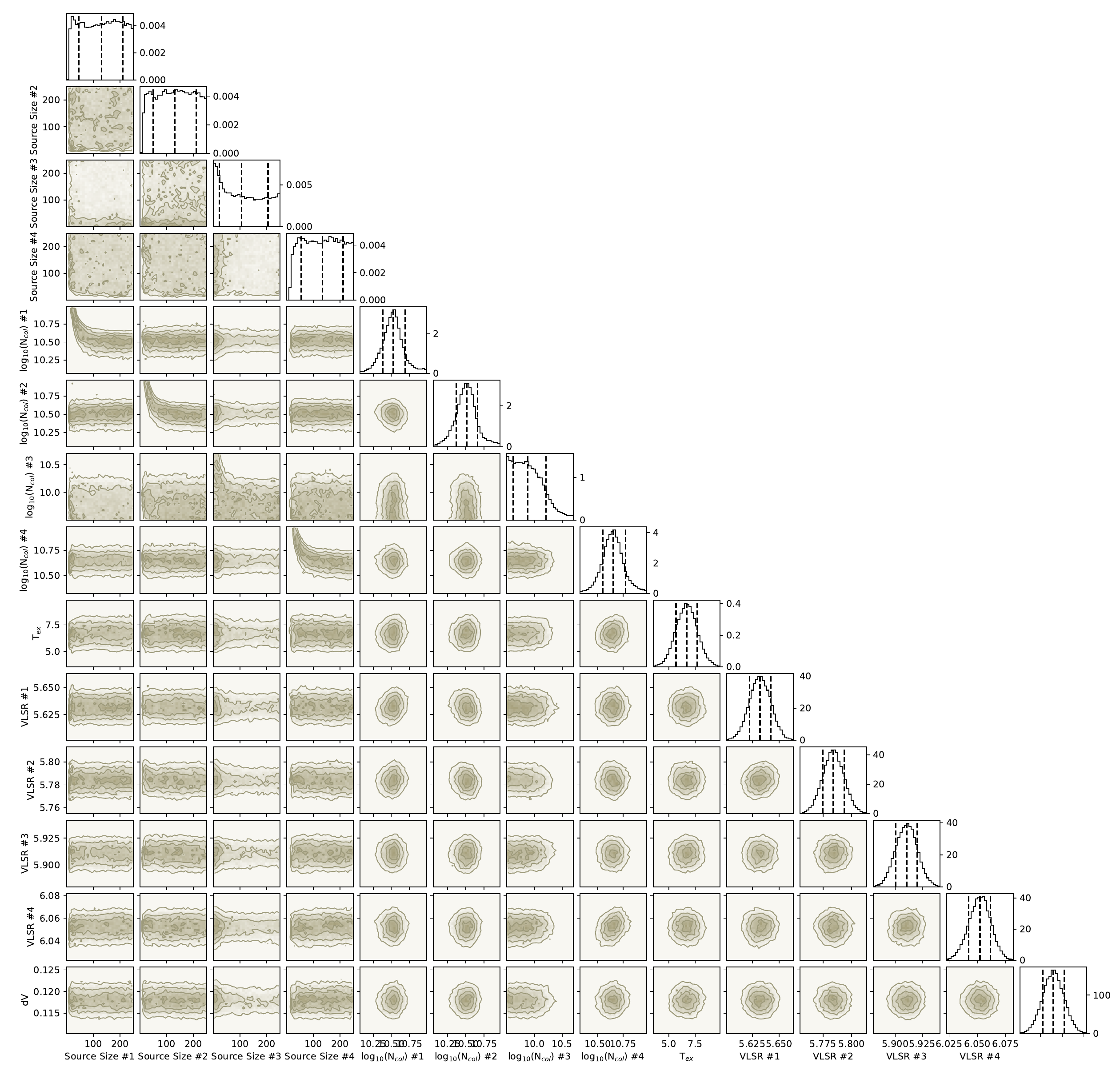}
\figsetgrpnote{The 16$^{th}$, 50$^{th}$, and 84$^{th}$ confidence intervals (corresponding to $\pm$1 sigma for a Gaussian posterior distribution) are shown as vertical lines. The contour lines are posterior probability levels, starting at $20\%$ of the maximum a posteriori estimate, with evenly spaced intervals of $20\%$ up to the peak density.}
\figsetgrpend

\figsetgrpstart
\figsetgrpnum{8.10}
\figsetgrptitle{Corner plot for HOCN.}
\figsetplot{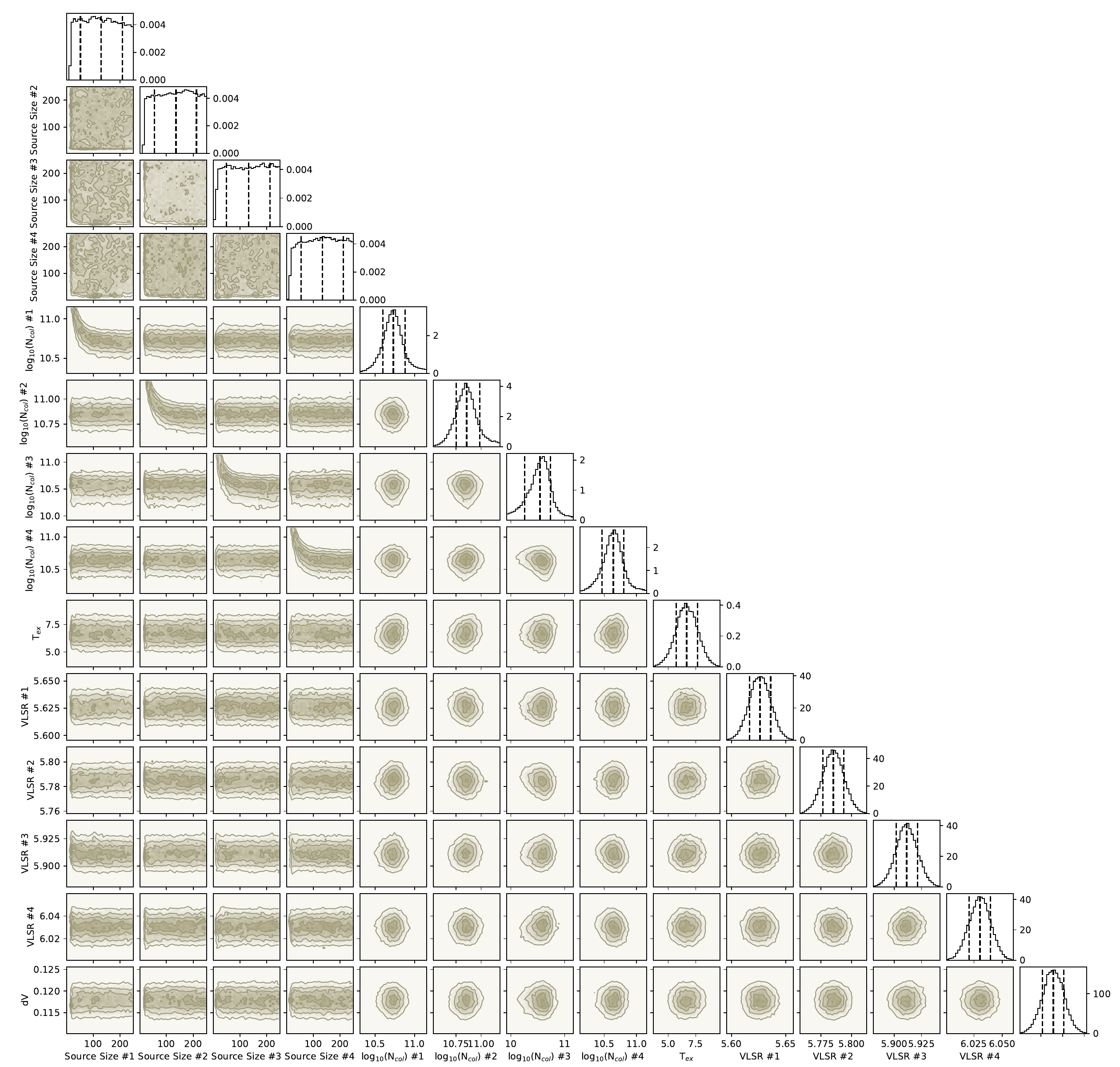}
\figsetgrpnote{The 16$^{th}$, 50$^{th}$, and 84$^{th}$ confidence intervals (corresponding to $\pm$1 sigma for a Gaussian posterior distribution) are shown as vertical lines. The contour lines are posterior probability levels, starting at $20\%$ of the maximum a posteriori estimate, with evenly spaced intervals of $20\%$ up to the peak density.}
\figsetgrpend

\figsetgrpstart
\figsetgrpnum{8.11}
\figsetgrptitle{Corner plot for HNCO.}
\figsetplot{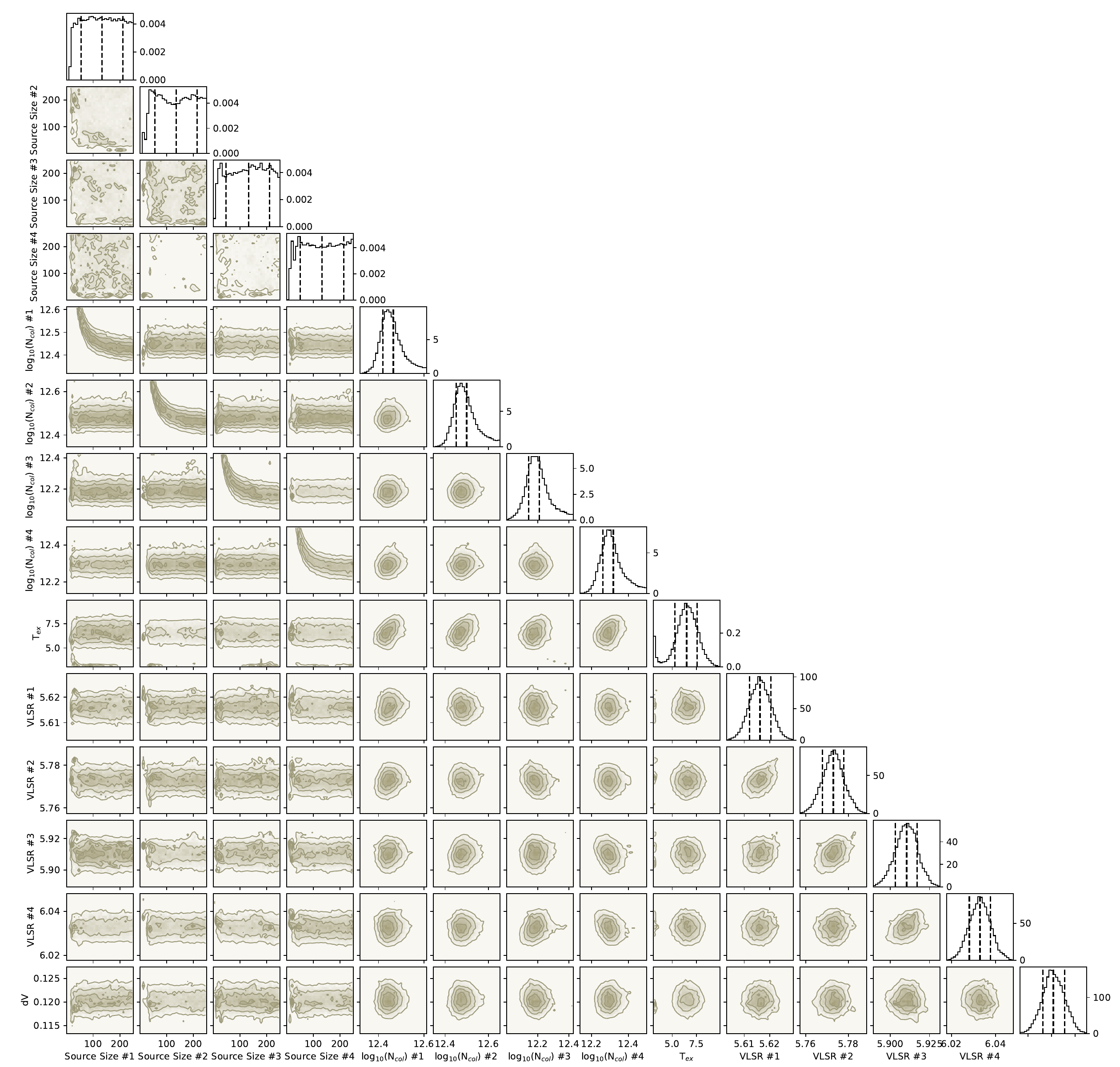}
\figsetgrpnote{The 16$^{th}$, 50$^{th}$, and 84$^{th}$ confidence intervals (corresponding to $\pm$1 sigma for a Gaussian posterior distribution) are shown as vertical lines. The contour lines are posterior probability levels, starting at $20\%$ of the maximum a posteriori estimate, with evenly spaced intervals of $20\%$ up to the peak density.}
\figsetgrpend

\figsetgrpstart
\figsetgrpnum{8.12}
\figsetgrptitle{Corner plot for HOCO$^{+}$.}
\figsetplot{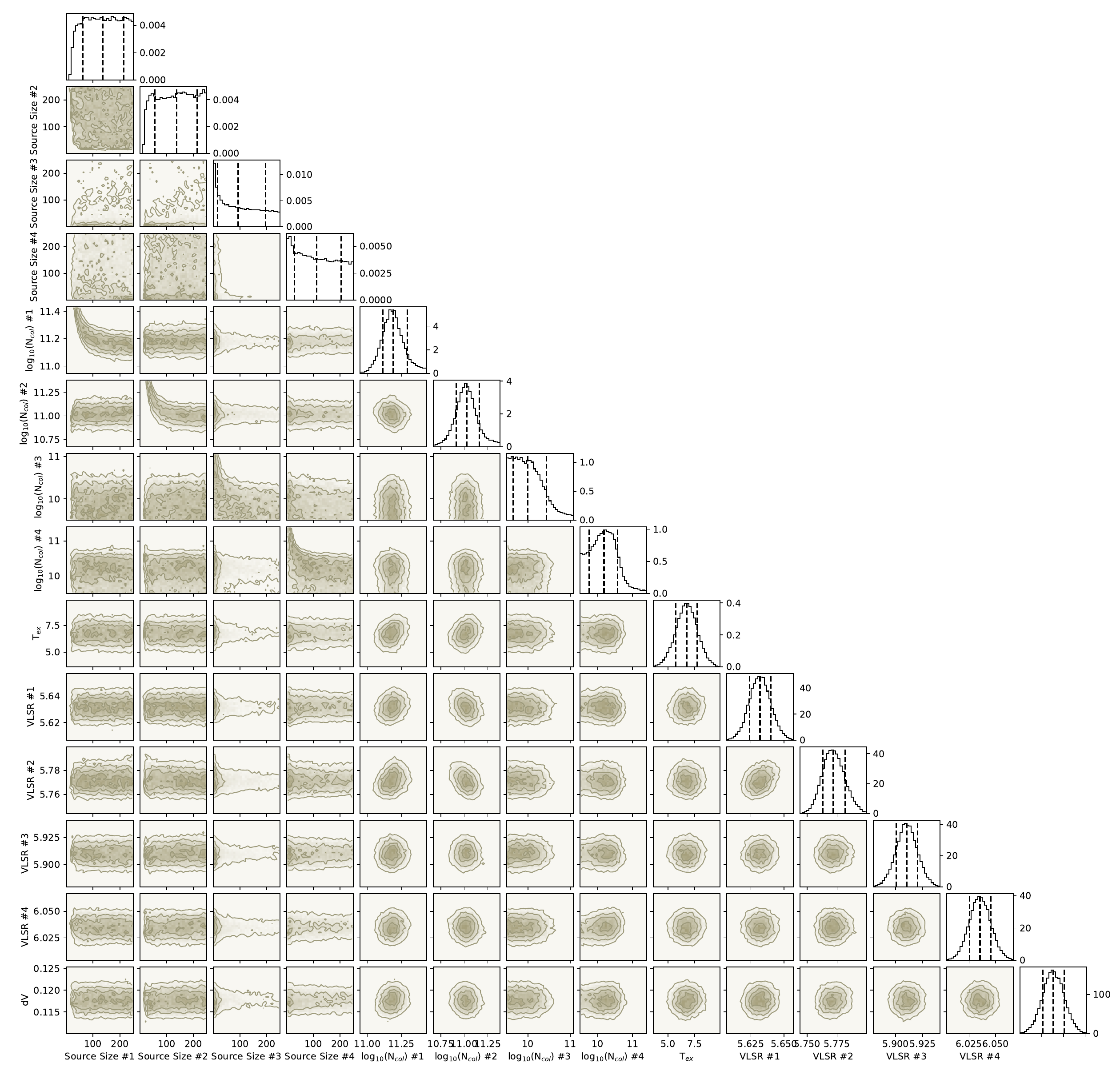}
\figsetgrpnote{The 16$^{th}$, 50$^{th}$, and 84$^{th}$ confidence intervals (corresponding to $\pm$1 sigma for a Gaussian posterior distribution) are shown as vertical lines. The contour lines are posterior probability levels, starting at $20\%$ of the maximum a posteriori estimate, with evenly spaced intervals of $20\%$ up to the peak density.}
\figsetgrpend

\figsetgrpstart
\figsetgrpnum{8.13}
\figsetgrptitle{Corner plot for $t$-HCOOH.}
\figsetplot{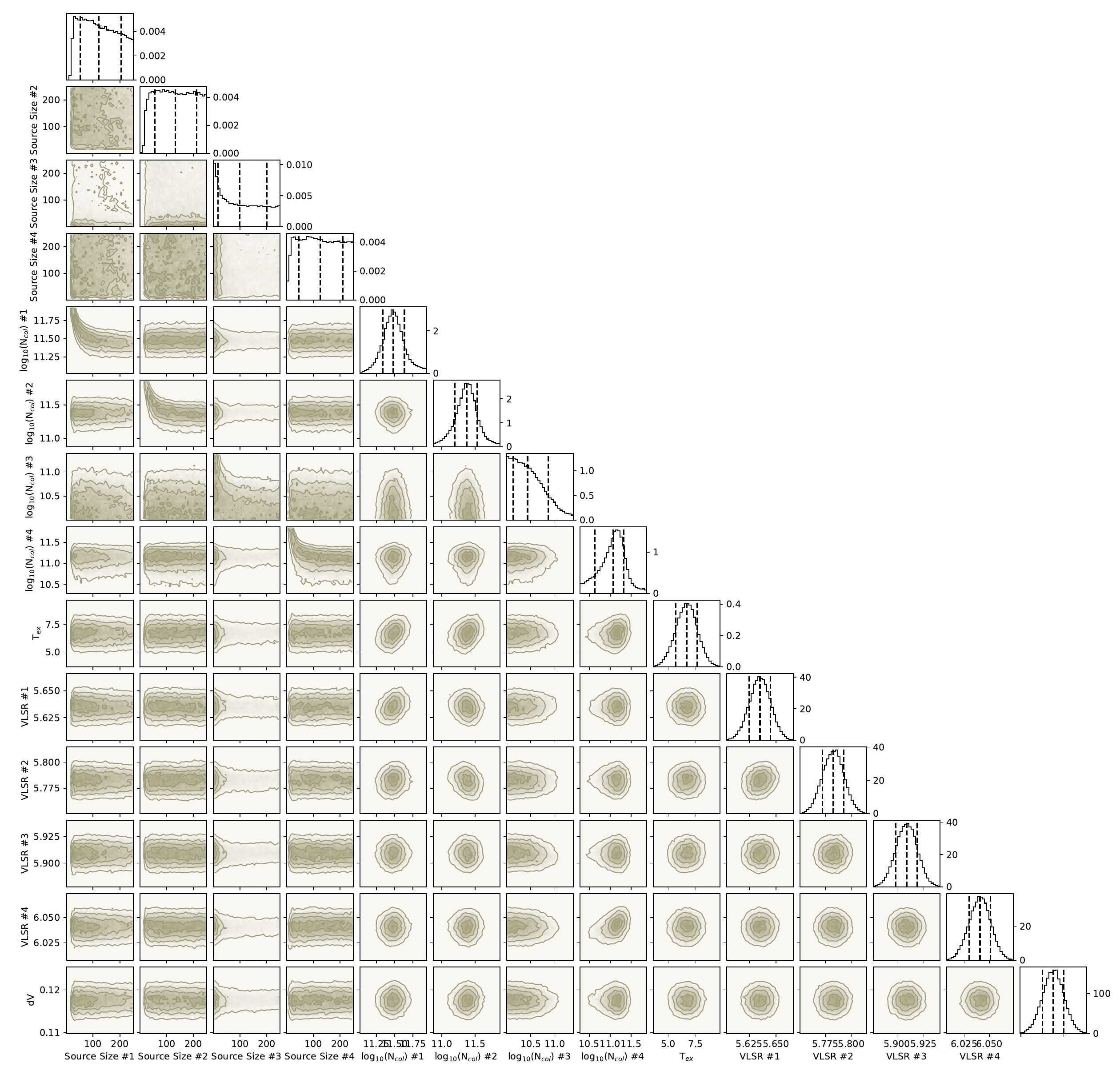}
\figsetgrpnote{The 16$^{th}$, 50$^{th}$, and 84$^{th}$ confidence intervals (corresponding to $\pm$1 sigma for a Gaussian posterior distribution) are shown as vertical lines. The contour lines are posterior probability levels, starting at $20\%$ of the maximum a posteriori estimate, with evenly spaced intervals of $20\%$ up to the peak density.}
\figsetgrpend

\figsetgrpstart
\figsetgrpnum{8.14}
\figsetgrptitle{Corner plot for C$_{4}$H$^{-}$.}
\figsetplot{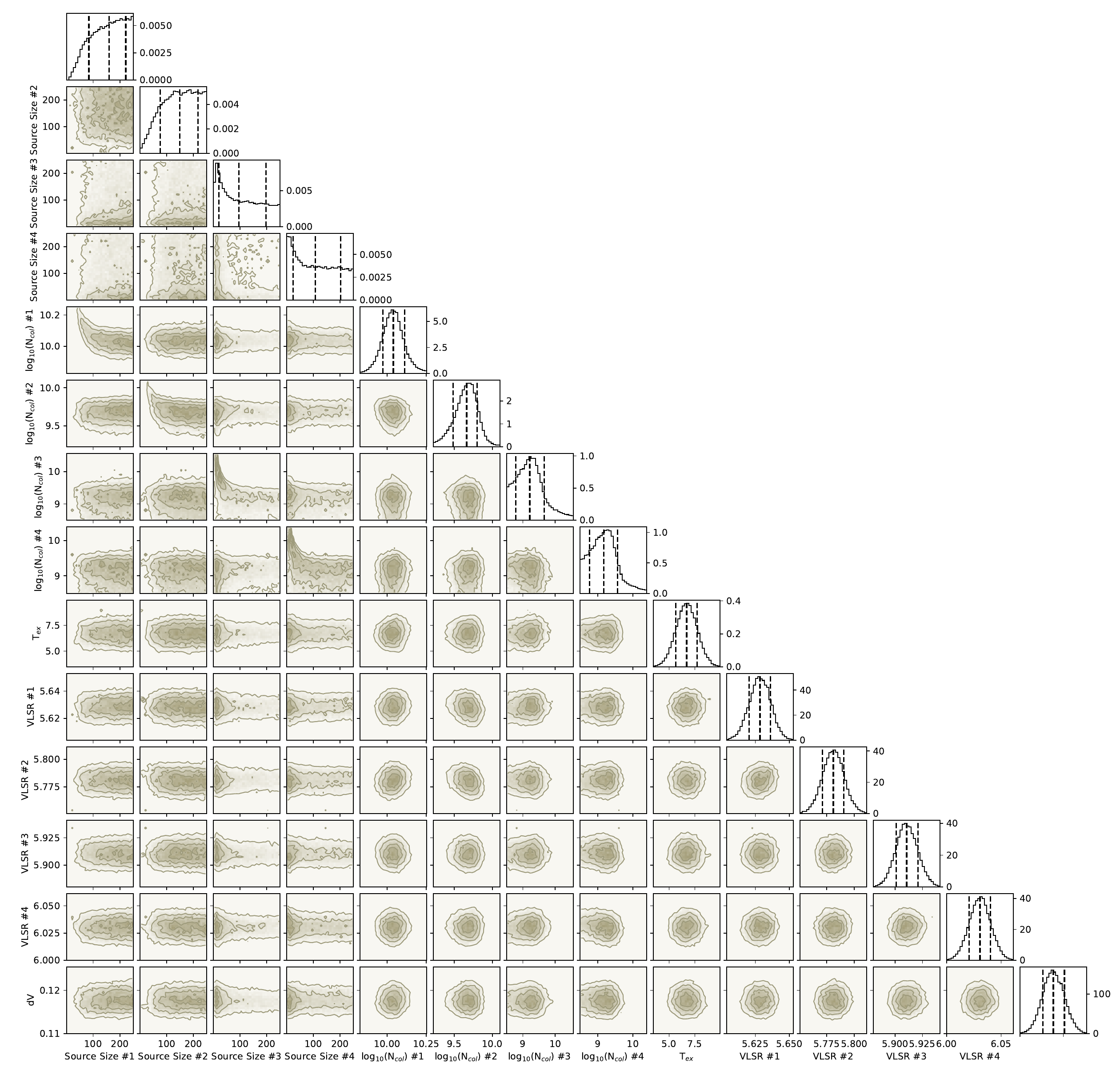}
\figsetgrpnote{The 16$^{th}$, 50$^{th}$, and 84$^{th}$ confidence intervals (corresponding to $\pm$1 sigma for a Gaussian posterior distribution) are shown as vertical lines. The contour lines are posterior probability levels, starting at $20\%$ of the maximum a posteriori estimate, with evenly spaced intervals of $20\%$ up to the peak density.}
\figsetgrpend

\figsetgrpstart
\figsetgrpnum{8.15}
\figsetgrptitle{Corner plot for C$_{4}$H.}
\figsetplot{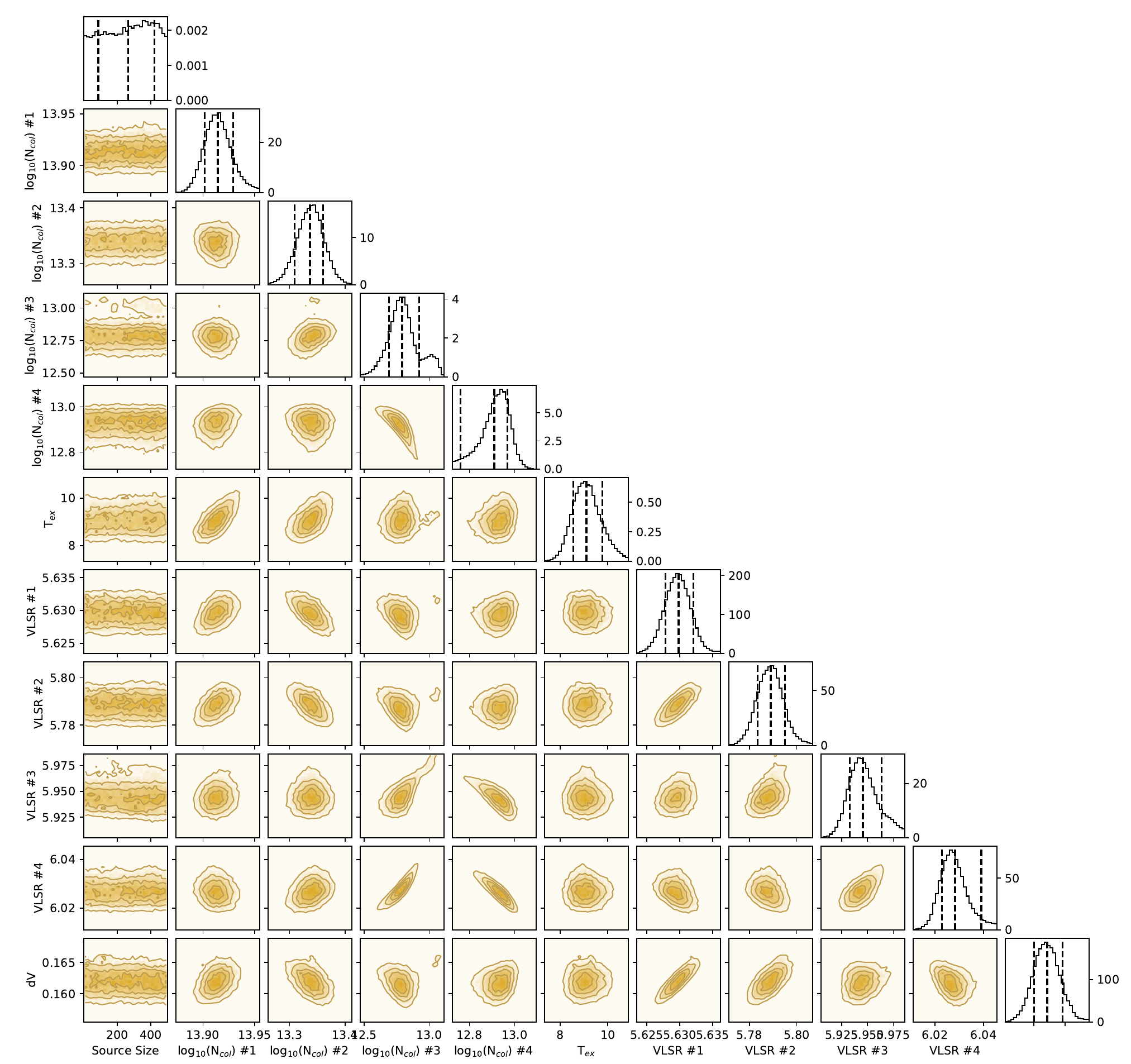}
\figsetgrpnote{The 16$^{th}$, 50$^{th}$, and 84$^{th}$ confidence intervals (corresponding to $\pm$1 sigma for a Gaussian posterior distribution) are shown as vertical lines. The contour lines are posterior probability levels, starting at $20\%$ of the maximum a posteriori estimate, with evenly spaced intervals of $20\%$ up to the peak density.}
\figsetgrpend

\figsetgrpstart
\figsetgrpnum{8.16}
\figsetgrptitle{Corner plot for C$_{3}$N.}
\figsetplot{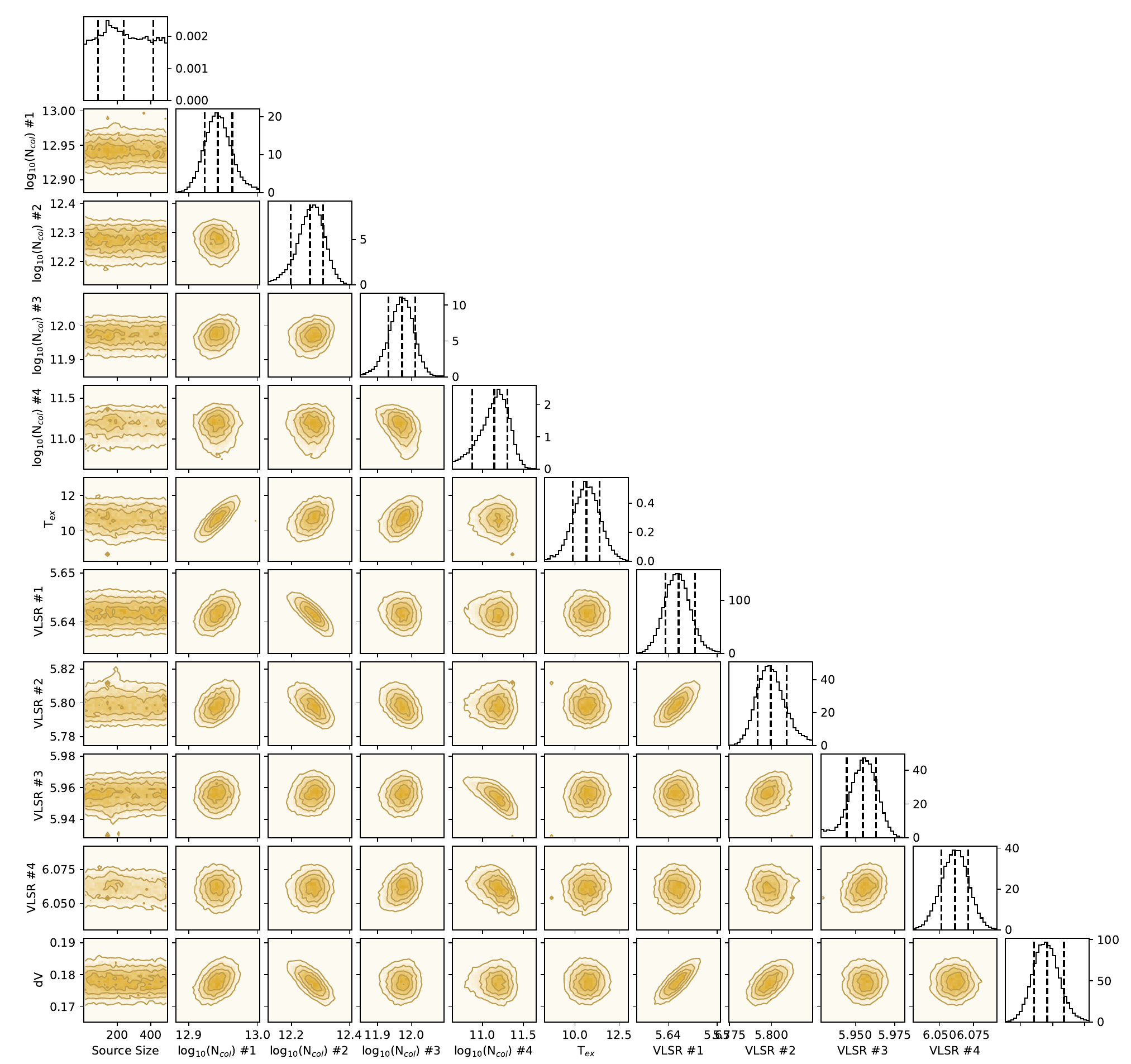}
\figsetgrpnote{The 16$^{th}$, 50$^{th}$, and 84$^{th}$ confidence intervals (corresponding to $\pm$1 sigma for a Gaussian posterior distribution) are shown as vertical lines. The contour lines are posterior probability levels, starting at $20\%$ of the maximum a posteriori estimate, with evenly spaced intervals of $20\%$ up to the peak density.}
\figsetgrpend

\figsetgrpstart
\figsetgrpnum{8.17}
\figsetgrptitle{Corner plot for HC$_{3}$N.}
\figsetplot{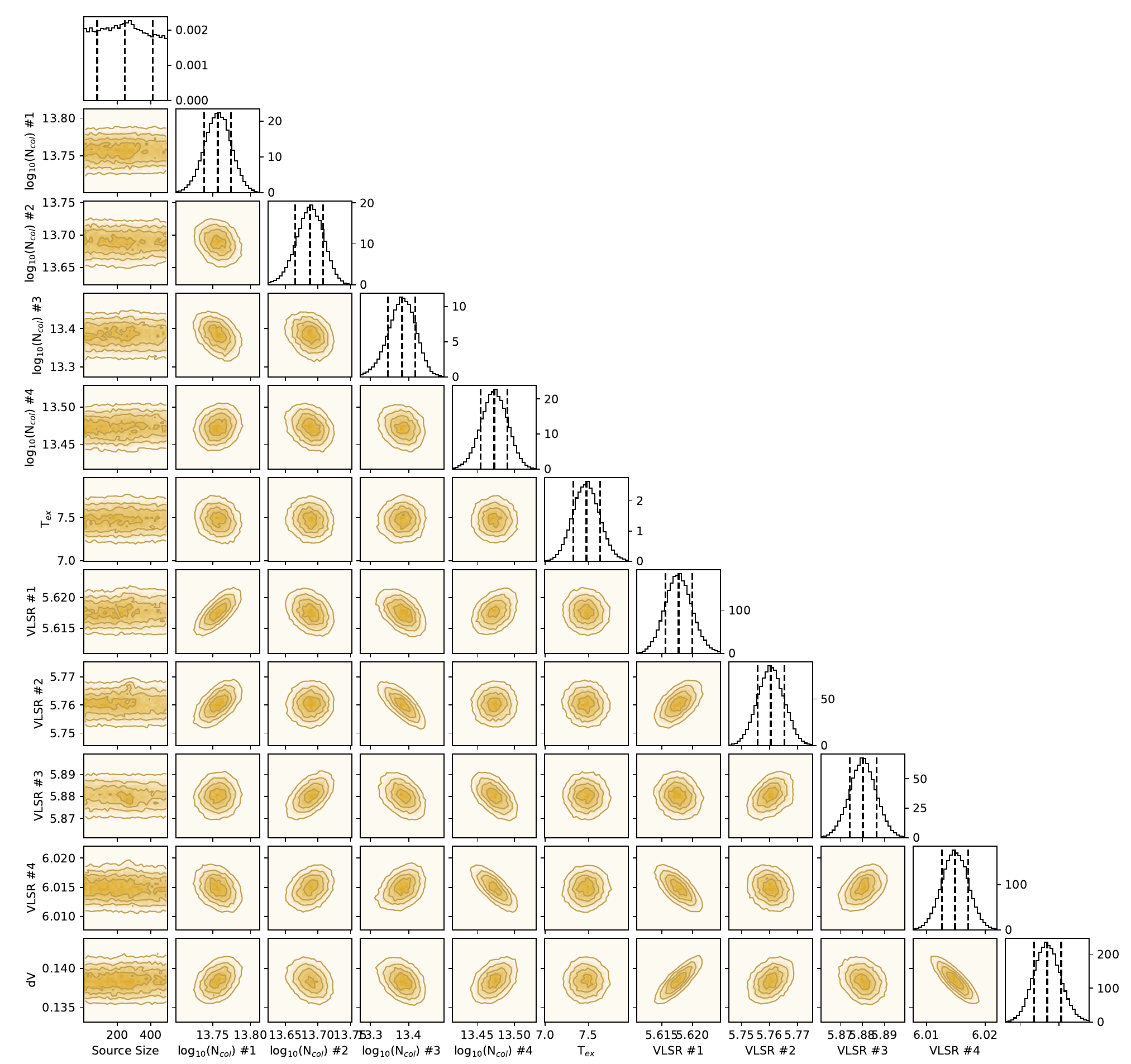}
\figsetgrpnote{The 16$^{th}$, 50$^{th}$, and 84$^{th}$ confidence intervals (corresponding to $\pm$1 sigma for a Gaussian posterior distribution) are shown as vertical lines. The contour lines are posterior probability levels, starting at $20\%$ of the maximum a posteriori estimate, with evenly spaced intervals of $20\%$ up to the peak density.}
\figsetgrpend

\figsetgrpstart
\figsetgrpnum{8.18}
\figsetgrptitle{Corner plot for HNC$_{3}$.}
\figsetplot{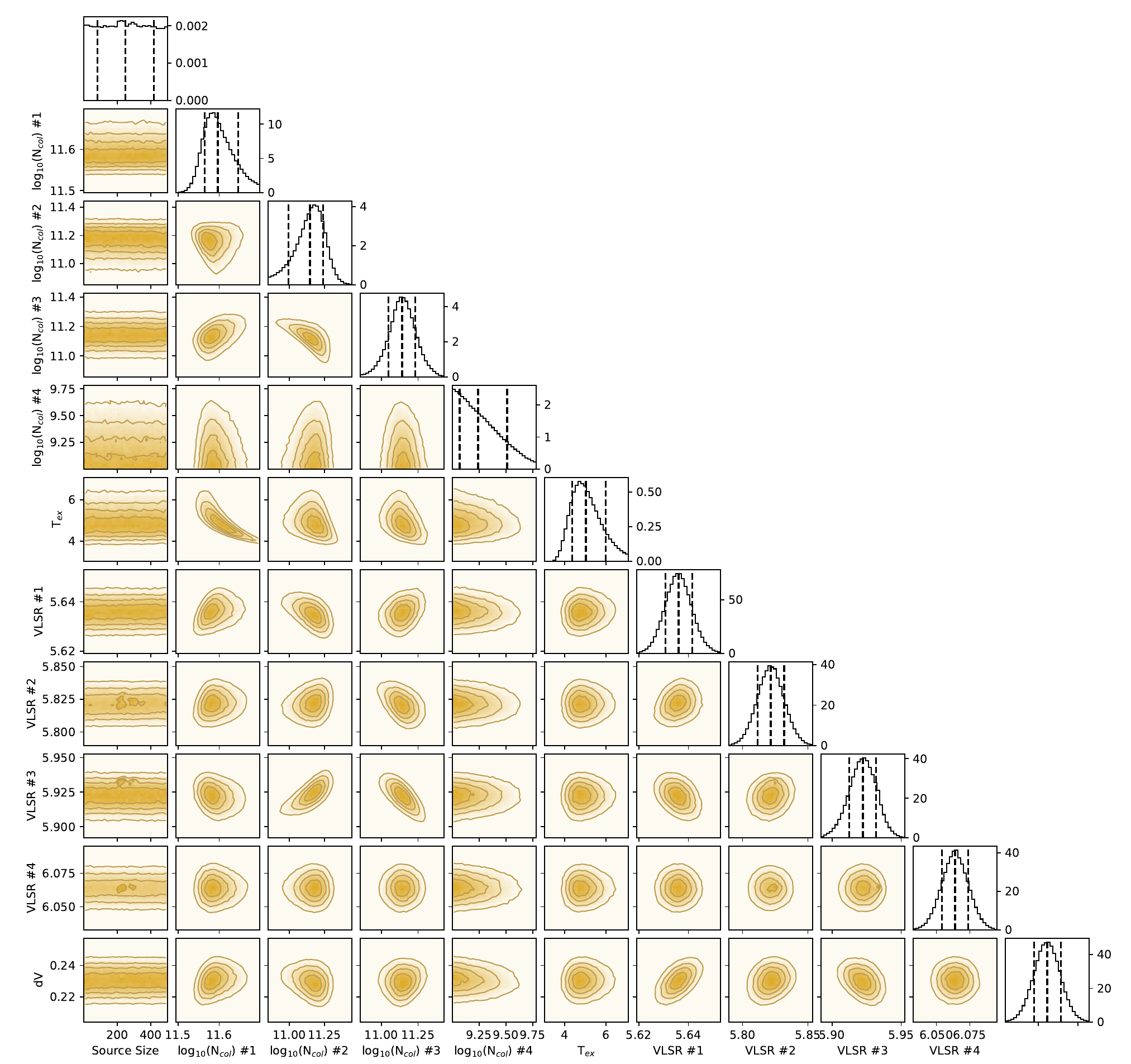}
\figsetgrpnote{The 16$^{th}$, 50$^{th}$, and 84$^{th}$ confidence intervals (corresponding to $\pm$1 sigma for a Gaussian posterior distribution) are shown as vertical lines. The contour lines are posterior probability levels, starting at $20\%$ of the maximum a posteriori estimate, with evenly spaced intervals of $20\%$ up to the peak density.}
\figsetgrpend

\figsetgrpstart
\figsetgrpnum{8.19}
\figsetgrptitle{Corner plot for HCCNC.}
\figsetplot{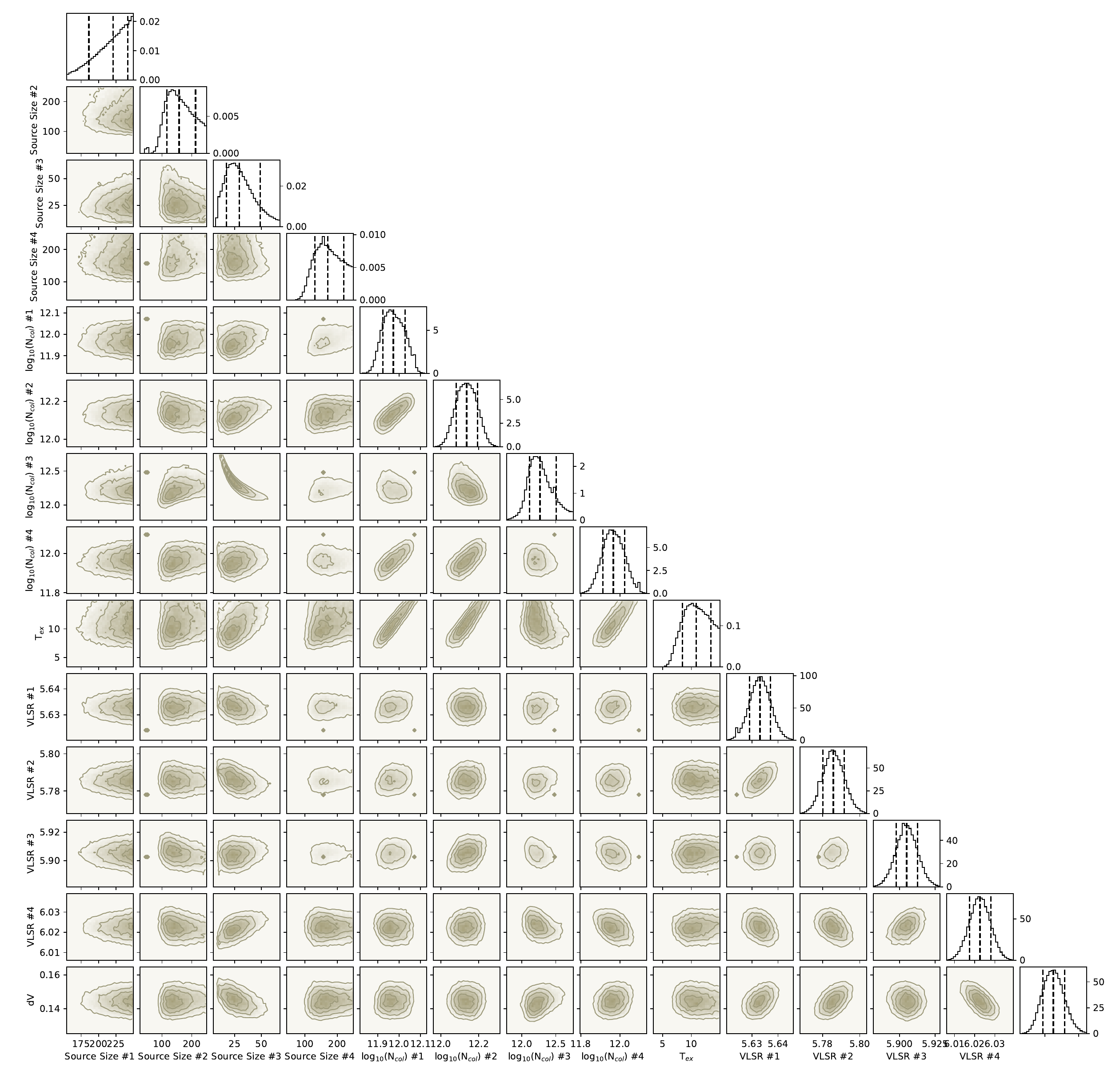}
\figsetgrpnote{The 16$^{th}$, 50$^{th}$, and 84$^{th}$ confidence intervals (corresponding to $\pm$1 sigma for a Gaussian posterior distribution) are shown as vertical lines. The contour lines are posterior probability levels, starting at $20\%$ of the maximum a posteriori estimate, with evenly spaced intervals of $20\%$ up to the peak density.}
\figsetgrpend

\figsetgrpstart
\figsetgrpnum{8.20}
\figsetgrptitle{Corner plot for C$_{3}$O.}
\figsetplot{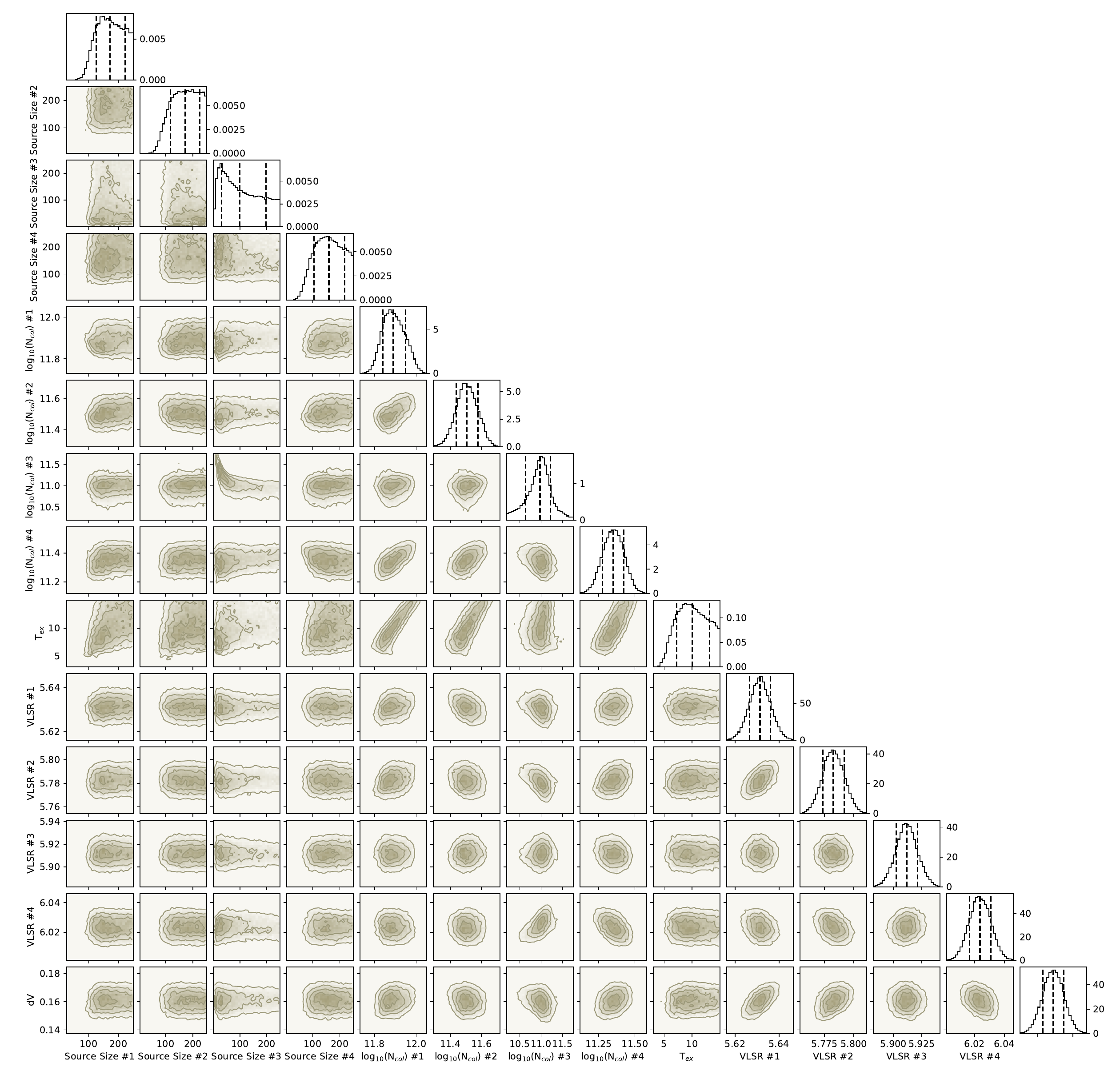}
\figsetgrpnote{The 16$^{th}$, 50$^{th}$, and 84$^{th}$ confidence intervals (corresponding to $\pm$1 sigma for a Gaussian posterior distribution) are shown as vertical lines. The contour lines are posterior probability levels, starting at $20\%$ of the maximum a posteriori estimate, with evenly spaced intervals of $20\%$ up to the peak density.}
\figsetgrpend

\figsetgrpstart
\figsetgrpnum{8.21}
\figsetgrptitle{Corner plot for HC$_{3}$NH$^{+}$.}
\figsetplot{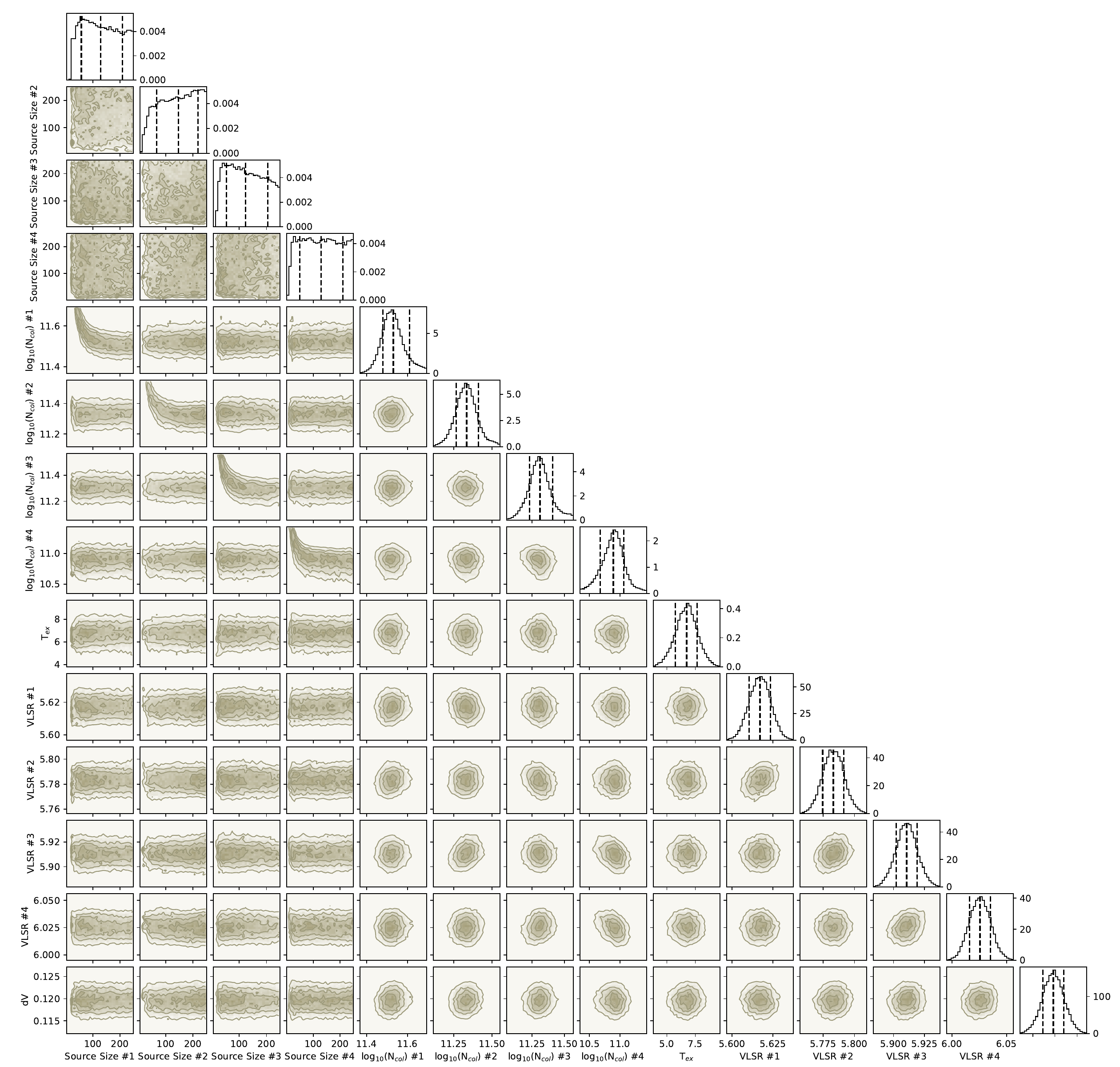}
\figsetgrpnote{The 16$^{th}$, 50$^{th}$, and 84$^{th}$ confidence intervals (corresponding to $\pm$1 sigma for a Gaussian posterior distribution) are shown as vertical lines. The contour lines are posterior probability levels, starting at $20\%$ of the maximum a posteriori estimate, with evenly spaced intervals of $20\%$ up to the peak density.}
\figsetgrpend

\figsetgrpstart
\figsetgrpnum{8.22}
\figsetgrptitle{Corner plot for C$_{2}$H$_{3}$CN.}
\figsetplot{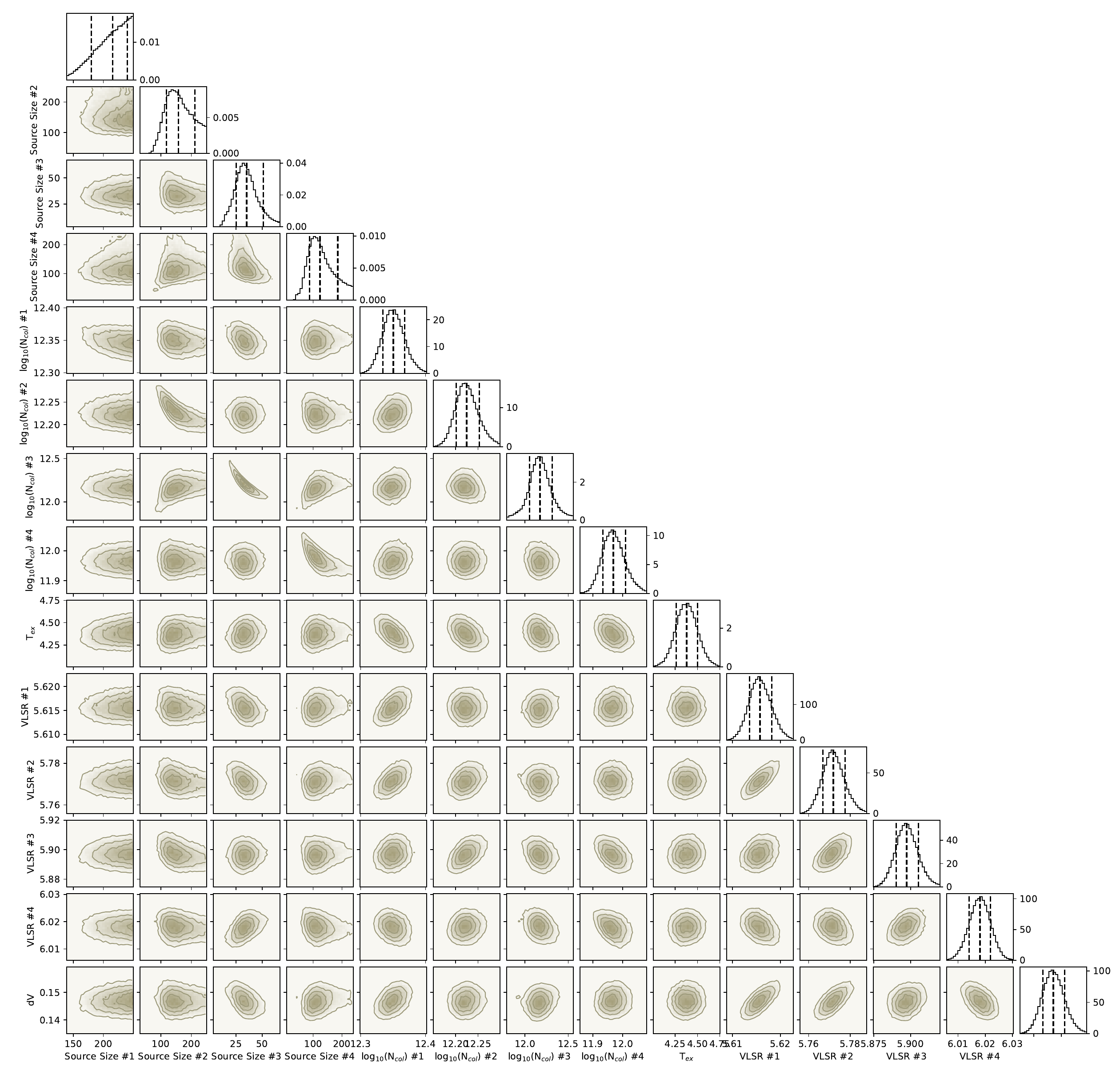}
\figsetgrpnote{The 16$^{th}$, 50$^{th}$, and 84$^{th}$ confidence intervals (corresponding to $\pm$1 sigma for a Gaussian posterior distribution) are shown as vertical lines. The contour lines are posterior probability levels, starting at $20\%$ of the maximum a posteriori estimate, with evenly spaced intervals of $20\%$ up to the peak density.}
\figsetgrpend

\figsetgrpstart
\figsetgrpnum{8.23}
\figsetgrptitle{Corner plot for NCCNH$^{+}$.}
\figsetplot{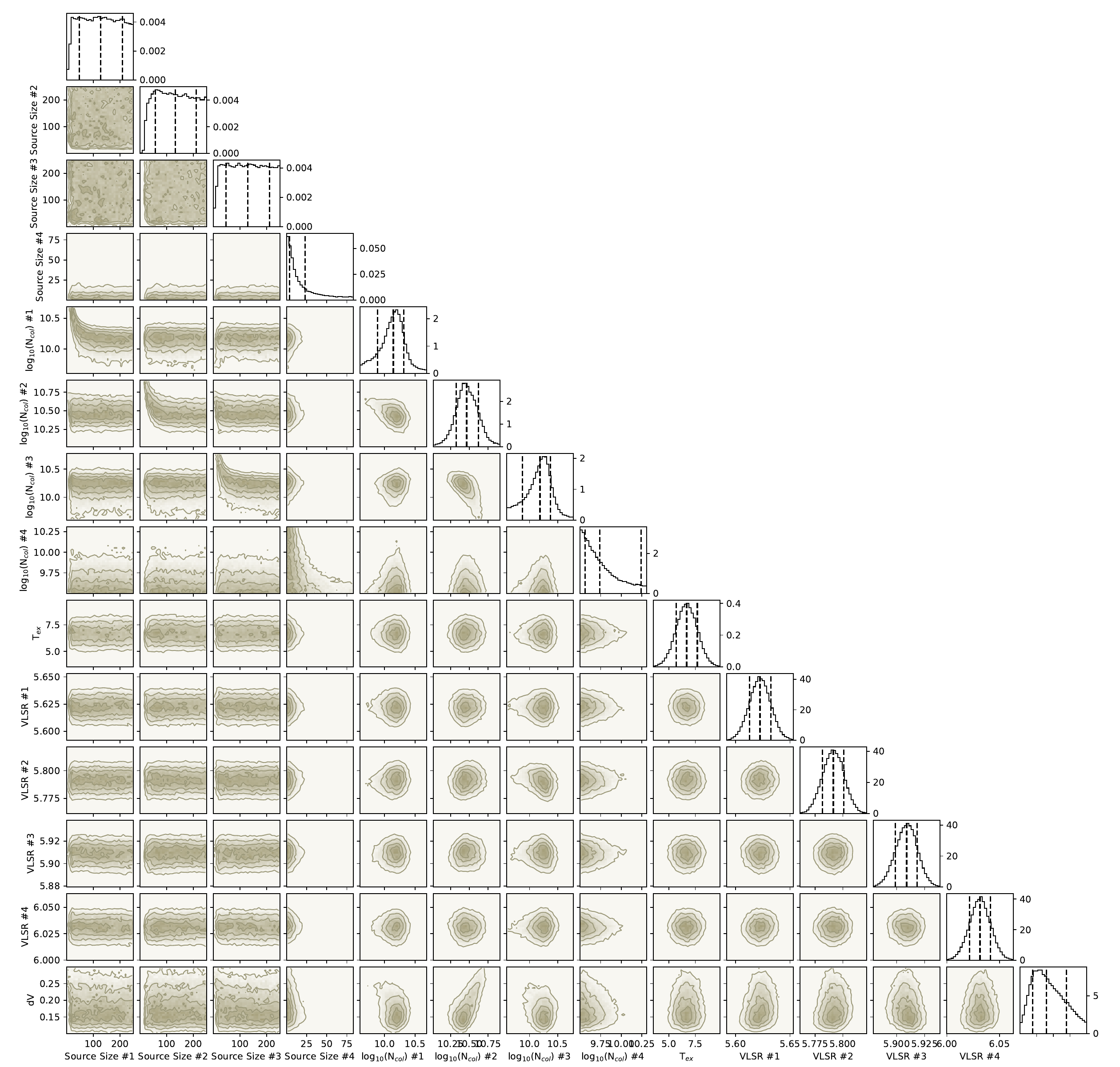}
\figsetgrpnote{The 16$^{th}$, 50$^{th}$, and 84$^{th}$ confidence intervals (corresponding to $\pm$1 sigma for a Gaussian posterior distribution) are shown as vertical lines. The contour lines are posterior probability levels, starting at $20\%$ of the maximum a posteriori estimate, with evenly spaced intervals of $20\%$ up to the peak density.}
\figsetgrpend

\figsetgrpstart
\figsetgrpnum{8.24}
\figsetgrptitle{Corner plot for HC$_{3}$O$^{+}$.}
\figsetplot{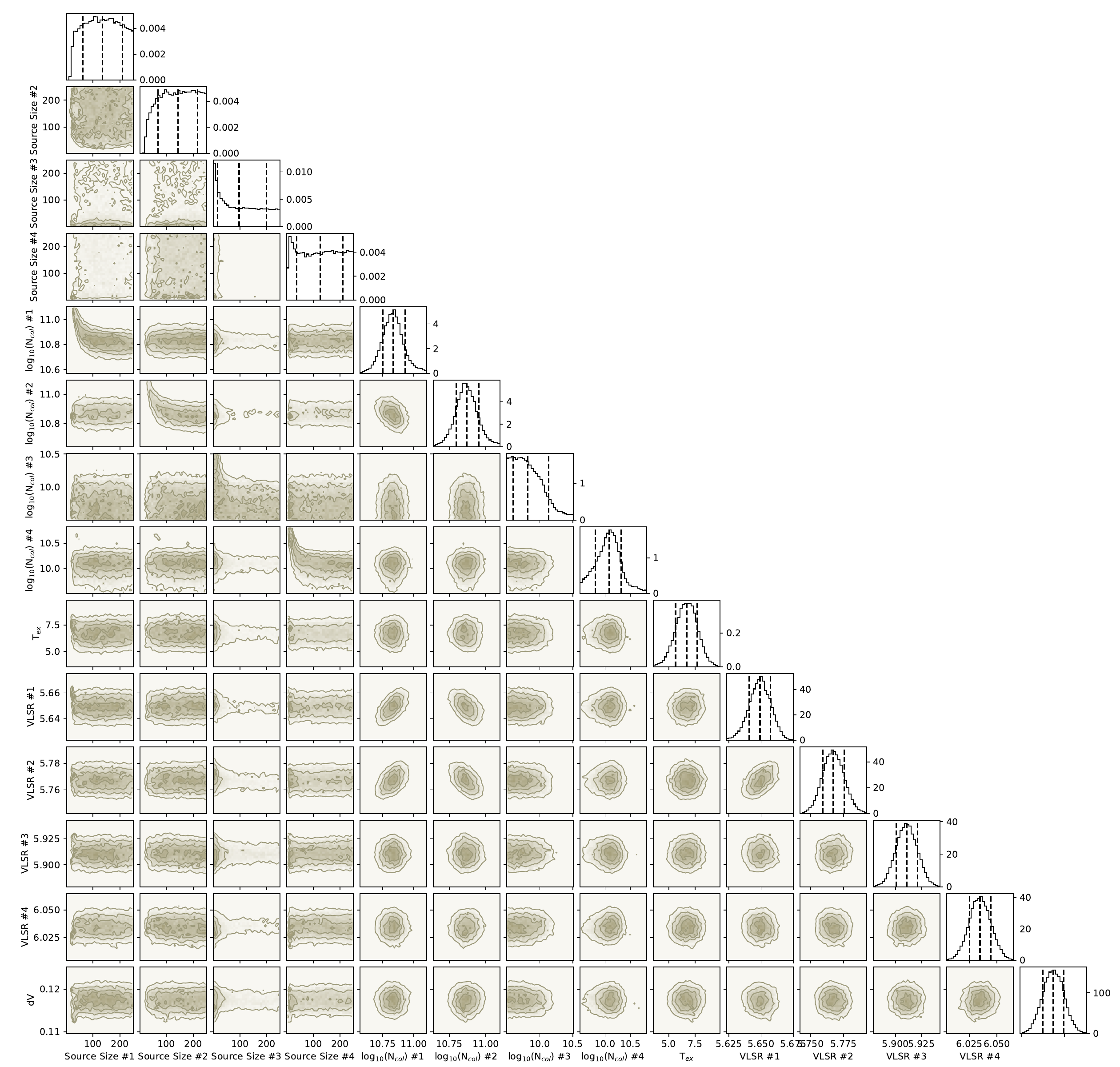}
\figsetgrpnote{The 16$^{th}$, 50$^{th}$, and 84$^{th}$ confidence intervals (corresponding to $\pm$1 sigma for a Gaussian posterior distribution) are shown as vertical lines. The contour lines are posterior probability levels, starting at $20\%$ of the maximum a posteriori estimate, with evenly spaced intervals of $20\%$ up to the peak density.}
\figsetgrpend

\figsetgrpstart
\figsetgrpnum{8.25}
\figsetgrptitle{Corner plot for $c$-H$_{2}$C$_{3}$O.}
\figsetplot{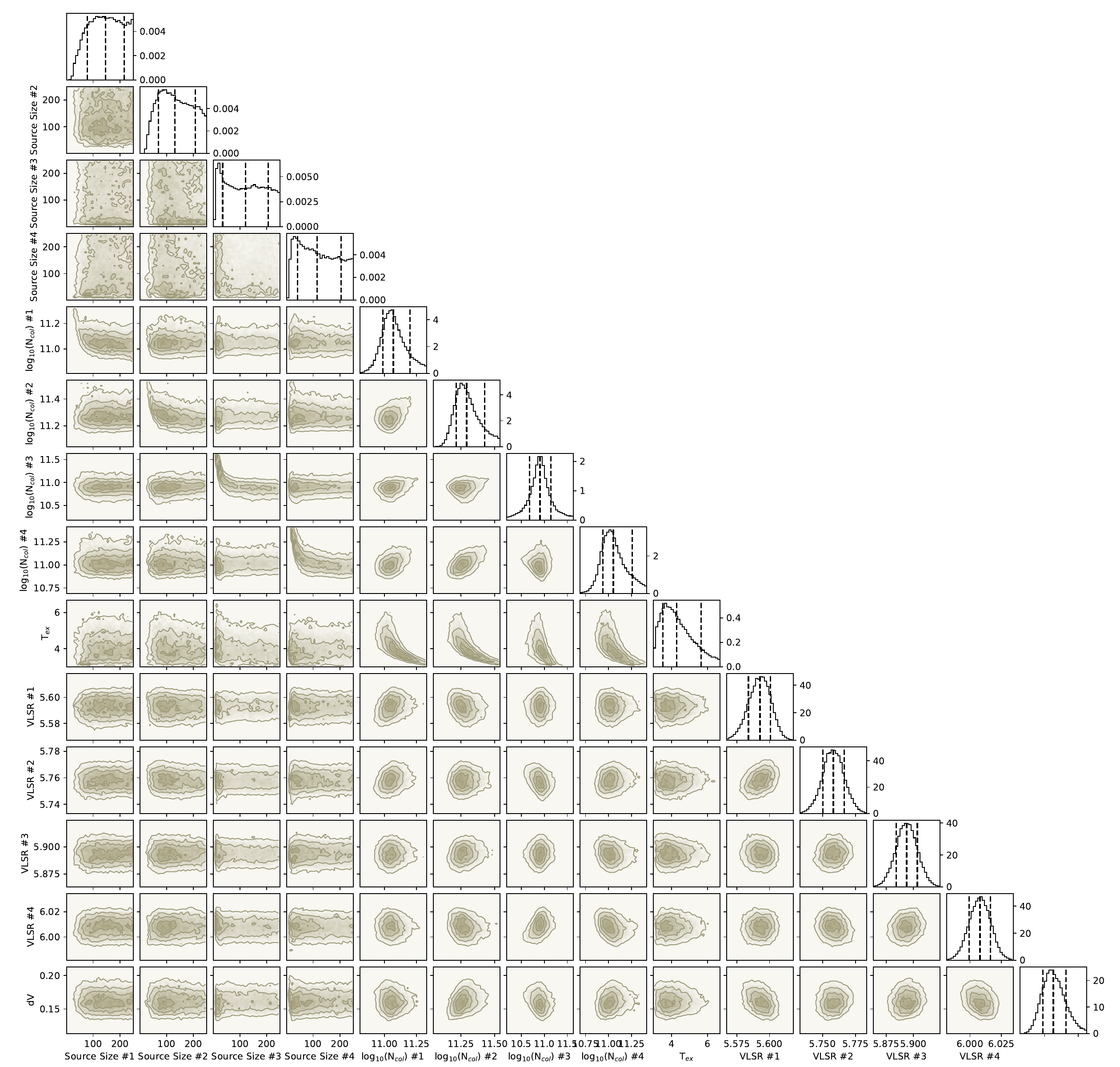}
\figsetgrpnote{The 16$^{th}$, 50$^{th}$, and 84$^{th}$ confidence intervals (corresponding to $\pm$1 sigma for a Gaussian posterior distribution) are shown as vertical lines. The contour lines are posterior probability levels, starting at $20\%$ of the maximum a posteriori estimate, with evenly spaced intervals of $20\%$ up to the peak density.}
\figsetgrpend

\figsetgrpstart
\figsetgrpnum{8.26}
\figsetgrptitle{Corner plot for HCCCHO.}
\figsetplot{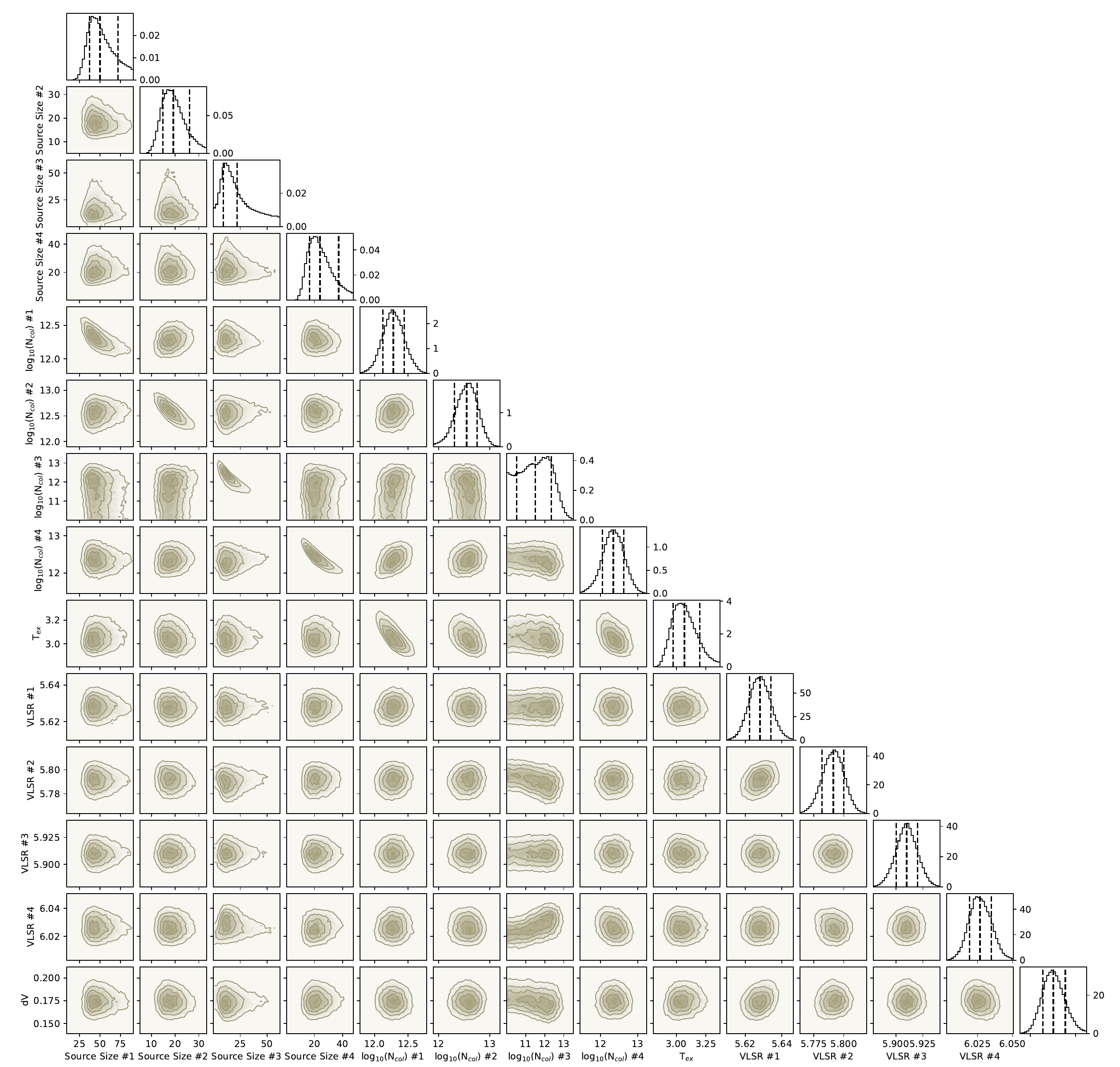}
\figsetgrpnote{The 16$^{th}$, 50$^{th}$, and 84$^{th}$ confidence intervals (corresponding to $\pm$1 sigma for a Gaussian posterior distribution) are shown as vertical lines. The contour lines are posterior probability levels, starting at $20\%$ of the maximum a posteriori estimate, with evenly spaced intervals of $20\%$ up to the peak density.}
\figsetgrpend

\figsetgrpstart
\figsetgrpnum{8.27}
\figsetgrptitle{Corner plot for trans-C$_{2}$H$_{3}$CHO.}
\figsetplot{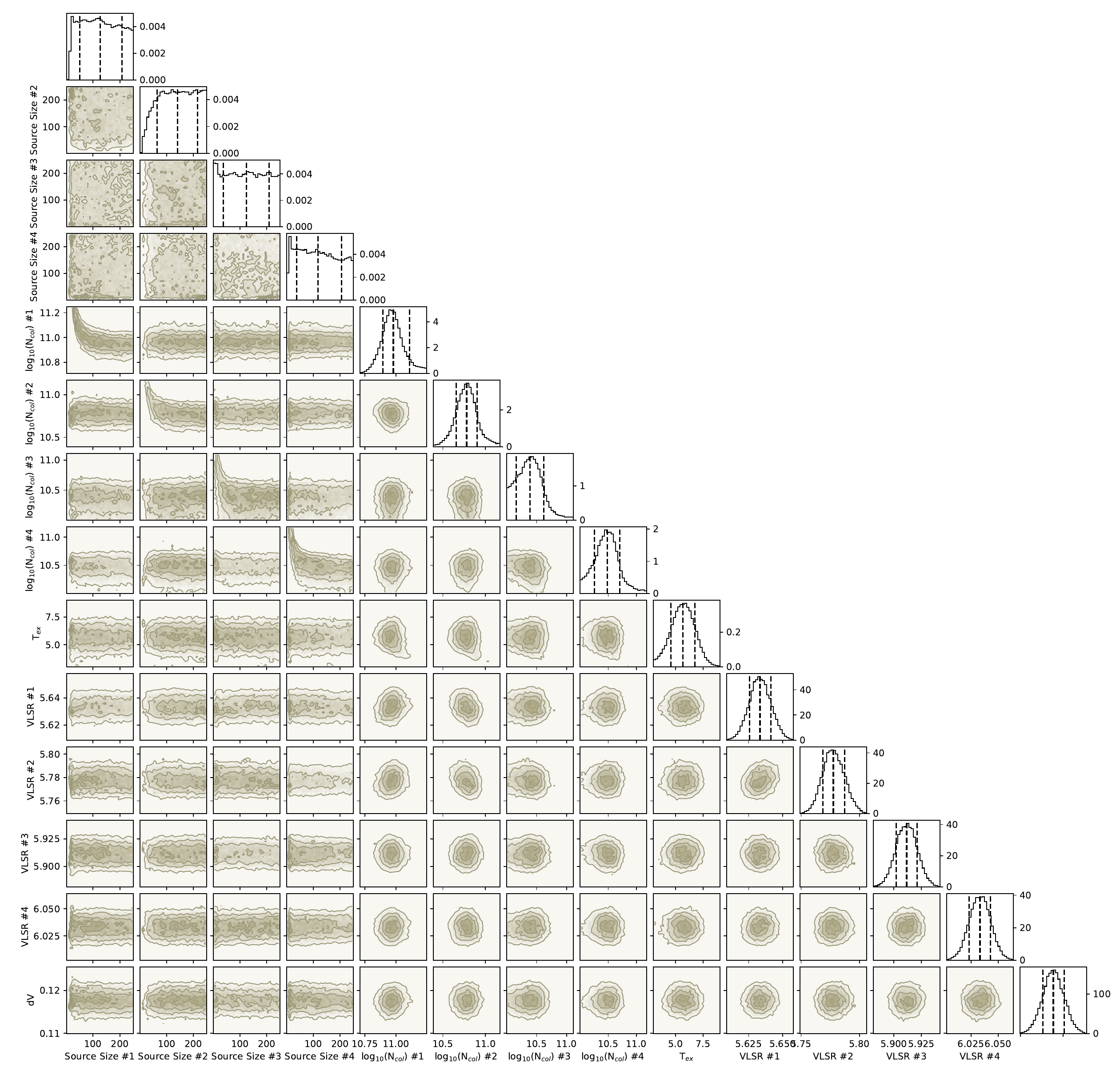}
\figsetgrpnote{The 16$^{th}$, 50$^{th}$, and 84$^{th}$ confidence intervals (corresponding to $\pm$1 sigma for a Gaussian posterior distribution) are shown as vertical lines. The contour lines are posterior probability levels, starting at $20\%$ of the maximum a posteriori estimate, with evenly spaced intervals of $20\%$ up to the peak density.}
\figsetgrpend

\figsetgrpstart
\figsetgrpnum{8.28}
\figsetgrptitle{Corner plot for HCCS.}
\figsetplot{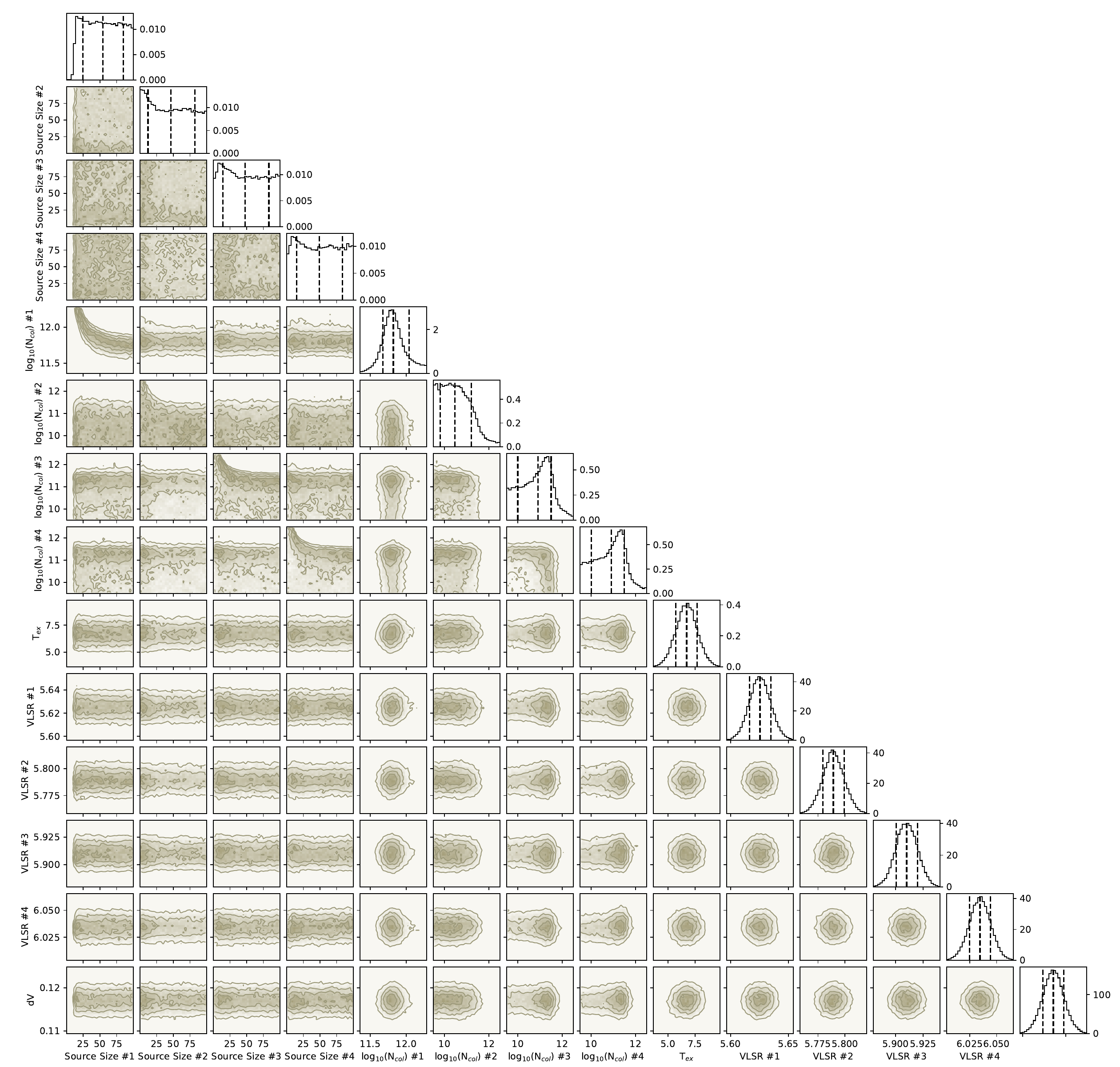}
\figsetgrpnote{The 16$^{th}$, 50$^{th}$, and 84$^{th}$ confidence intervals (corresponding to $\pm$1 sigma for a Gaussian posterior distribution) are shown as vertical lines. The contour lines are posterior probability levels, starting at $20\%$ of the maximum a posteriori estimate, with evenly spaced intervals of $20\%$ up to the peak density.}
\figsetgrpend

\figsetgrpstart
\figsetgrpnum{8.29}
\figsetgrptitle{Corner plot for HCCS$^{+}$.}
\figsetplot{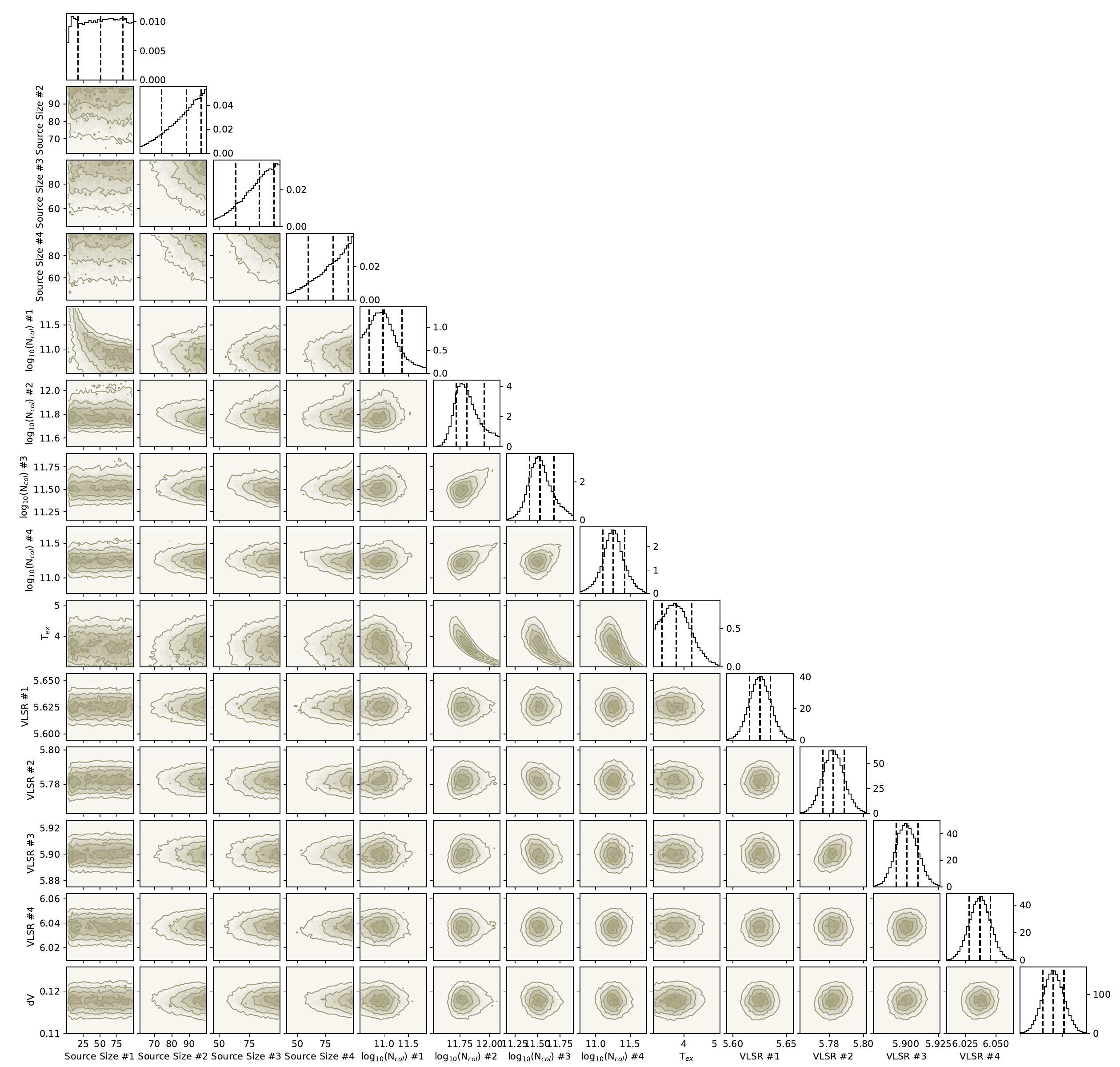}
\figsetgrpnote{The 16$^{th}$, 50$^{th}$, and 84$^{th}$ confidence intervals (corresponding to $\pm$1 sigma for a Gaussian posterior distribution) are shown as vertical lines. The contour lines are posterior probability levels, starting at $20\%$ of the maximum a posteriori estimate, with evenly spaced intervals of $20\%$ up to the peak density.}
\figsetgrpend

\figsetgrpstart
\figsetgrpnum{8.30}
\figsetgrptitle{Corner plot for C$_{5}$H.}
\figsetplot{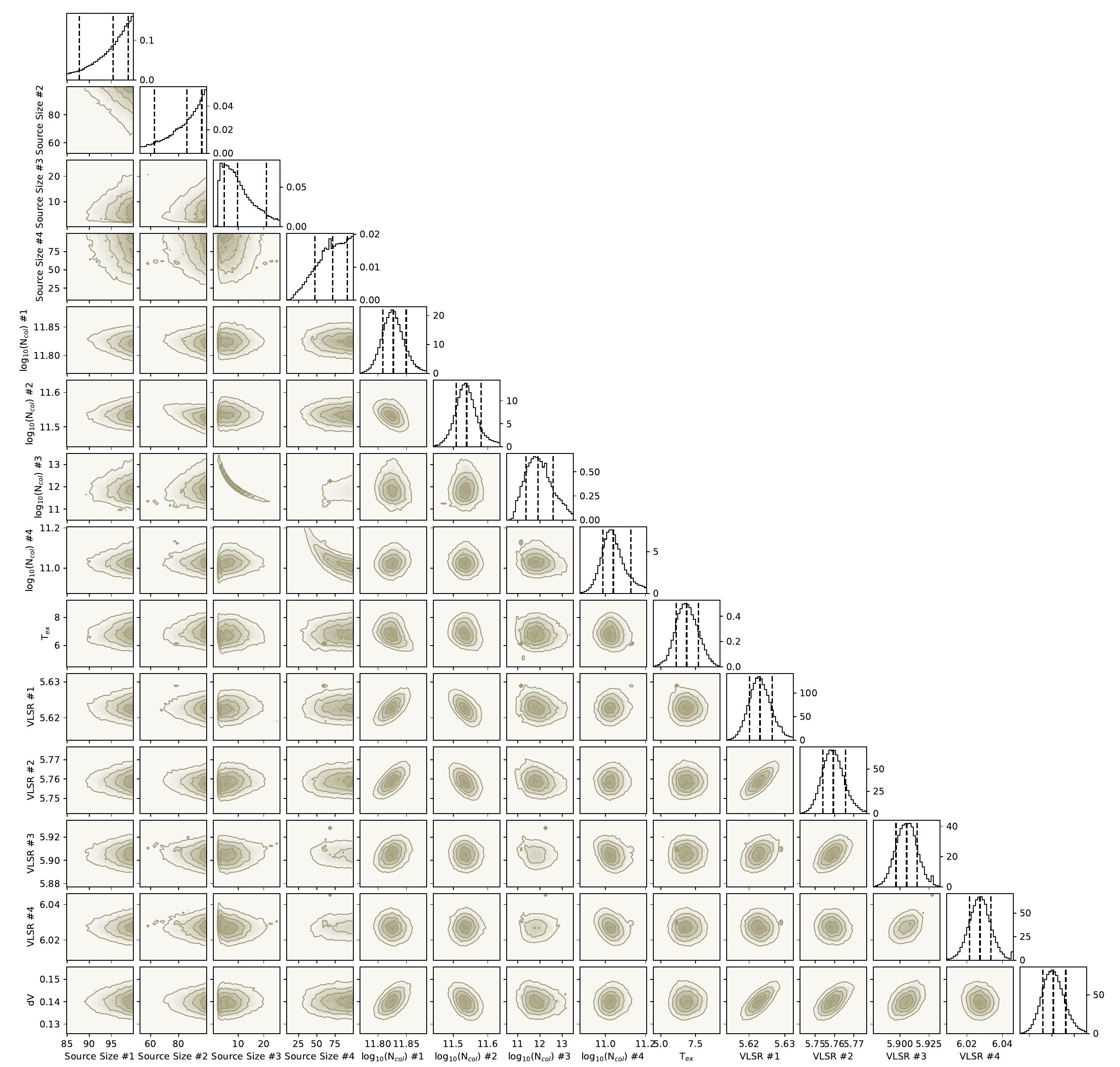}
\figsetgrpnote{The 16$^{th}$, 50$^{th}$, and 84$^{th}$ confidence intervals (corresponding to $\pm$1 sigma for a Gaussian posterior distribution) are shown as vertical lines. The contour lines are posterior probability levels, starting at $20\%$ of the maximum a posteriori estimate, with evenly spaced intervals of $20\%$ up to the peak density.}
\figsetgrpend

\figsetgrpstart
\figsetgrpnum{8.31}
\figsetgrptitle{Corner plot for $l$-HC$_{4}$N.}
\figsetplot{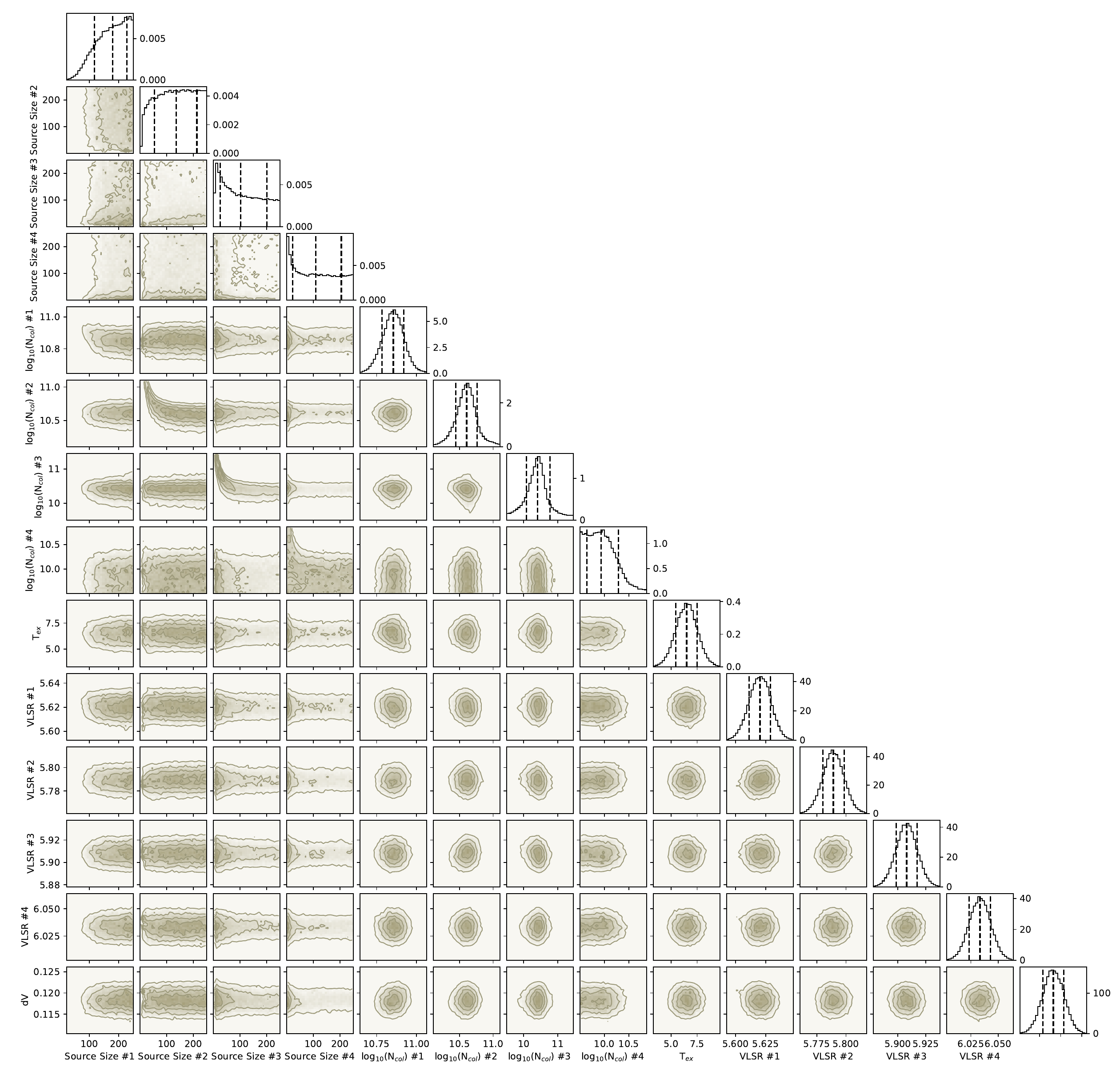}
\figsetgrpnote{The 16$^{th}$, 50$^{th}$, and 84$^{th}$ confidence intervals (corresponding to $\pm$1 sigma for a Gaussian posterior distribution) are shown as vertical lines. The contour lines are posterior probability levels, starting at $20\%$ of the maximum a posteriori estimate, with evenly spaced intervals of $20\%$ up to the peak density.}
\figsetgrpend

\figsetgrpstart
\figsetgrpnum{8.32}
\figsetgrptitle{Corner plot for H$_{2}$C$_{3}$HCCH.}
\figsetplot{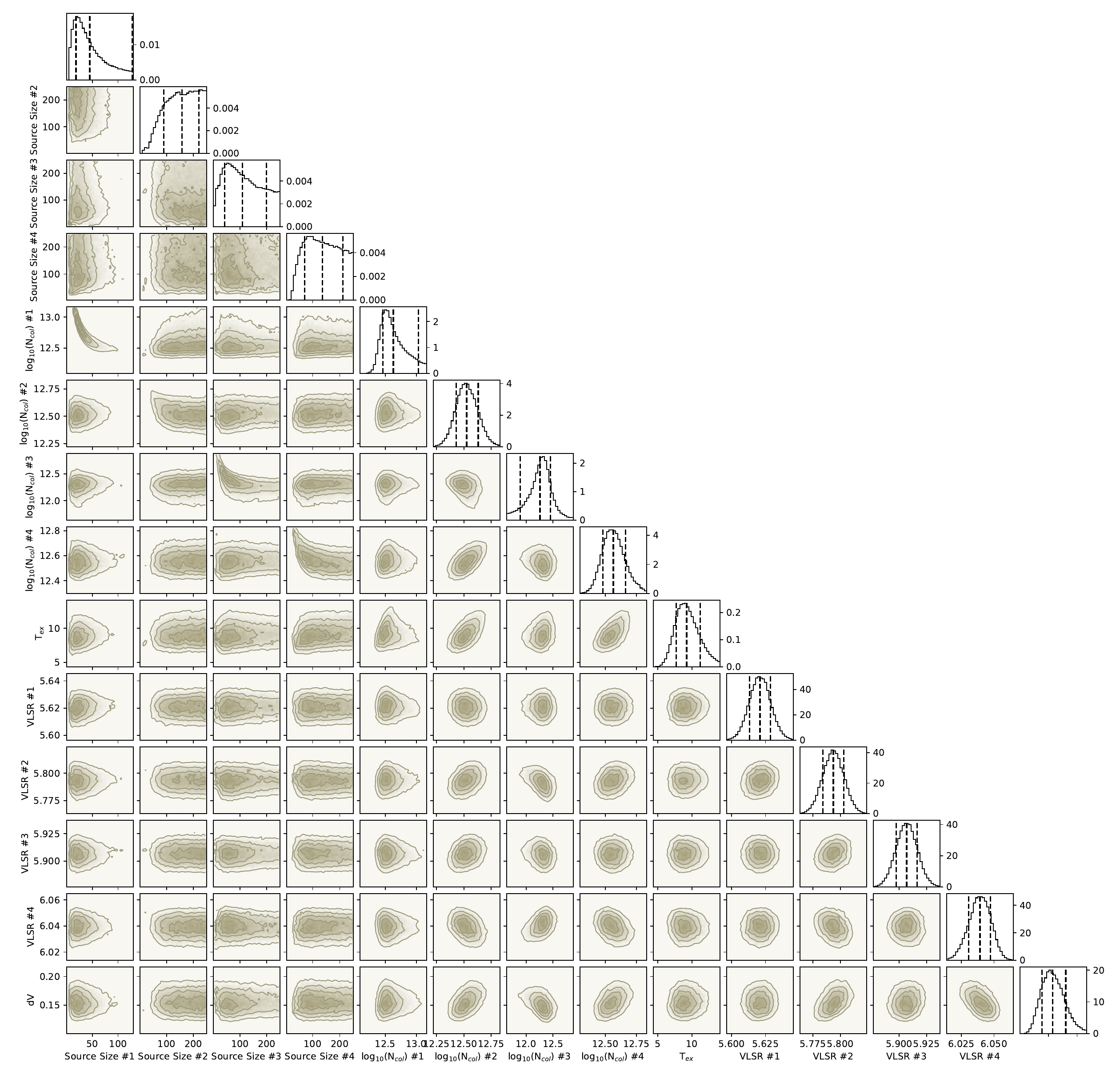}
\figsetgrpnote{The 16$^{th}$, 50$^{th}$, and 84$^{th}$ confidence intervals (corresponding to $\pm$1 sigma for a Gaussian posterior distribution) are shown as vertical lines. The contour lines are posterior probability levels, starting at $20\%$ of the maximum a posteriori estimate, with evenly spaced intervals of $20\%$ up to the peak density.}
\figsetgrpend

\figsetgrpstart
\figsetgrpnum{8.33}
\figsetgrptitle{Corner plot for CH$_{3}$C$_{4}$H $A$.}
\figsetplot{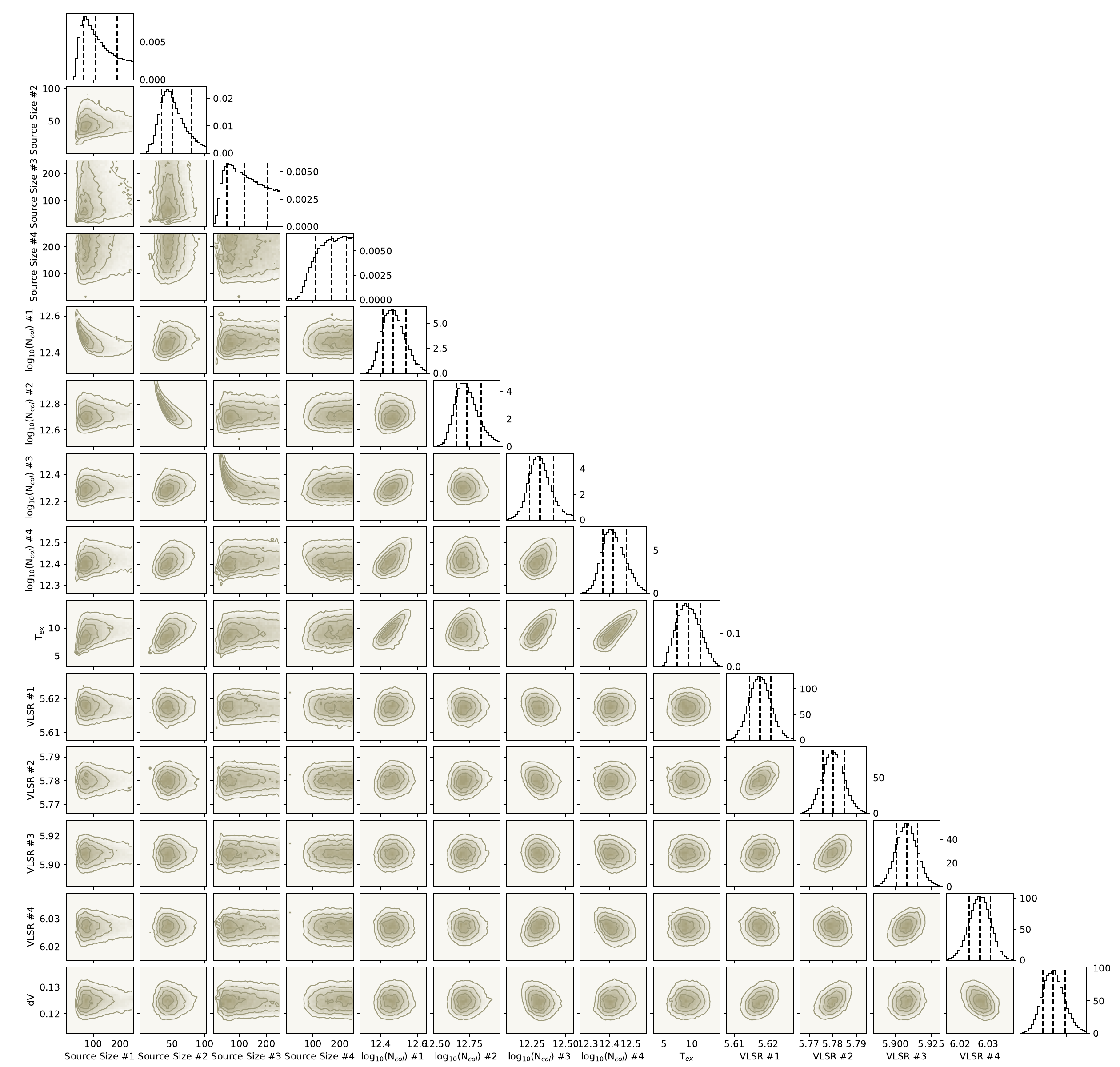}
\figsetgrpnote{The 16$^{th}$, 50$^{th}$, and 84$^{th}$ confidence intervals (corresponding to $\pm$1 sigma for a Gaussian posterior distribution) are shown as vertical lines. The contour lines are posterior probability levels, starting at $20\%$ of the maximum a posteriori estimate, with evenly spaced intervals of $20\%$ up to the peak density.}
\figsetgrpend

\figsetgrpstart
\figsetgrpnum{8.34}
\figsetgrptitle{Corner plot for CH$_{3}$C$_{4}$H $E$.}
\figsetplot{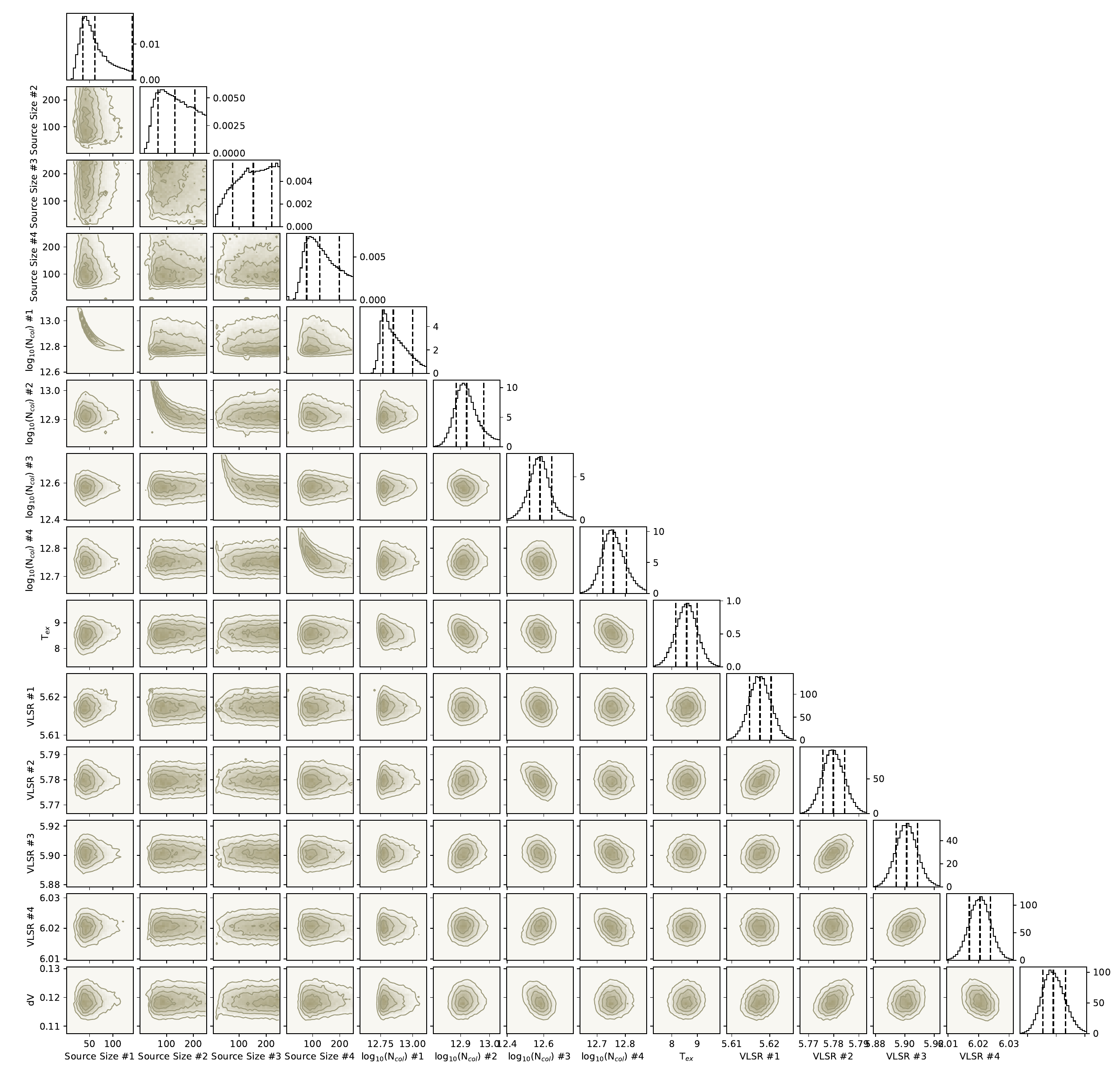}
\figsetgrpnote{The 16$^{th}$, 50$^{th}$, and 84$^{th}$ confidence intervals (corresponding to $\pm$1 sigma for a Gaussian posterior distribution) are shown as vertical lines. The contour lines are posterior probability levels, starting at $20\%$ of the maximum a posteriori estimate, with evenly spaced intervals of $20\%$ up to the peak density.}
\figsetgrpend

\figsetgrpstart
\figsetgrpnum{8.35}
\figsetgrptitle{Corner plot for H$_{2}$CCCHCN.}
\figsetplot{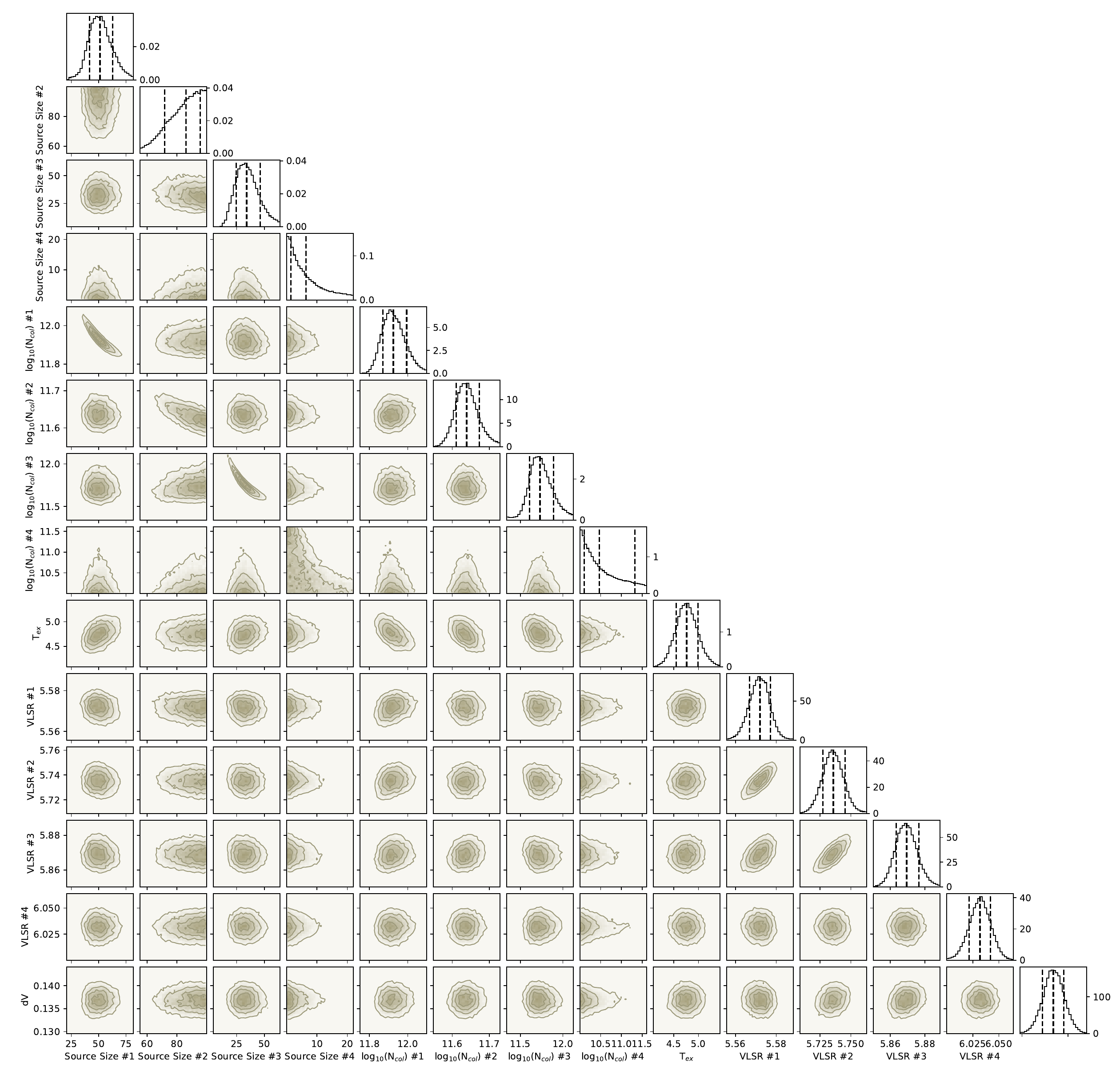}
\figsetgrpnote{The 16$^{th}$, 50$^{th}$, and 84$^{th}$ confidence intervals (corresponding to $\pm$1 sigma for a Gaussian posterior distribution) are shown as vertical lines. The contour lines are posterior probability levels, starting at $20\%$ of the maximum a posteriori estimate, with evenly spaced intervals of $20\%$ up to the peak density.}
\figsetgrpend

\figsetgrpstart
\figsetgrpnum{8.36}
\figsetgrptitle{Corner plot for HCCCH$_{2}$CN.}
\figsetplot{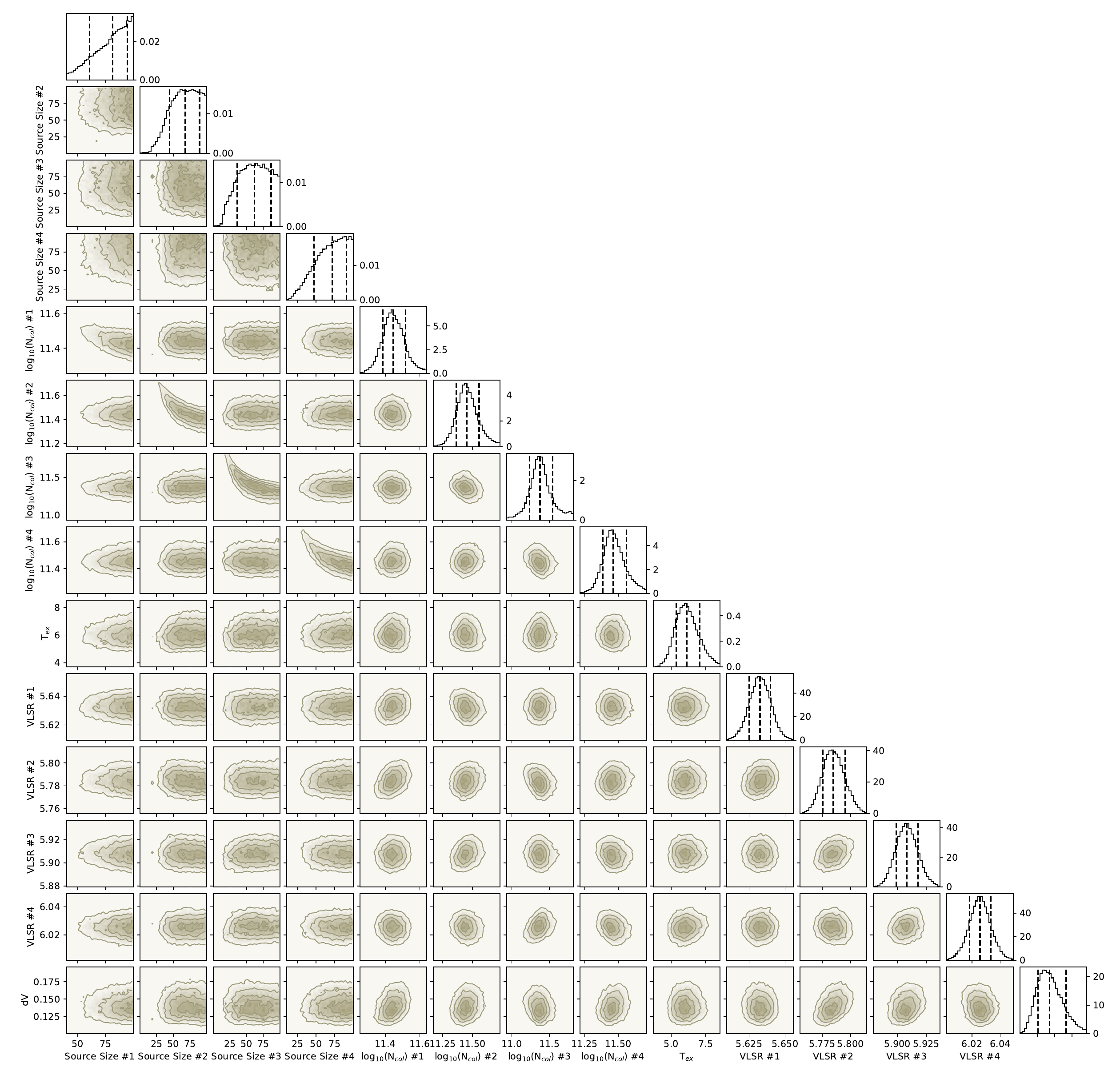}
\figsetgrpnote{The 16$^{th}$, 50$^{th}$, and 84$^{th}$ confidence intervals (corresponding to $\pm$1 sigma for a Gaussian posterior distribution) are shown as vertical lines. The contour lines are posterior probability levels, starting at $20\%$ of the maximum a posteriori estimate, with evenly spaced intervals of $20\%$ up to the peak density.}
\figsetgrpend

\figsetgrpstart
\figsetgrpnum{8.37}
\figsetgrptitle{Corner plot for CH$_{3}$C$_{3}$N $A$.}
\figsetplot{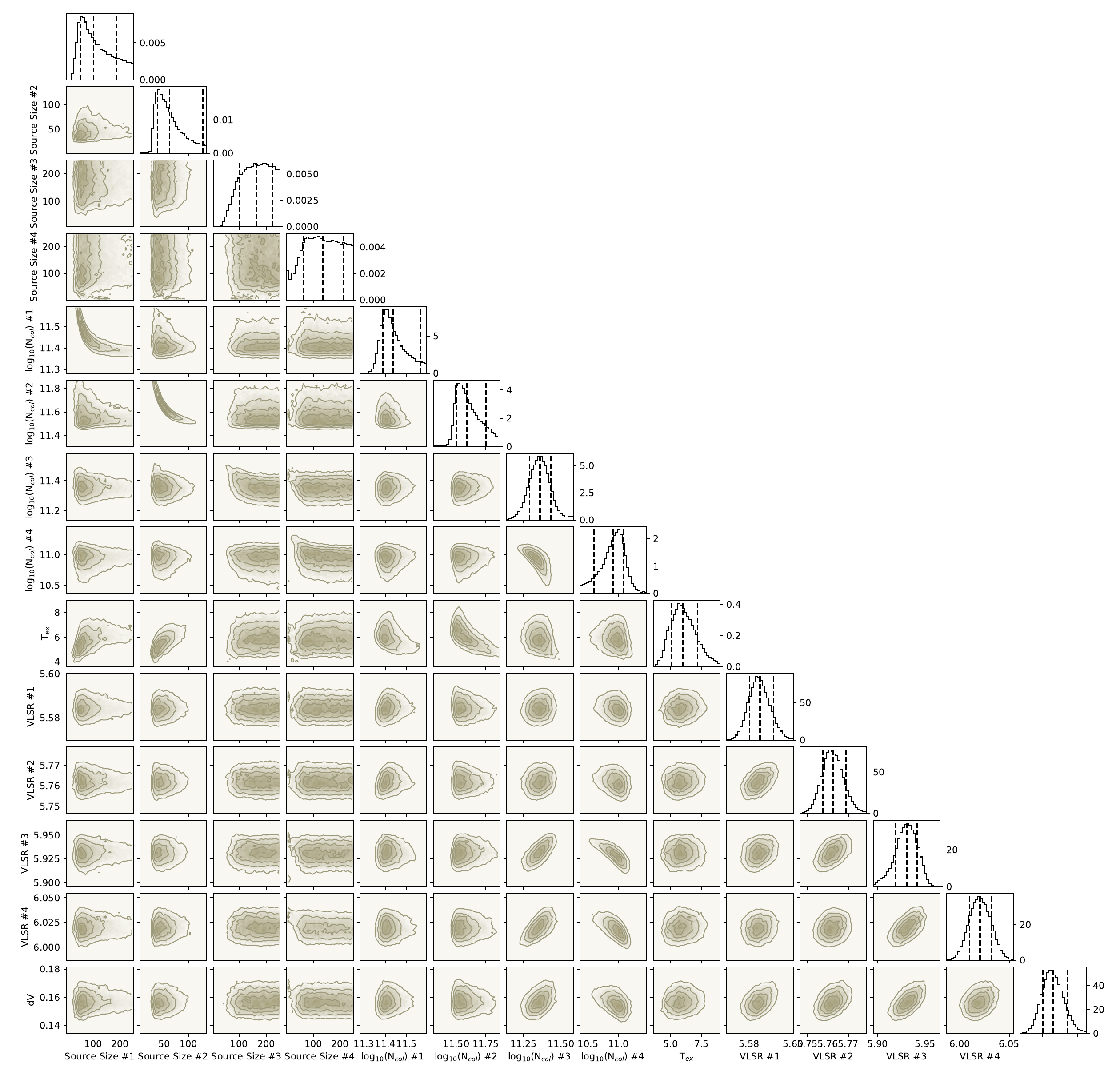}
\figsetgrpnote{The 16$^{th}$, 50$^{th}$, and 84$^{th}$ confidence intervals (corresponding to $\pm$1 sigma for a Gaussian posterior distribution) are shown as vertical lines. The contour lines are posterior probability levels, starting at $20\%$ of the maximum a posteriori estimate, with evenly spaced intervals of $20\%$ up to the peak density.}
\figsetgrpend

\figsetgrpstart
\figsetgrpnum{8.38}
\figsetgrptitle{Corner plot for CH$_{3}$C$_{3}$N $E$.}
\figsetplot{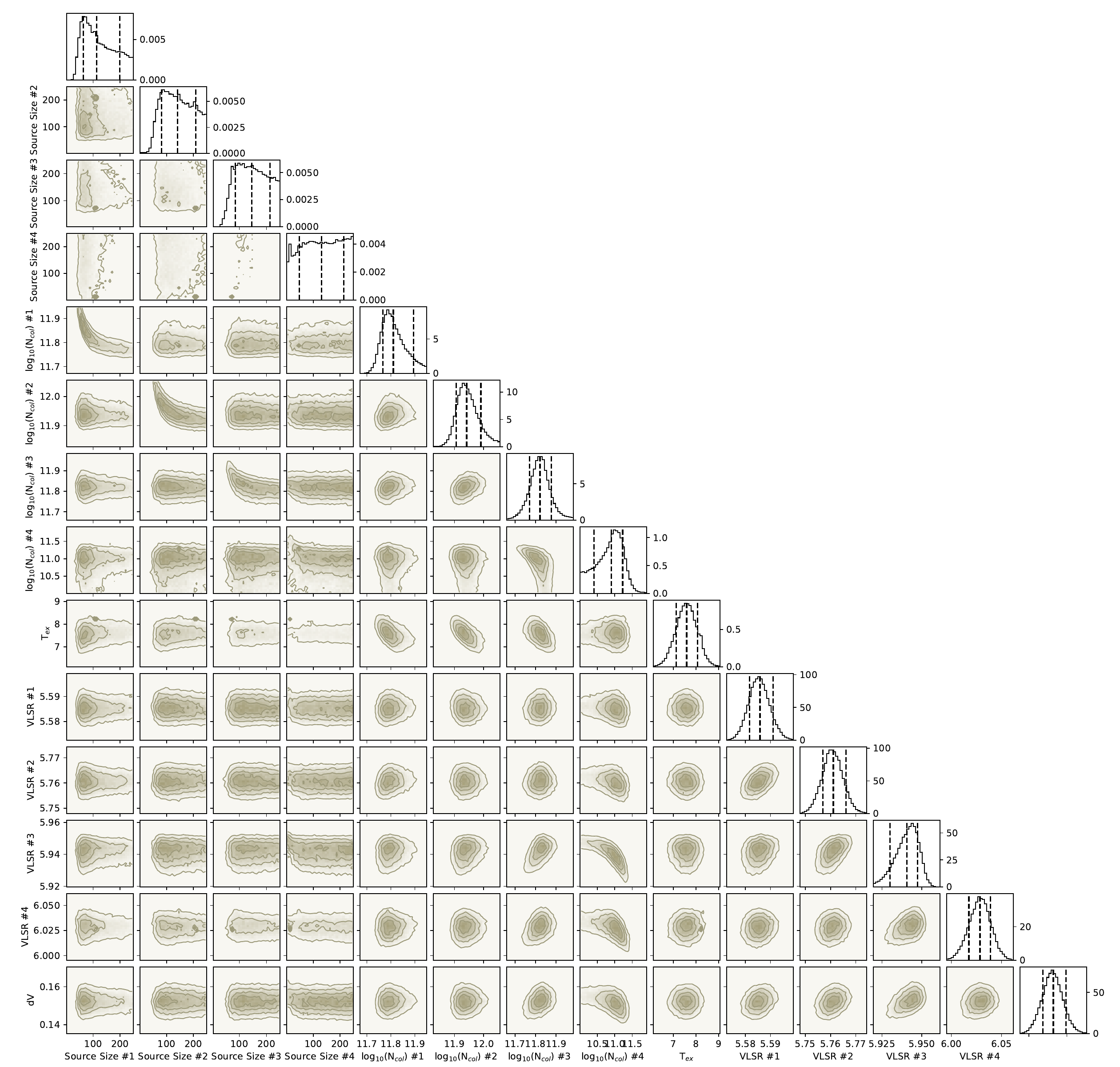}
\figsetgrpnote{The 16$^{th}$, 50$^{th}$, and 84$^{th}$ confidence intervals (corresponding to $\pm$1 sigma for a Gaussian posterior distribution) are shown as vertical lines. The contour lines are posterior probability levels, starting at $20\%$ of the maximum a posteriori estimate, with evenly spaced intervals of $20\%$ up to the peak density.}
\figsetgrpend

\figsetgrpstart
\figsetgrpnum{8.39}
\figsetgrptitle{Corner plot for C$_{3}$S.}
\figsetplot{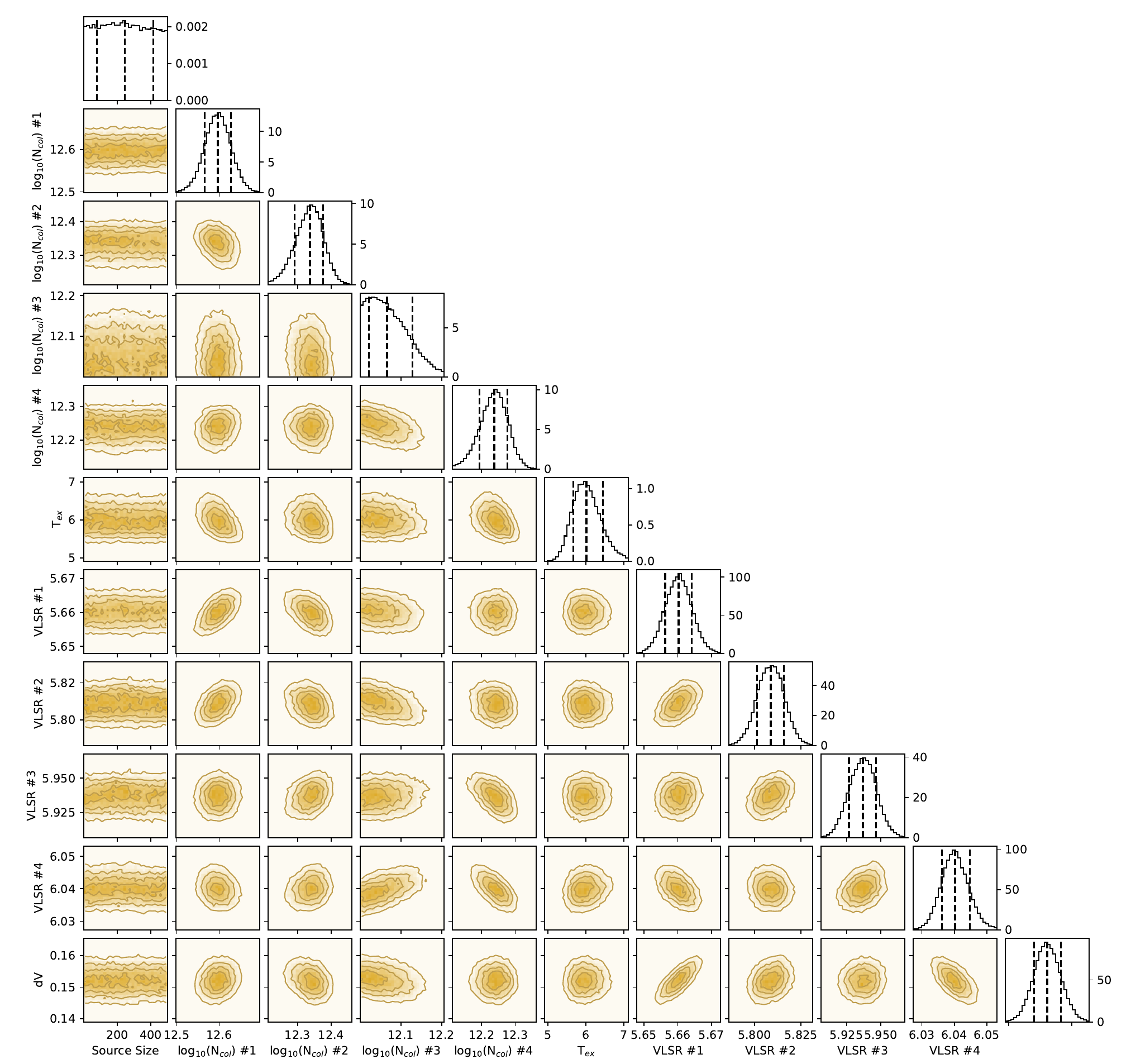}
\figsetgrpnote{The 16$^{th}$, 50$^{th}$, and 84$^{th}$ confidence intervals (corresponding to $\pm$1 sigma for a Gaussian posterior distribution) are shown as vertical lines. The contour lines are posterior probability levels, starting at $20\%$ of the maximum a posteriori estimate, with evenly spaced intervals of $20\%$ up to the peak density.}
\figsetgrpend

\figsetgrpstart
\figsetgrpnum{8.40}
\figsetgrptitle{Corner plot for H$_{2}$C$_{3}$S.}
\figsetplot{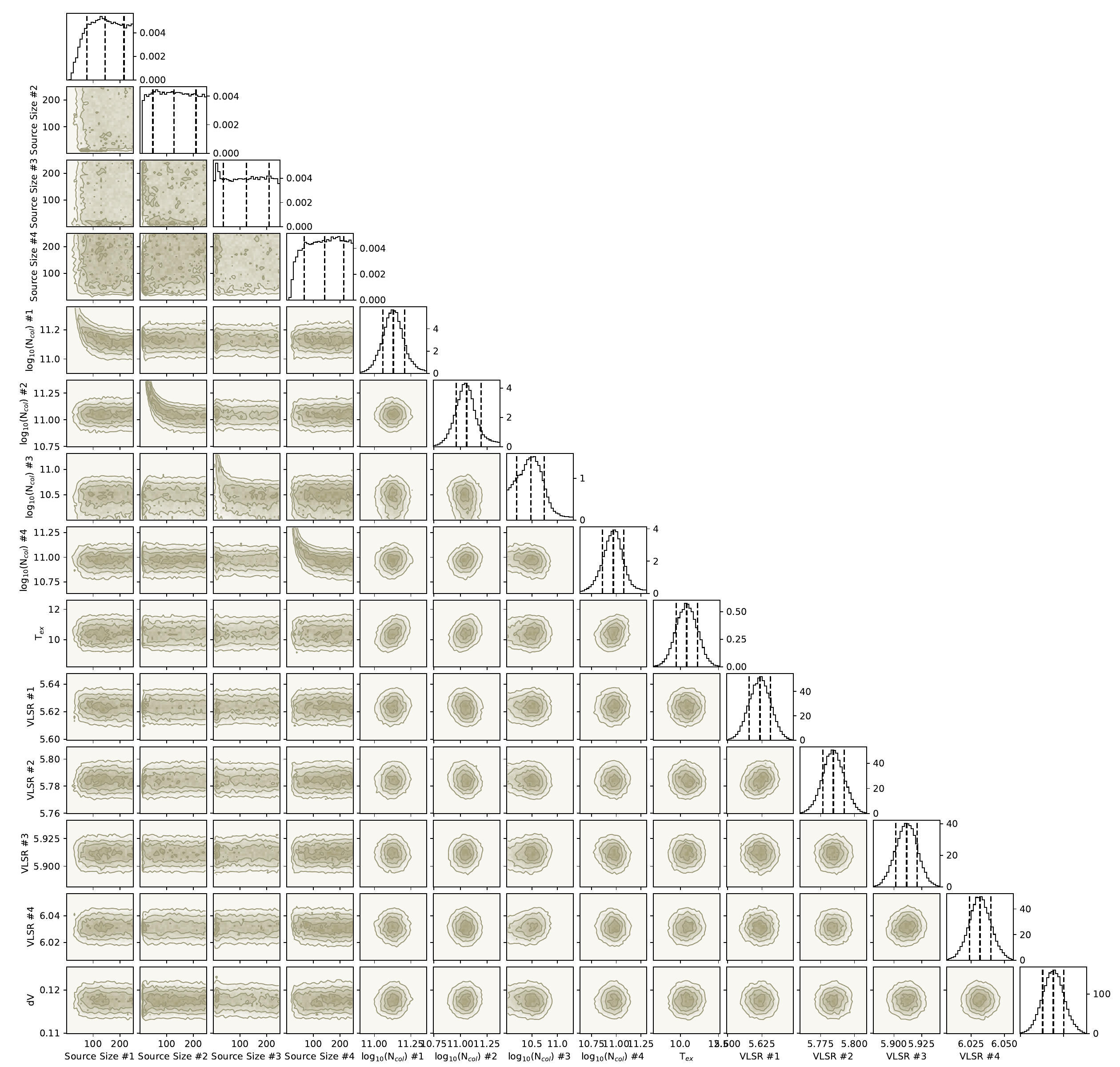}
\figsetgrpnote{The 16$^{th}$, 50$^{th}$, and 84$^{th}$ confidence intervals (corresponding to $\pm$1 sigma for a Gaussian posterior distribution) are shown as vertical lines. The contour lines are posterior probability levels, starting at $20\%$ of the maximum a posteriori estimate, with evenly spaced intervals of $20\%$ up to the peak density.}
\figsetgrpend

\figsetgrpstart
\figsetgrpnum{8.41}
\figsetgrptitle{Corner plot for $c$-H$_{2}$C$_{3}$S.}
\figsetplot{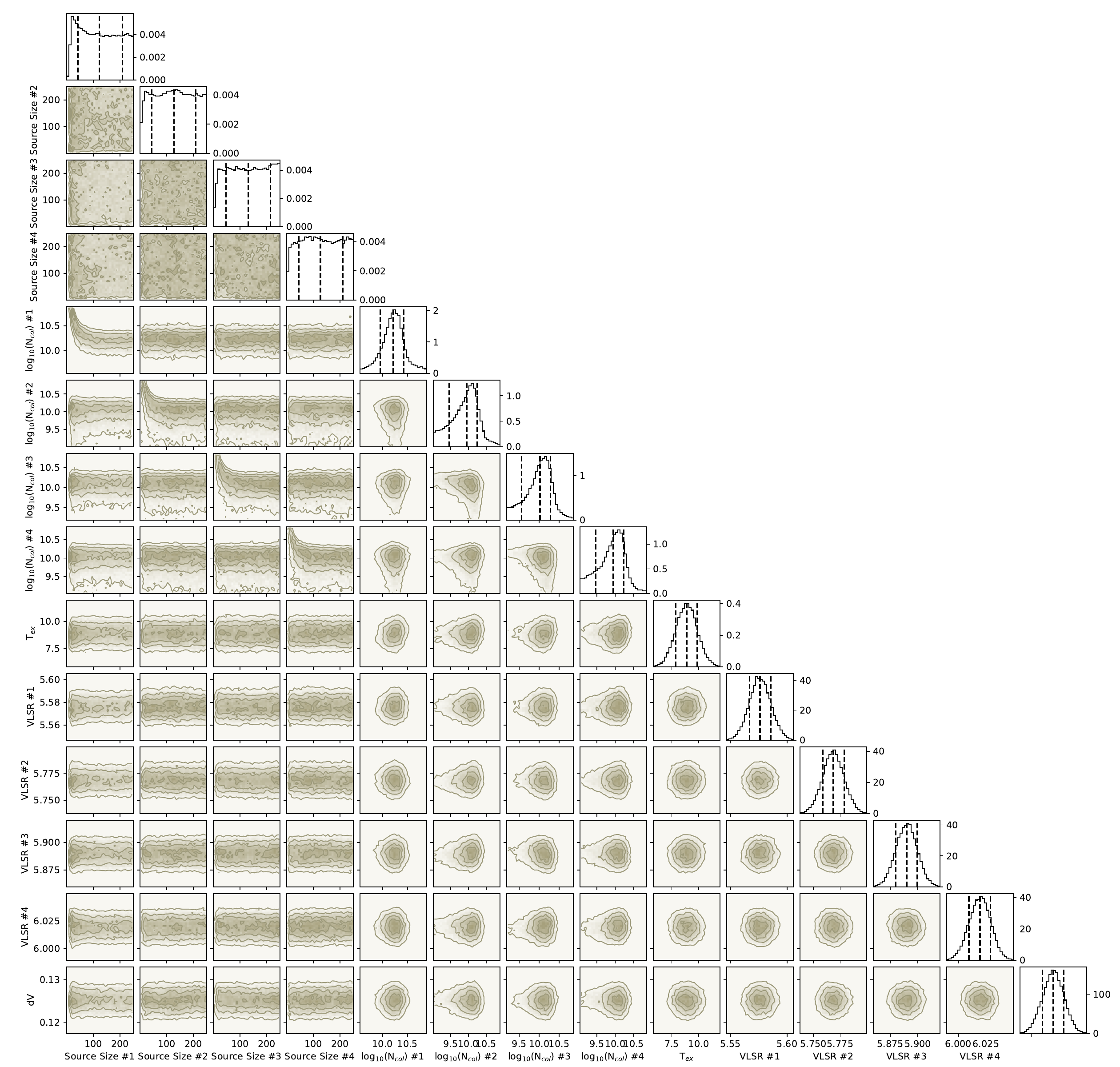}
\figsetgrpnote{The 16$^{th}$, 50$^{th}$, and 84$^{th}$ confidence intervals (corresponding to $\pm$1 sigma for a Gaussian posterior distribution) are shown as vertical lines. The contour lines are posterior probability levels, starting at $20\%$ of the maximum a posteriori estimate, with evenly spaced intervals of $20\%$ up to the peak density.}
\figsetgrpend

\figsetgrpstart
\figsetgrpnum{8.42}
\figsetgrptitle{Corner plot for C$_{6}$H$^{-}$.}
\figsetplot{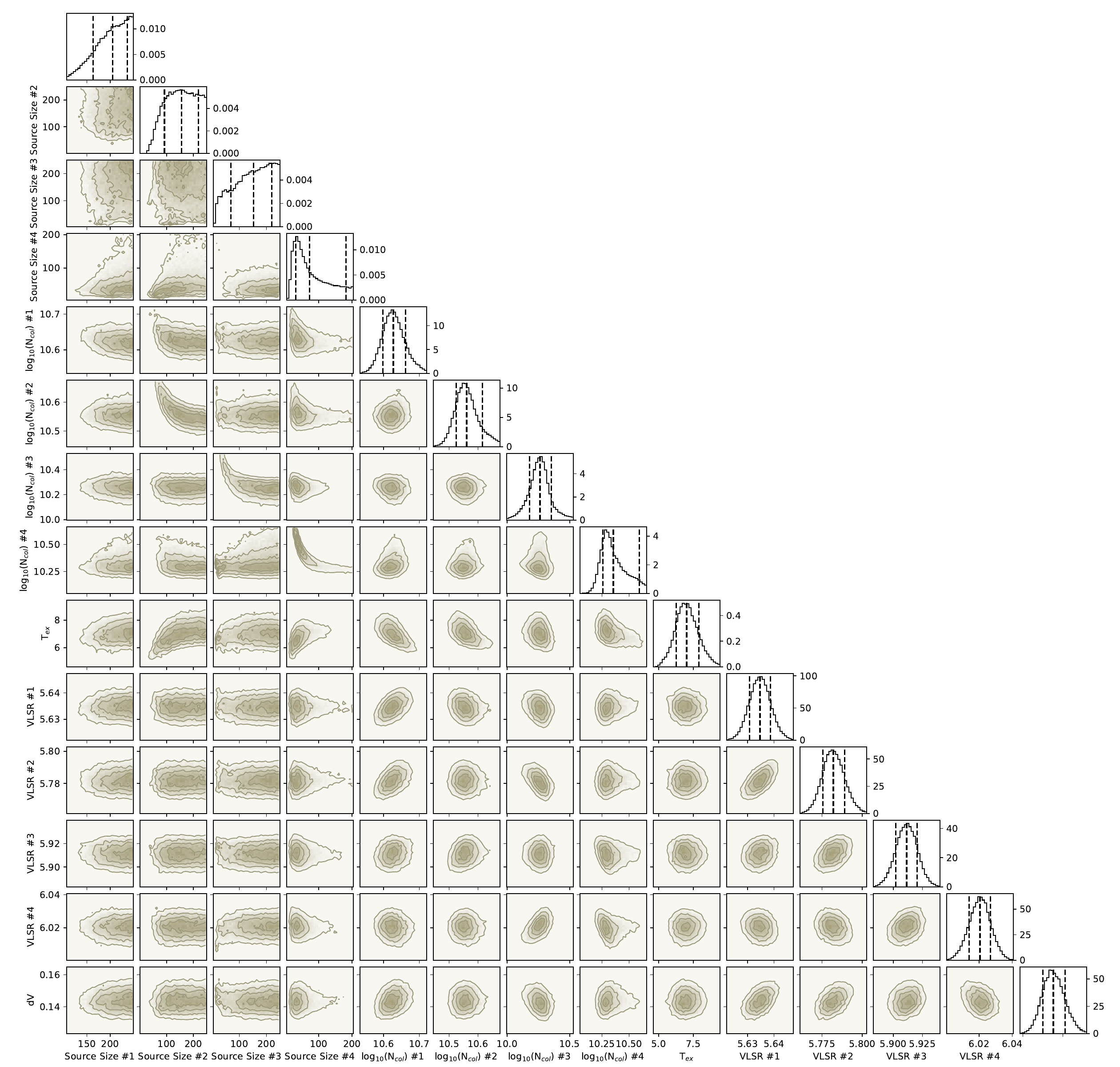}
\figsetgrpnote{The 16$^{th}$, 50$^{th}$, and 84$^{th}$ confidence intervals (corresponding to $\pm$1 sigma for a Gaussian posterior distribution) are shown as vertical lines. The contour lines are posterior probability levels, starting at $20\%$ of the maximum a posteriori estimate, with evenly spaced intervals of $20\%$ up to the peak density.}
\figsetgrpend

\figsetgrpstart
\figsetgrpnum{8.43}
\figsetgrptitle{Corner plot for C$_{6}$H.}
\figsetplot{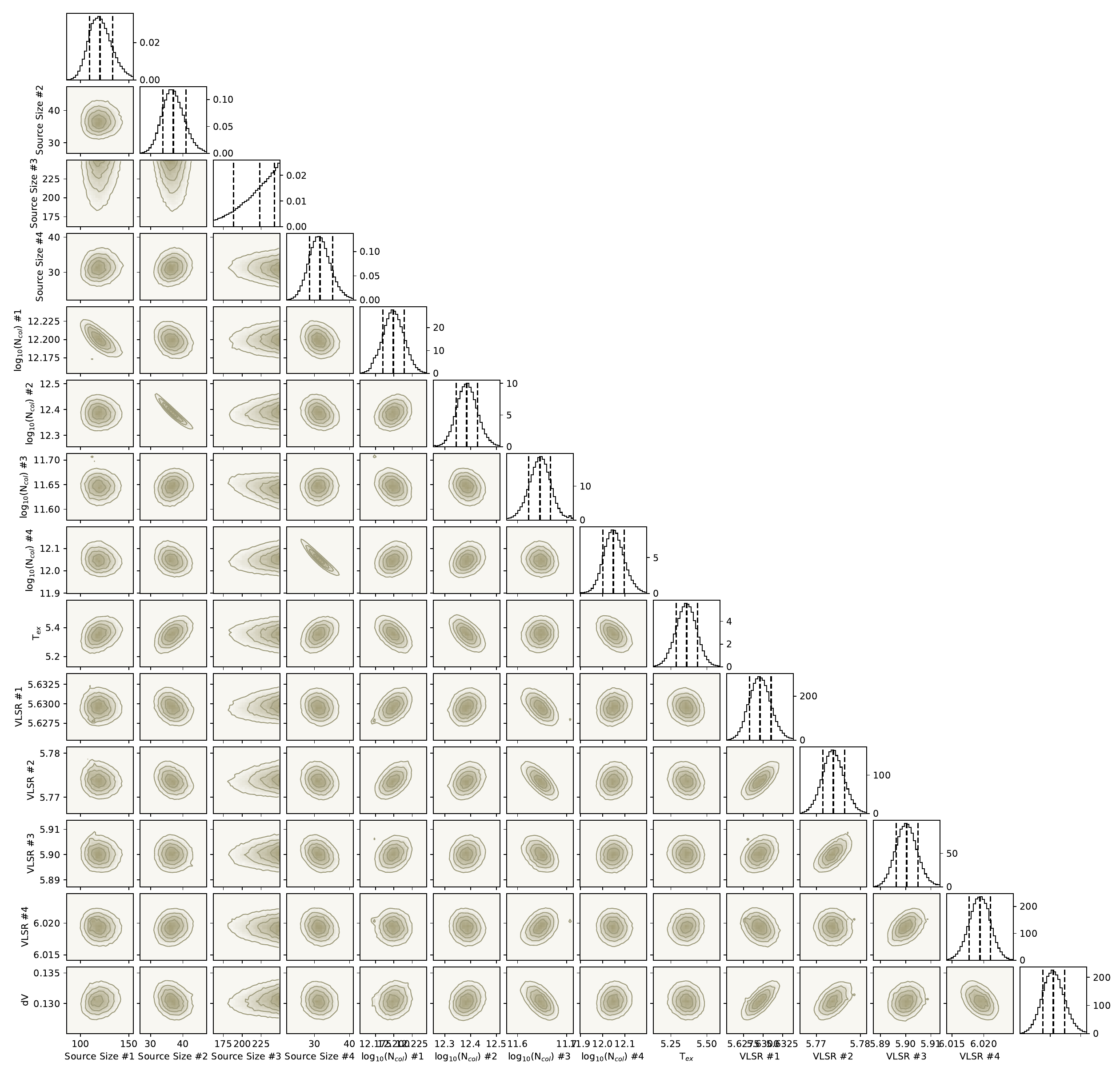}
\figsetgrpnote{The 16$^{th}$, 50$^{th}$, and 84$^{th}$ confidence intervals (corresponding to $\pm$1 sigma for a Gaussian posterior distribution) are shown as vertical lines. The contour lines are posterior probability levels, starting at $20\%$ of the maximum a posteriori estimate, with evenly spaced intervals of $20\%$ up to the peak density.}
\figsetgrpend

\figsetgrpstart
\figsetgrpnum{8.44}
\figsetgrptitle{Corner plot for C$_{5}$N.}
\figsetplot{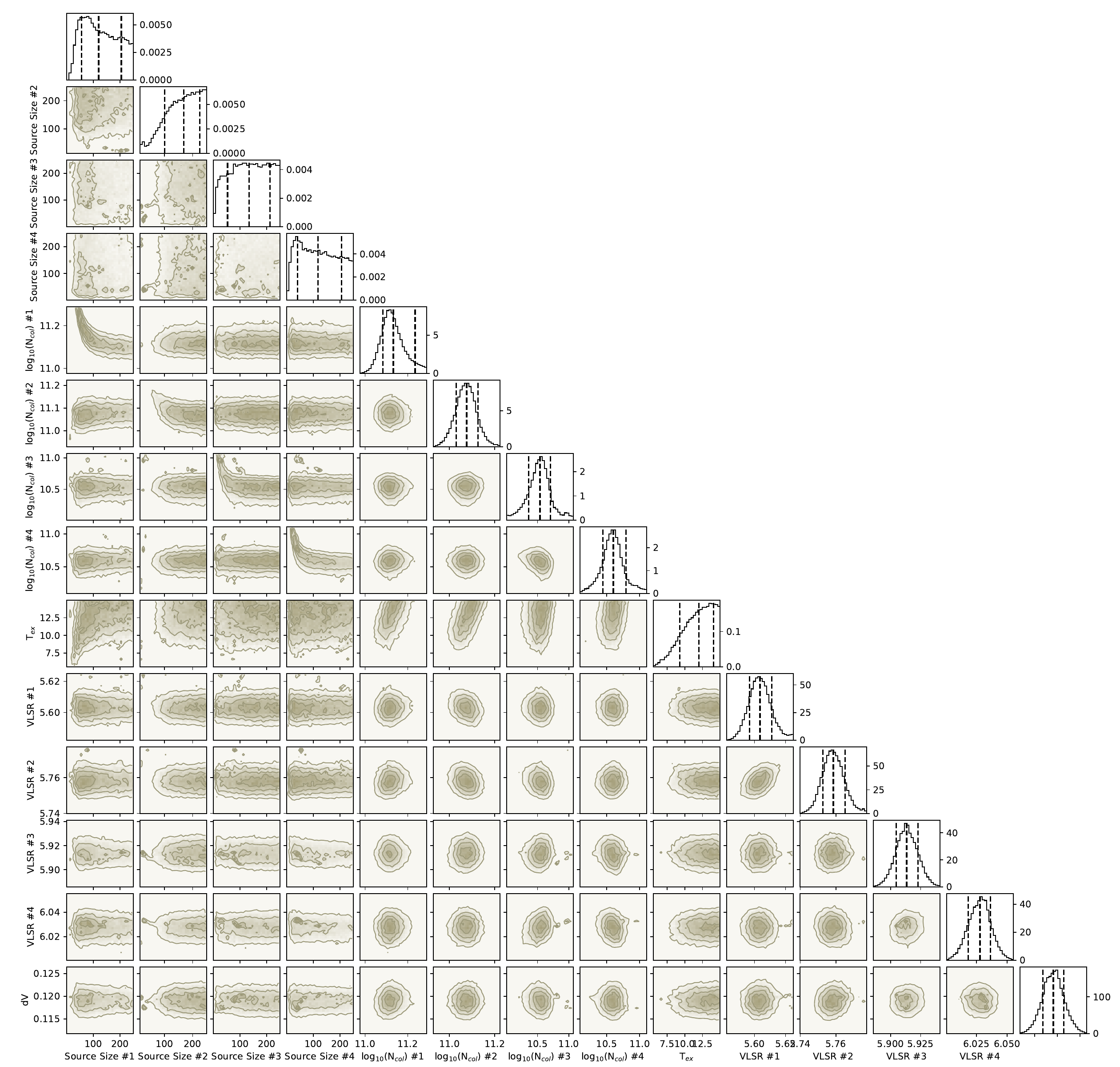}
\figsetgrpnote{The 16$^{th}$, 50$^{th}$, and 84$^{th}$ confidence intervals (corresponding to $\pm$1 sigma for a Gaussian posterior distribution) are shown as vertical lines. The contour lines are posterior probability levels, starting at $20\%$ of the maximum a posteriori estimate, with evenly spaced intervals of $20\%$ up to the peak density.}
\figsetgrpend

\figsetgrpstart
\figsetgrpnum{8.45}
\figsetgrptitle{Corner plot for $l$-C$_{6}$H$_{2}$.}
\figsetplot{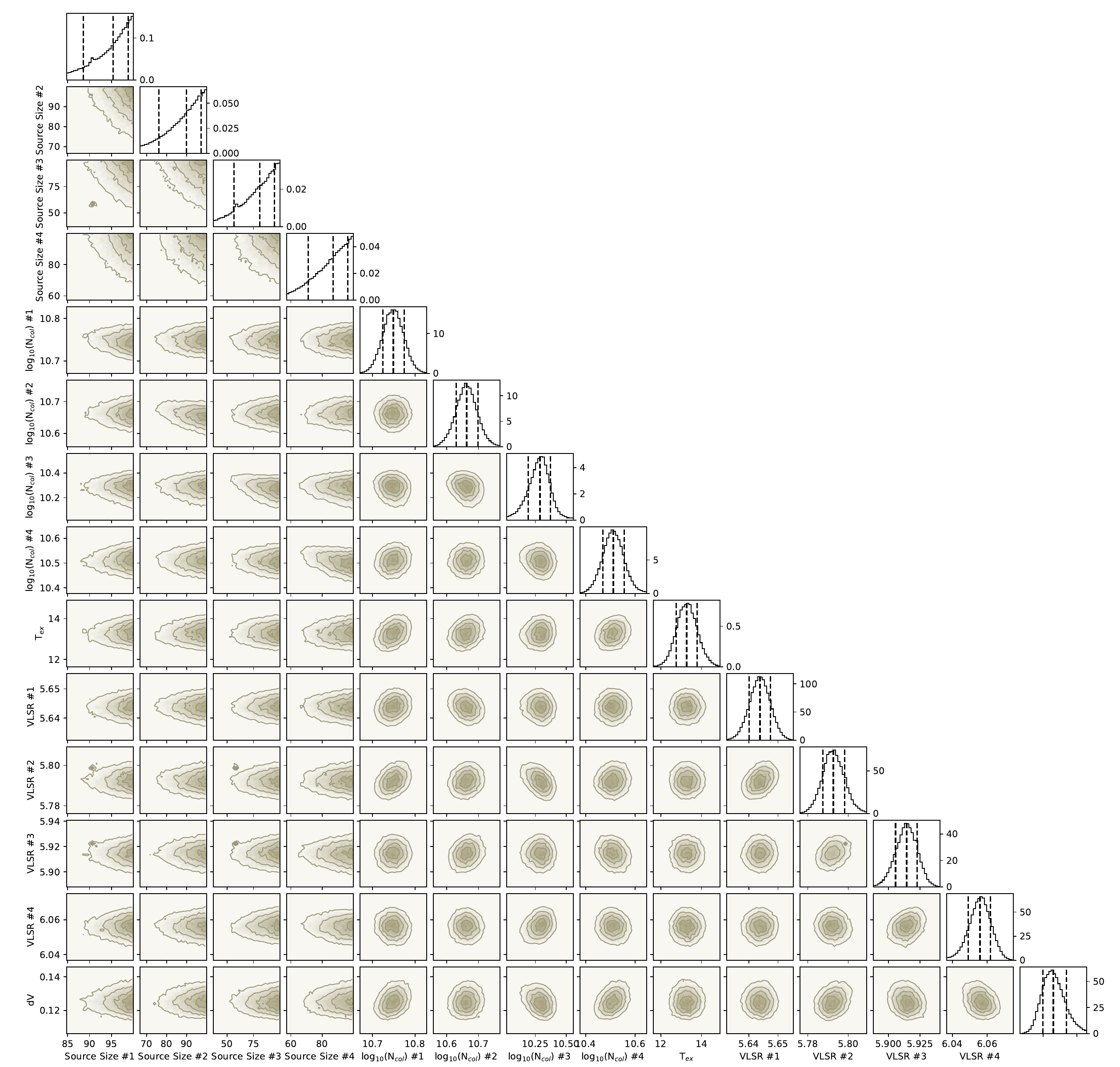}
\figsetgrpnote{The 16$^{th}$, 50$^{th}$, and 84$^{th}$ confidence intervals (corresponding to $\pm$1 sigma for a Gaussian posterior distribution) are shown as vertical lines. The contour lines are posterior probability levels, starting at $20\%$ of the maximum a posteriori estimate, with evenly spaced intervals of $20\%$ up to the peak density.}
\figsetgrpend

\figsetgrpstart
\figsetgrpnum{8.46}
\figsetgrptitle{Corner plot for C$_{5}$N$^{-}$.}
\figsetplot{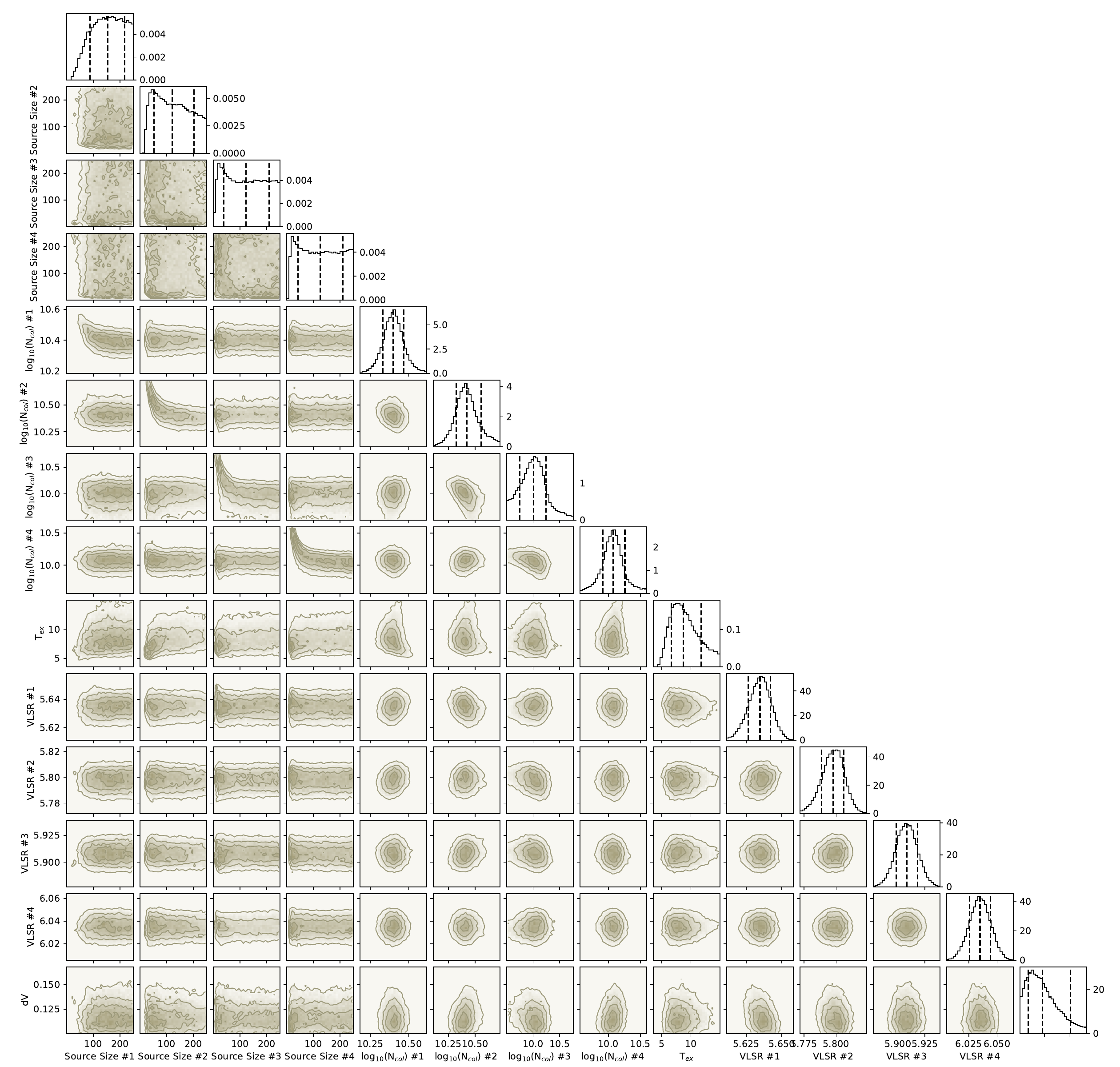}
\figsetgrpnote{The 16$^{th}$, 50$^{th}$, and 84$^{th}$ confidence intervals (corresponding to $\pm$1 sigma for a Gaussian posterior distribution) are shown as vertical lines. The contour lines are posterior probability levels, starting at $20\%$ of the maximum a posteriori estimate, with evenly spaced intervals of $20\%$ up to the peak density.}
\figsetgrpend

\figsetgrpstart
\figsetgrpnum{8.47}
\figsetgrptitle{Corner plot for HC$_{4}$NC.}
\figsetplot{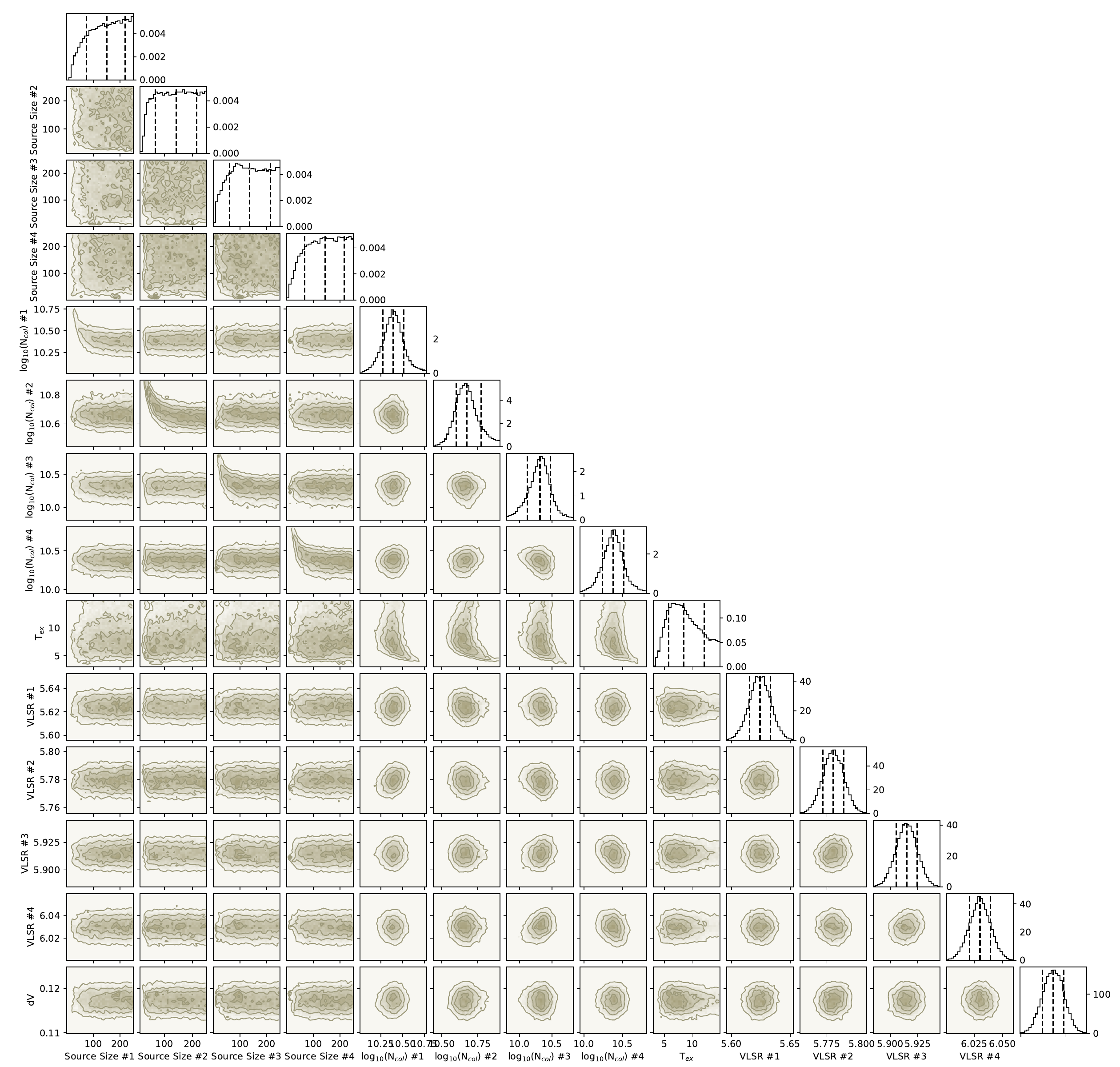}
\figsetgrpnote{The 16$^{th}$, 50$^{th}$, and 84$^{th}$ confidence intervals (corresponding to $\pm$1 sigma for a Gaussian posterior distribution) are shown as vertical lines. The contour lines are posterior probability levels, starting at $20\%$ of the maximum a posteriori estimate, with evenly spaced intervals of $20\%$ up to the peak density.}
\figsetgrpend

\figsetgrpstart
\figsetgrpnum{8.48}
\figsetgrptitle{Corner plot for HC$_{5}$N.}
\figsetplot{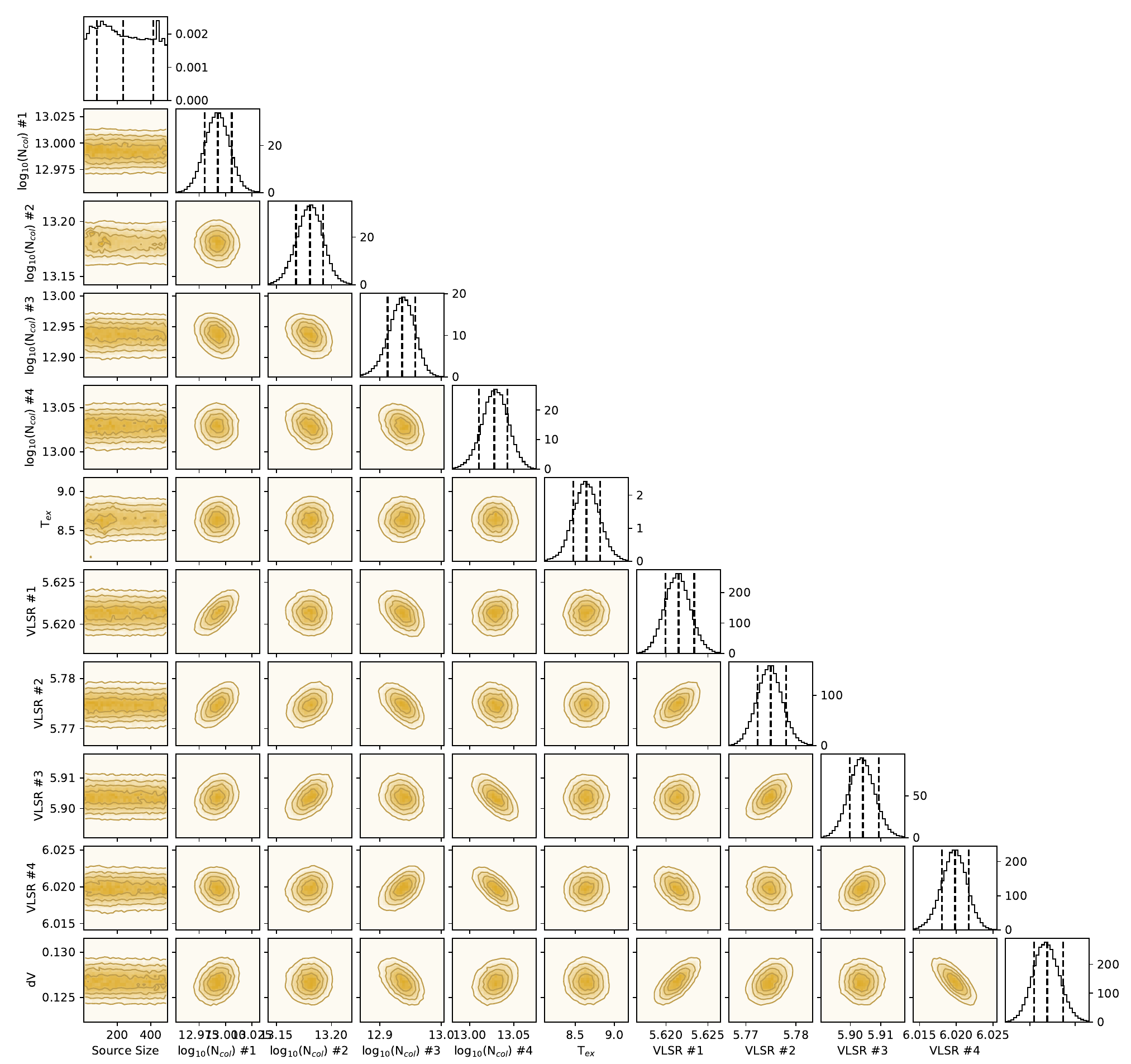}
\figsetgrpnote{The 16$^{th}$, 50$^{th}$, and 84$^{th}$ confidence intervals (corresponding to $\pm$1 sigma for a Gaussian posterior distribution) are shown as vertical lines. The contour lines are posterior probability levels, starting at $20\%$ of the maximum a posteriori estimate, with evenly spaced intervals of $20\%$ up to the peak density.}
\figsetgrpend

\figsetgrpstart
\figsetgrpnum{8.49}
\figsetgrptitle{Corner plot for C$_{2}$H$_{3}$C$_{3}$N.}
\figsetplot{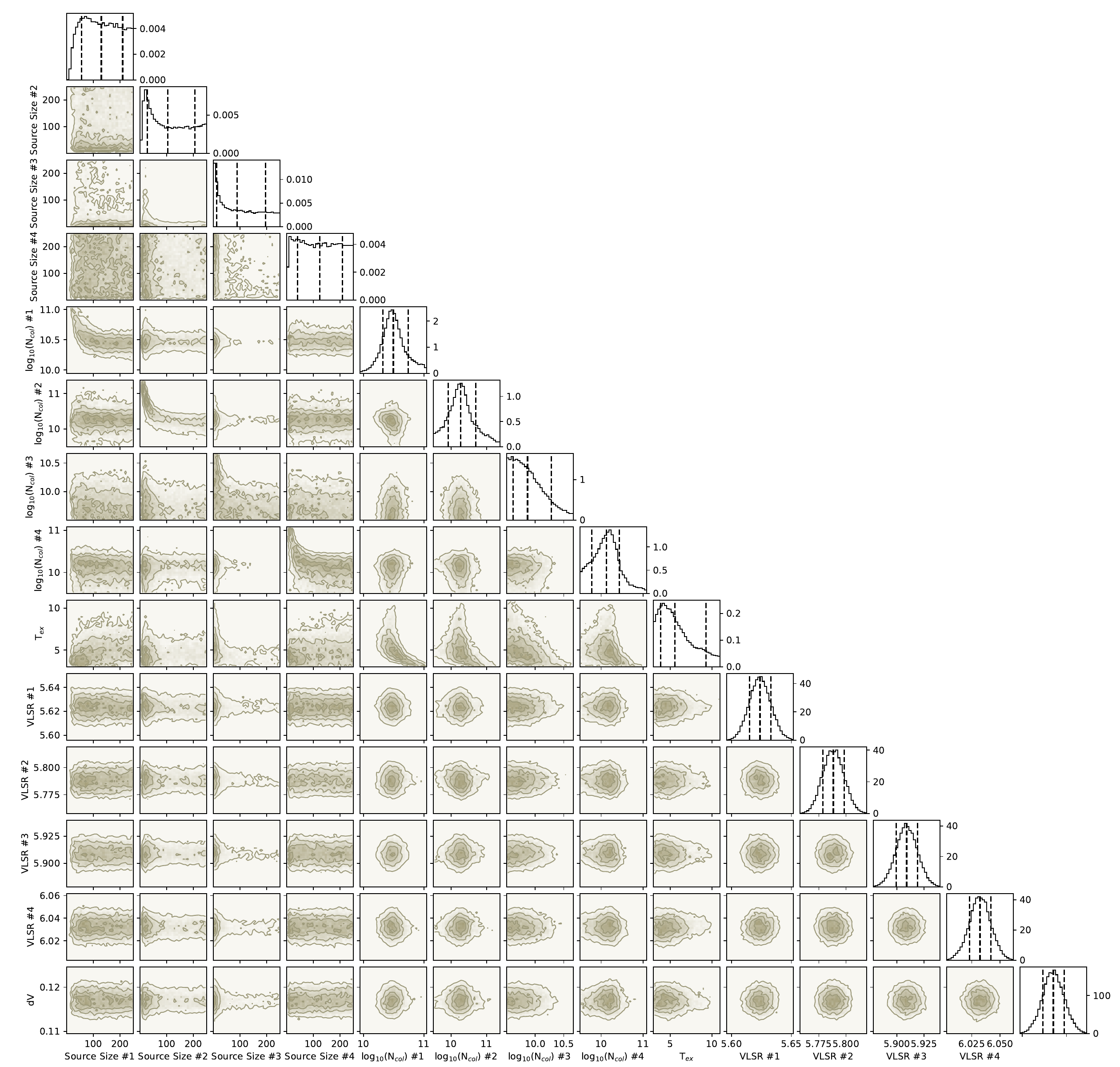}
\figsetgrpnote{The 16$^{th}$, 50$^{th}$, and 84$^{th}$ confidence intervals (corresponding to $\pm$1 sigma for a Gaussian posterior distribution) are shown as vertical lines. The contour lines are posterior probability levels, starting at $20\%$ of the maximum a posteriori estimate, with evenly spaced intervals of $20\%$ up to the peak density.}
\figsetgrpend

\figsetgrpstart
\figsetgrpnum{8.50}
\figsetgrptitle{Corner plot for HC$_{5}$O.}
\figsetplot{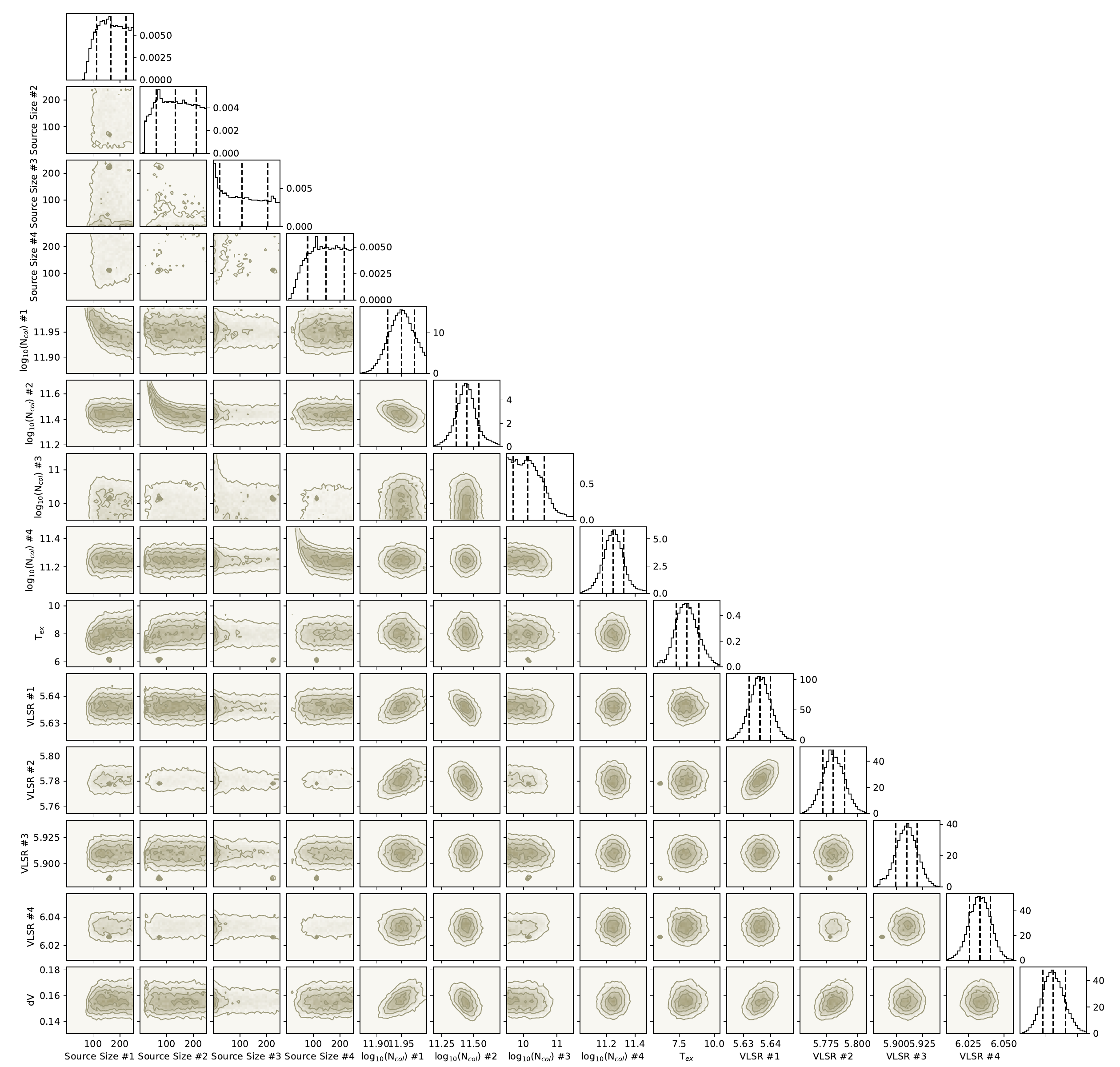}
\figsetgrpnote{The 16$^{th}$, 50$^{th}$, and 84$^{th}$ confidence intervals (corresponding to $\pm$1 sigma for a Gaussian posterior distribution) are shown as vertical lines. The contour lines are posterior probability levels, starting at $20\%$ of the maximum a posteriori estimate, with evenly spaced intervals of $20\%$ up to the peak density.}
\figsetgrpend

\figsetgrpstart
\figsetgrpnum{8.51}
\figsetgrptitle{Corner plot for NC$_{4}$NH$^{+}$.}
\figsetplot{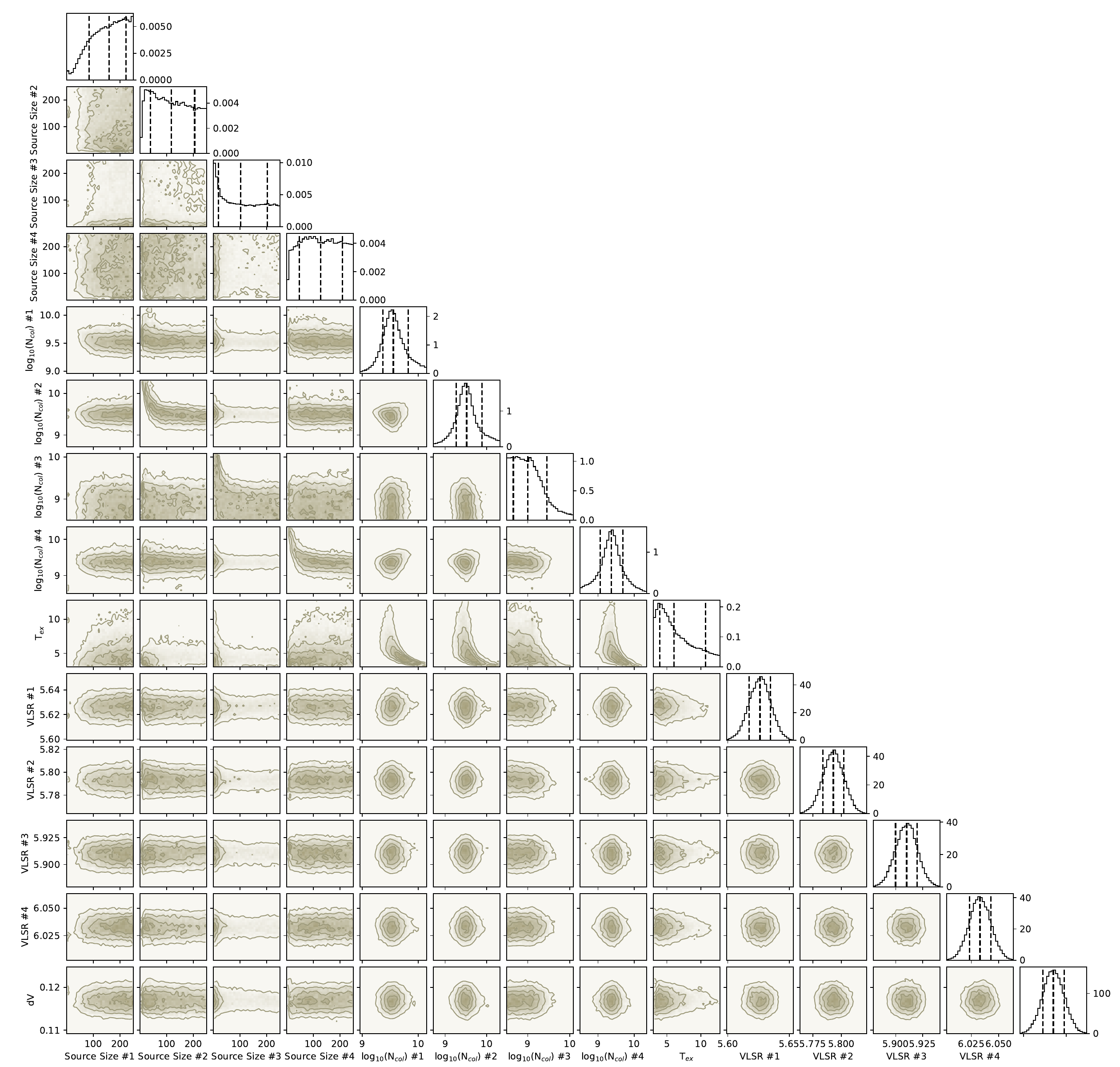}
\figsetgrpnote{The 16$^{th}$, 50$^{th}$, and 84$^{th}$ confidence intervals (corresponding to $\pm$1 sigma for a Gaussian posterior distribution) are shown as vertical lines. The contour lines are posterior probability levels, starting at $20\%$ of the maximum a posteriori estimate, with evenly spaced intervals of $20\%$ up to the peak density.}
\figsetgrpend

\figsetgrpstart
\figsetgrpnum{8.52}
\figsetgrptitle{Corner plot for E-1-C$_{4}$H$_{5}$CN.}
\figsetplot{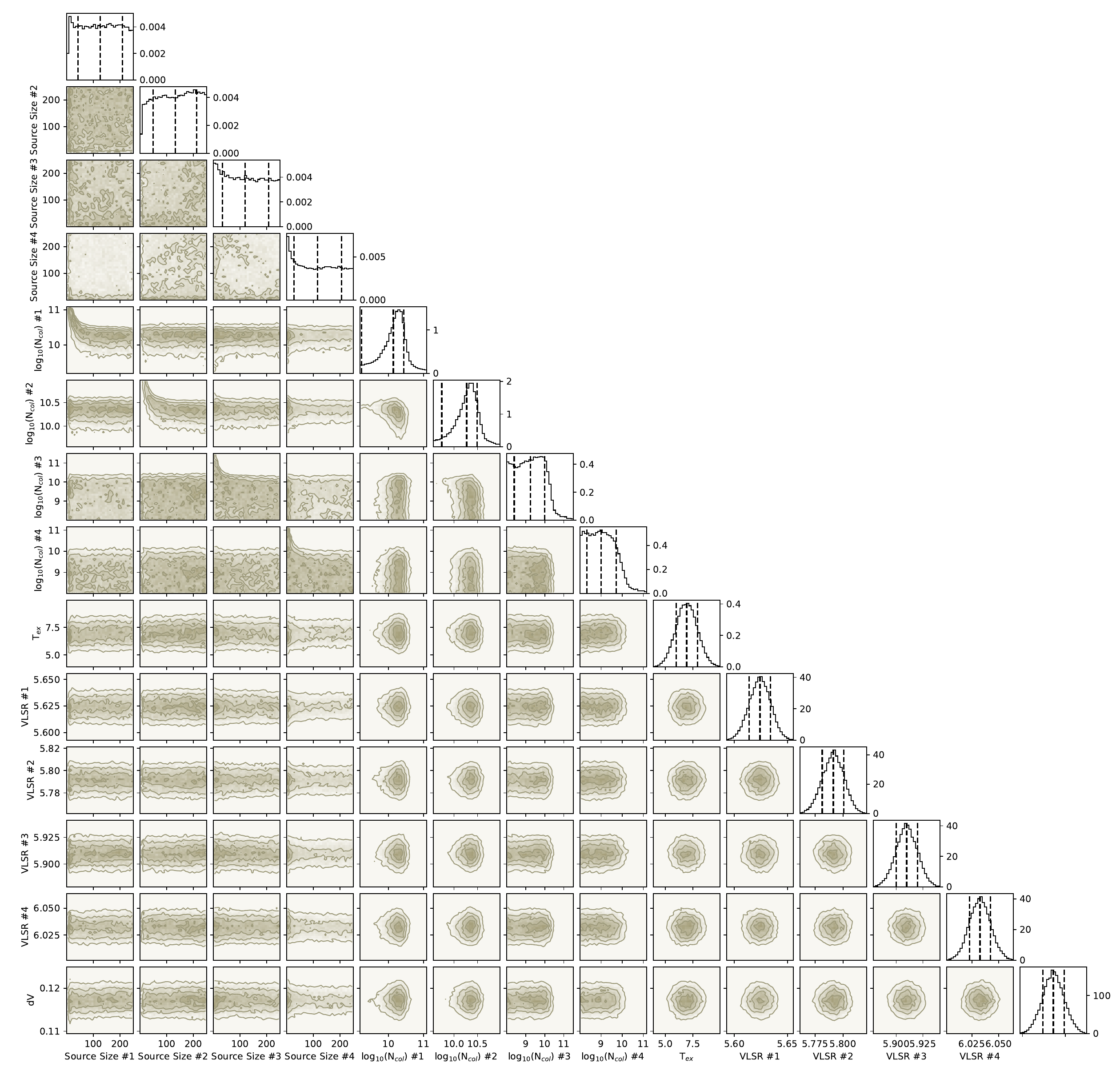}
\figsetgrpnote{The 16$^{th}$, 50$^{th}$, and 84$^{th}$ confidence intervals (corresponding to $\pm$1 sigma for a Gaussian posterior distribution) are shown as vertical lines. The contour lines are posterior probability levels, starting at $20\%$ of the maximum a posteriori estimate, with evenly spaced intervals of $20\%$ up to the peak density.}
\figsetgrpend

\figsetgrpstart
\figsetgrpnum{8.53}
\figsetgrptitle{Corner plot for CH$_{3}$C$_{6}$H $A$.}
\figsetplot{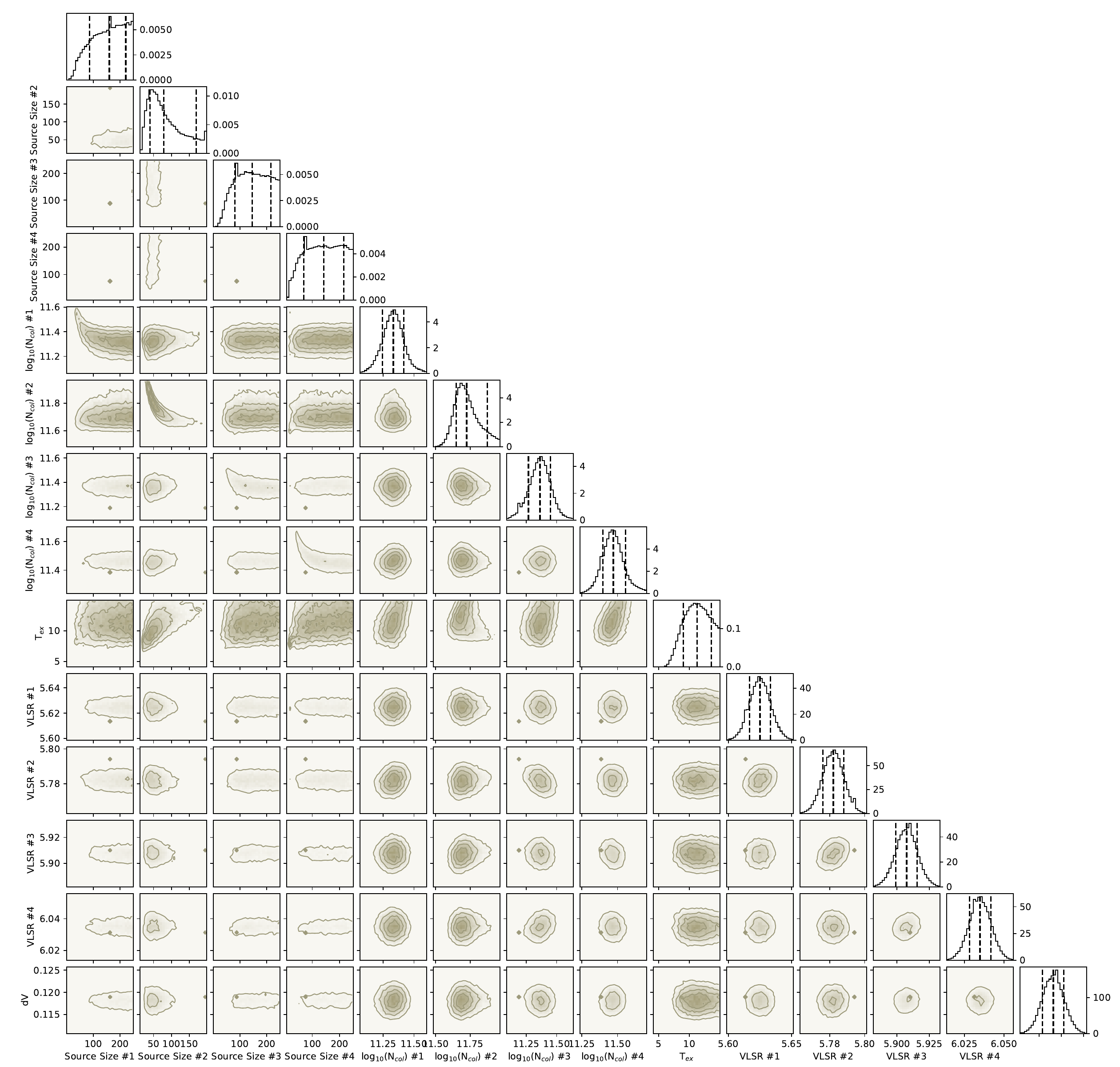}
\figsetgrpnote{The 16$^{th}$, 50$^{th}$, and 84$^{th}$ confidence intervals (corresponding to $\pm$1 sigma for a Gaussian posterior distribution) are shown as vertical lines. The contour lines are posterior probability levels, starting at $20\%$ of the maximum a posteriori estimate, with evenly spaced intervals of $20\%$ up to the peak density.}
\figsetgrpend

\figsetgrpstart
\figsetgrpnum{8.54}
\figsetgrptitle{Corner plot for CH$_{3}$C$_{6}$H $E$.}
\figsetplot{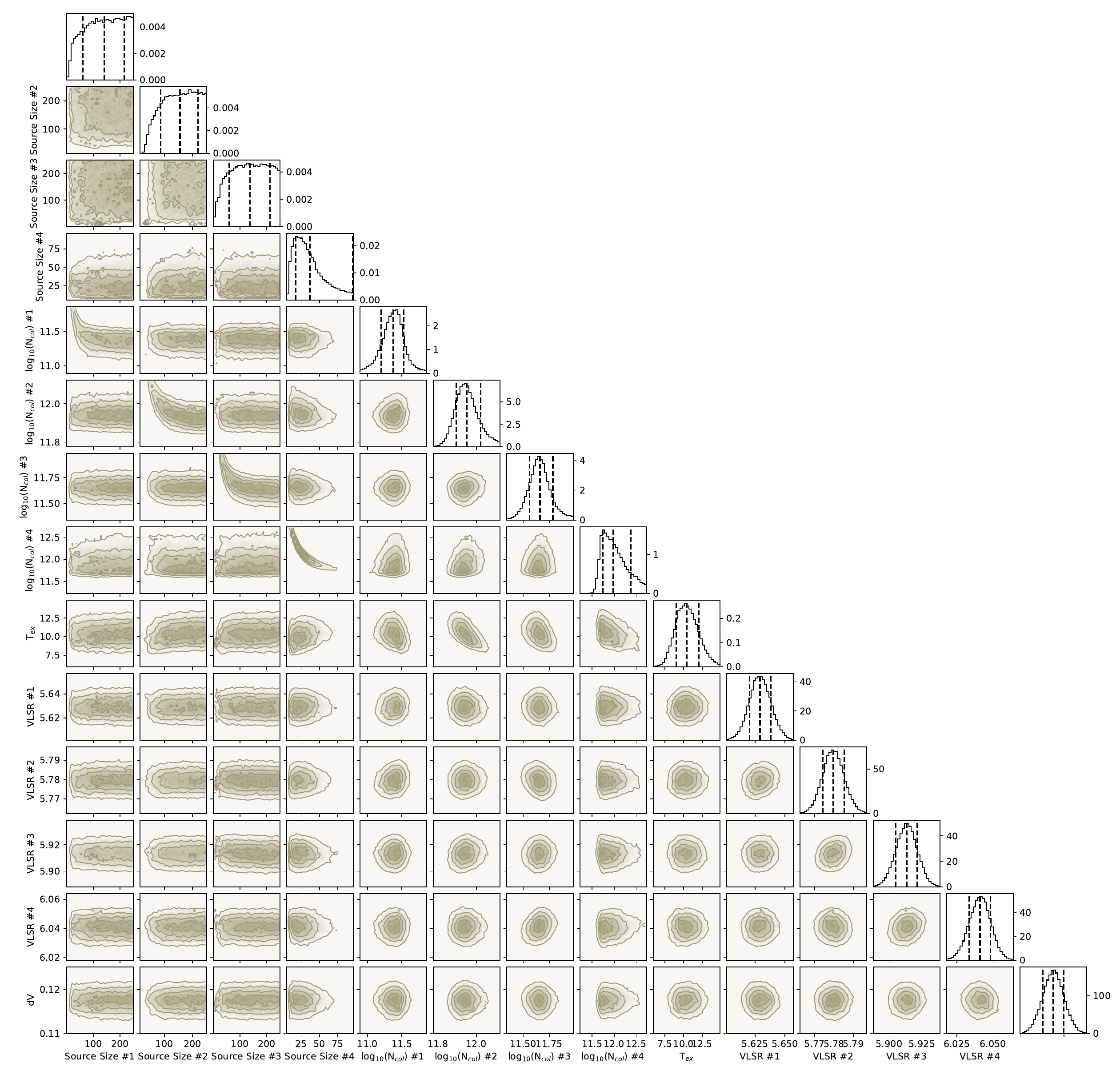}
\figsetgrpnote{The 16$^{th}$, 50$^{th}$, and 84$^{th}$ confidence intervals (corresponding to $\pm$1 sigma for a Gaussian posterior distribution) are shown as vertical lines. The contour lines are posterior probability levels, starting at $20\%$ of the maximum a posteriori estimate, with evenly spaced intervals of $20\%$ up to the peak density.}
\figsetgrpend

\figsetgrpstart
\figsetgrpnum{8.55}
\figsetgrptitle{Corner plot for CH$_{3}$C$_{5}$N $A$.}
\figsetplot{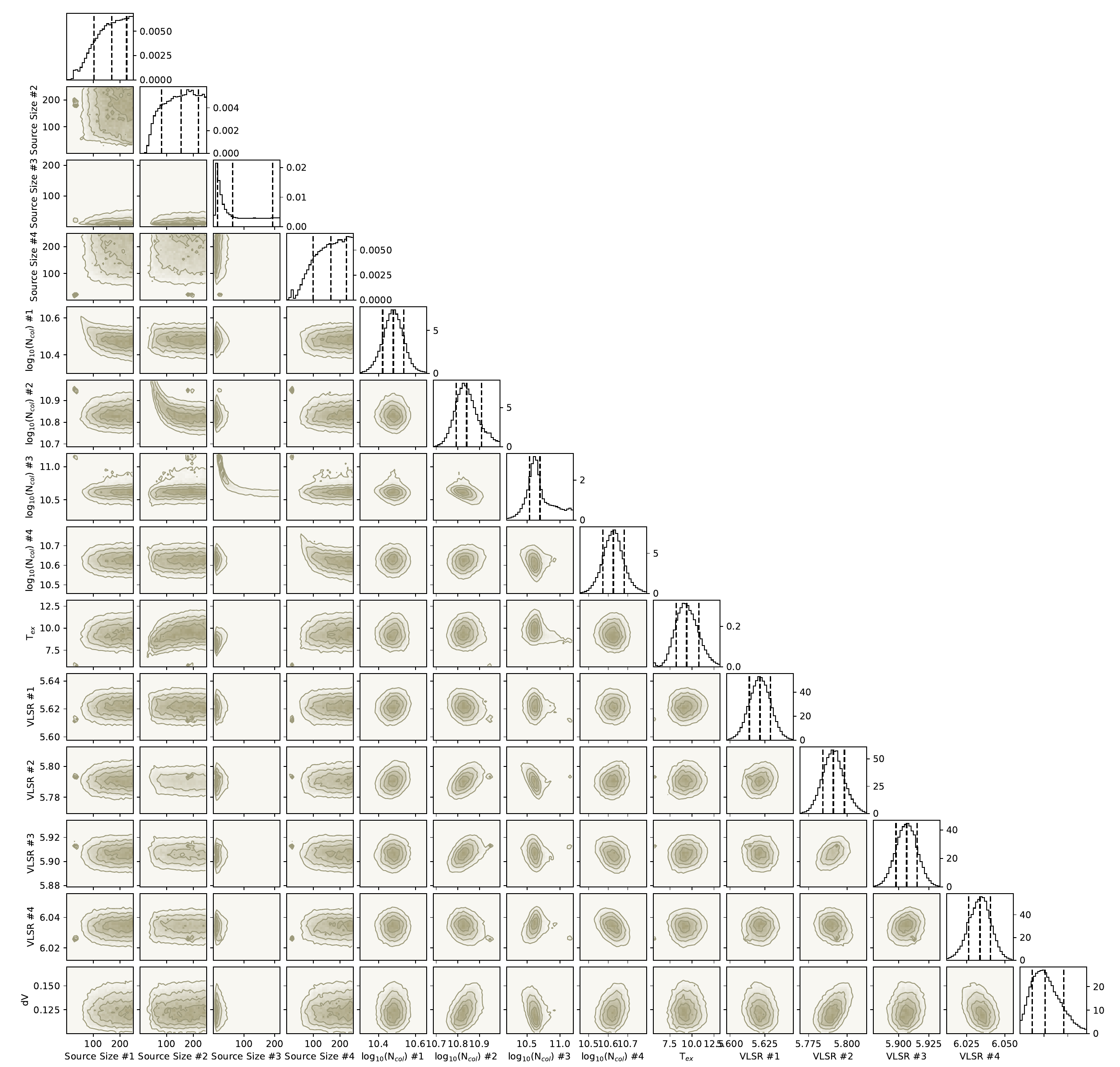}
\figsetgrpnote{The 16$^{th}$, 50$^{th}$, and 84$^{th}$ confidence intervals (corresponding to $\pm$1 sigma for a Gaussian posterior distribution) are shown as vertical lines. The contour lines are posterior probability levels, starting at $20\%$ of the maximum a posteriori estimate, with evenly spaced intervals of $20\%$ up to the peak density.}
\figsetgrpend

\figsetgrpstart
\figsetgrpnum{8.56}
\figsetgrptitle{Corner plot for CH$_{3}$C$_{5}$N $E$.}
\figsetplot{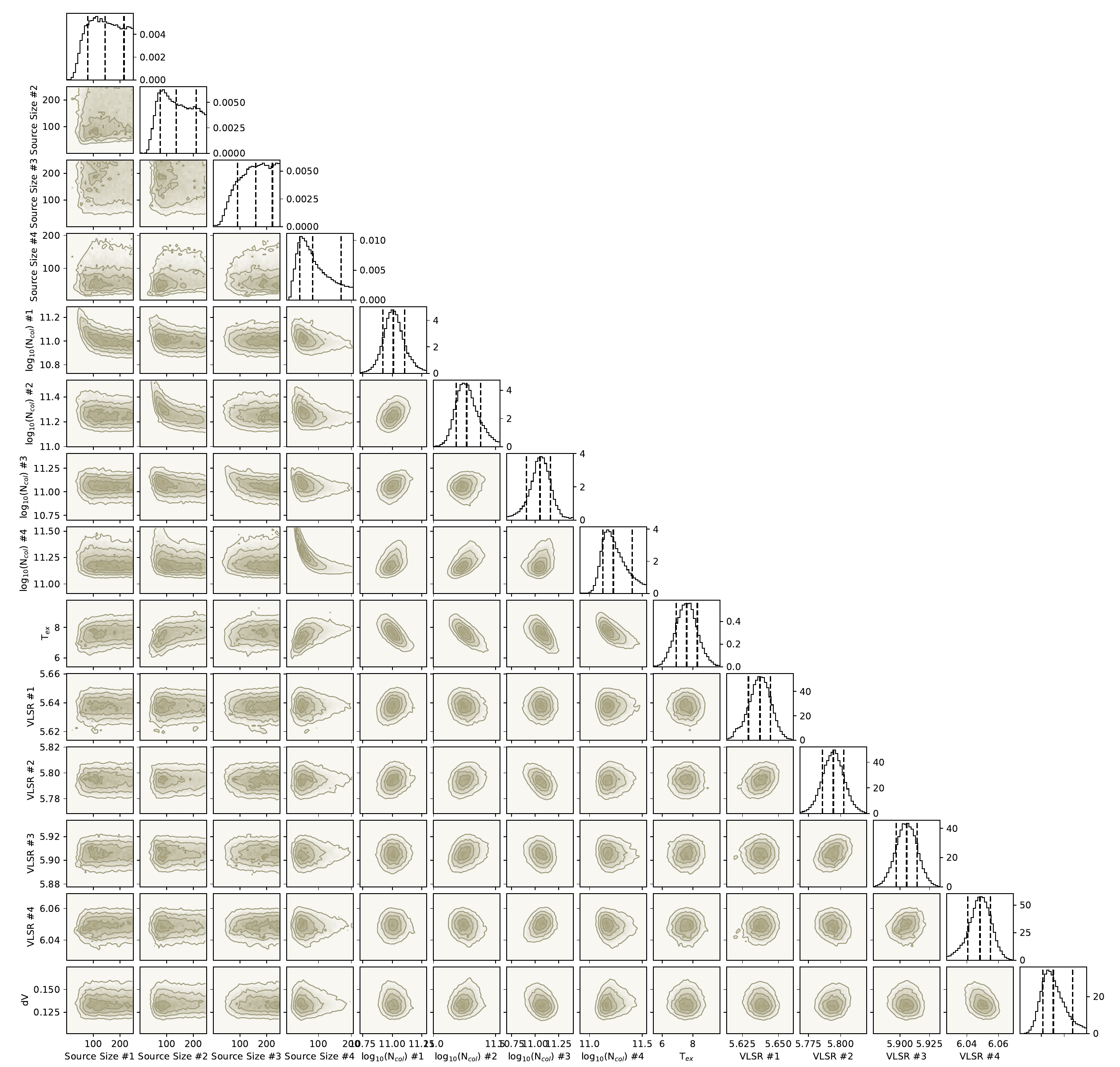}
\figsetgrpnote{The 16$^{th}$, 50$^{th}$, and 84$^{th}$ confidence intervals (corresponding to $\pm$1 sigma for a Gaussian posterior distribution) are shown as vertical lines. The contour lines are posterior probability levels, starting at $20\%$ of the maximum a posteriori estimate, with evenly spaced intervals of $20\%$ up to the peak density.}
\figsetgrpend

\figsetgrpstart
\figsetgrpnum{8.57}
\figsetgrptitle{Corner plot for $c$-1-C$_{5}$H$_{5}$CN.}
\figsetplot{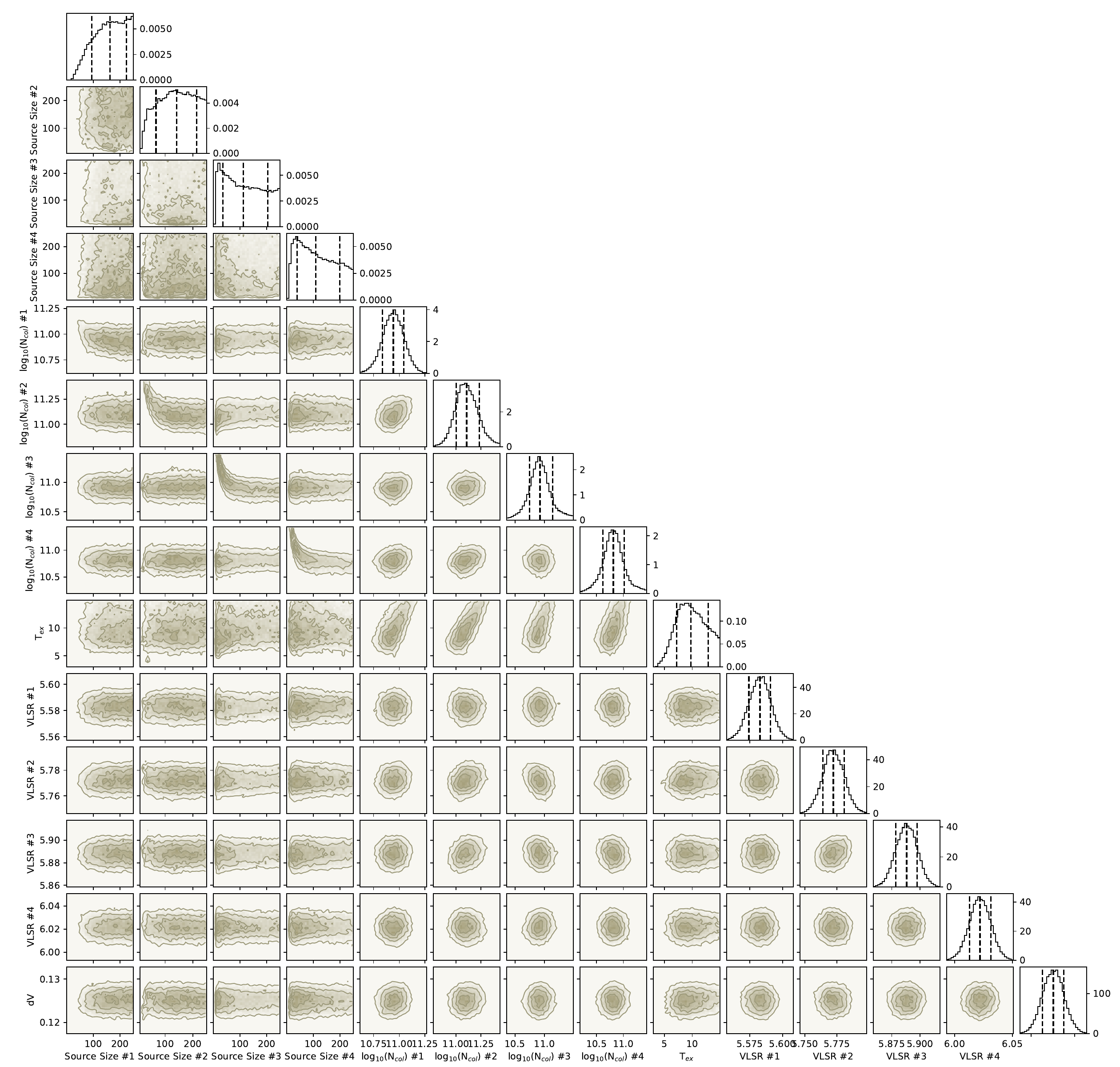}
\figsetgrpnote{The 16$^{th}$, 50$^{th}$, and 84$^{th}$ confidence intervals (corresponding to $\pm$1 sigma for a Gaussian posterior distribution) are shown as vertical lines. The contour lines are posterior probability levels, starting at $20\%$ of the maximum a posteriori estimate, with evenly spaced intervals of $20\%$ up to the peak density.}
\figsetgrpend

\figsetgrpstart
\figsetgrpnum{8.58}
\figsetgrptitle{Corner plot for $c$-2-C$_{5}$H$_{5}$CN.}
\figsetplot{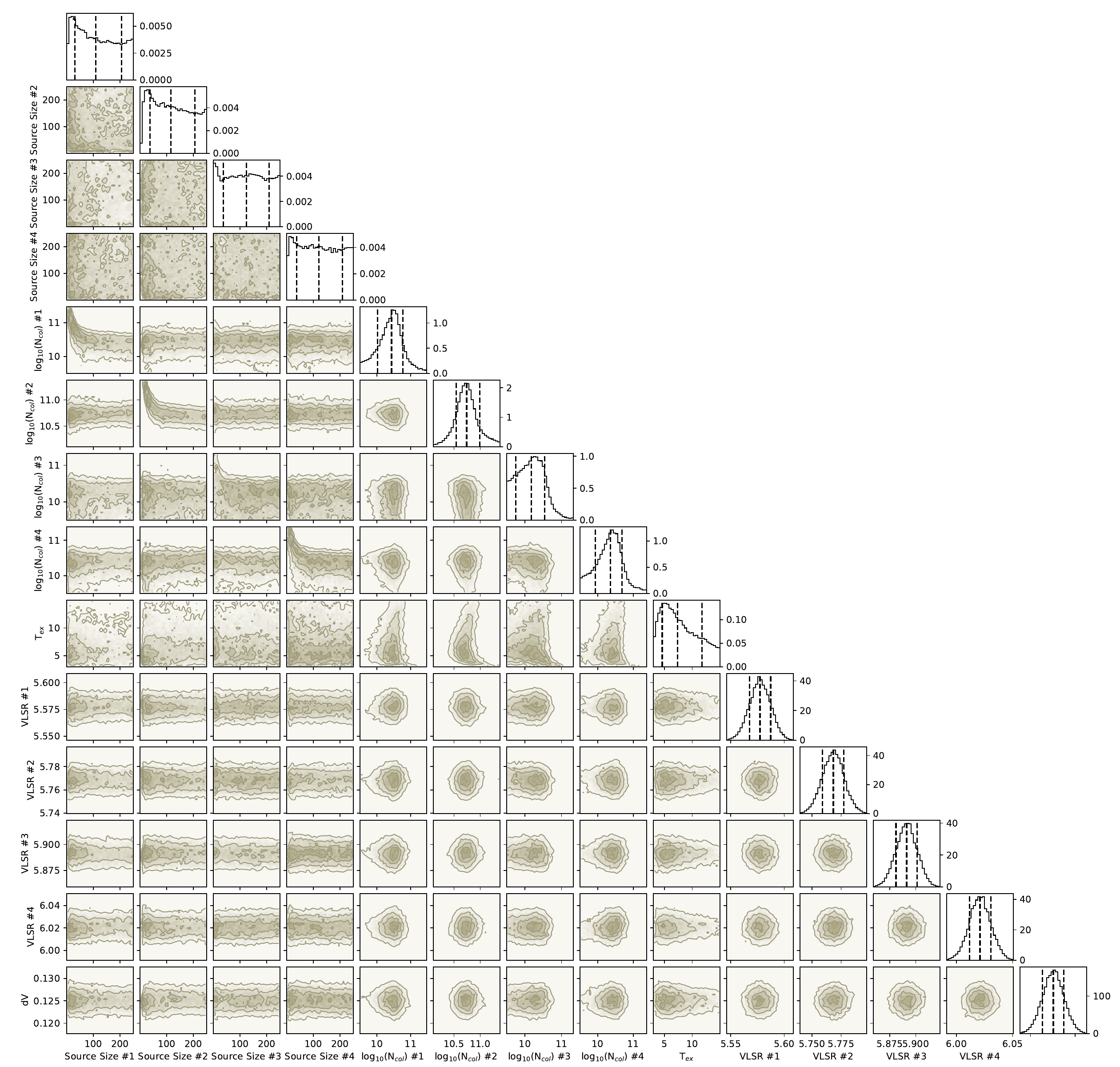}
\figsetgrpnote{The 16$^{th}$, 50$^{th}$, and 84$^{th}$ confidence intervals (corresponding to $\pm$1 sigma for a Gaussian posterior distribution) are shown as vertical lines. The contour lines are posterior probability levels, starting at $20\%$ of the maximum a posteriori estimate, with evenly spaced intervals of $20\%$ up to the peak density.}
\figsetgrpend

\figsetgrpstart
\figsetgrpnum{8.59}
\figsetgrptitle{Corner plot for C$_{5}$S.}
\figsetplot{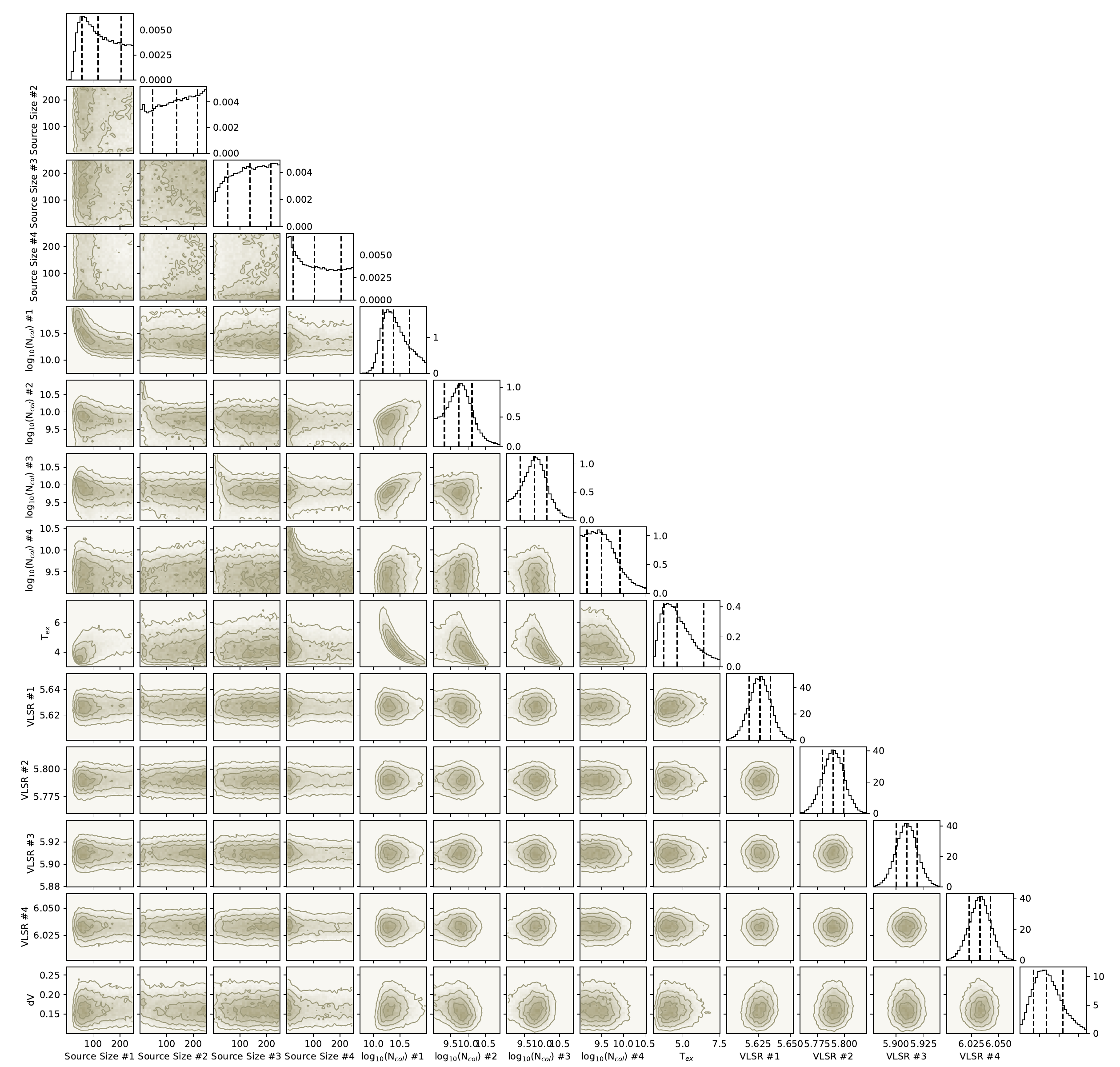}
\figsetgrpnote{The 16$^{th}$, 50$^{th}$, and 84$^{th}$ confidence intervals (corresponding to $\pm$1 sigma for a Gaussian posterior distribution) are shown as vertical lines. The contour lines are posterior probability levels, starting at $20\%$ of the maximum a posteriori estimate, with evenly spaced intervals of $20\%$ up to the peak density.}
\figsetgrpend

\figsetgrpstart
\figsetgrpnum{8.60}
\figsetgrptitle{Corner plot for C$_{8}$H.}
\figsetplot{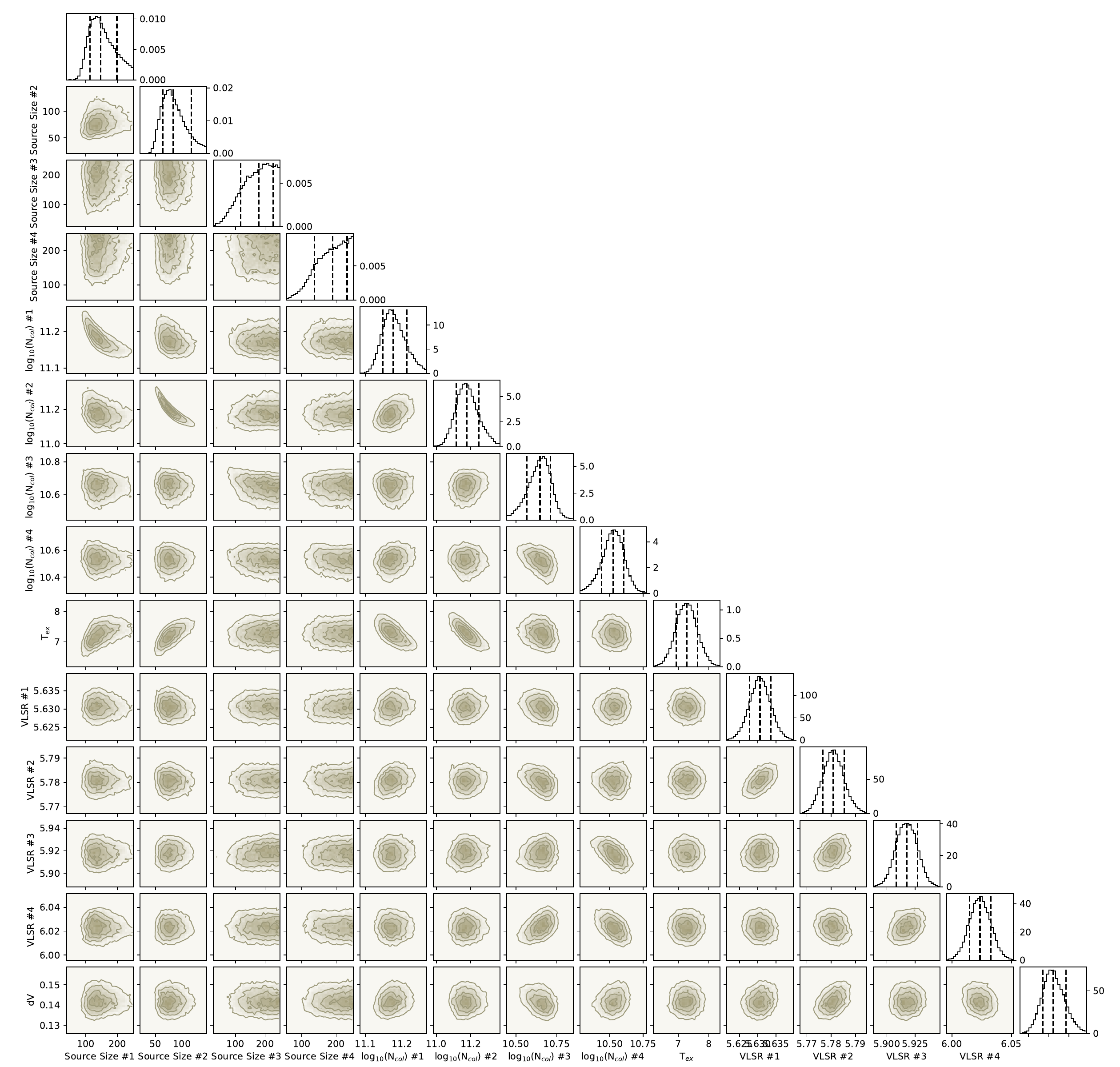}
\figsetgrpnote{The 16$^{th}$, 50$^{th}$, and 84$^{th}$ confidence intervals (corresponding to $\pm$1 sigma for a Gaussian posterior distribution) are shown as vertical lines. The contour lines are posterior probability levels, starting at $20\%$ of the maximum a posteriori estimate, with evenly spaced intervals of $20\%$ up to the peak density.}
\figsetgrpend

\figsetgrpstart
\figsetgrpnum{8.61}
\figsetgrptitle{Corner plot for C$_{8}$H$^{-}$.}
\figsetplot{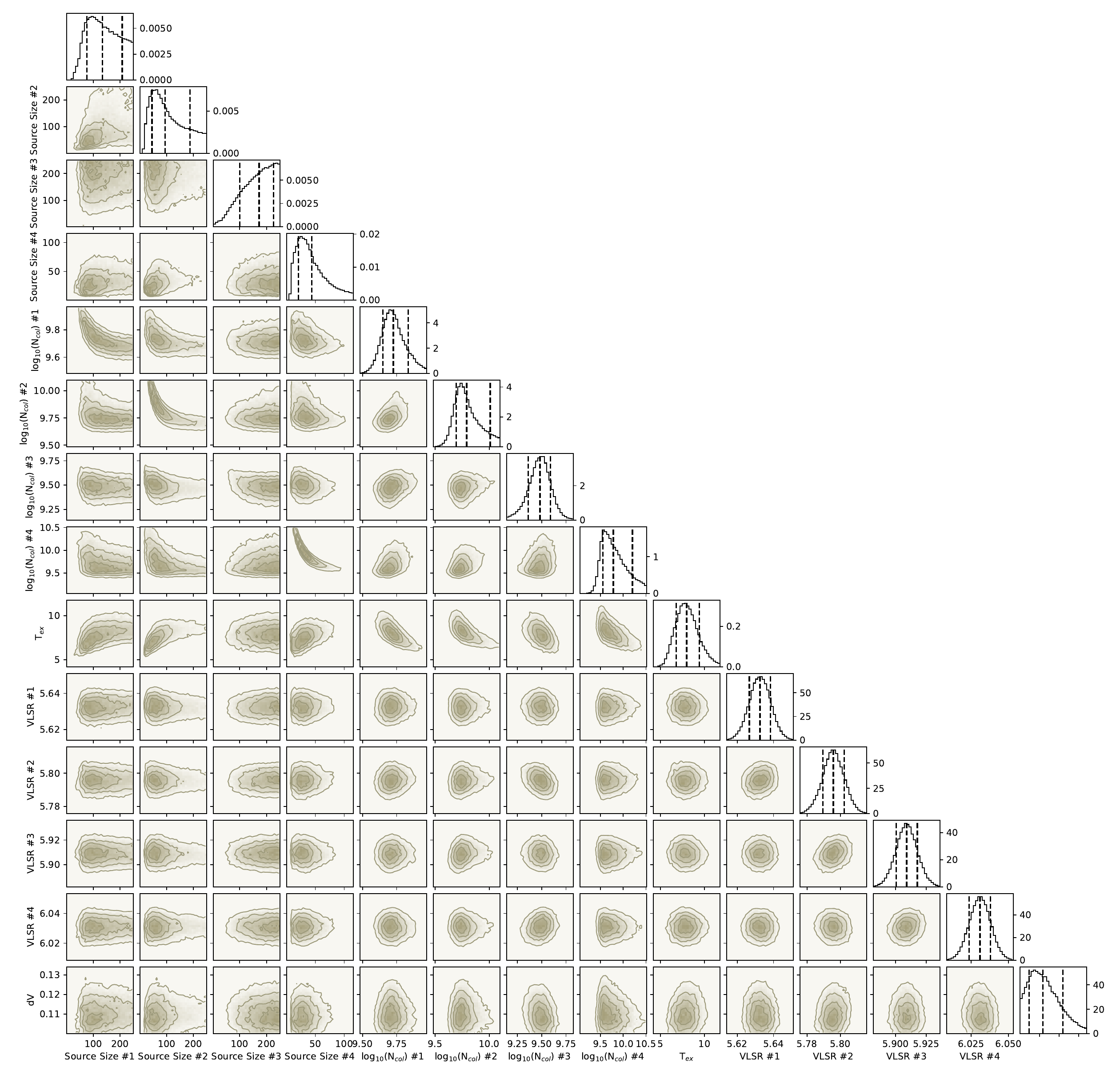}
\figsetgrpnote{The 16$^{th}$, 50$^{th}$, and 84$^{th}$ confidence intervals (corresponding to $\pm$1 sigma for a Gaussian posterior distribution) are shown as vertical lines. The contour lines are posterior probability levels, starting at $20\%$ of the maximum a posteriori estimate, with evenly spaced intervals of $20\%$ up to the peak density.}
\figsetgrpend

\figsetgrpstart
\figsetgrpnum{8.62}
\figsetgrptitle{Corner plot for C$_{7}$N$^{-}$.}
\figsetplot{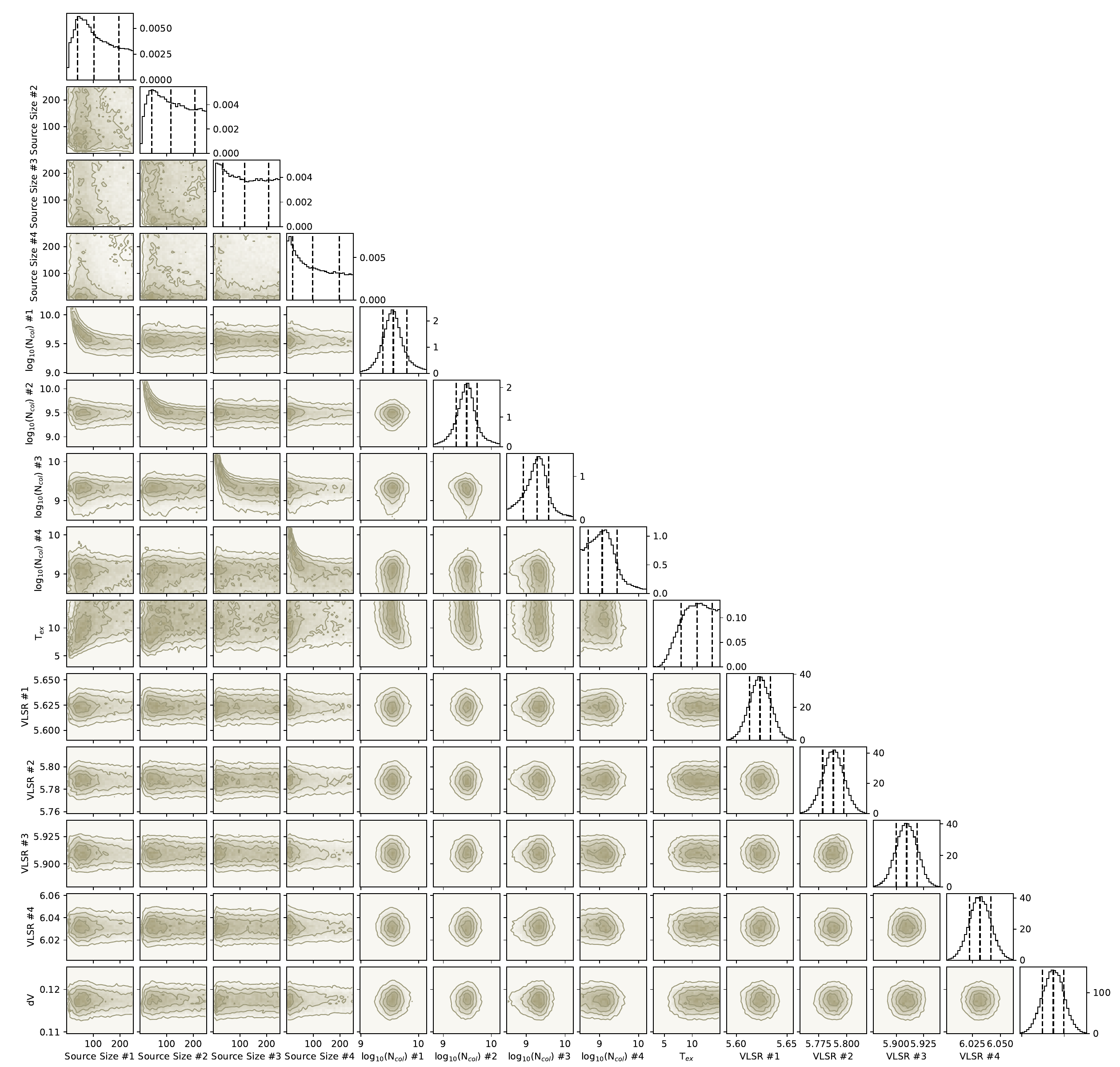}
\figsetgrpnote{The 16$^{th}$, 50$^{th}$, and 84$^{th}$ confidence intervals (corresponding to $\pm$1 sigma for a Gaussian posterior distribution) are shown as vertical lines. The contour lines are posterior probability levels, starting at $20\%$ of the maximum a posteriori estimate, with evenly spaced intervals of $20\%$ up to the peak density.}
\figsetgrpend

\figsetgrpstart
\figsetgrpnum{8.63}
\figsetgrptitle{Corner plot for HC$_{7}$N.}
\figsetplot{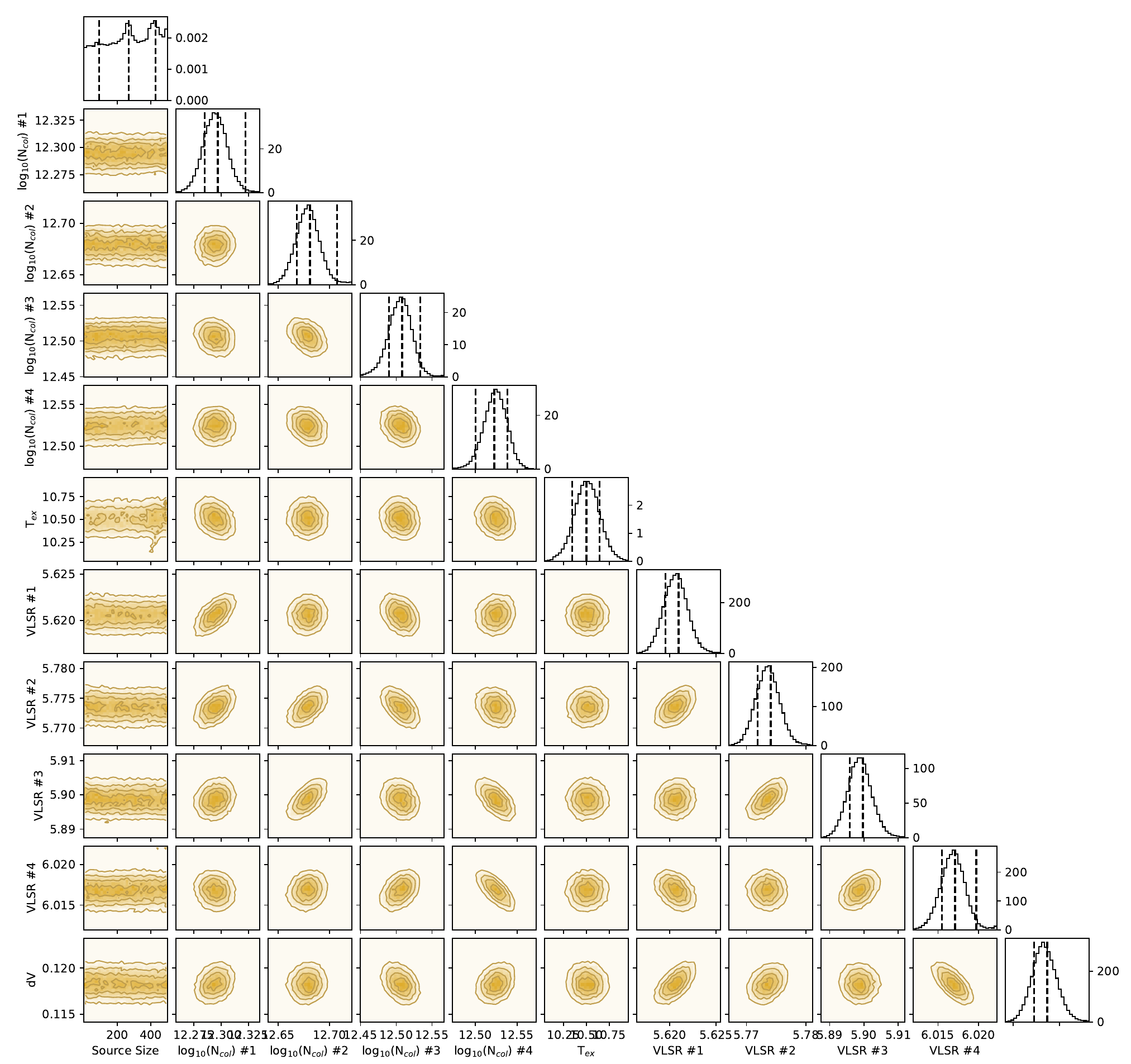}
\figsetgrpnote{The 16$^{th}$, 50$^{th}$, and 84$^{th}$ confidence intervals (corresponding to $\pm$1 sigma for a Gaussian posterior distribution) are shown as vertical lines. The contour lines are posterior probability levels, starting at $20\%$ of the maximum a posteriori estimate, with evenly spaced intervals of $20\%$ up to the peak density.}
\figsetgrpend

\figsetgrpstart
\figsetgrpnum{8.64}
\figsetgrptitle{Corner plot for HC$_{7}$NH$^{+}$.}
\figsetplot{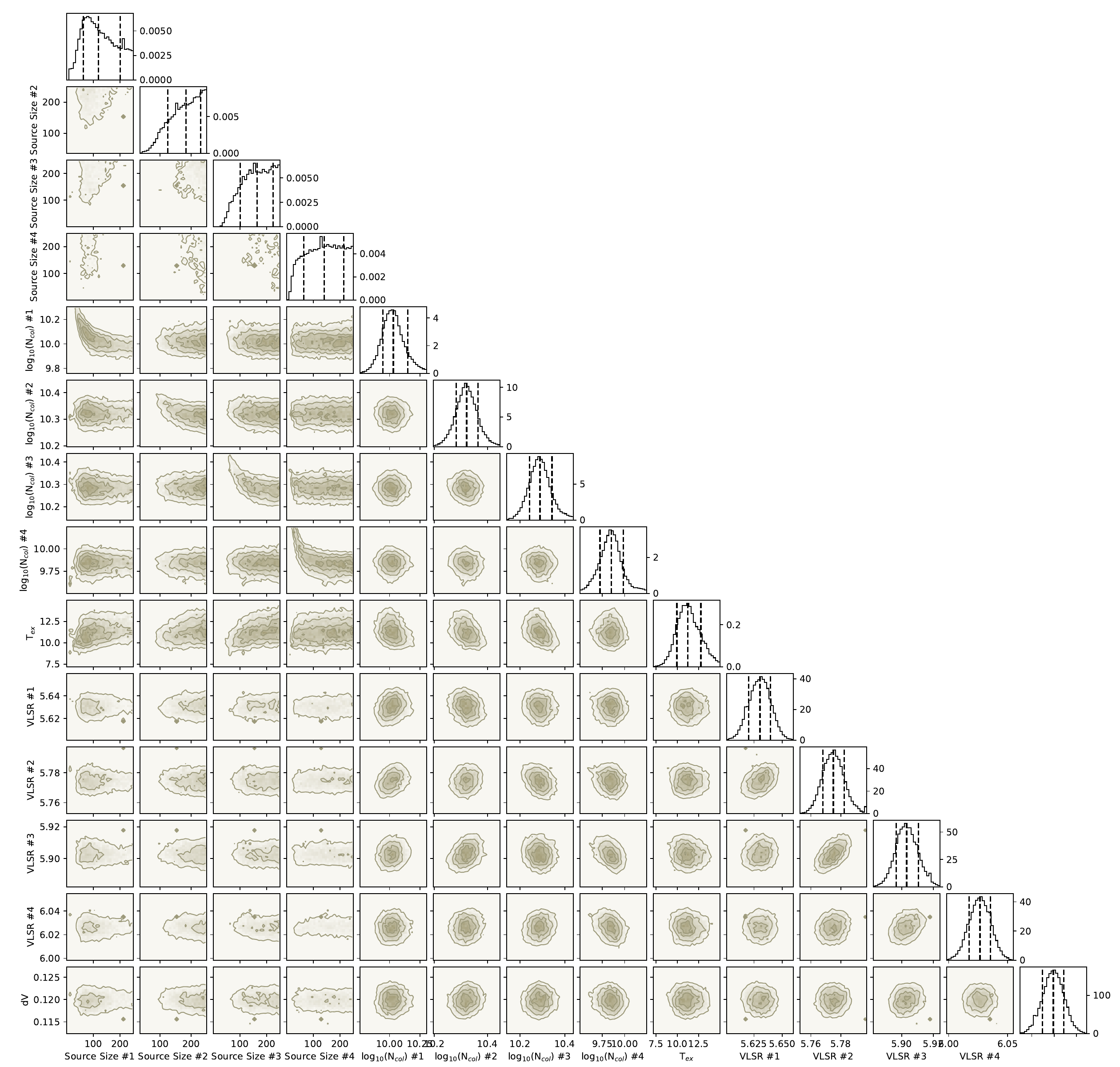}
\figsetgrpnote{The 16$^{th}$, 50$^{th}$, and 84$^{th}$ confidence intervals (corresponding to $\pm$1 sigma for a Gaussian posterior distribution) are shown as vertical lines. The contour lines are posterior probability levels, starting at $20\%$ of the maximum a posteriori estimate, with evenly spaced intervals of $20\%$ up to the peak density.}
\figsetgrpend

\figsetgrpstart
\figsetgrpnum{8.65}
\figsetgrptitle{Corner plot for HC$_{7}$O.}
\figsetplot{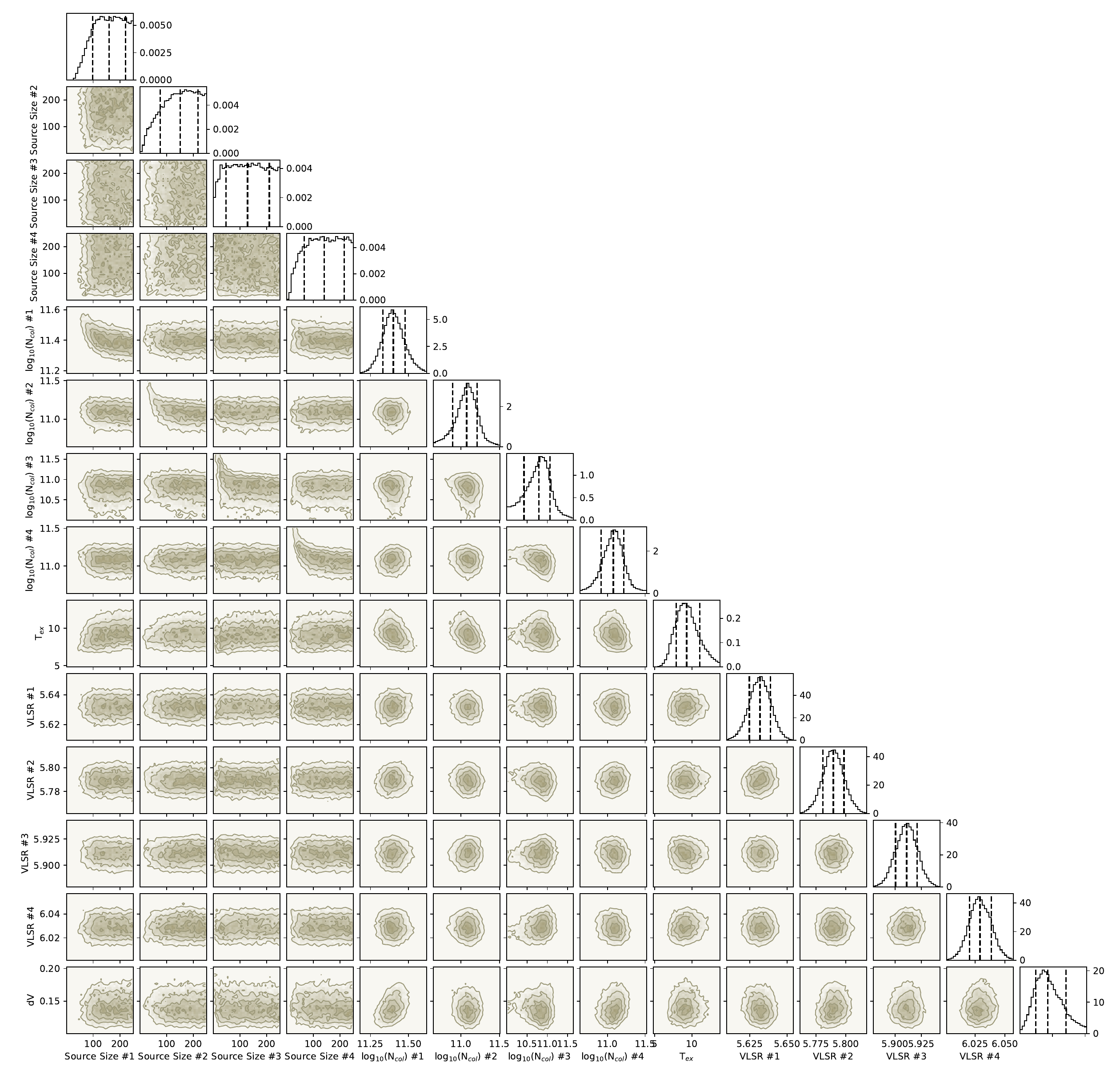}
\figsetgrpnote{The 16$^{th}$, 50$^{th}$, and 84$^{th}$ confidence intervals (corresponding to $\pm$1 sigma for a Gaussian posterior distribution) are shown as vertical lines. The contour lines are posterior probability levels, starting at $20\%$ of the maximum a posteriori estimate, with evenly spaced intervals of $20\%$ up to the peak density.}
\figsetgrpend

\figsetgrpstart
\figsetgrpnum{8.66}
\figsetgrptitle{Corner plot for $c$-C$_{6}$H$_{5}$CN.}
\figsetplot{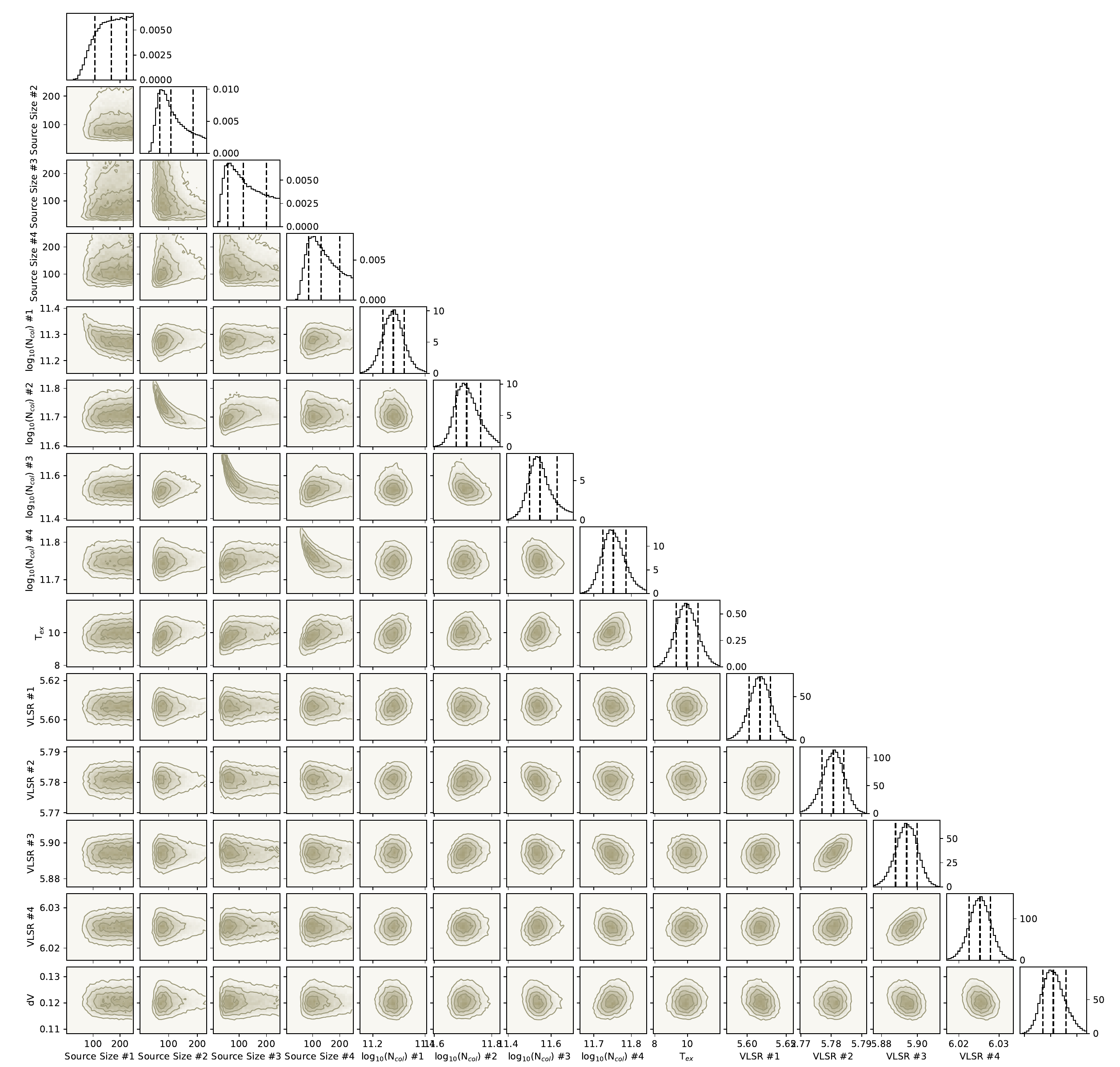}
\figsetgrpnote{The 16$^{th}$, 50$^{th}$, and 84$^{th}$ confidence intervals (corresponding to $\pm$1 sigma for a Gaussian posterior distribution) are shown as vertical lines. The contour lines are posterior probability levels, starting at $20\%$ of the maximum a posteriori estimate, with evenly spaced intervals of $20\%$ up to the peak density.}
\figsetgrpend

\figsetgrpstart
\figsetgrpnum{8.67}
\figsetgrptitle{Corner plot for CH$_{3}$C$_{7}$N $A$.}
\figsetplot{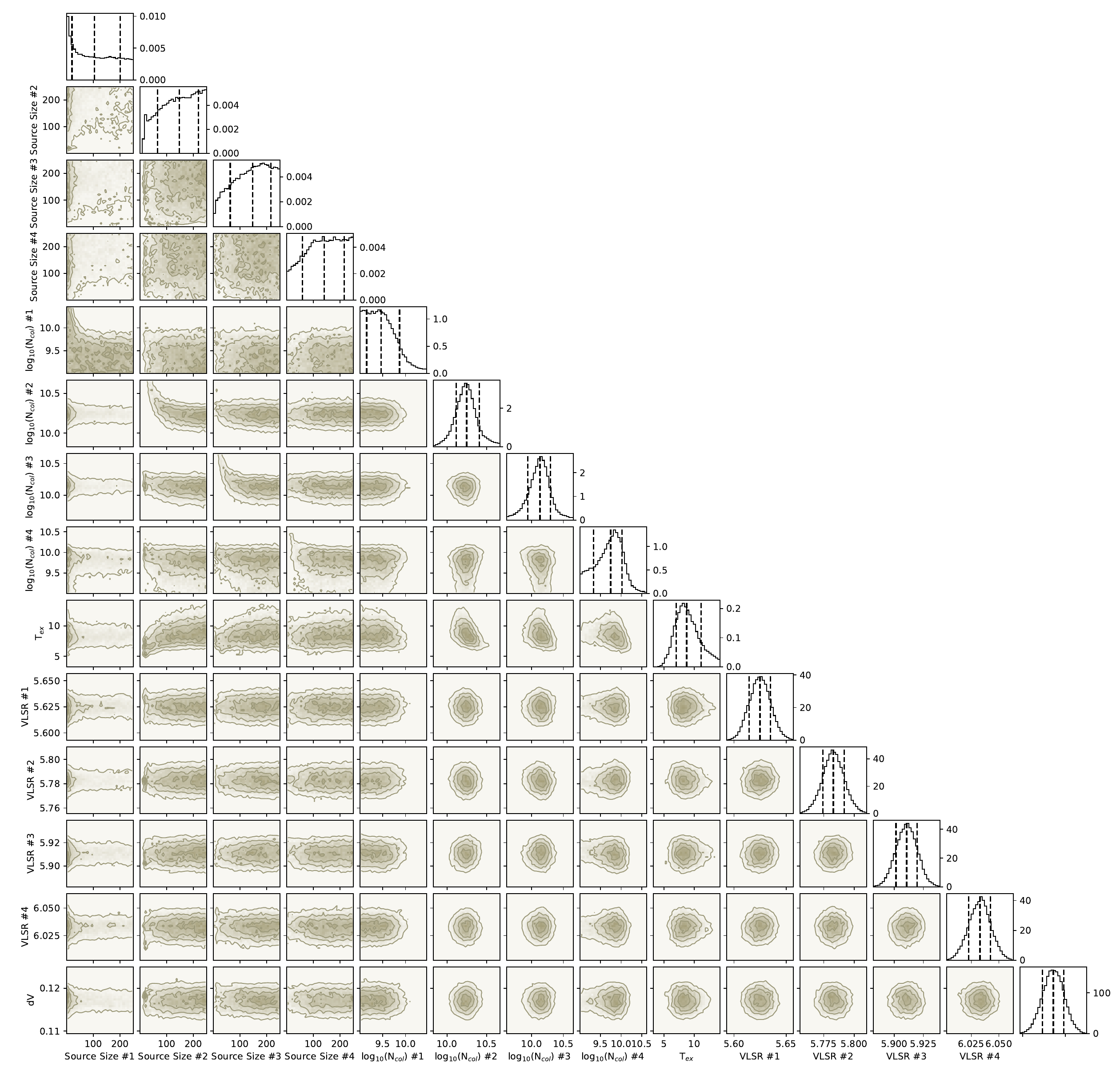}
\figsetgrpnote{The 16$^{th}$, 50$^{th}$, and 84$^{th}$ confidence intervals (corresponding to $\pm$1 sigma for a Gaussian posterior distribution) are shown as vertical lines. The contour lines are posterior probability levels, starting at $20\%$ of the maximum a posteriori estimate, with evenly spaced intervals of $20\%$ up to the peak density.}
\figsetgrpend

\figsetgrpstart
\figsetgrpnum{8.68}
\figsetgrptitle{Corner plot for CH$_{3}$C$_{7}$N $E$.}
\figsetplot{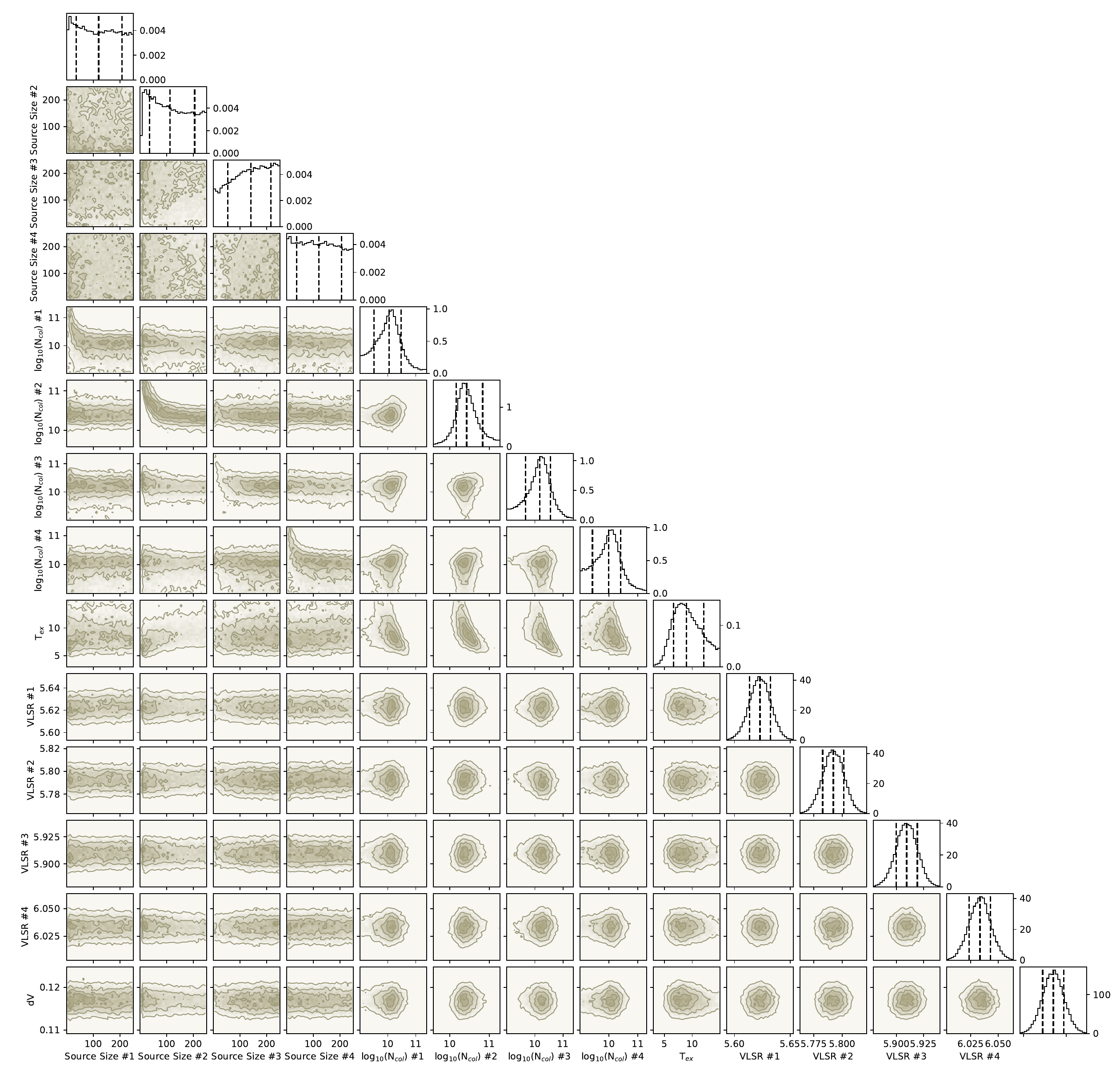}
\figsetgrpnote{The 16$^{th}$, 50$^{th}$, and 84$^{th}$ confidence intervals (corresponding to $\pm$1 sigma for a Gaussian posterior distribution) are shown as vertical lines. The contour lines are posterior probability levels, starting at $20\%$ of the maximum a posteriori estimate, with evenly spaced intervals of $20\%$ up to the peak density.}
\figsetgrpend

\figsetgrpstart
\figsetgrpnum{8.69}
\figsetgrptitle{Corner plot for C$_{9}$H$_{8}$.}
\figsetplot{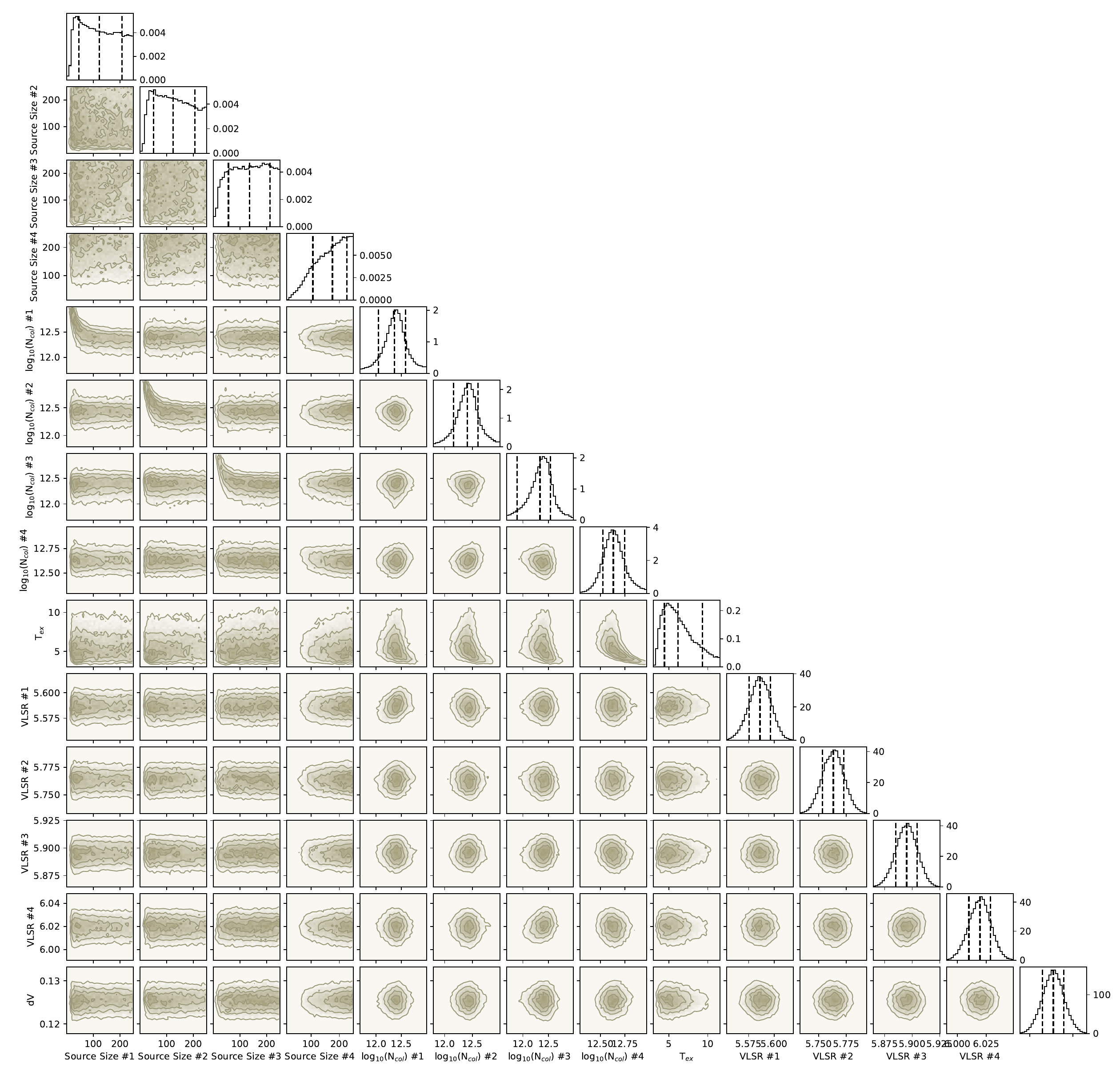}
\figsetgrpnote{The 16$^{th}$, 50$^{th}$, and 84$^{th}$ confidence intervals (corresponding to $\pm$1 sigma for a Gaussian posterior distribution) are shown as vertical lines. The contour lines are posterior probability levels, starting at $20\%$ of the maximum a posteriori estimate, with evenly spaced intervals of $20\%$ up to the peak density.}
\figsetgrpend

\figsetgrpstart
\figsetgrpnum{8.70}
\figsetgrptitle{Corner plot for C$_{10}$H$^{-}$.}
\figsetplot{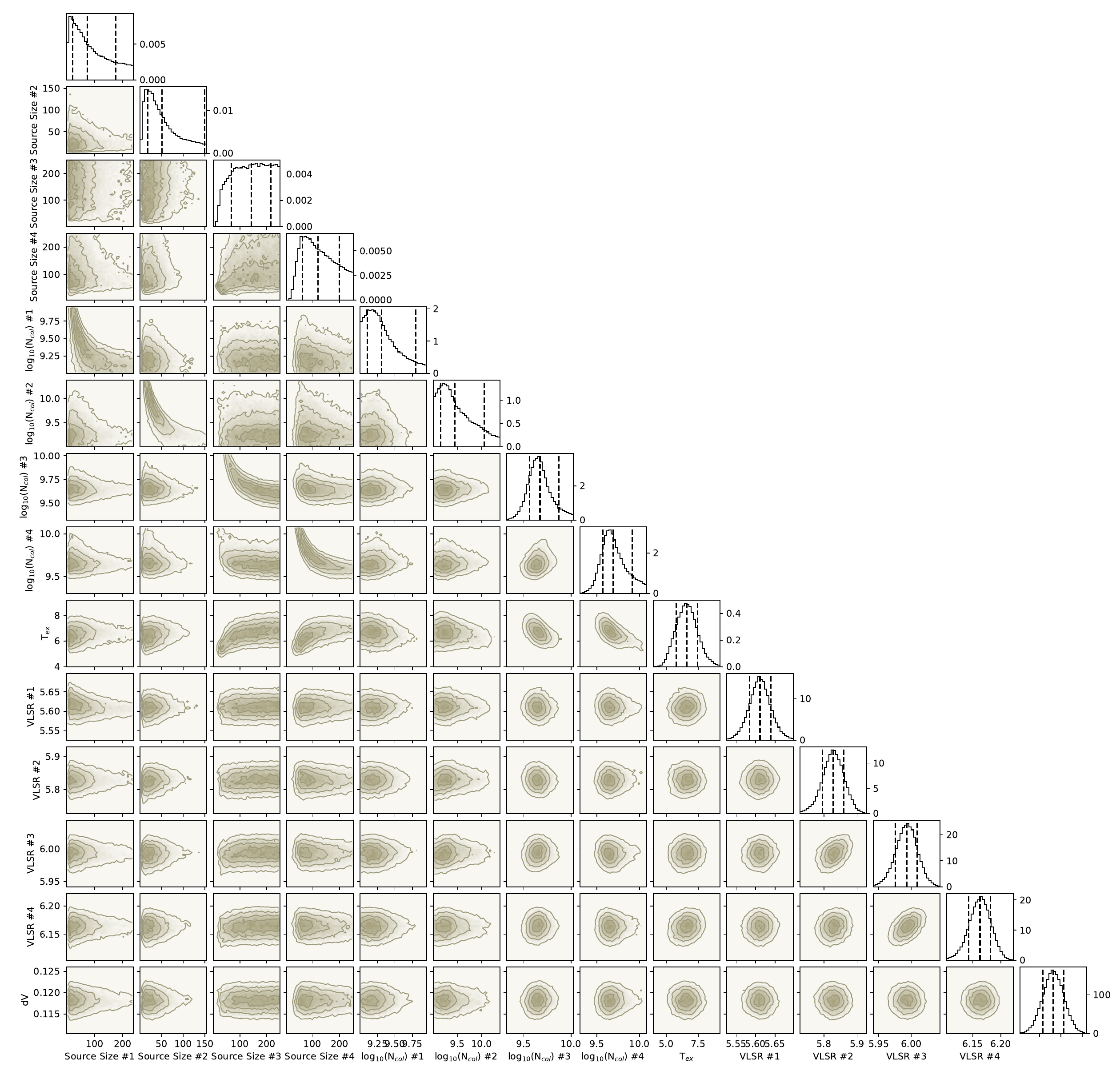}
\figsetgrpnote{The 16$^{th}$, 50$^{th}$, and 84$^{th}$ confidence intervals (corresponding to $\pm$1 sigma for a Gaussian posterior distribution) are shown as vertical lines. The contour lines are posterior probability levels, starting at $20\%$ of the maximum a posteriori estimate, with evenly spaced intervals of $20\%$ up to the peak density.}
\figsetgrpend

\figsetgrpstart
\figsetgrpnum{8.71}
\figsetgrptitle{Corner plot for HC$_{9}$N.}
\figsetplot{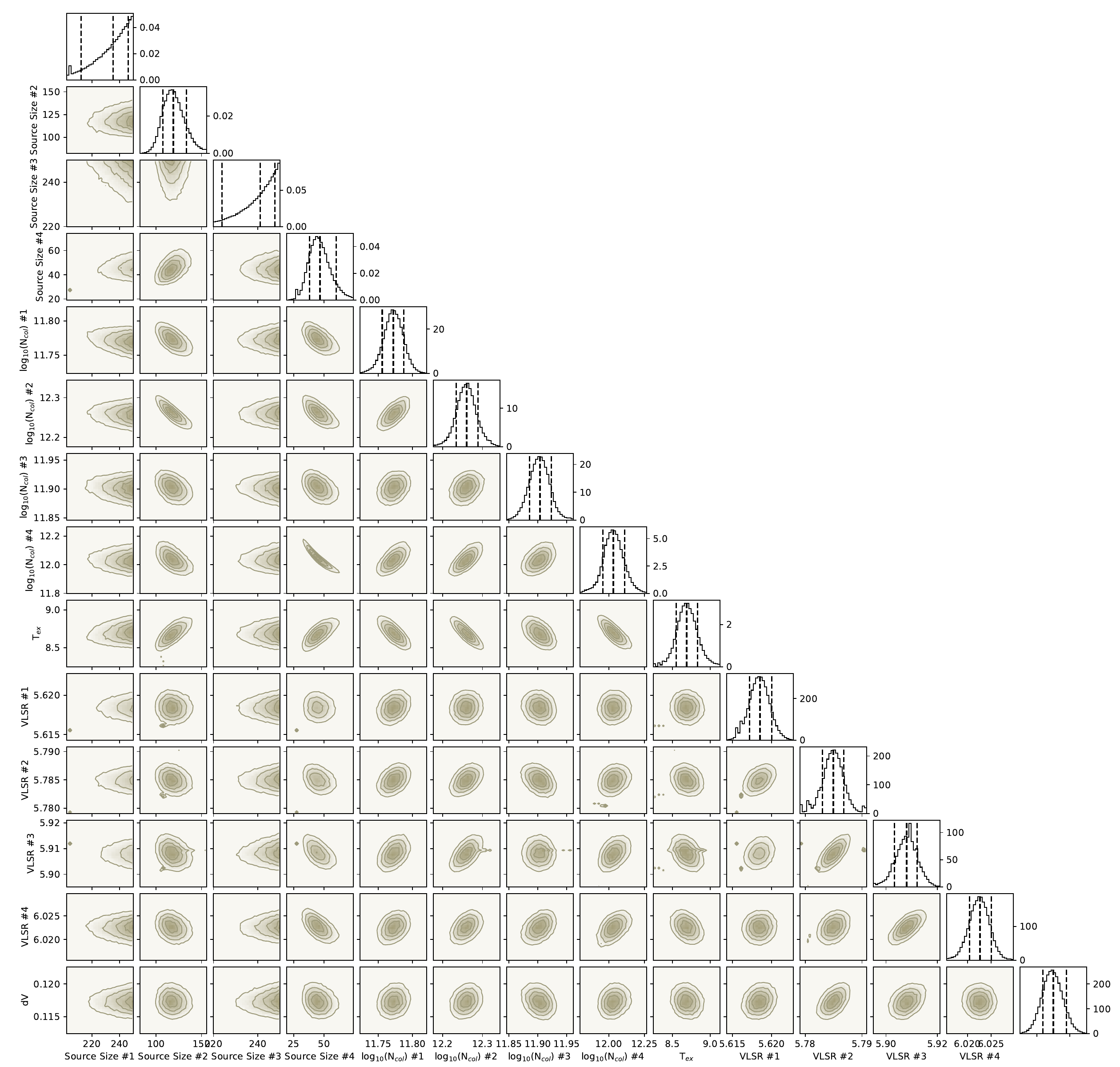}
\figsetgrpnote{The 16$^{th}$, 50$^{th}$, and 84$^{th}$ confidence intervals (corresponding to $\pm$1 sigma for a Gaussian posterior distribution) are shown as vertical lines. The contour lines are posterior probability levels, starting at $20\%$ of the maximum a posteriori estimate, with evenly spaced intervals of $20\%$ up to the peak density.}
\figsetgrpend

\figsetgrpstart
\figsetgrpnum{8.72}
\figsetgrptitle{Corner plot for 2-C$_{9}$H$_{7}$CN.}
\figsetplot{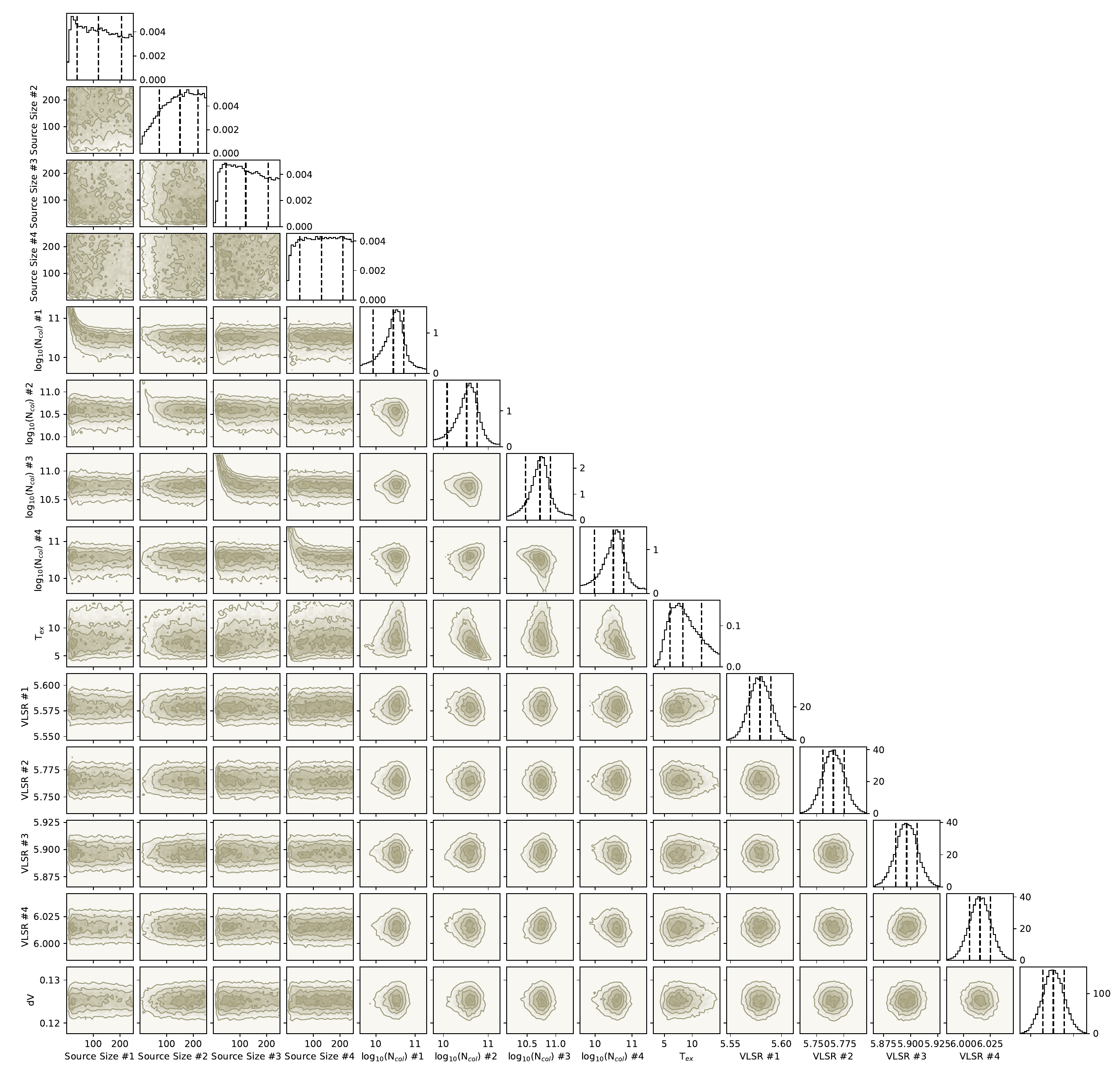}
\figsetgrpnote{The 16$^{th}$, 50$^{th}$, and 84$^{th}$ confidence intervals (corresponding to $\pm$1 sigma for a Gaussian posterior distribution) are shown as vertical lines. The contour lines are posterior probability levels, starting at $20\%$ of the maximum a posteriori estimate, with evenly spaced intervals of $20\%$ up to the peak density.}
\figsetgrpend

\figsetgrpstart
\figsetgrpnum{8.73}
\figsetgrptitle{Corner plot for HC$_{11}$N.}
\figsetplot{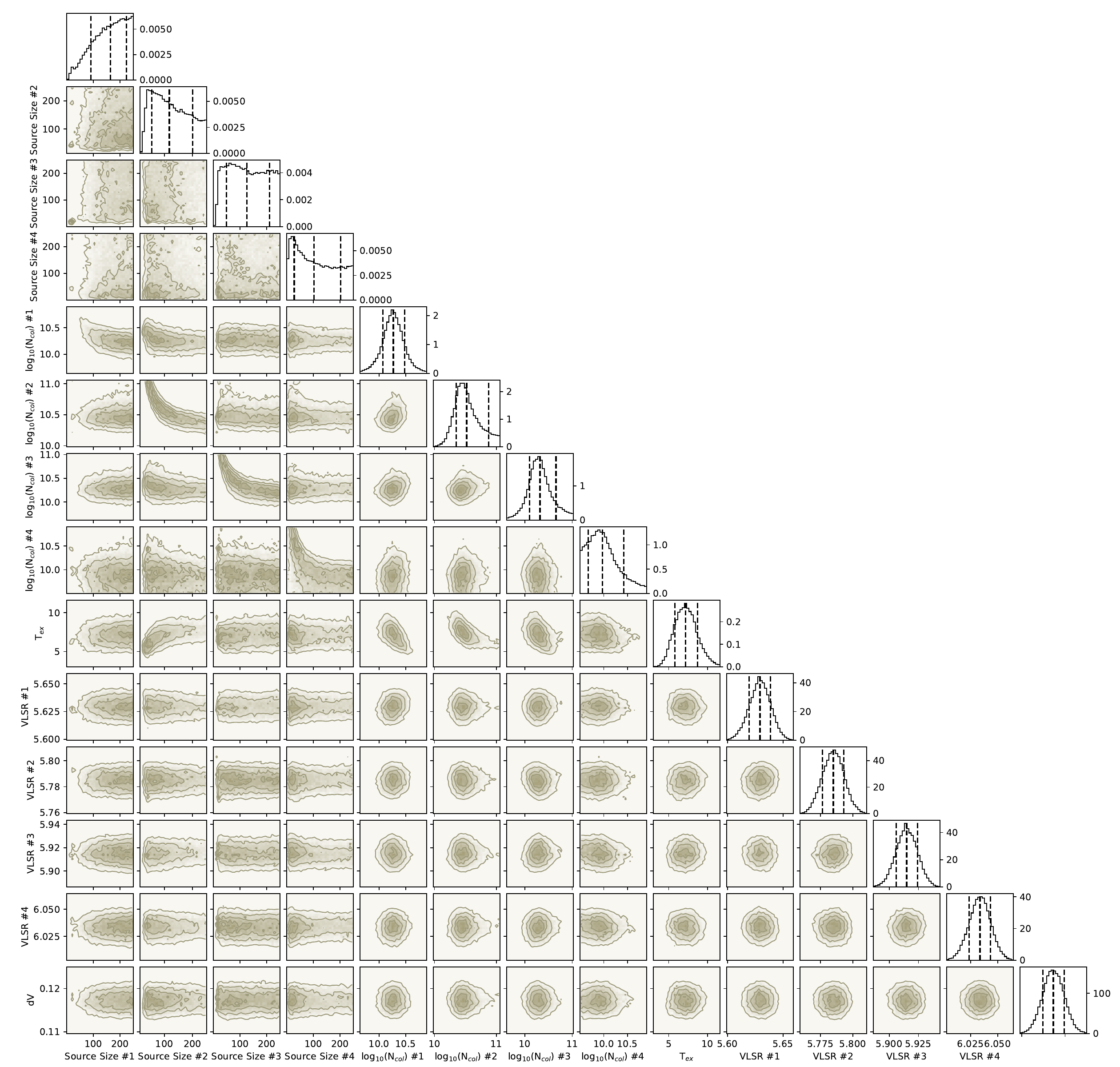}
\figsetgrpnote{The 16$^{th}$, 50$^{th}$, and 84$^{th}$ confidence intervals (corresponding to $\pm$1 sigma for a Gaussian posterior distribution) are shown as vertical lines. The contour lines are posterior probability levels, starting at $20\%$ of the maximum a posteriori estimate, with evenly spaced intervals of $20\%$ up to the peak density.}
\figsetgrpend

\figsetgrpstart
\figsetgrpnum{8.74}
\figsetgrptitle{Corner plot for 1-C$_{10}$H$_{7}$CN.}
\figsetplot{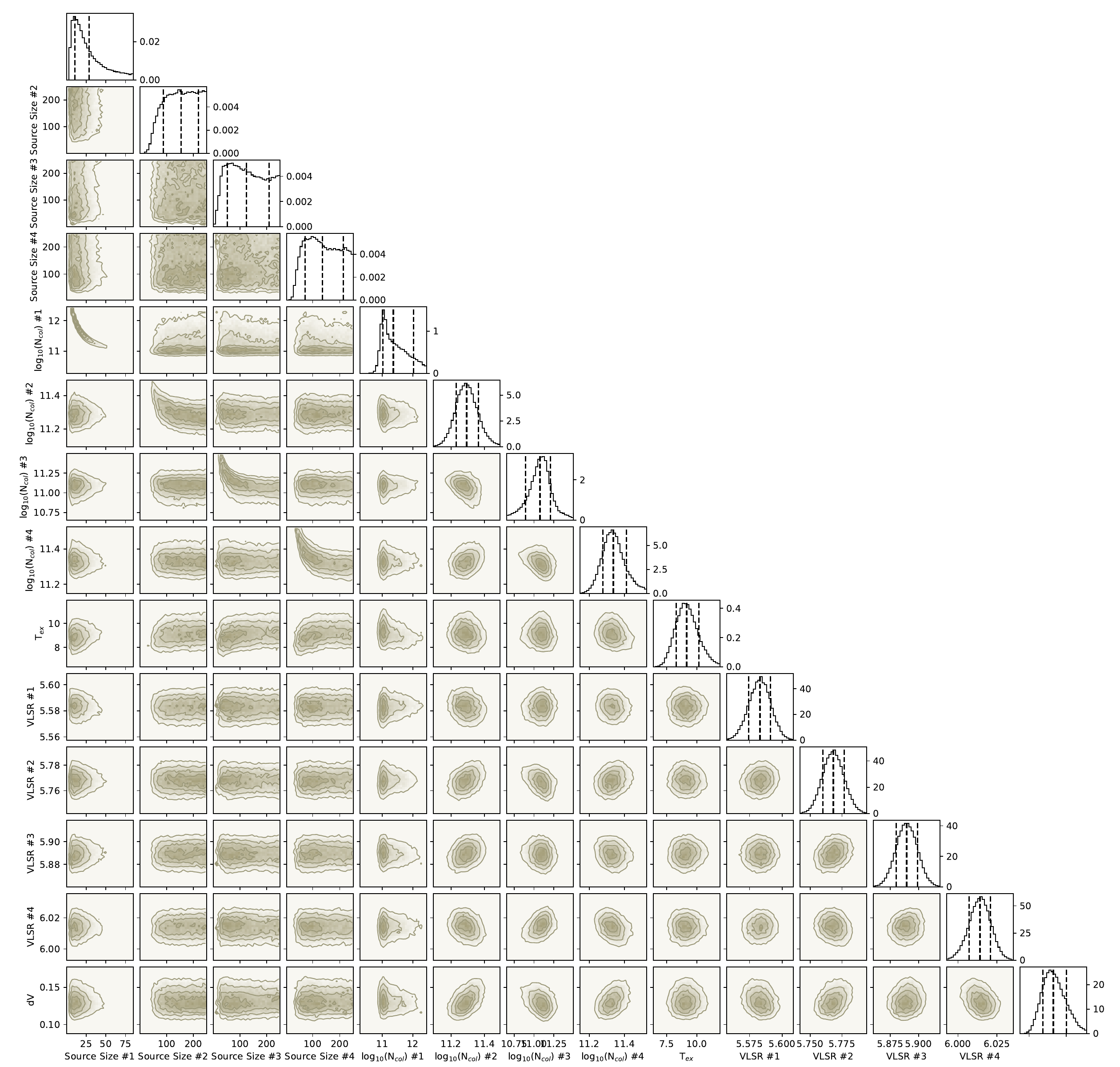}
\figsetgrpnote{The 16$^{th}$, 50$^{th}$, and 84$^{th}$ confidence intervals (corresponding to $\pm$1 sigma for a Gaussian posterior distribution) are shown as vertical lines. The contour lines are posterior probability levels, starting at $20\%$ of the maximum a posteriori estimate, with evenly spaced intervals of $20\%$ up to the peak density.}
\figsetgrpend

\figsetgrpstart
\figsetgrpnum{8.75}
\figsetgrptitle{Corner plot for 2-C$_{10}$H$_{7}$CN.}
\figsetplot{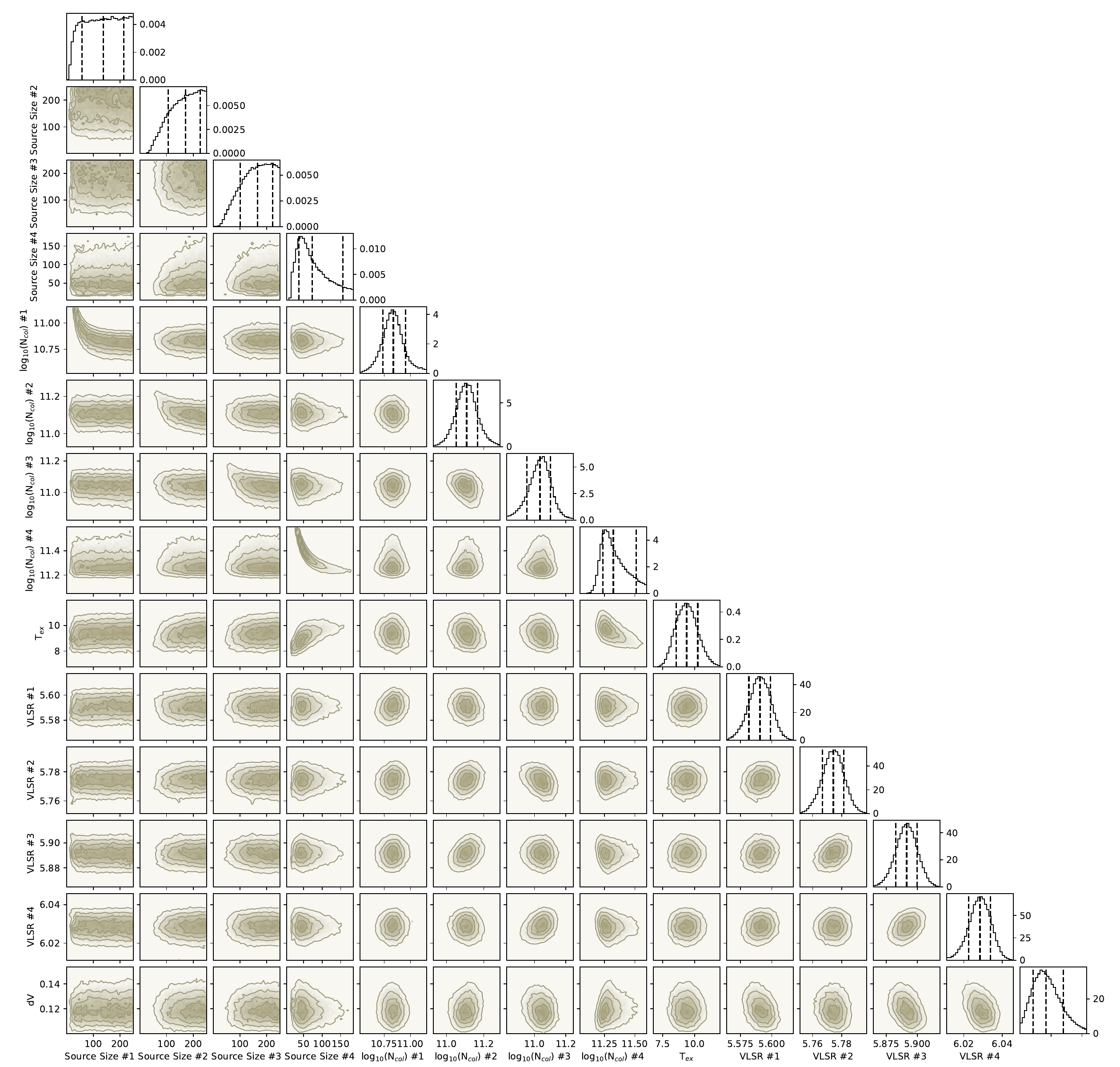}
\figsetgrpnote{The 16$^{th}$, 50$^{th}$, and 84$^{th}$ confidence intervals (corresponding to $\pm$1 sigma for a Gaussian posterior distribution) are shown as vertical lines. The contour lines are posterior probability levels, starting at $20\%$ of the maximum a posteriori estimate, with evenly spaced intervals of $20\%$ up to the peak density.}
\figsetgrpend

\figsetgrpstart
\figsetgrpnum{8.76}
\figsetgrptitle{Corner plot for 1-C$_{12}$H$_{7}$CN.}
\figsetplot{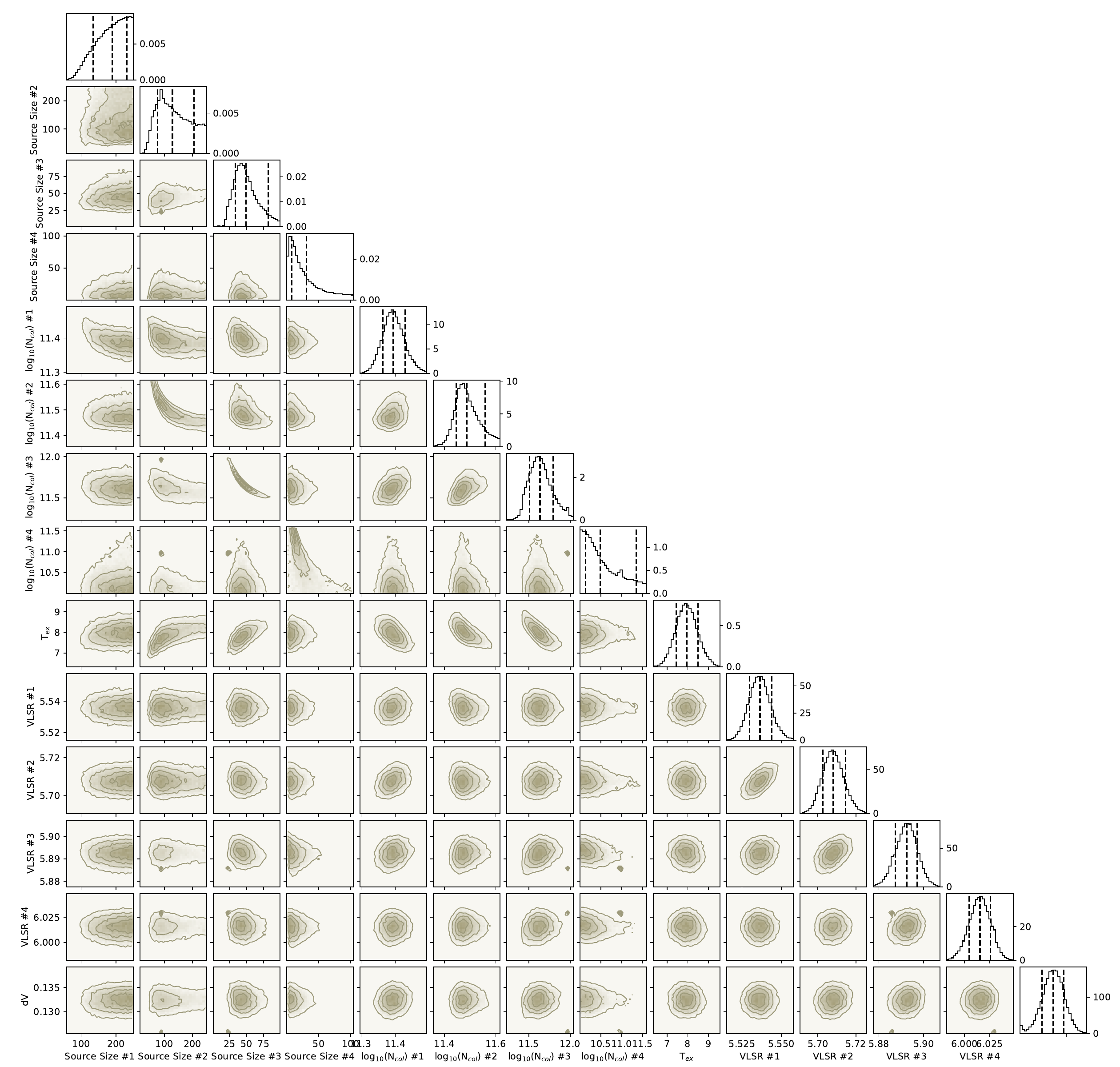}
\figsetgrpnote{The 16$^{th}$, 50$^{th}$, and 84$^{th}$ confidence intervals (corresponding to $\pm$1 sigma for a Gaussian posterior distribution) are shown as vertical lines. The contour lines are posterior probability levels, starting at $20\%$ of the maximum a posteriori estimate, with evenly spaced intervals of $20\%$ up to the peak density.}
\figsetgrpend

\figsetgrpstart
\figsetgrpnum{8.77}
\figsetgrptitle{Corner plot for 5-C$_{12}$H$_{7}$CN.}
\figsetplot{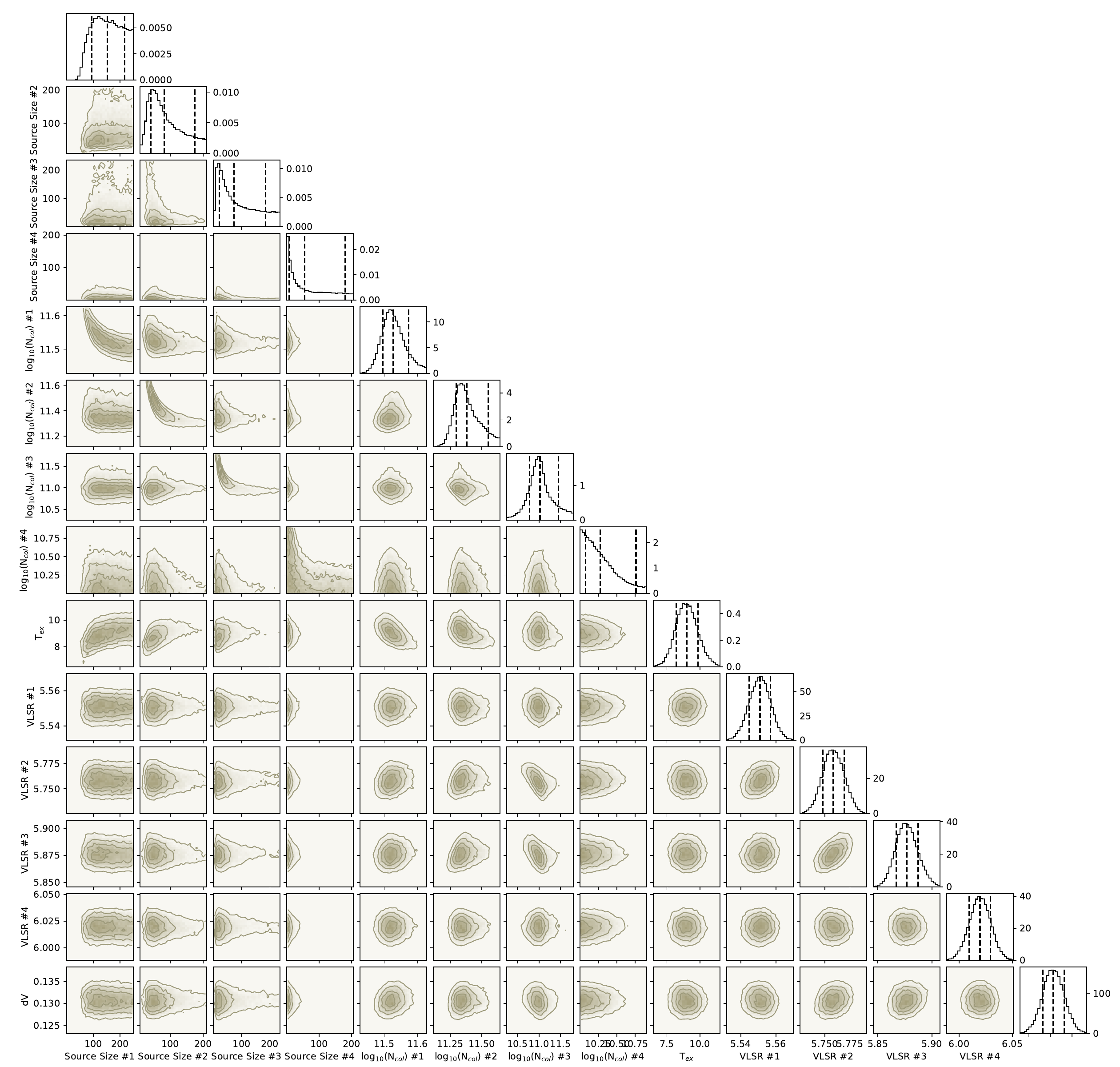}
\figsetgrpnote{The 16$^{th}$, 50$^{th}$, and 84$^{th}$ confidence intervals (corresponding to $\pm$1 sigma for a Gaussian posterior distribution) are shown as vertical lines. The contour lines are posterior probability levels, starting at $20\%$ of the maximum a posteriori estimate, with evenly spaced intervals of $20\%$ up to the peak density.}
\figsetgrpend

\figsetgrpstart
\figsetgrpnum{8.78}
\figsetgrptitle{Corner plot for 1-C$_{16}$H$_{9}$CN.}
\figsetplot{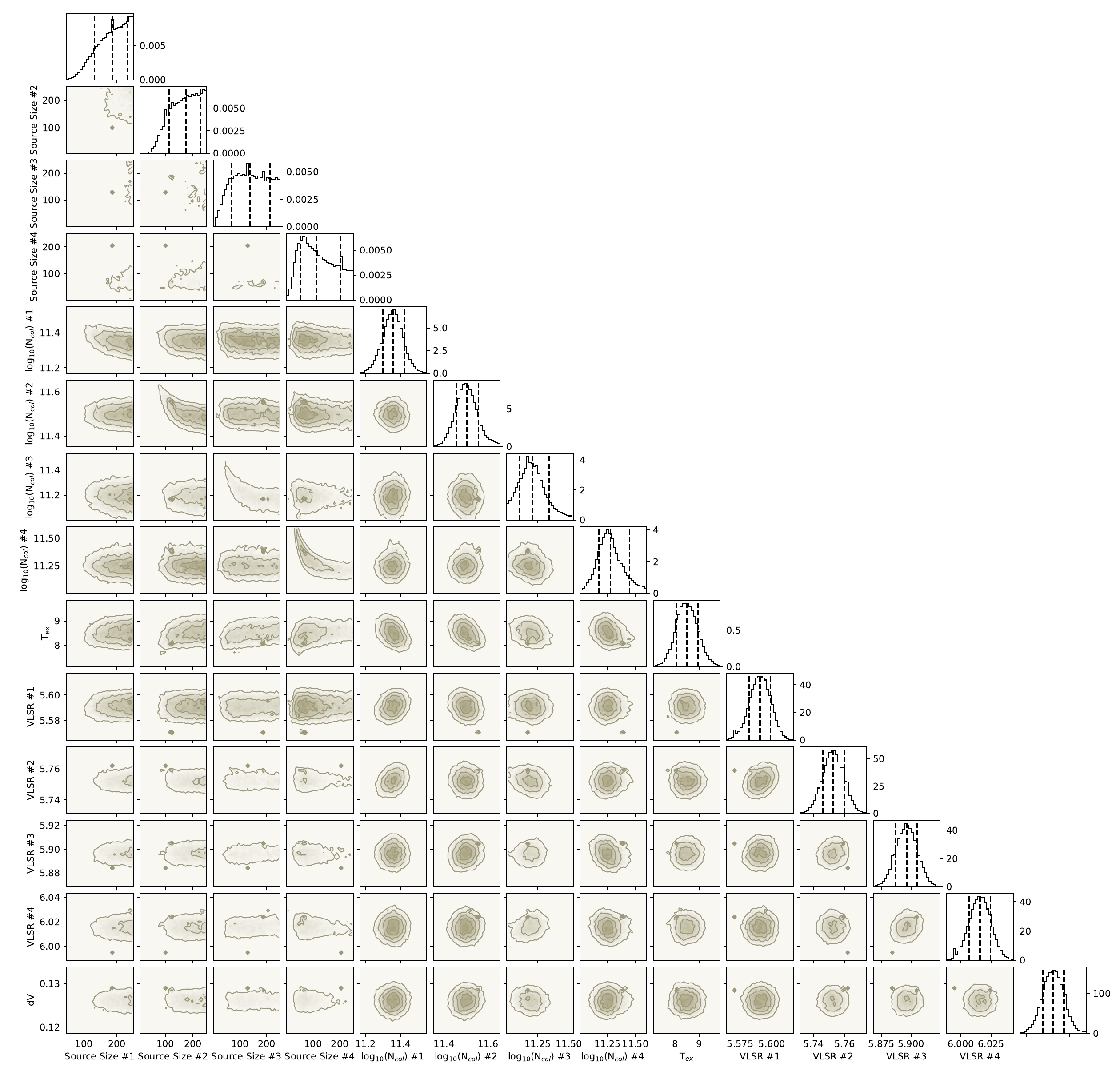}
\figsetgrpnote{The 16$^{th}$, 50$^{th}$, and 84$^{th}$ confidence intervals (corresponding to $\pm$1 sigma for a Gaussian posterior distribution) are shown as vertical lines. The contour lines are posterior probability levels, starting at $20\%$ of the maximum a posteriori estimate, with evenly spaced intervals of $20\%$ up to the peak density.}
\figsetgrpend

\figsetgrpstart
\figsetgrpnum{8.79}
\figsetgrptitle{Corner plot for 2-C$_{16}$H$_{9}$CN.}
\figsetplot{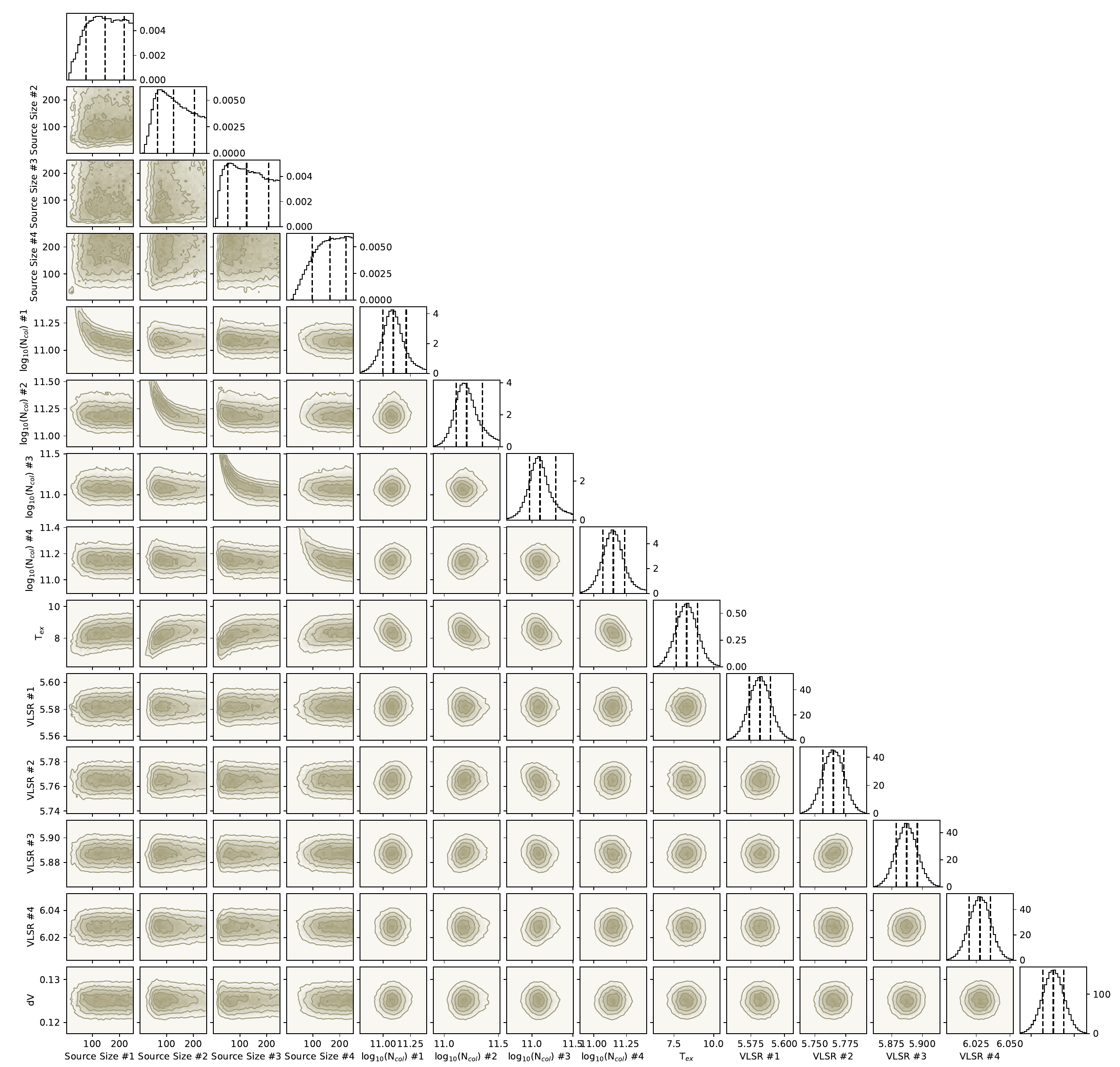}
\figsetgrpnote{The 16$^{th}$, 50$^{th}$, and 84$^{th}$ confidence intervals (corresponding to $\pm$1 sigma for a Gaussian posterior distribution) are shown as vertical lines. The contour lines are posterior probability levels, starting at $20\%$ of the maximum a posteriori estimate, with evenly spaced intervals of $20\%$ up to the peak density.}
\figsetgrpend

\figsetgrpstart
\figsetgrpnum{8.80}
\figsetgrptitle{Corner plot for 4-C$_{16}$H$_{9}$CN.}
\figsetplot{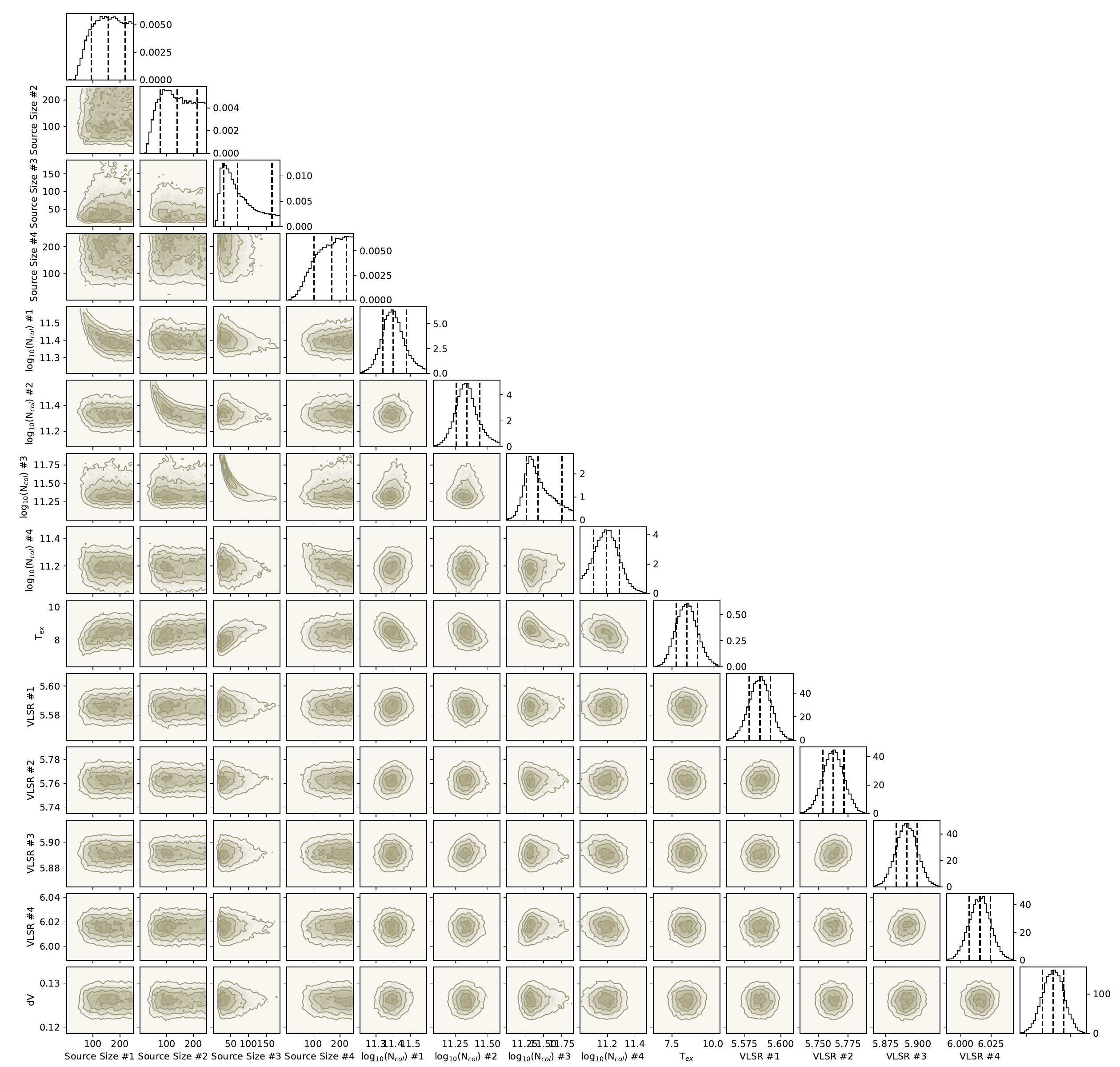}
\figsetgrpnote{The 16$^{th}$, 50$^{th}$, and 84$^{th}$ confidence intervals (corresponding to $\pm$1 sigma for a Gaussian posterior distribution) are shown as vertical lines. The contour lines are posterior probability levels, starting at $20\%$ of the maximum a posteriori estimate, with evenly spaced intervals of $20\%$ up to the peak density.}
\figsetgrpend

\figsetgrpstart
\figsetgrpnum{8.81}
\figsetgrptitle{Corner plot for $^{13}$CCCCH.}
\figsetplot{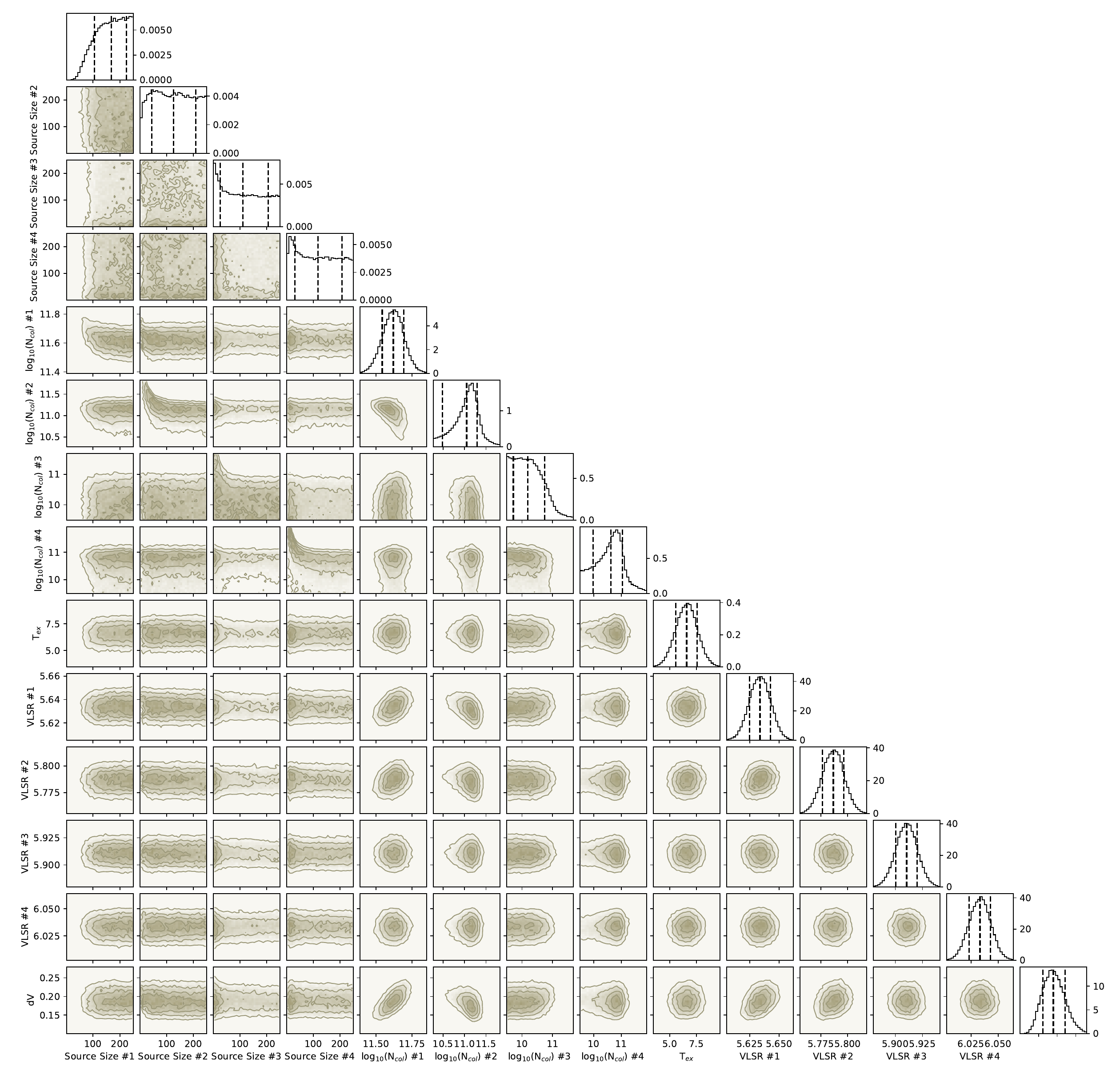}
\figsetgrpnote{The 16$^{th}$, 50$^{th}$, and 84$^{th}$ confidence intervals (corresponding to $\pm$1 sigma for a Gaussian posterior distribution) are shown as vertical lines. The contour lines are posterior probability levels, starting at $20\%$ of the maximum a posteriori estimate, with evenly spaced intervals of $20\%$ up to the peak density.}
\figsetgrpend

\figsetgrpstart
\figsetgrpnum{8.82}
\figsetgrptitle{Corner plot for C$^{13}$CCCH.}
\figsetplot{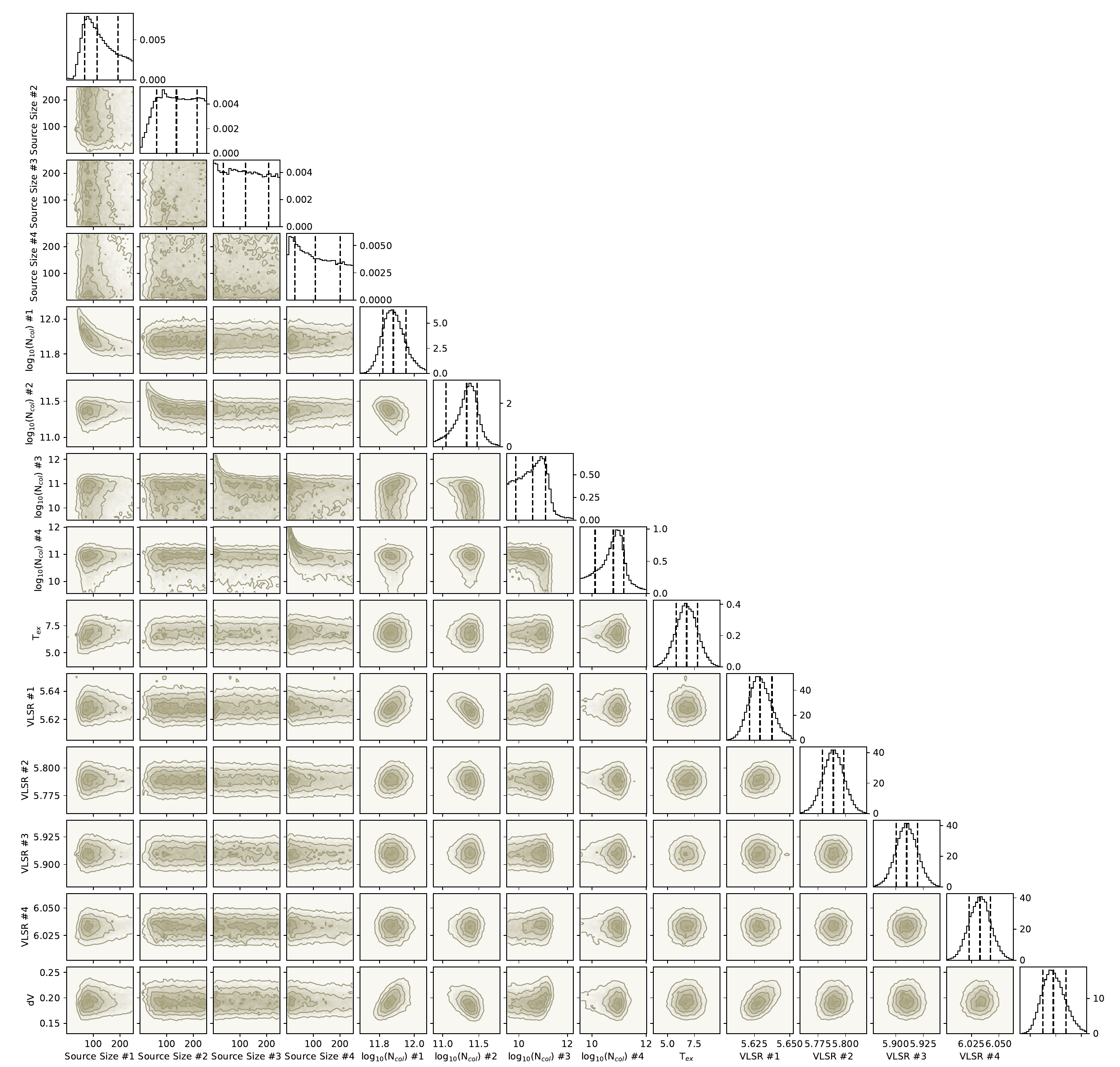}
\figsetgrpnote{The 16$^{th}$, 50$^{th}$, and 84$^{th}$ confidence intervals (corresponding to $\pm$1 sigma for a Gaussian posterior distribution) are shown as vertical lines. The contour lines are posterior probability levels, starting at $20\%$ of the maximum a posteriori estimate, with evenly spaced intervals of $20\%$ up to the peak density.}
\figsetgrpend

\figsetgrpstart
\figsetgrpnum{8.83}
\figsetgrptitle{Corner plot for CC$^{13}$CCH.}
\figsetplot{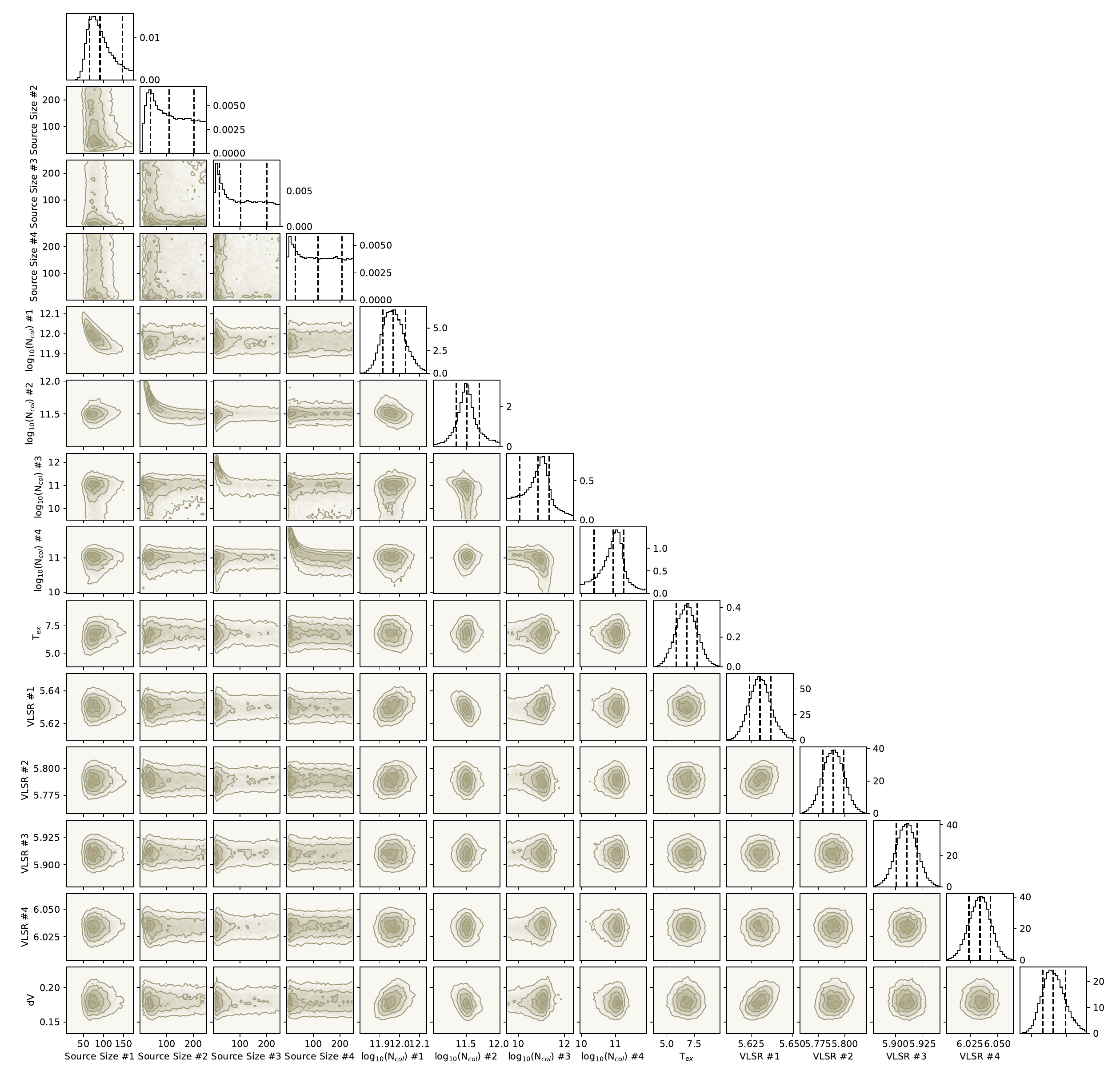}
\figsetgrpnote{The 16$^{th}$, 50$^{th}$, and 84$^{th}$ confidence intervals (corresponding to $\pm$1 sigma for a Gaussian posterior distribution) are shown as vertical lines. The contour lines are posterior probability levels, starting at $20\%$ of the maximum a posteriori estimate, with evenly spaced intervals of $20\%$ up to the peak density.}
\figsetgrpend

\figsetgrpstart
\figsetgrpnum{8.84}
\figsetgrptitle{Corner plot for CCC$^{13}$CH.}
\figsetplot{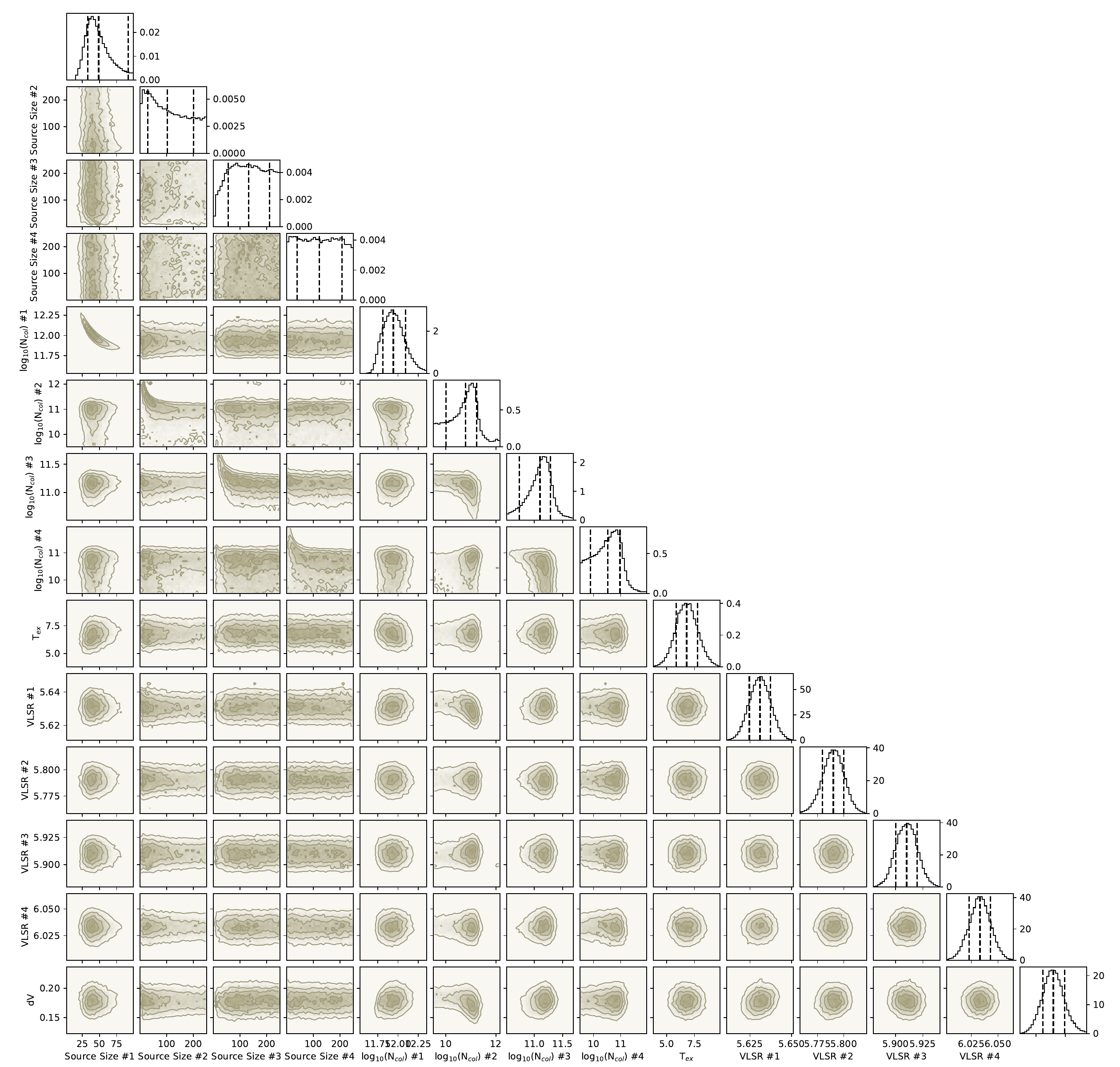}
\figsetgrpnote{The 16$^{th}$, 50$^{th}$, and 84$^{th}$ confidence intervals (corresponding to $\pm$1 sigma for a Gaussian posterior distribution) are shown as vertical lines. The contour lines are posterior probability levels, starting at $20\%$ of the maximum a posteriori estimate, with evenly spaced intervals of $20\%$ up to the peak density.}
\figsetgrpend

\figsetgrpstart
\figsetgrpnum{8.85}
\figsetgrptitle{Corner plot for H$^{13}$CCCN.}
\figsetplot{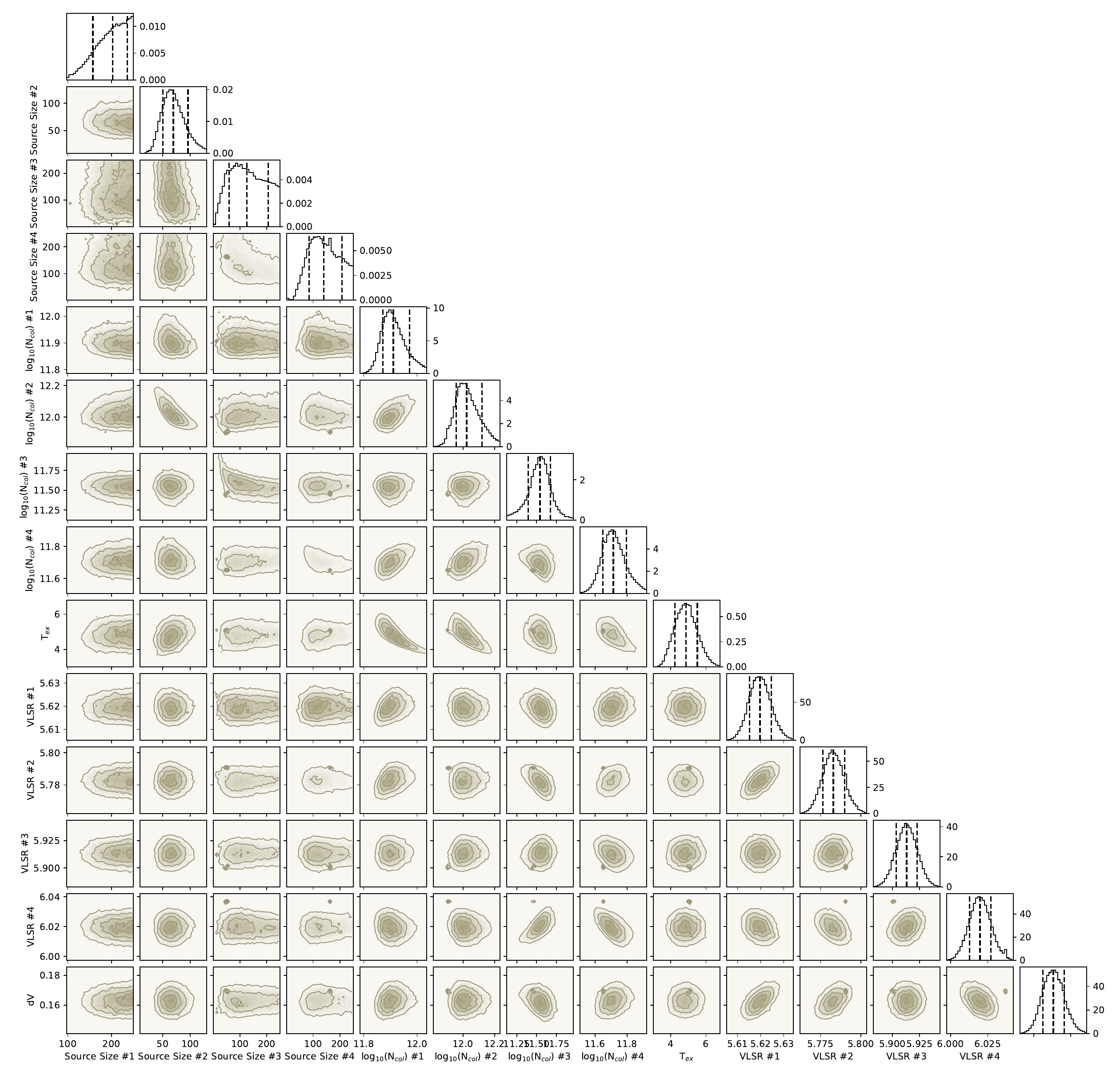}
\figsetgrpnote{The 16$^{th}$, 50$^{th}$, and 84$^{th}$ confidence intervals (corresponding to $\pm$1 sigma for a Gaussian posterior distribution) are shown as vertical lines. The contour lines are posterior probability levels, starting at $20\%$ of the maximum a posteriori estimate, with evenly spaced intervals of $20\%$ up to the peak density.}
\figsetgrpend

\figsetgrpstart
\figsetgrpnum{8.86}
\figsetgrptitle{Corner plot for HC$^{13}$CCN.}
\figsetplot{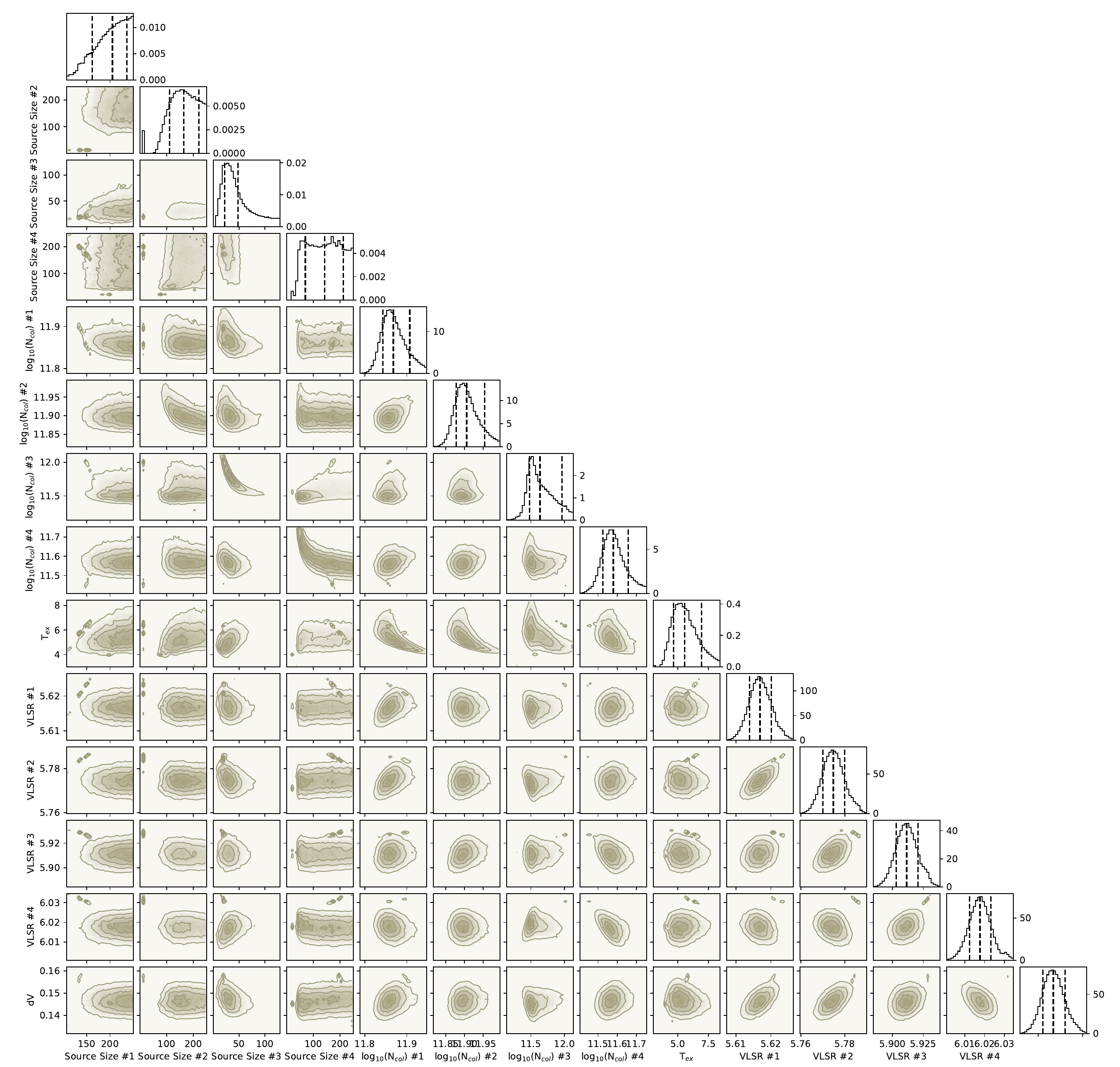}
\figsetgrpnote{The 16$^{th}$, 50$^{th}$, and 84$^{th}$ confidence intervals (corresponding to $\pm$1 sigma for a Gaussian posterior distribution) are shown as vertical lines. The contour lines are posterior probability levels, starting at $20\%$ of the maximum a posteriori estimate, with evenly spaced intervals of $20\%$ up to the peak density.}
\figsetgrpend

\figsetgrpstart
\figsetgrpnum{8.87}
\figsetgrptitle{Corner plot for HCC$^{13}$CN.}
\figsetplot{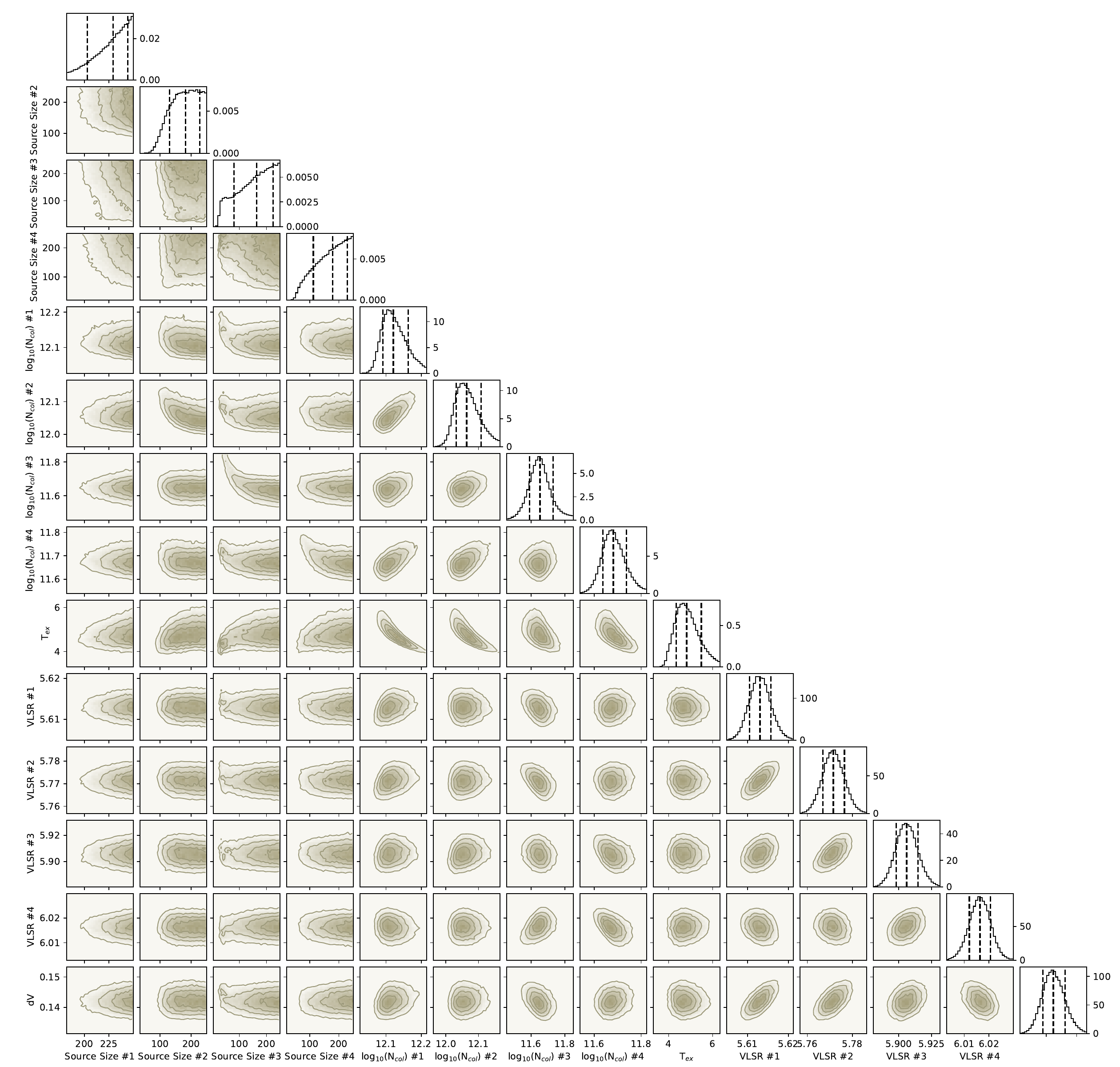}
\figsetgrpnote{The 16$^{th}$, 50$^{th}$, and 84$^{th}$ confidence intervals (corresponding to $\pm$1 sigma for a Gaussian posterior distribution) are shown as vertical lines. The contour lines are posterior probability levels, starting at $20\%$ of the maximum a posteriori estimate, with evenly spaced intervals of $20\%$ up to the peak density.}
\figsetgrpend

\figsetgrpstart
\figsetgrpnum{8.88}
\figsetgrptitle{Corner plot for C$^{13}$CS.}
\figsetplot{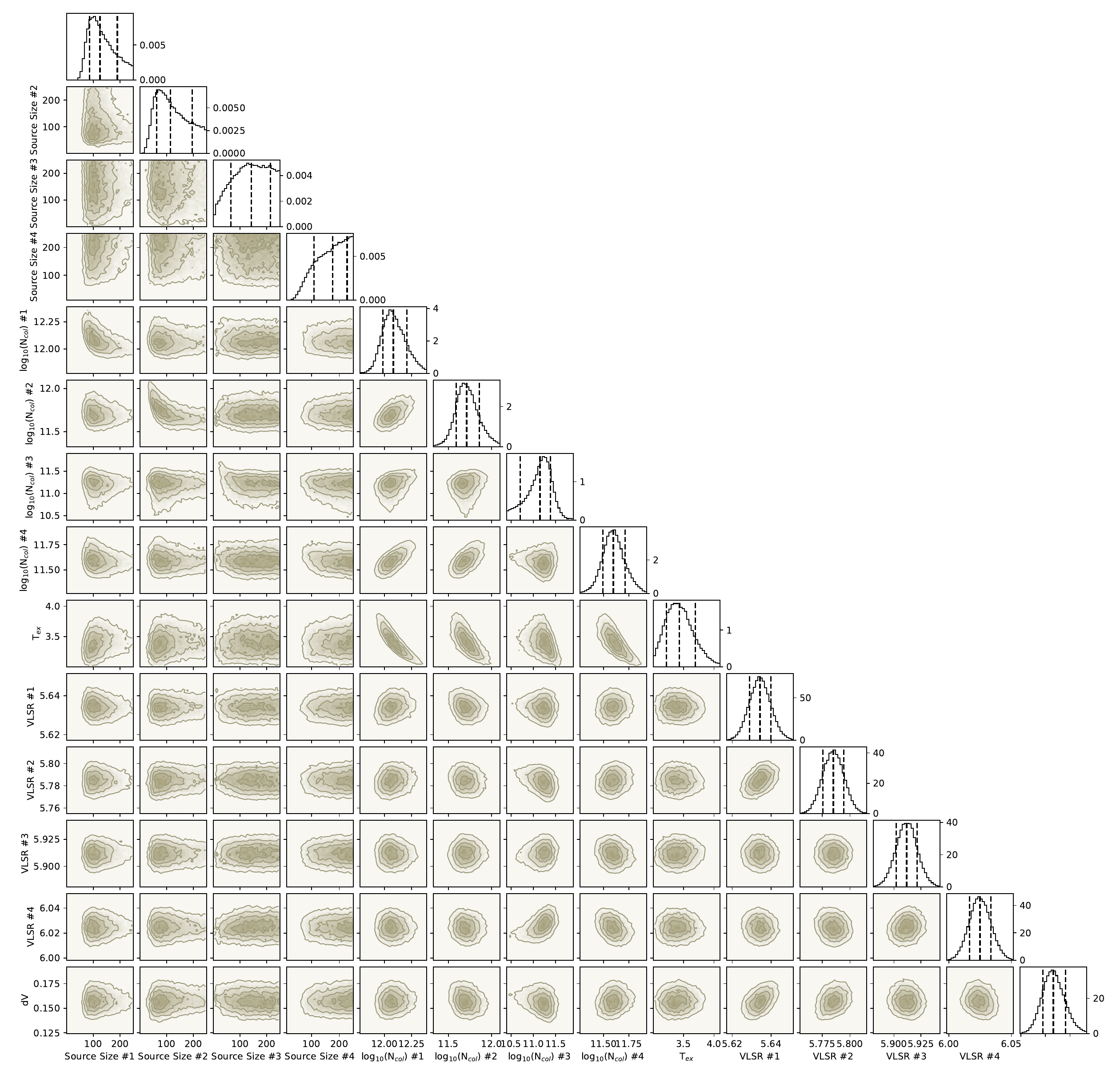}
\figsetgrpnote{The 16$^{th}$, 50$^{th}$, and 84$^{th}$ confidence intervals (corresponding to $\pm$1 sigma for a Gaussian posterior distribution) are shown as vertical lines. The contour lines are posterior probability levels, starting at $20\%$ of the maximum a posteriori estimate, with evenly spaced intervals of $20\%$ up to the peak density.}
\figsetgrpend

\figsetgrpstart
\figsetgrpnum{8.89}
\figsetgrptitle{Corner plot for H$^{13}$CC$_{4}$N.}
\figsetplot{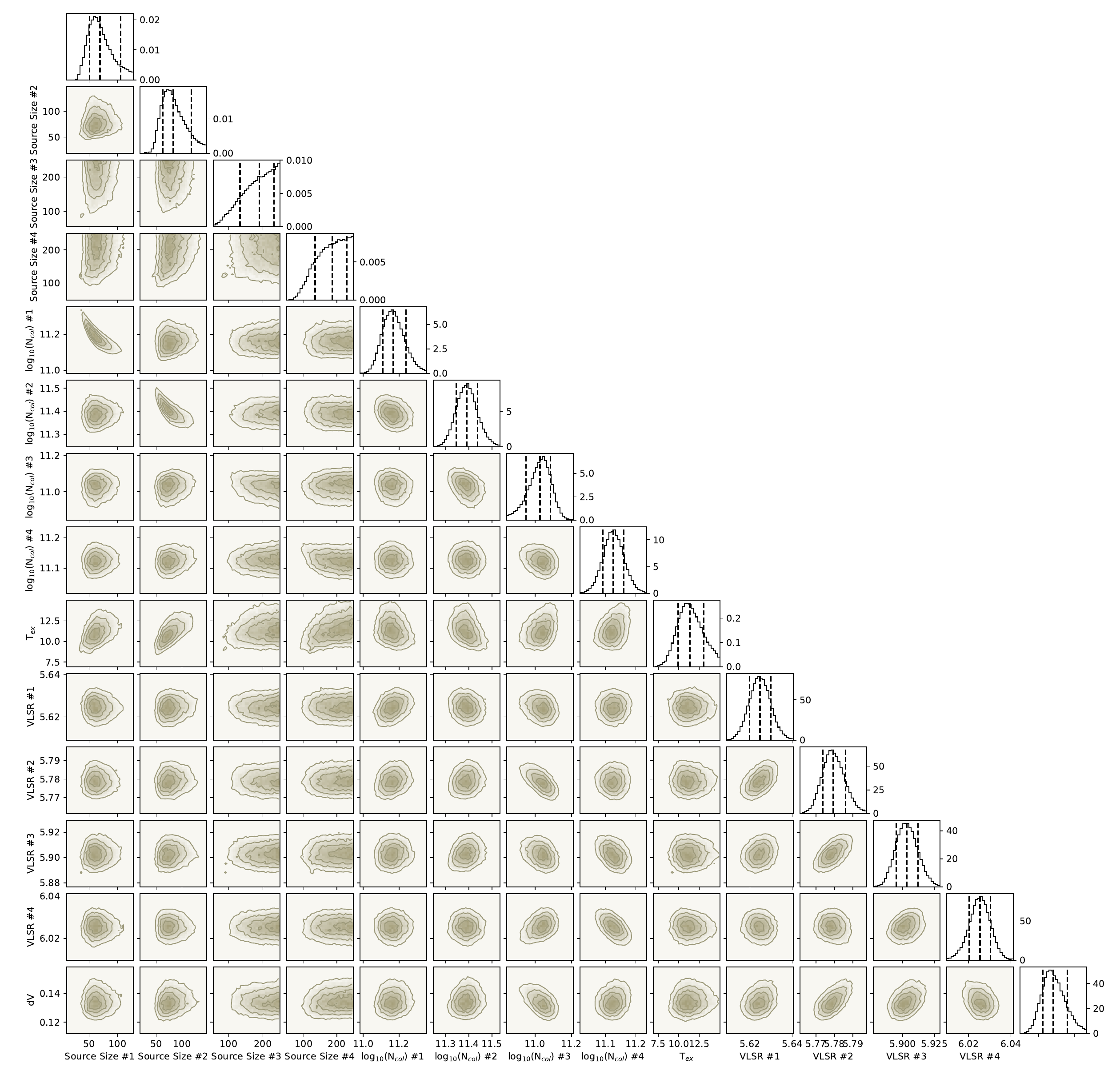}
\figsetgrpnote{The 16$^{th}$, 50$^{th}$, and 84$^{th}$ confidence intervals (corresponding to $\pm$1 sigma for a Gaussian posterior distribution) are shown as vertical lines. The contour lines are posterior probability levels, starting at $20\%$ of the maximum a posteriori estimate, with evenly spaced intervals of $20\%$ up to the peak density.}
\figsetgrpend

\figsetgrpstart
\figsetgrpnum{8.90}
\figsetgrptitle{Corner plot for HC$^{13}$CC$_{3}$N.}
\figsetplot{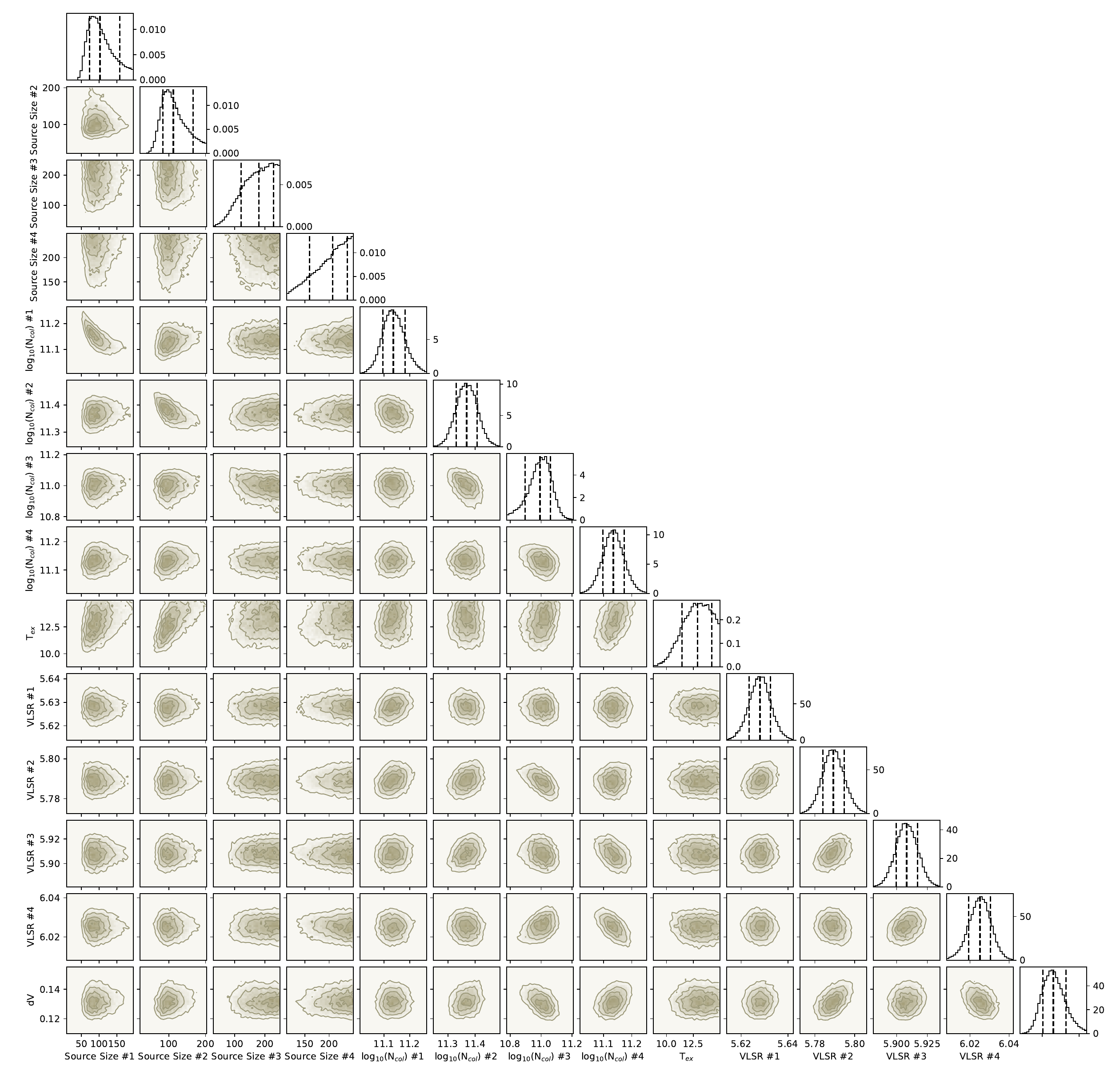}
\figsetgrpnote{The 16$^{th}$, 50$^{th}$, and 84$^{th}$ confidence intervals (corresponding to $\pm$1 sigma for a Gaussian posterior distribution) are shown as vertical lines. The contour lines are posterior probability levels, starting at $20\%$ of the maximum a posteriori estimate, with evenly spaced intervals of $20\%$ up to the peak density.}
\figsetgrpend

\figsetgrpstart
\figsetgrpnum{8.91}
\figsetgrptitle{Corner plot for HC$_{2}$$^{13}$CC$_{2}$N.}
\figsetplot{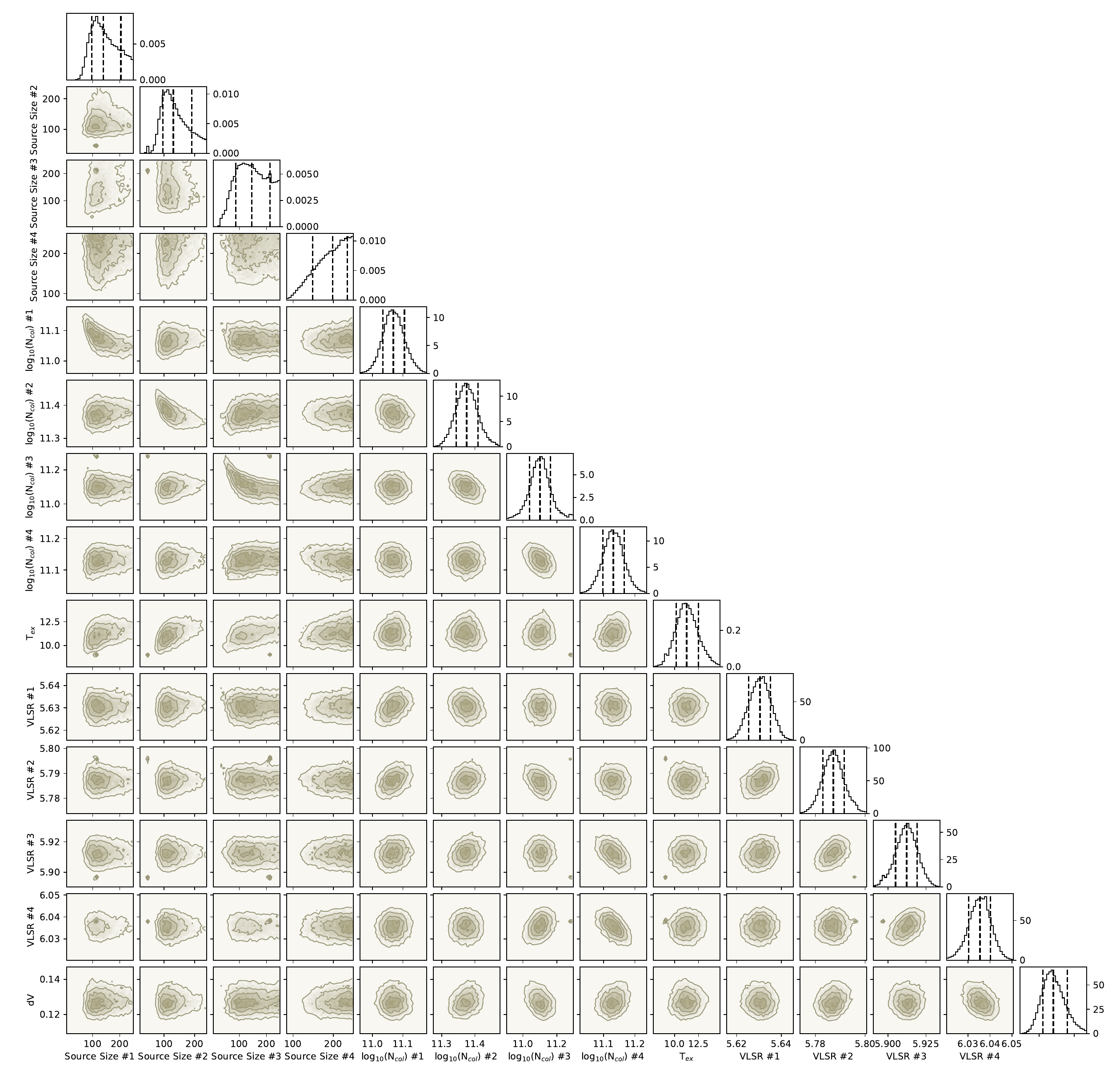}
\figsetgrpnote{The 16$^{th}$, 50$^{th}$, and 84$^{th}$ confidence intervals (corresponding to $\pm$1 sigma for a Gaussian posterior distribution) are shown as vertical lines. The contour lines are posterior probability levels, starting at $20\%$ of the maximum a posteriori estimate, with evenly spaced intervals of $20\%$ up to the peak density.}
\figsetgrpend

\figsetgrpstart
\figsetgrpnum{8.92}
\figsetgrptitle{Corner plot for HC$_{3}$$^{13}$CCN.}
\figsetplot{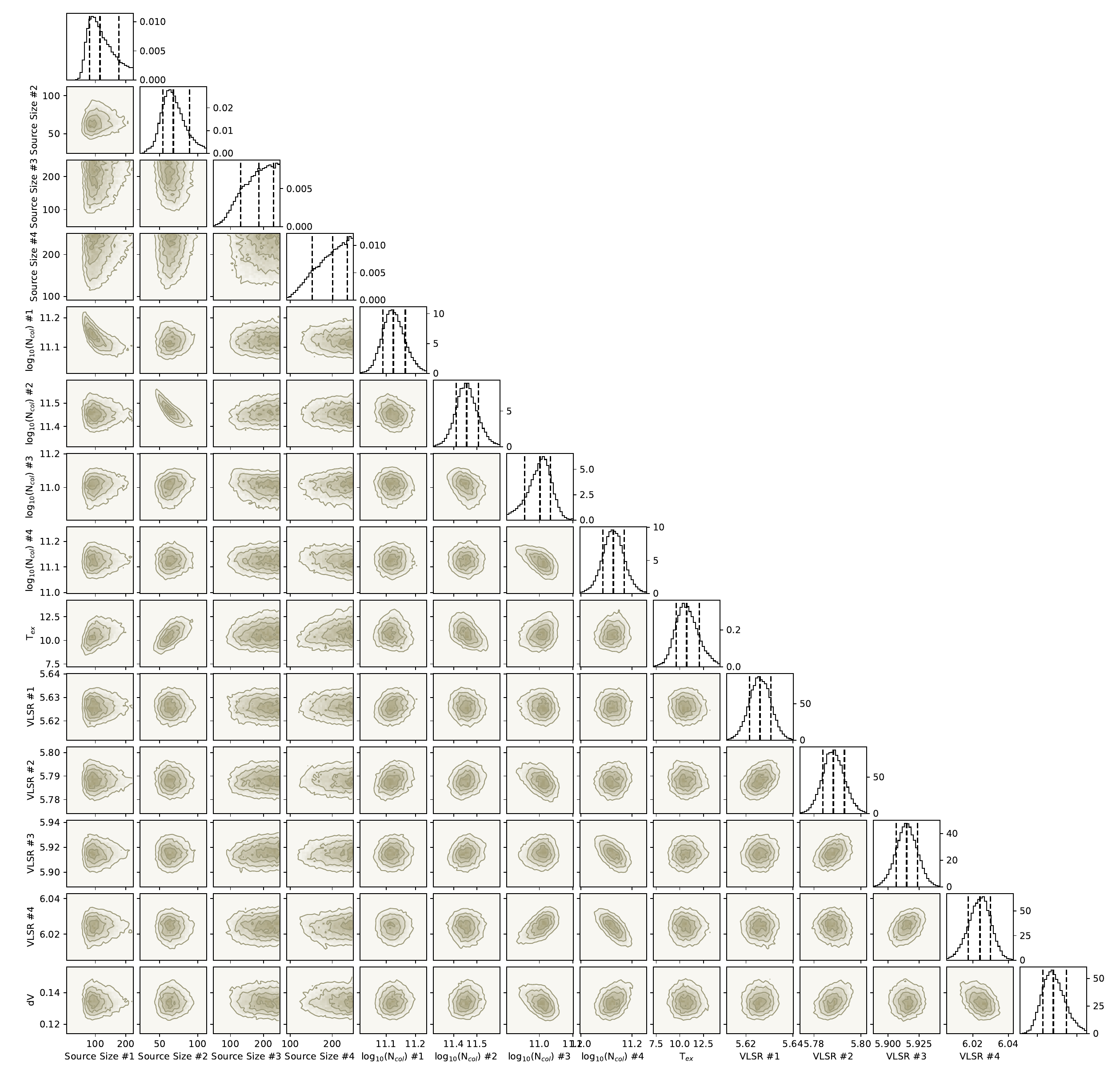}
\figsetgrpnote{The 16$^{th}$, 50$^{th}$, and 84$^{th}$ confidence intervals (corresponding to $\pm$1 sigma for a Gaussian posterior distribution) are shown as vertical lines. The contour lines are posterior probability levels, starting at $20\%$ of the maximum a posteriori estimate, with evenly spaced intervals of $20\%$ up to the peak density.}
\figsetgrpend

\figsetgrpstart
\figsetgrpnum{8.93}
\figsetgrptitle{Corner plot for HC$_{4}$$^{13}$CN.}
\figsetplot{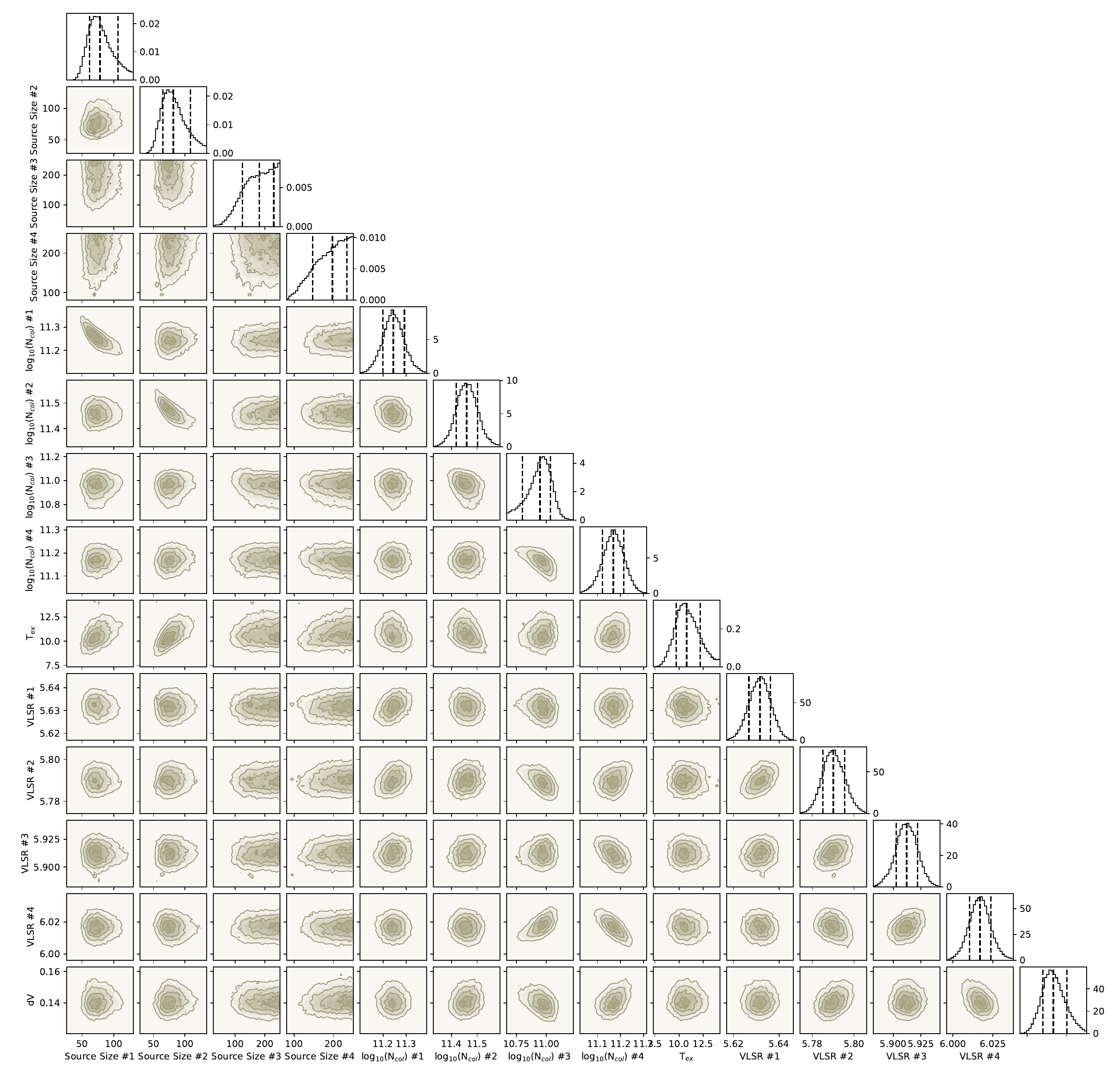}
\figsetgrpnote{The 16$^{th}$, 50$^{th}$, and 84$^{th}$ confidence intervals (corresponding to $\pm$1 sigma for a Gaussian posterior distribution) are shown as vertical lines. The contour lines are posterior probability levels, starting at $20\%$ of the maximum a posteriori estimate, with evenly spaced intervals of $20\%$ up to the peak density.}
\figsetgrpend

\figsetgrpstart
\figsetgrpnum{8.94}
\figsetgrptitle{Corner plot for H$^{13}$CC$_{6}$N.}
\figsetplot{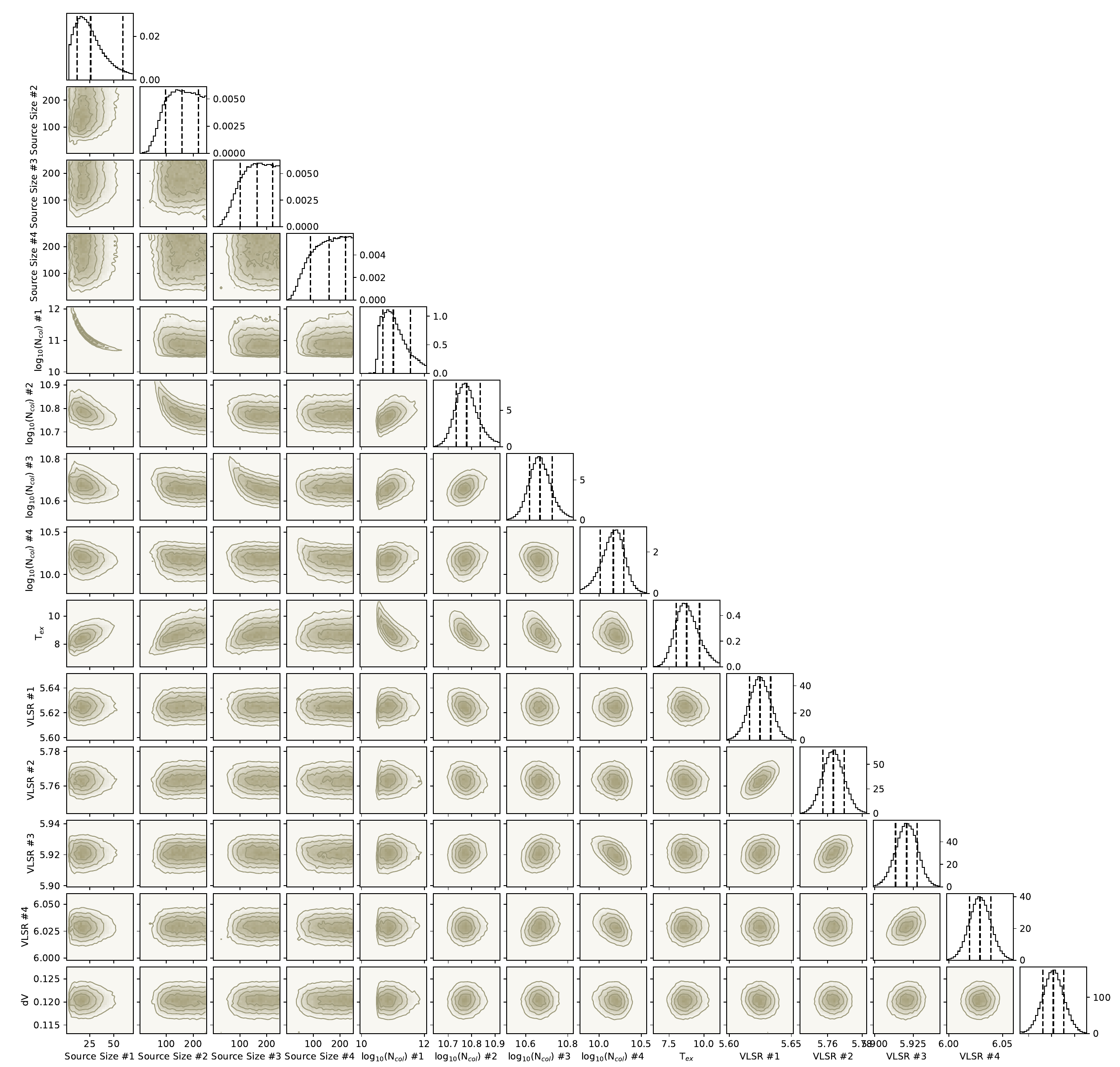}
\figsetgrpnote{The 16$^{th}$, 50$^{th}$, and 84$^{th}$ confidence intervals (corresponding to $\pm$1 sigma for a Gaussian posterior distribution) are shown as vertical lines. The contour lines are posterior probability levels, starting at $20\%$ of the maximum a posteriori estimate, with evenly spaced intervals of $20\%$ up to the peak density.}
\figsetgrpend

\figsetgrpstart
\figsetgrpnum{8.95}
\figsetgrptitle{Corner plot for HC$^{13}$CC$_{5}$N.}
\figsetplot{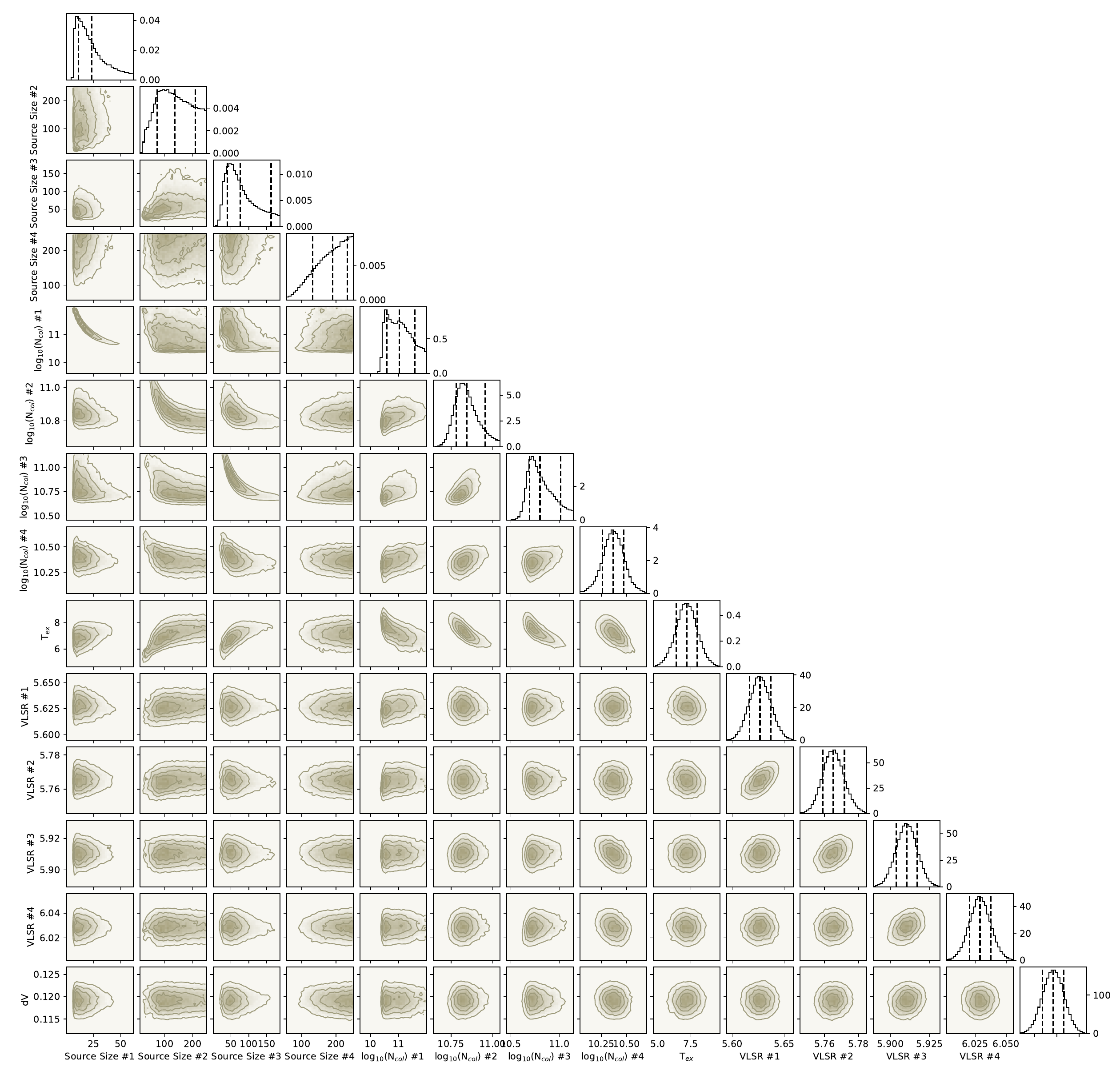}
\figsetgrpnote{The 16$^{th}$, 50$^{th}$, and 84$^{th}$ confidence intervals (corresponding to $\pm$1 sigma for a Gaussian posterior distribution) are shown as vertical lines. The contour lines are posterior probability levels, starting at $20\%$ of the maximum a posteriori estimate, with evenly spaced intervals of $20\%$ up to the peak density.}
\figsetgrpend

\figsetgrpstart
\figsetgrpnum{8.96}
\figsetgrptitle{Corner plot for HC$_{2}$$^{13}$CC$_{4}$N.}
\figsetplot{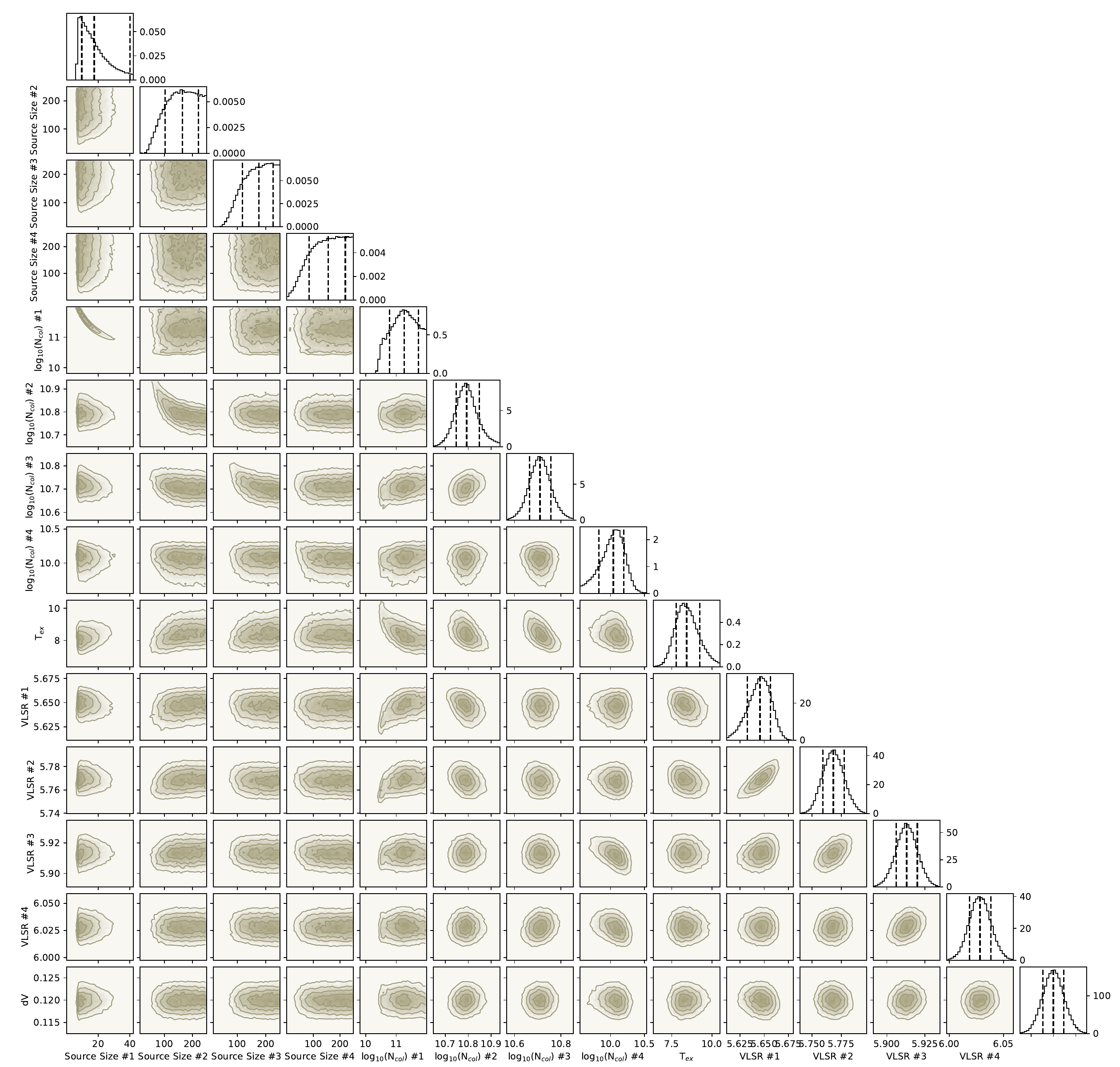}
\figsetgrpnote{The 16$^{th}$, 50$^{th}$, and 84$^{th}$ confidence intervals (corresponding to $\pm$1 sigma for a Gaussian posterior distribution) are shown as vertical lines. The contour lines are posterior probability levels, starting at $20\%$ of the maximum a posteriori estimate, with evenly spaced intervals of $20\%$ up to the peak density.}
\figsetgrpend

\figsetgrpstart
\figsetgrpnum{8.97}
\figsetgrptitle{Corner plot for HC$_{3}$$^{13}$CC$_{3}$N.}
\figsetplot{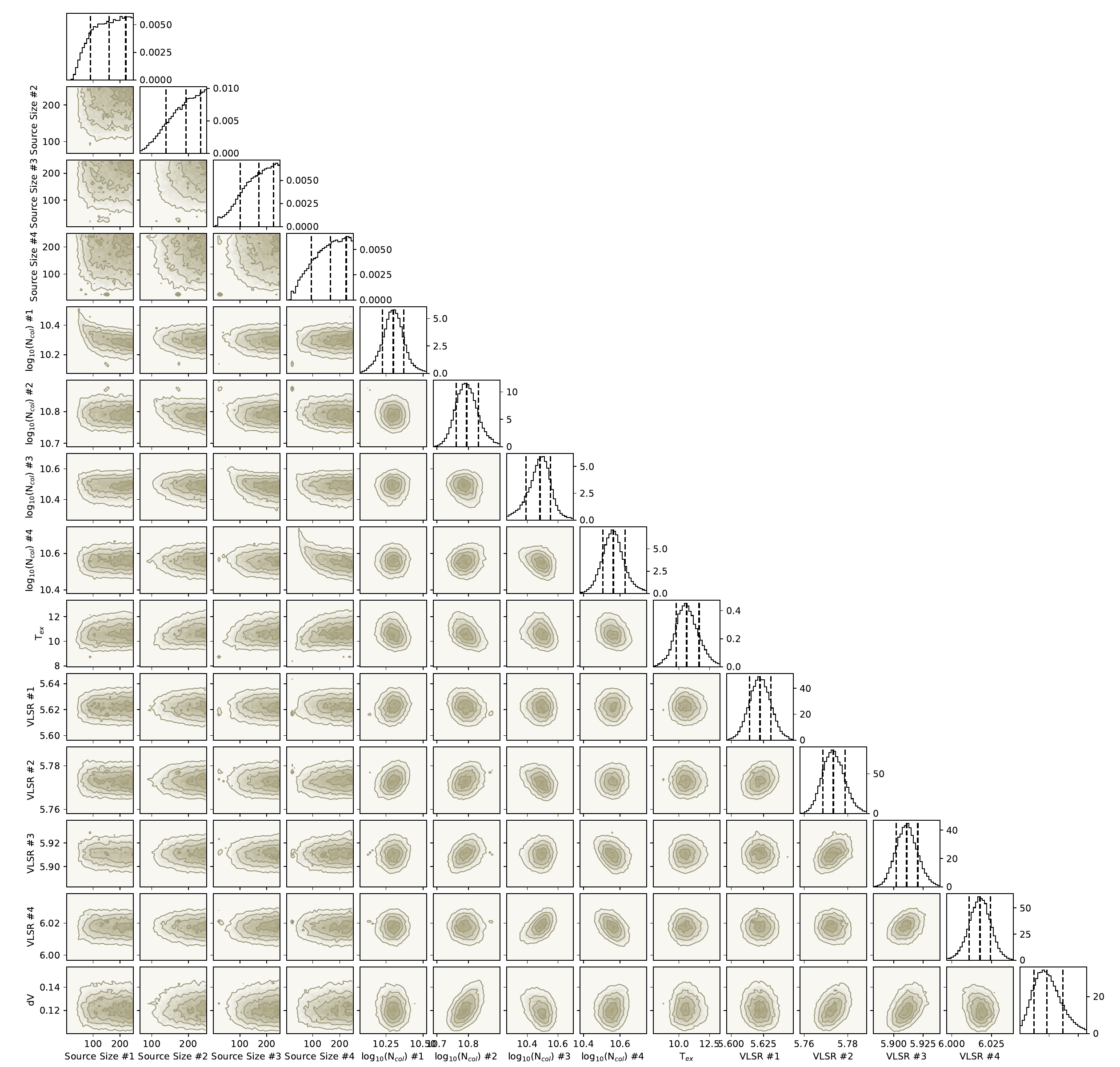}
\figsetgrpnote{The 16$^{th}$, 50$^{th}$, and 84$^{th}$ confidence intervals (corresponding to $\pm$1 sigma for a Gaussian posterior distribution) are shown as vertical lines. The contour lines are posterior probability levels, starting at $20\%$ of the maximum a posteriori estimate, with evenly spaced intervals of $20\%$ up to the peak density.}
\figsetgrpend

\figsetgrpstart
\figsetgrpnum{8.98}
\figsetgrptitle{Corner plot for HC$_{4}$$^{13}$CC$_{2}$N.}
\figsetplot{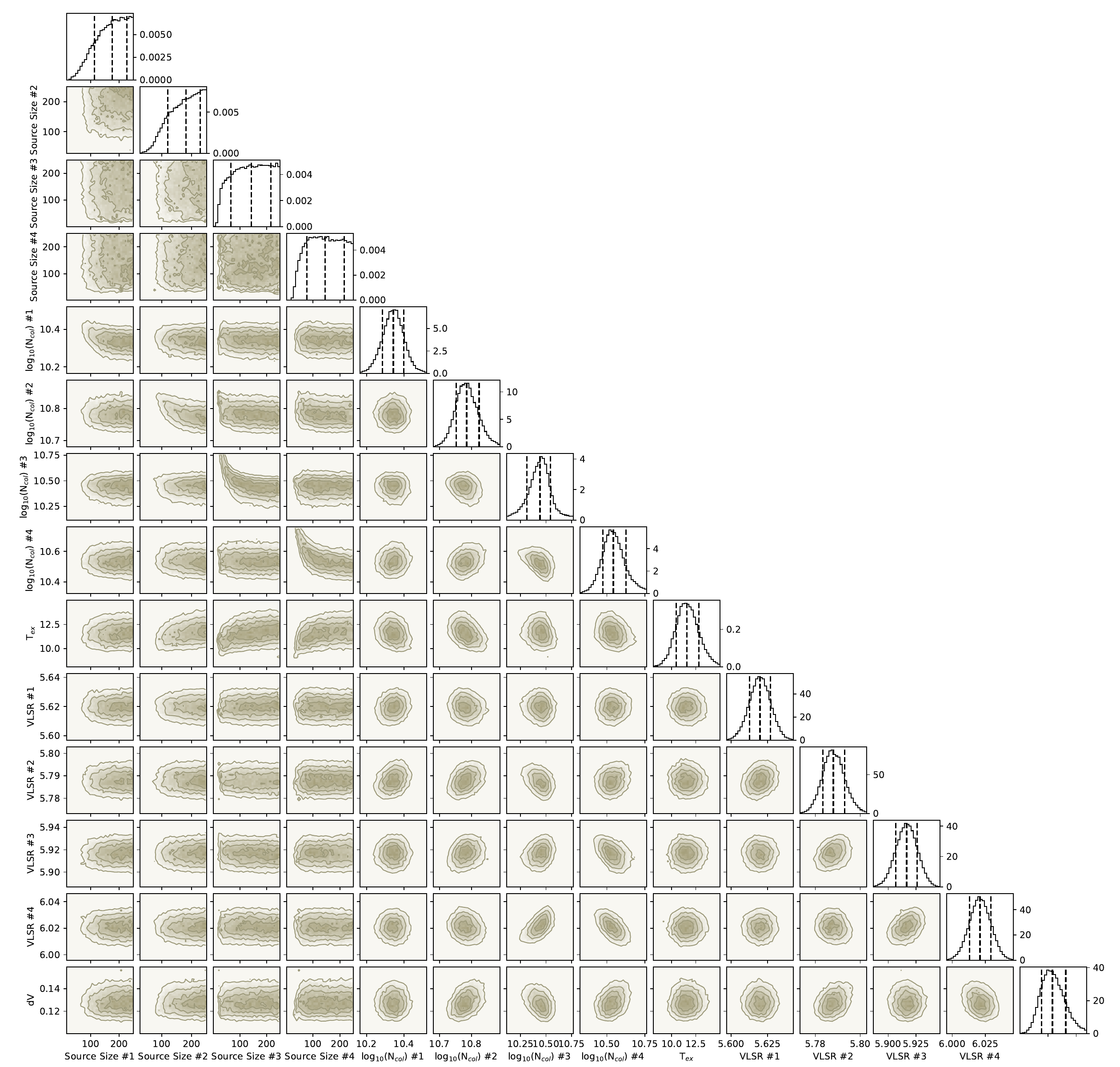}
\figsetgrpnote{The 16$^{th}$, 50$^{th}$, and 84$^{th}$ confidence intervals (corresponding to $\pm$1 sigma for a Gaussian posterior distribution) are shown as vertical lines. The contour lines are posterior probability levels, starting at $20\%$ of the maximum a posteriori estimate, with evenly spaced intervals of $20\%$ up to the peak density.}
\figsetgrpend

\figsetgrpstart
\figsetgrpnum{8.99}
\figsetgrptitle{Corner plot for HC$_{5}$$^{13}$CCN.}
\figsetplot{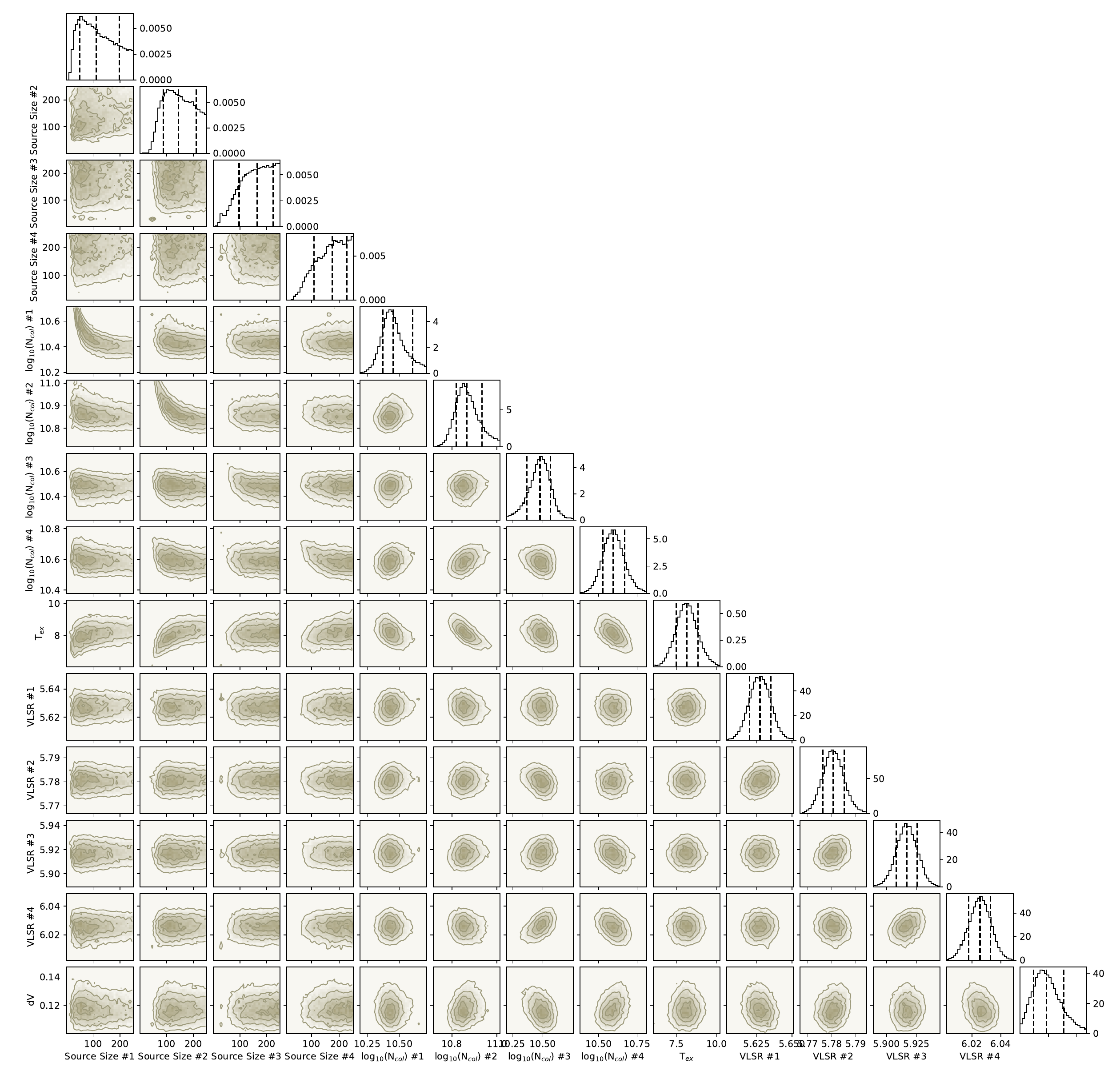}
\figsetgrpnote{The 16$^{th}$, 50$^{th}$, and 84$^{th}$ confidence intervals (corresponding to $\pm$1 sigma for a Gaussian posterior distribution) are shown as vertical lines. The contour lines are posterior probability levels, starting at $20\%$ of the maximum a posteriori estimate, with evenly spaced intervals of $20\%$ up to the peak density.}
\figsetgrpend

\figsetgrpstart
\figsetgrpnum{8.100}
\figsetgrptitle{Corner plot for HC$_{6}$$^{13}$CN.}
\figsetplot{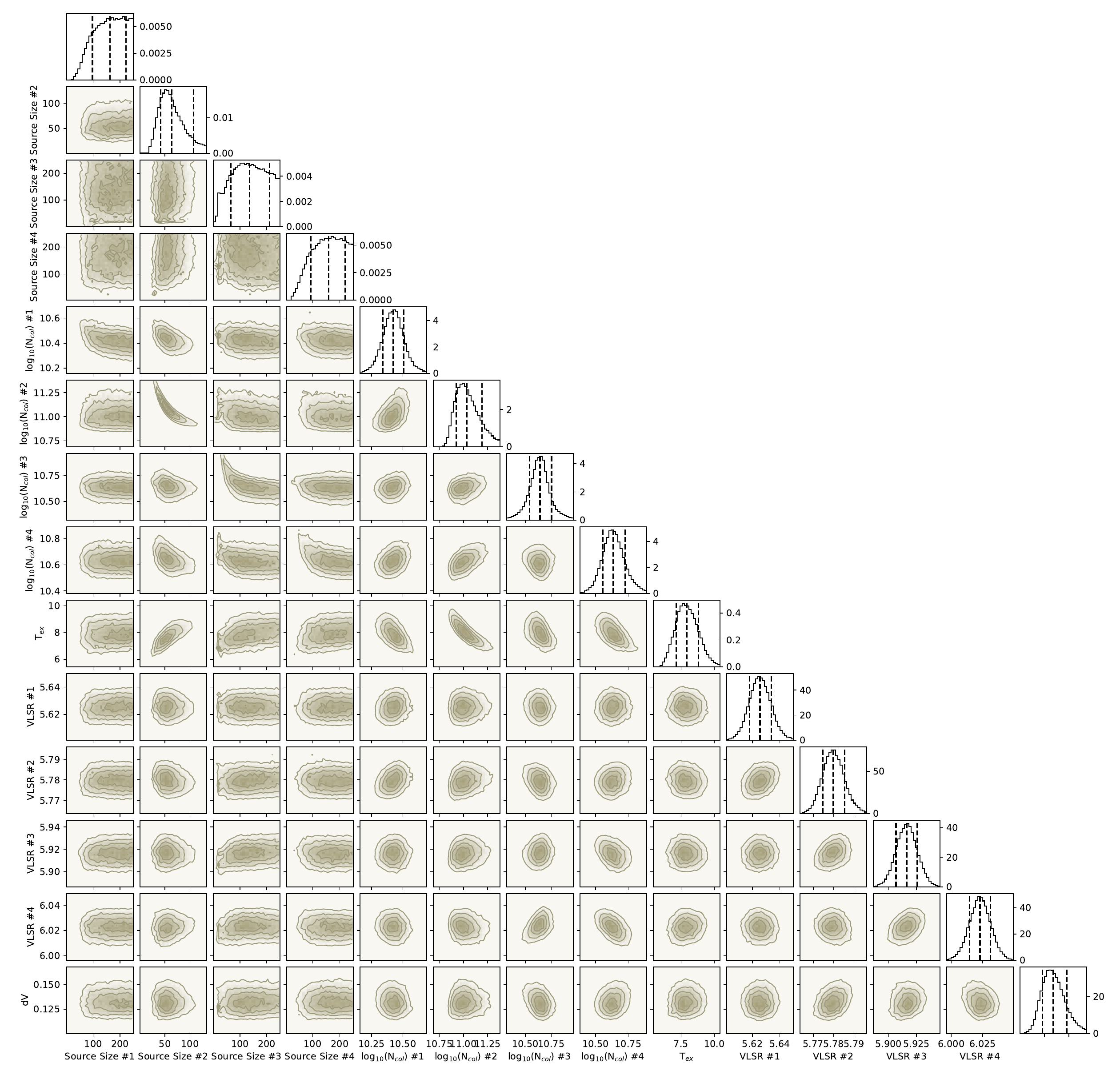}
\figsetgrpnote{The 16$^{th}$, 50$^{th}$, and 84$^{th}$ confidence intervals (corresponding to $\pm$1 sigma for a Gaussian posterior distribution) are shown as vertical lines. The contour lines are posterior probability levels, starting at $20\%$ of the maximum a posteriori estimate, with evenly spaced intervals of $20\%$ up to the peak density.}
\figsetgrpend

\figsetgrpstart
\figsetgrpnum{8.101}
\figsetgrptitle{Corner plot for HDCS.}
\figsetplot{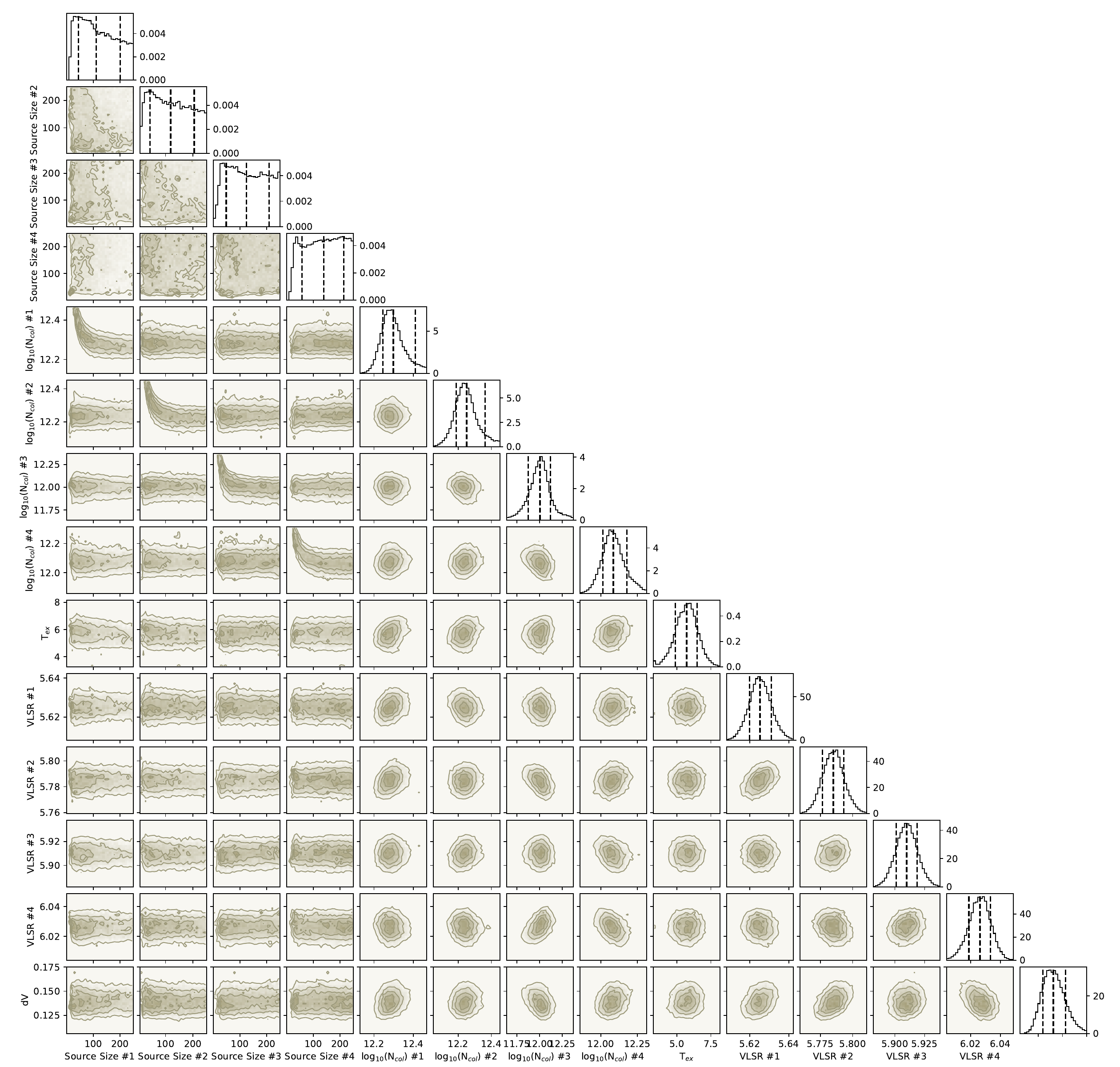}
\figsetgrpnote{The 16$^{th}$, 50$^{th}$, and 84$^{th}$ confidence intervals (corresponding to $\pm$1 sigma for a Gaussian posterior distribution) are shown as vertical lines. The contour lines are posterior probability levels, starting at $20\%$ of the maximum a posteriori estimate, with evenly spaced intervals of $20\%$ up to the peak density.}
\figsetgrpend

\figsetgrpstart
\figsetgrpnum{8.102}
\figsetgrptitle{Corner plot for C$_{4}$D.}
\figsetplot{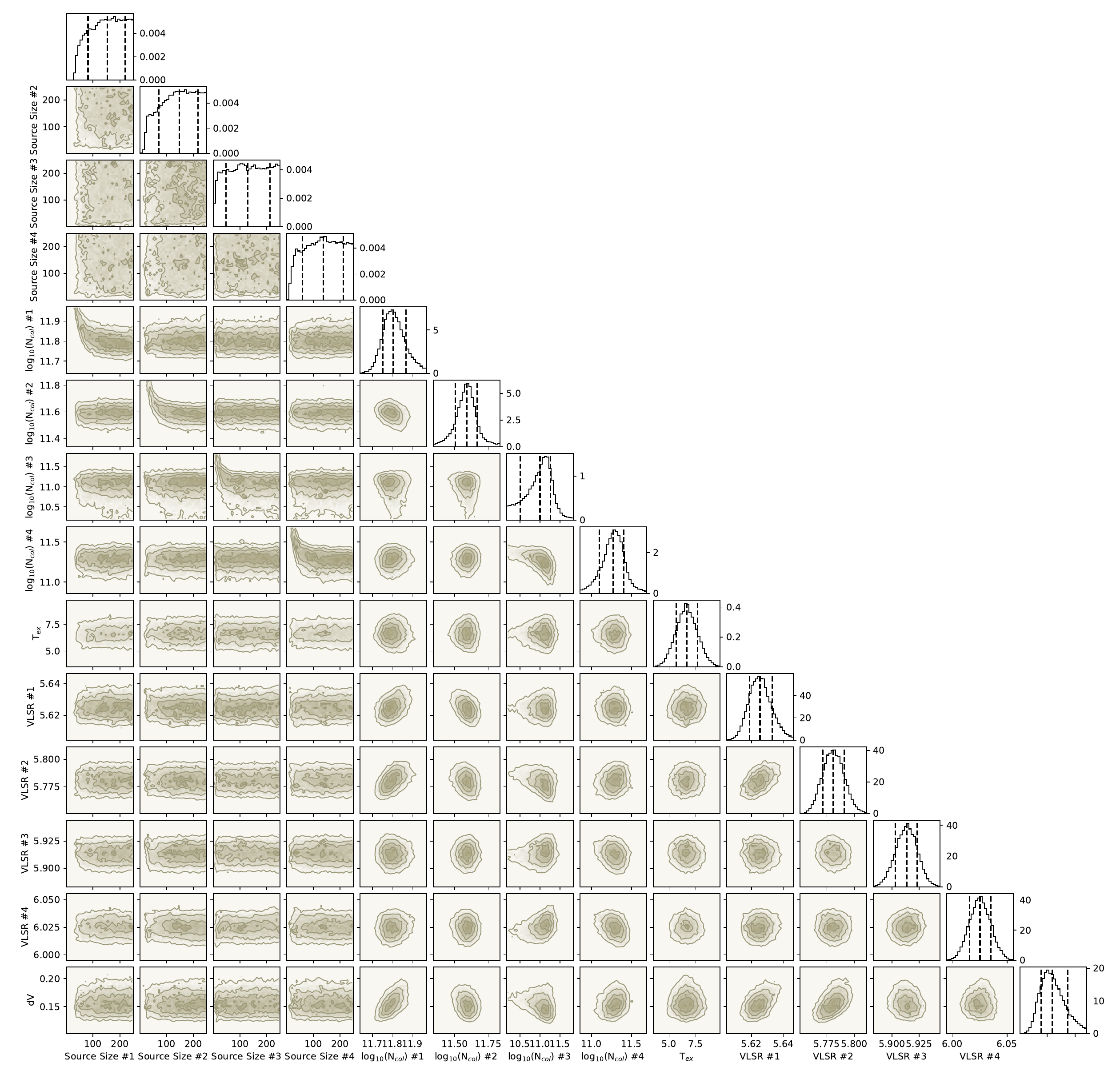}
\figsetgrpnote{The 16$^{th}$, 50$^{th}$, and 84$^{th}$ confidence intervals (corresponding to $\pm$1 sigma for a Gaussian posterior distribution) are shown as vertical lines. The contour lines are posterior probability levels, starting at $20\%$ of the maximum a posteriori estimate, with evenly spaced intervals of $20\%$ up to the peak density.}
\figsetgrpend

\figsetgrpstart
\figsetgrpnum{8.103}
\figsetgrptitle{Corner plot for DC$_{3}$N.}
\figsetplot{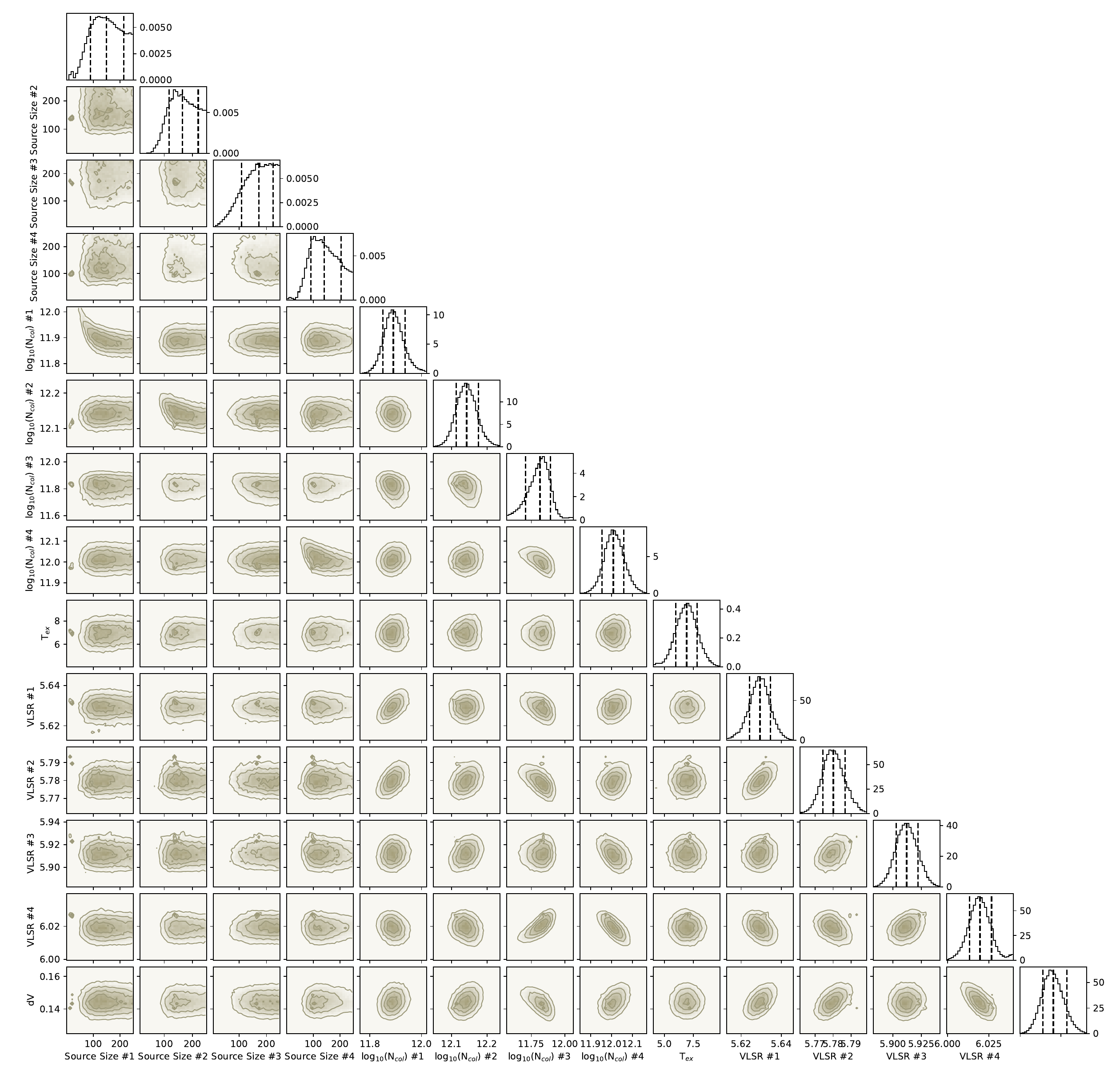}
\figsetgrpnote{The 16$^{th}$, 50$^{th}$, and 84$^{th}$ confidence intervals (corresponding to $\pm$1 sigma for a Gaussian posterior distribution) are shown as vertical lines. The contour lines are posterior probability levels, starting at $20\%$ of the maximum a posteriori estimate, with evenly spaced intervals of $20\%$ up to the peak density.}
\figsetgrpend

\figsetgrpstart
\figsetgrpnum{8.104}
\figsetgrptitle{Corner plot for CH$_{2}$DC$_{4}$H.}
\figsetplot{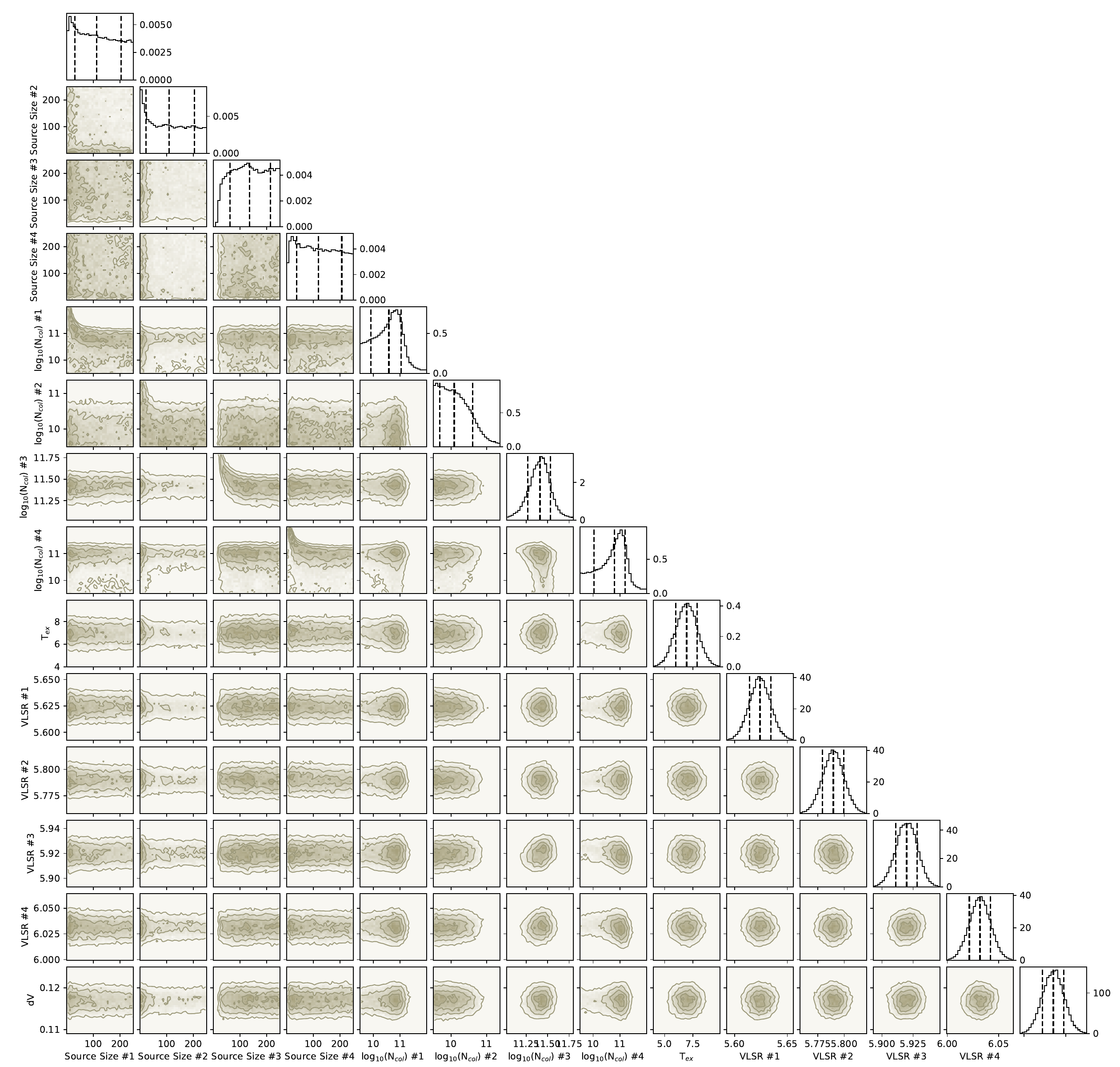}
\figsetgrpnote{The 16$^{th}$, 50$^{th}$, and 84$^{th}$ confidence intervals (corresponding to $\pm$1 sigma for a Gaussian posterior distribution) are shown as vertical lines. The contour lines are posterior probability levels, starting at $20\%$ of the maximum a posteriori estimate, with evenly spaced intervals of $20\%$ up to the peak density.}
\figsetgrpend

\figsetgrpstart
\figsetgrpnum{8.105}
\figsetgrptitle{Corner plot for CH$_{2}$DC$_{3}$N.}
\figsetplot{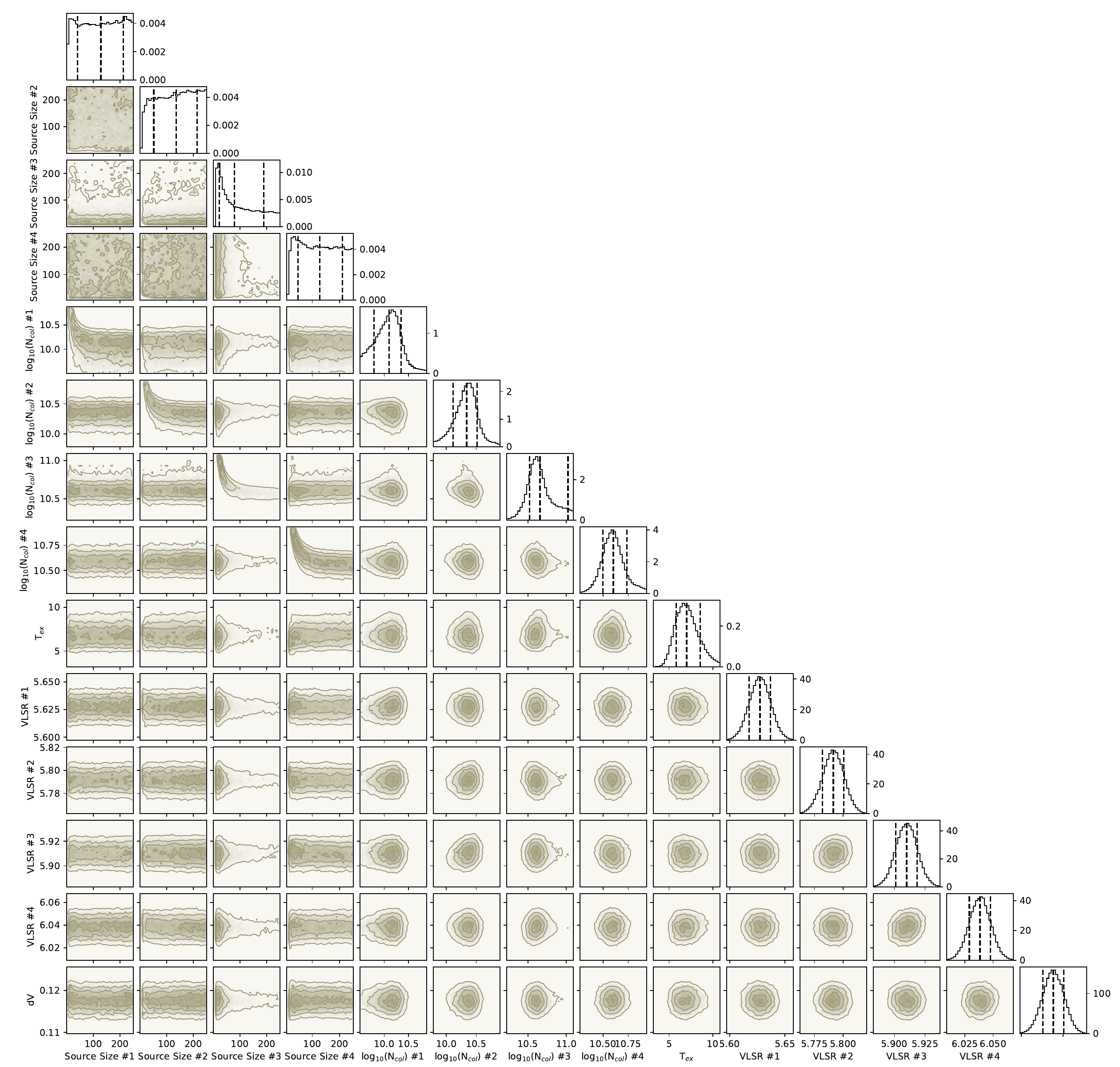}
\figsetgrpnote{The 16$^{th}$, 50$^{th}$, and 84$^{th}$ confidence intervals (corresponding to $\pm$1 sigma for a Gaussian posterior distribution) are shown as vertical lines. The contour lines are posterior probability levels, starting at $20\%$ of the maximum a posteriori estimate, with evenly spaced intervals of $20\%$ up to the peak density.}
\figsetgrpend

\figsetgrpstart
\figsetgrpnum{8.106}
\figsetgrptitle{Corner plot for DC$_{5}$N.}
\figsetplot{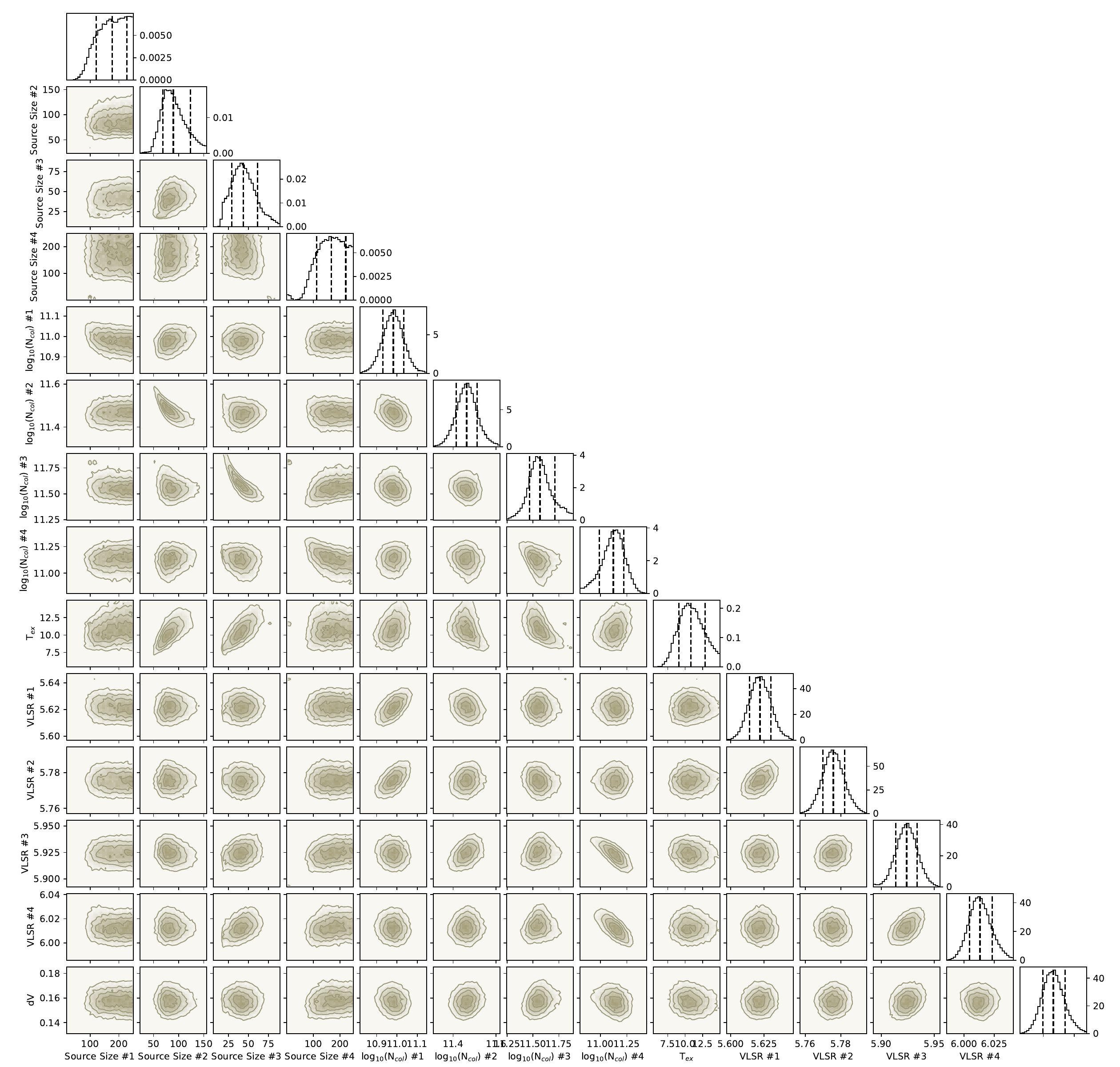}
\figsetgrpnote{The 16$^{th}$, 50$^{th}$, and 84$^{th}$ confidence intervals (corresponding to $\pm$1 sigma for a Gaussian posterior distribution) are shown as vertical lines. The contour lines are posterior probability levels, starting at $20\%$ of the maximum a posteriori estimate, with evenly spaced intervals of $20\%$ up to the peak density.}
\figsetgrpend

\figsetgrpstart
\figsetgrpnum{8.107}
\figsetgrptitle{Corner plot for DC$_{7}$N.}
\figsetplot{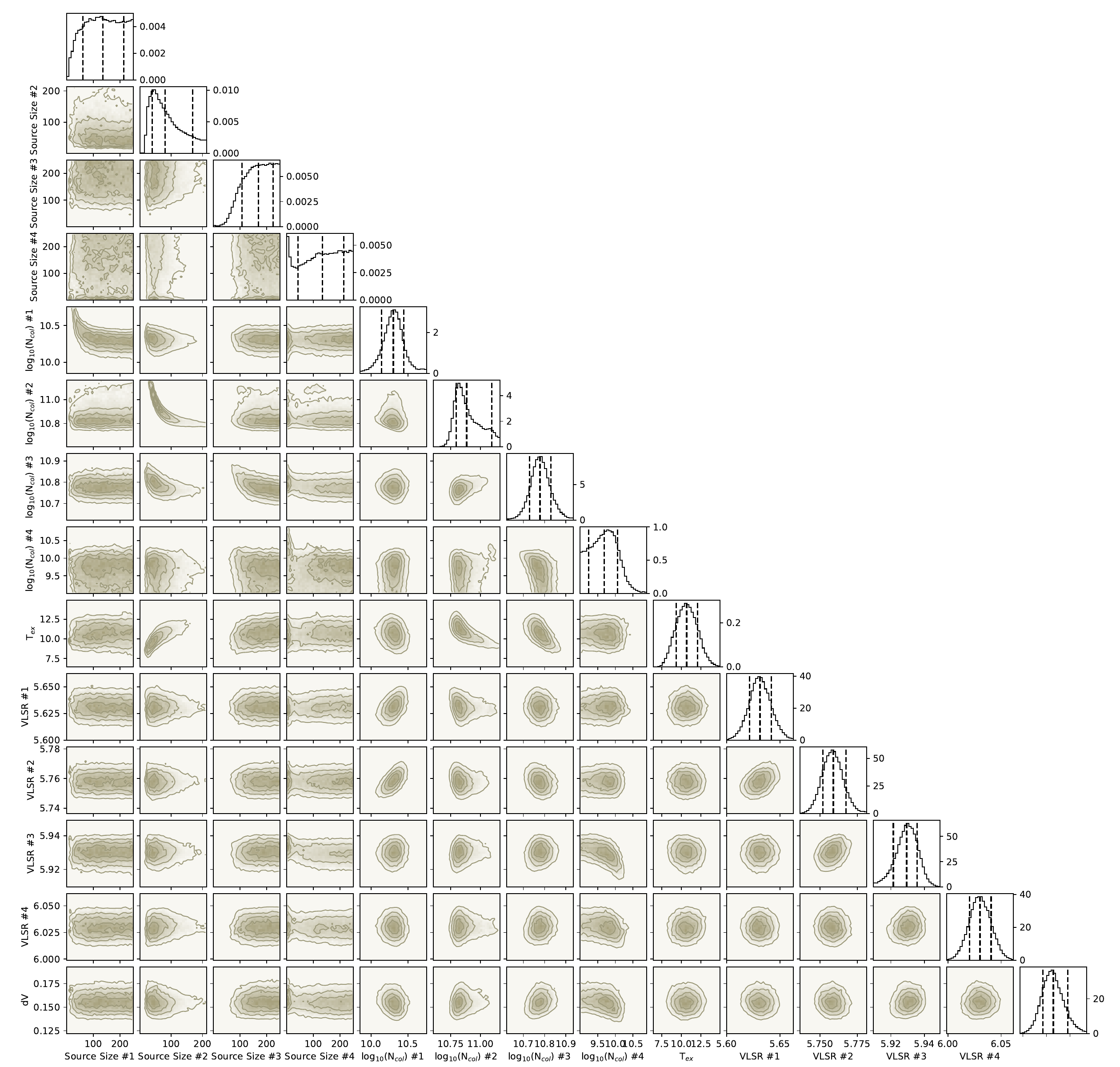}
\figsetgrpnote{The 16$^{th}$, 50$^{th}$, and 84$^{th}$ confidence intervals (corresponding to $\pm$1 sigma for a Gaussian posterior distribution) are shown as vertical lines. The contour lines are posterior probability levels, starting at $20\%$ of the maximum a posteriori estimate, with evenly spaced intervals of $20\%$ up to the peak density.}
\figsetgrpend

\figsetend

\begin{figure}
\plotone{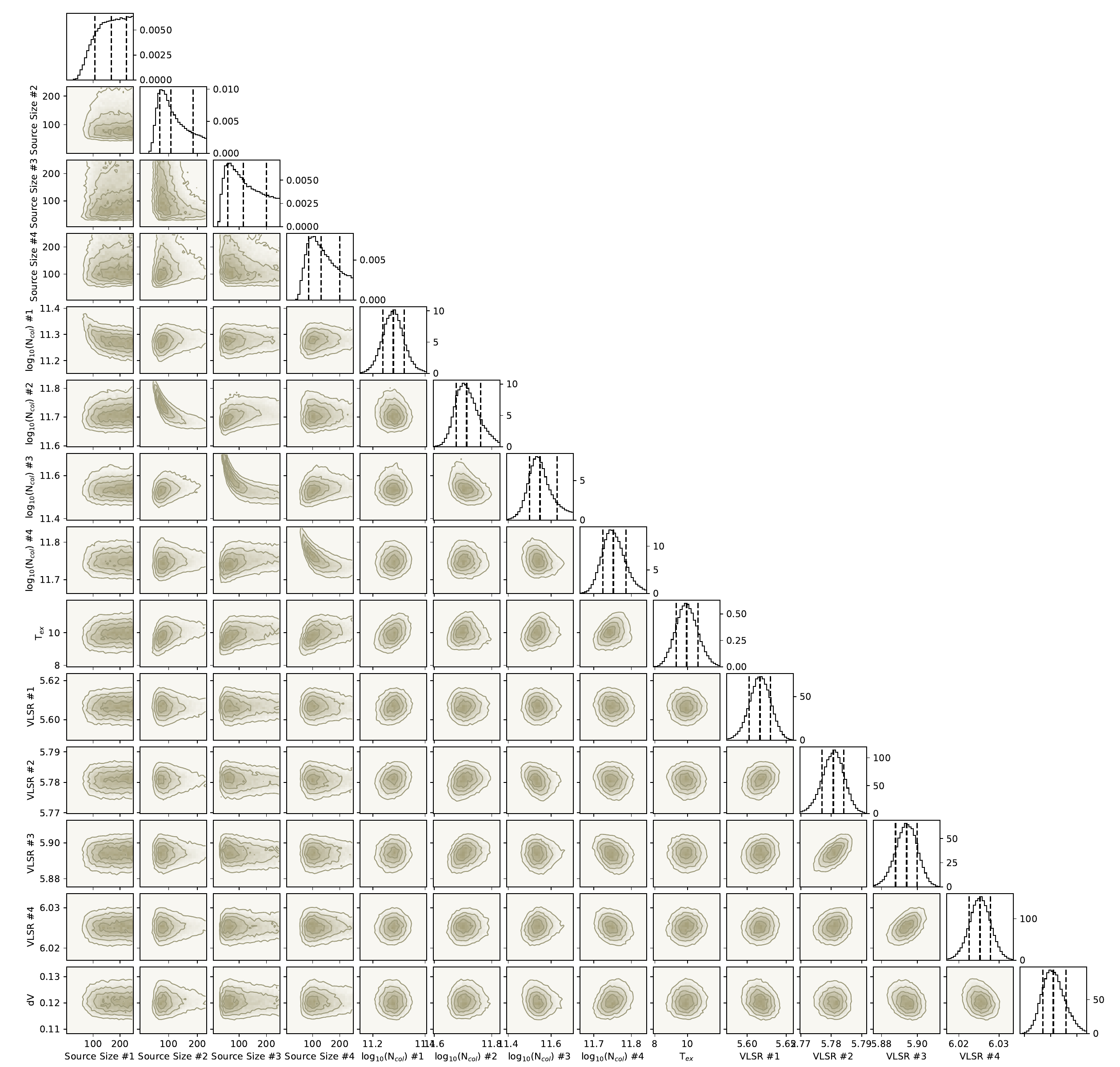}
\caption{\label{fig:corner_c-c6h5cn} Parameter covariances and marginalized posterior distributions for the MCMC fit of the $c$-\ce{C6H5CN} transitions in GOTHAM DR. The 16$^{th}$, 50$^{th}$, and 84$^{th}$ confidence intervals (corresponding to $\pm$1 sigma for a Gaussian posterior distribution) are shown as vertical lines. The contour lines are posterior probability levels, starting at $20\%$ of the maximum a posteriori estimate, with evenly spaced intervals of $20\%$ up to the peak density. (The complete figure set for 102 molecular species (107 images) is available in the online journal)}
\end{figure}